\documentclass[11pt]{article}

\usepackage[utf8]{inputenc}
\usepackage[margin=1in]{geometry}
\usepackage[round]{natbib}
\usepackage{amsmath,amsthm,amssymb}
\IfFileExists{bbm.sty}{\usepackage{bbm}}{\usepackage{dsfont}}
\providecommand{\mathbbm}[1]{\mathds{#1}}

\usepackage{graphicx}
\usepackage{float}
\usepackage{microtype}
\usepackage{setspace}
\usepackage{tabularx}
\usepackage{subcaption}
\usepackage{nicefrac}
\usepackage{etoolbox}
\usepackage{xcolor}
\usepackage{tikz}
\usepackage{hyperref}
\usepackage{url}
\definecolor{linkgray}{gray}{0.45}
\hypersetup{
  colorlinks=true,
  linkcolor=black,
  citecolor=black,
  urlcolor=linkgray,
}
\usepackage[english]{isodate}
\usepackage{listings}
\usepackage{tcolorbox}
\usepackage{booktabs}
\usepackage{longtable}
\usepackage{colortbl}
\usepackage{arydshln}
\tcbuselibrary{skins,breakable}

\usepackage{kpfonts}
\graphicspath{{./Images/}}

\newcommand{\ordonly}{\mathrm{ord}}

\DeclareMathOperator{\diag}{diag}

\newcommand{\cset}{\mathbb{K}}% closure set: a set of chokepoints shut together

\newtcolorbox{keybox}[1][]{colback=blue!5,colframe=blue!50!black,fonttitle=\bfseries,title={#1},breakable}
\lstdefinestyle{Rstyle}{language=R,basicstyle=\ttfamily\small,keywordstyle=\color{blue!70},commentstyle=\color{gray},stringstyle=\color{orange!80!black},breaklines=true,frame=single,numbers=left,numberstyle=\tiny\color{gray}}

\theoremstyle{plain}

\newtheorem{proposition}{Proposition}
\newtheorem{lemma}{Lemma}

\newtheorem{assumption}{Assumption}

\theoremstyle{remark}

\theoremstyle{plain}
\newtheorem{definition}{Definition}

\title{\textbf{Macroeconomic Risks from Maritime Trade Disruptions}}

\author{Vipin P. Veetil\thanks{Economics Area, Indian Institute of Management Kozhikode.} \and Fathimath S. Vemmarath\thanks{Indian Statistical Institute, Kolkata.}}

\date{\small\printdayoff\today}

\usepackage{adjustbox}

\begin{document}
% --- arXiv build -------------------------------------------------------------
% \maketitle disables the title commands after its first use. Keep them, so
% that each online appendix can set its own title block as it does when it is
% compiled as a separate document.
\makeatletter
\let\OAkeep@maketitle\maketitle  \let\OAkeep@@maketitle\@maketitle
\let\OAkeep@title\title          \let\OAkeep@author\author
\let\OAkeep@date\date            \let\OAkeep@and\and
\let\OAkeep@thanks\thanks
\newcommand{\OArestoretitle}{%
  \global\let\maketitle\OAkeep@maketitle  \global\let\@maketitle\OAkeep@@maketitle
  \global\let\title\OAkeep@title          \global\let\author\OAkeep@author
  \global\let\date\OAkeep@date            \global\let\and\OAkeep@and
  \global\let\thanks\OAkeep@thanks}
\makeatother
% -----------------------------------------------------------------------------
\maketitle

\begin{abstract}
\setstretch{1.3}
This paper develops a model of maritime chokepoint
closures in which interrupting a shipping
passage produces losses that are not measured, or even bounded, by the
value of the trade that transits it. The losses stem from disruptions to the flow of intermediate inputs that are complementary in downstream production. They travel through the world trade network, which itself rewires around the closed passage. Re-matching displaced trade on the buyer and seller sides of the market, which the model carries out as rationing at fixed prices, limits the damage but does not eliminate it. Across countries, the incidence of these losses is heavy-tailed, and it reaches economies that sell nothing through the passage, some of which buy little through it either. The two
ends of a severed corridor lose unequally. At the typical corridor the economies that mainly sell through it lose more than twice as much as those that mainly buy through it, and at the largest energy gate between four and five times as much, as economic geography funnels commodity-concentrated
sellers through a single passage while their buyers re-source.
Economic geography decides how the loss from closing several chokepoints together compares with the sum of the losses from closing each alone. A shipment stopped at the first of two gates in series cannot be stopped again at the second, so such gates duplicate one another, whereas gates in parallel take different cargo for the most part, so their losses add. Which is why the Middle East scenario costs about a tenth less than the sum of its single closures, while the East Asian and Russia--Europe scenarios cost that sum to within three percent.

\vspace{0.5cm}
\end{abstract}

\newpage
\setstretch{1.3}
\section{Introduction}

A quarter millennium ago, Adam \citeauthor{smith1776} placed at the foundation of economics the
proposition that the division of labor is limited by the extent of the market. \citet[p.~529]{young1928increasing} thought it ``one of the most illuminating and fruitful generalisations'' in the whole literature of economics. It illuminates when supplied with the right empirical material. One such `empirical material', perhaps the most significant of all, is economic geography. For what integrates markets or hinders their coalescence is the cost, safety, and ease of transportation. The great but treacherous Himalayas and the ever so pliant Mediterranean are not neutral topologies when it comes to division of labor and extension of the market.  \citet{smith1776} was cognizant of how much his proposition about the origins of human prosperity depended on economic geography. He writes that ``as by means of water-carriage a more extensive market is opened to every sort of industry than what land-carriage alone can afford it, so it is upon the sea-coast, and along the banks of navigable rivers, that industry of every kind naturally begins to subdivide and improve itself.'' In some senses, the world economy of our day is Smith's observation made planetary. The greater part of international trade travels by sea. What the ships carry is no
longer principally wine and wool for final consumption but the intermediate inputs of
a division of labor stretched across continents, crude for refineries, feedstocks
for chemical works, ores for smelters, components for assembly lines. The sea, however, does not spread this commerce evenly over its
surface. Economic calculus funnels it through a small number of narrow
passages, a canal here, a strait there, which means that the division of
labor of an entire planet is made to pass, at certain points, through
channels a few miles wide.

But the geography that makes possible this truly global division of labor carries within it grave threats to the very system it fosters. The canals that link some of the world's deepest manufacturing chains are far more susceptible to closure than thousands of nautical miles of open ocean.   In March 2021, for instance, a single container ship ran aground in the Suez Canal and for six days blocked a passage that carries something close to one-eighth of world seaborne trade. Two years later drones
and missiles in the Bab-el-Mandeb emptied the Red Sea of container traffic
and pushed the Asia--Europe trades around the Cape of Good Hope. In the same year a drought drew down the reservoirs that feed the Panama Canal and rationed its locks. As for the Strait of Hormuz, a war-risk premium has hung over it for nearly forty years, and in early 2026 the risk behind that premium finally arrived, with recorded transits falling by 96 percent between February and March, to fewer than three ships a day. Each of these episodes was geographically minute, a canal, a strait, a
few hundred miles of contested water. But each generated economic consequences
out of all proportion to its geography. The reason is that a maritime
chokepoint is not a place where goods are merely made costlier or delayed. It is the needle's eye through which the supply of intermediate inputs to
the world's producers must pass, the narrowest point of Smith's extended
market.  In an economy without
a production network, where a severed shipment is simply a shipment lost, the
loss is the value of the trade that passes through the chokepoint. In an economy whose producers are bound
to one another by intermediate inputs, it is not. A Korean refinery cannot
run on the crude it does not receive, and a German chemical works cannot run on
the Asian feedstocks that do not arrive. The conventional gauge counts the
severed shipment once, at the strait. The economy though loses it again at every
downstream producer for whom it was a necessary input, and may not lose it at
all where a surviving supplier can be found. Put differently, the map from the
trade that passes through a chokepoint to the macroeconomic loss from its closure is drawn by the world trade network and the substitutabilities therein.

Economic geography shapes not just the shock that propagates through the world trade network but also the structure of propagation itself. Once a strait is closed, buyers re-source from the suppliers they can still reach, and severed suppliers push idle output toward the customers that remain. The world trade network itself rewires, with the rewiring shaped by
the locational specifics of the detours around the strait that closed and
of the routes that remain open. The matter will perhaps be better understood by taking an aerial view of things. The world trade table, with its myriad flows of goods between
country-sectors, has beneath it an economic geography which makes these
flows possible. A change in the underlying economic geography, whether temporary or permanent, is not a shock that travels through the world trade network but one that transforms the network itself. The cost of closing one of these chokepoints, and the way that cost falls across nations, is therefore a comparison of two flow equilibria. One is the trade-table fixed point consistent with the old geography, and the other is the fixed point consistent with the new. We must emphasize that the equilibrium the economy settles into afterward is not the benchmark equilibrium with the exposed links deleted. For the pre-shock flows are generated by the very configuration the closure dismantles, and cannot serve as a sufficient statistic for the loss their disappearance sets in motion. Which is precisely why the value of the trade that crosses the passage is not a good measure of the damage to the world economy due to its closure, for the trade table to which it belonged is no longer operational.

\begin{figure}[H]
\centering
\includegraphics[width=\linewidth]{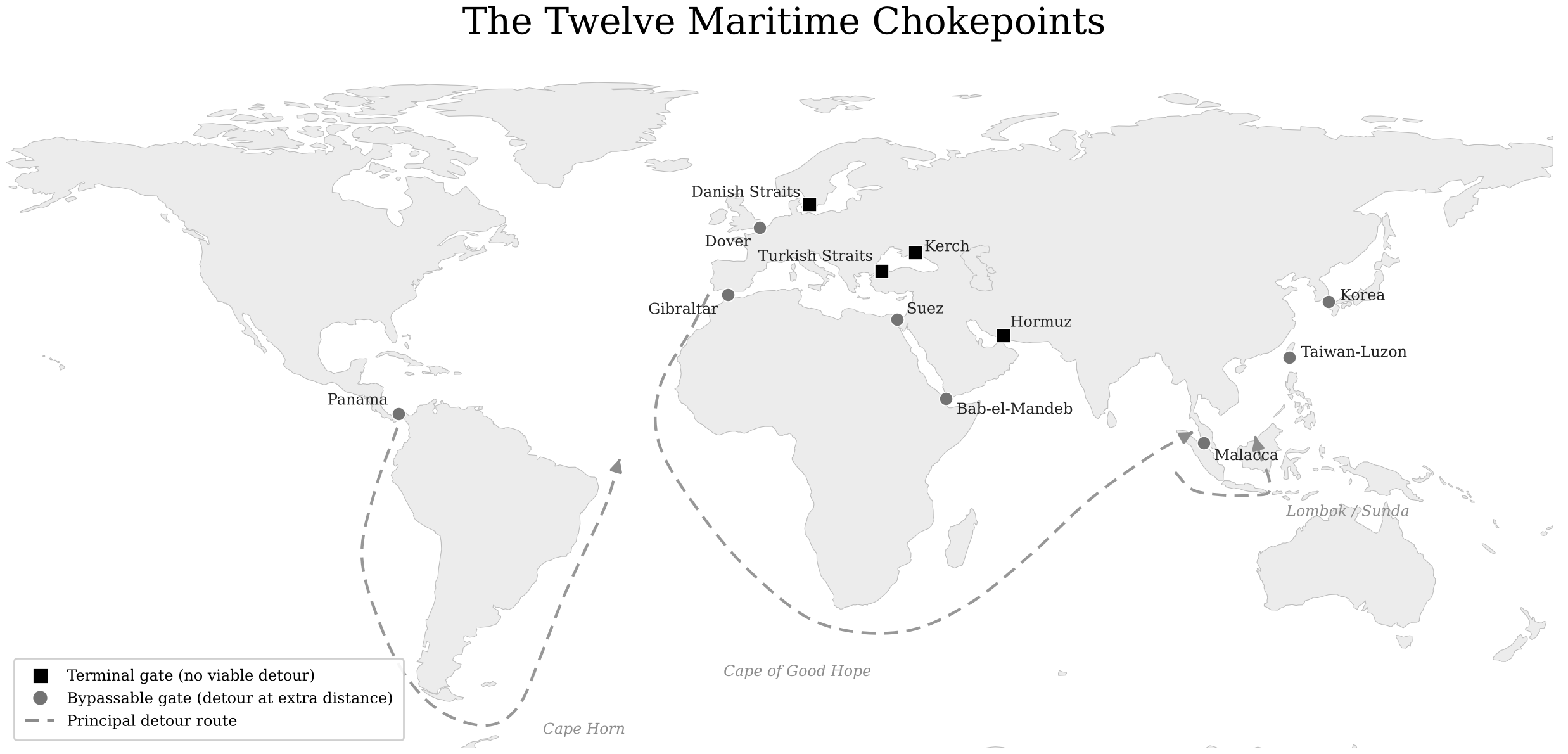}
\caption{The twelve maritime chokepoints studied in the paper. Squares are terminal
gates, the sole sea exit of an enclosed basin, which admit no detour.
Circles are bypassable gates, which a longer voyage can circumvent. Dashed
arcs are the three principal detours: the Cape of Good Hope for the
Suez--Bab-el-Mandeb--Gibraltar corridor, Cape Horn for Panama, and the
Lombok and Sunda Straits for Malacca.}
\label{fig:chokepoint_map}
\end{figure}

We develop a model to compute geography-consistent equilibria that arise from different maritime chokepoint closures, including joint closures. We concern ourselves with twelve major maritime chokepoints: the Straits
of Hormuz, Malacca, Gibraltar, and Dover, the Turkish and Danish Straits,
the Korea and Kerch Straits, the Suez and Panama Canals, Bab-el-Mandeb,
and the Taiwan--Luzon Corridor (Figure~\ref{fig:chokepoint_map}). The world trade network we work with is that of flows of intermediate inputs among 4{,}350
country-sectors, i.e., 87 economies at 50 sectors.\footnote{We study only the losses that arise from the disruption of the flow of intermediate inputs and from its propagation through production. A closure also stops the goods bound for households, governments, and investors. We do not consider those stoppages, holding final demand at its benchmark throughout. The reason is not a lack of data, since the input--output table we use records the origin and the destination of final goods as it does of intermediate inputs. The reason is that a model of production can follow a shipment between two producers at both of its ends, the seller losing an outlet and the buyer an input, whereas it can follow a shipment to a final user at the seller's end alone. The losses we report are to that extent a lower bound. Online Appendix~\ref{app:final_goods} gives a rough count of what the stopped final goods would add to them.} Historically, a chokepoint closure has seldom been complete. Neither nature nor international politics has, as a rule, been so heedless of its dire consequences.\footnote{What still moves through a closed passage has differed from one episode to the next and from one cargo to another, the tankers sailing on through a Red Sea the container lines had abandoned in 2023--24. For the determined, sustained closures the paper studies we take a severe and uniform baseline, in which each closed link retains a tenth of its benchmark seaborne flow.}  Our model, therefore, implements a chokepoint closure as a parametric reduction in throughput. What the model returns is a pattern of exposure that is heavy-tailed across countries, lopsided between the exporting and importing ends of each severed corridor, and nearly additive across gates, except where two gates would take the same cargo.

Where a longer voyage around the obstacle exists (Lombok and Sunda for Malacca, the Cape of Good Hope for Suez), part of each severed shipment can be salvaged at a cost that rises with the detour's share of the voyage. A terminal gate such as Hormuz, the sole exit of an enclosed basin, admits no such salvage. Beyond such rerouting, buyers and sellers also seek to redirect some of their lost trades toward partners still reachable under the new geography. Because prices are fixed in the short run, this redirection takes the form of rationing rather than of price adjustment. We implement it proportionally: a buyer that loses a supplier turns to the suppliers it can still reach in proportion to what it already bought from each, a seller that loses a buyer turns likewise to the buyers it can still reach, and what still passes through the closed passage is held fixed.  We assume that the re-matching of displaced trade to new buyers and sellers is neither free nor
unbounded. More specifically, a fraction of the displaced mass is
unrecoverable within the disruption horizon.

Not all trade between two economies moves by sea, even where a sea route joins them. In our benchmark year Russian gas reached Germany through pipelines, Chinese goods reached Kazakhstan by rail, and Taiwanese semiconductors left the island by air. The closing of a strait stops none of these. We therefore count as exposed to a chokepoint only that part of each trade flow which moves by sea. The part that moves by pipeline, road, rail, or air stays in place at its benchmark value. We do not let that part grow to take over the cargo which a closed passage strands, since pipelines, railways, and air freight run at a capacity that is given in the short run.

The model is not complete without a specification of how the fall in the production of one good affects that of another. We assume that input combinations in pre-closure equilibrium were optimal given their prices. Since ours is a short-run model, we assume that prices are fixed and that, therefore, optimal input combinations remain unchanged. Which means that a maritime chokepoint closure generates a decline in output not
merely because the quantities of inputs available to producers fall but
also because their proportions move away from the optimal. We do not impose a global production function but assume that, whatever these functions may be, they are locally well approximated by a CES curvature.

Needless to say, the decline in production due to the scale and the proportions in which inputs are available propagates through the world trade network. The post-closure fixed point is the one at which this propagation completes itself within the trade network consistent with the new geography. In fact, this percolation of production losses, along with the rewiring of the trade network itself, is the reason why the loss in world output due to the closing of a maritime trade route need not even be monotone in the value of the trade that passes through it. In principle, a country that trades little through a chokepoint can be devastated by its closure, while a country that trades a great deal through it remains barely touched.

That said, what can happen in principle can differ from that which occurs in reality. We therefore discipline the model parameters and inputs with empirical regularities. The buyer-seller trade table is the OECD's Inter-Country Input--Output table extended with GTAP to differentiate the rest-of-world composite. The cost of rerouting goods through other waterways is measured from sailing distances on the world's shipping routes. The share of each trade flow that moves by sea is measured from customs records of trade by mode of transport. The share of trade that still moves through a closed passage is measured from the record of past closures. The ability to re-match trade to other buyers and sellers is taken from studies of how firms recovered from past supply disruptions. The local CES curvature, which sets how much output is lost when inputs arrive in suboptimal proportions, is taken from the sizable literature
that estimates it from natural experiments.

Closing the Strait of
Hormuz alone removes production worth about 2.8 percent of world GDP, in a computation that credits no bypass around the strait. Crediting the Gulf crude pipelines that the 2026 closure mobilized lowers that figure by no more than a half, and the Emirati and Saudi losses by about as much.\footnote{The Saudi East--West pipeline and the Emirati pipeline to Fujairah end at ports outside the strait, and during the 2026 closure the two carried about 2 million barrels a day more than before, an eighth of the crude that crossed the strait before the closure. The credited computation places no ceiling on what the two pipelines can carry, so we read its loss as a floor on the cost of the closure with the pipelines in use and the baseline loss as its ceiling (Online Appendix~\ref{app:hormuz_pipeline}).} The loss is borne very unevenly. The Gulf exporters behind the strait lose double-digit shares of their own GDP, an order of magnitude more than the world figure. Those shares stay double-digit once the bypass is credited, but most of the world then loses less than 1 percent. The tail of GDP losses is not confined to the Gulf. Korea, for instance, loses production worth about 10 percent of its GDP to the same closure (the Gulf supplies most of the crude oil and gas on which its refineries and power plants
run). About four fifths of what the closure costs the world begins with producers that go short of the inputs it withholds, and only a fifth with the orders it removes. Going short of inputs weighs nearly as much at every other gate, where its share of the world loss runs from about two thirds to three quarters. Behind the other gates the deepest losses are smaller, about 5 percent of GDP for Turkey and Kazakhstan when the Turkish Straits close. A small number of economies lose a large share of GDP under their single most binding closure, but the great majority face only modest losses.

Perhaps the most intriguing of our findings is an exporter--importer asymmetry in the losses that come from maritime chokepoint closures. Nearly every economy both sells and buys through a given passage, so by the `exporting side' of a passage we mean the economies whose sales through it exceed their purchases through it, and by the `importing side' those whose purchases exceed their sales. We find that the economies on the exporting side lose more, in proportion to their GDP, than those on the importing side at every gate, more than twice as much at the typical gate and between four and five times as much at Hormuz, the largest of the energy gates.\footnote{The asymmetry is measured on trade in intermediate inputs, the only trade the model carries. On a rough count of the final goods that each closure stops, the exporting side still loses more than the importing side at all twelve gates (Online Appendix~\ref{app:final_goods}).} Nothing in the model tilts the losses that way. The rationing holds the residual that still transits the closed passage fixed for buyers and sellers alike, and re-matches the remaining free flows under a common unrecovered share. Put differently, buyers can redirect lost demand to surviving same-sector suppliers, and sellers can redirect idle output to surviving same-sector buyers, on identical terms. What is tilted is the way in which geography is intertwined with the benchmark trade network. Chokepoint suppliers are often commodity-concentrated exporters funneled through one corridor: Gulf energy through Hormuz, Caspian and Black Sea exports through the Turkish Straits, and Baltic energy and materials through the Danish Straits. Once the corridor closes, they have few off-corridor same-sector buyers onto which displaced output can be placed. Importers, by contrast, are usually more diversified across suppliers and do not find themselves
behind narrow geographies.\footnote{Exporters lose because the closure physically cuts them off from the demand of the buyers beyond the closed passage, and the output that they cannot place elsewhere is written off. Importers lose because re-matching cannot fully reconstitute the input mix their technologies were built around. Japan's Hormuz loss, for instance, arrives even though Japan sells almost nothing through the strait.}

Single chokepoint closures, however informative, tell us little about the tail of
macroeconomic risk, because closing a chokepoint is sometimes a political act whose
geography follows the reach of a regional power and not the count of
waterways. We therefore close each region's gates together, in three scenarios: a Middle East scenario taking Hormuz, Bab-el-Mandeb, and Suez, an East Asian
scenario taking Malacca, the Taiwan--Luzon Corridor, and the Korea
Strait, and a Russia--Europe scenario taking the Turkish Straits, the
Danish Straits, and the Kerch Strait. Each scenario has a public record.
Iran threatened Hormuz for years and shut it at the end of February 2026, China has
rehearsed the blockade of sea space around Taiwan in large-scale
exercises since 2022, and Russia announced in 2023 that it would treat
ships bound for Ukrainian ports as potential military targets.\footnote{Online Appendix~\ref{app:scenariopol} documents the record gate by gate, with dates and sources.}

The joint closure scenarios are unambiguous in their ranking. The Middle
East scenario is by far the worst, costing the world economy about 3 percent of GDP. The Russia--Europe scenario follows at about 0.5 percent, and the East Asian scenario at about 0.4 percent. Set against what their gates
cost when closed one at a time and in isolation, the three scenarios part ways.
The Middle East scenario costs about a tenth less than the sum of its three single closures, but each of the other two falls short of that sum by less than 3 percent.
What decides the difference is how the closed gates sit on the routes they cut.

Gates in series lie
one behind another on a single voyage, so a shipment must clear both.
Cutting the first gate already stops the flow, and cutting the second gate stops
little that still moves, so a joint closure of gates in series costs less than the sum of its single closures. Bab-el-Mandeb and Suez lie one behind the other on the Red Sea route and
share the Cape of Good Hope detour, so the second gate disrupts nothing the first gate has not. The Middle East scenario therefore pays once for a severance that the sum of its single closures counts twice. Gates in parallel have detours of their own, and for the most part they take
different cargo. Russia's western gates are separate
sea outlets: the Turkish Straits from the Black Sea and the Danish Straits
from the Baltic. Few shipments cross both. East Asia's gates are
doors to the same producers: the Taiwan--Luzon Corridor onto the
Pacific, Malacca onto the Indian Ocean, and the Korea Strait onto the
Japan--Korea and Russian Far-East trades. Much of the traffic through Malacca goes on
through the corridor, but the two gates have separate detours, each of which saves most of a
cargo, so little is lost twice. The losses of such gates add.

The losses we report are not forecasts of what any actual closure will cost. They are the losses from determined and sustained closures, computed without the strategic reserves, the bypass pipelines, and the substitution that higher prices induce, all of which soften a closure in practice. We offer them as a map of where the extreme risks of the world's maritime geography lie, and of who bears them.

\subsection{Related literature}
We are not the first to study the economic consequences of maritime closures, but we study counterfactual ones where most of the existing
work estimates realized ones. Its richest strand treats past closures and
openings as natural experiments, from \citet{feyrer2021distance} on
the 1967--75 Suez closure to \citet{hugot_umana2017tradecosts} on the
openings of Suez and Panama, \citet{heiland2025tradefromspace} on the
2016 Panama Canal expansion, and \citet{qu_etal2024dynamic} on the 2021
Suez blockage. Together they show that maritime geography matters for welfare
well beyond the closed region. Closest in spirit to our paper are
\citet{kaenzig_raghavan2026supply}, who estimate the aggregate response to
realized shocks centered on Suez and Panama, and \citet{meza_etal2022lng},
whose counterfactual resembles ours but covers LNG
alone.\footnote{See also \citet{bodenschatz_etal2025germany}, who measure Germany's dependence on six chokepoints across products, sectors, and partners by the share of its trade that each chokepoint carries. \citet{verschuur_lumma_hall2025systemic} value the trade that disruptions at $24$ chokepoints put at risk worldwide, and the direct losses from delays, rerouting, insurance premiums, and higher freight costs.} Closest in
method to us is \citet{verschuur_koks_hall_2022}, who assign bilateral trade to
ports, remove the port-routed flows from a multi-regional input--output
matrix, and recompute Leontief and Ghosh inverses. Their recomputation is the precedent for ours, since once geography removes a link it is the modified matrix, and no longer the benchmark matrix, that carries the shock from one producer to the next. Nor are we the first to make trade depend on the physical network that carries it, for in quantitative spatial models the route a shipment takes and the cost of moving it are themselves outcomes of the infrastructure network \citep{allen_arkolakis_2022}. Our contribution is accordingly narrower than putting physical geography inside an economic model. What we do is let a closure decide not only which trades are severed but also how much of each severed trade a detour salvages and how the displaced remainder is re-matched among the buyers and sellers that can still reach one another. Which means that the operator that governs the propagation of the shock depends on the closure in each of these respects. Put simply, we consider not only the complementarity between inputs, which takes the study of production networks beyond the theorem of \citet{hulten1978growth}, but also the changes in the very structure of the network, which a Leontief inverse computed on the benchmark table holds fixed.

Hulten's theorem is the natural benchmark for a counterfactual that runs through the input--output table.
By the theorem, in an efficient economy the
first-order effect of a small productivity shock on aggregate output is the
affected producer's sales share, and the network reduces to one weighted
number. A chokepoint closure is instead a discrete link-feasibility shock, so its cost need not equal a local derivative. Low short-run substitution,
gross complementarity, input specificity, and capacity constraints make the
nonlinear terms large
\citep{atalay2017,boehm_flaaen_2019,barrot_sauvagnat_2016,baqaee_farhi_2019hulten}.
Our results are accordingly a beyond-Hulten phenomenon of the kind
\citet{baqaee_farhi_2019hulten} and \citet{baqaee_farhi_supply_2022}
characterize. What is central for the risk that a country bears is the
concentration of its exposure on a single gate. A country has most to fear from the closure that cuts the inputs it can least replace, whatever its average exposure.\footnote{Such
a closure can also be deliberate, which connects the exercise to the
literature on economic statecraft and weaponized interdependence. The
exporter--importer asymmetry we find is a
quantitative instance of the influence effect that
\citet{hirschman1945national} placed at the center of the politics of trade. \citet{keohane_nye_1977} make asymmetric interdependence a source of power, and \citet{baldwin_1985} counts trade measures among the instruments of statecraft. \citet{farrell_newman_2019} develop an account of coercion through central positions in economic networks, and \citet{clayton_maggiori_schreger_2024coercion} a theory of the coercion that hegemons exert by controlling the supply of key inputs. Those two accounts treat financial, informational, and technological chokepoints, where ours restores
the term to its original physical referent.}
That concentration places systemic risk in the topology of the world trade network rather than in its nodes and flows alone. The measures of systemic risk in finance make the same move, judging an institution by its tie to the distress of the whole system rather than by its own risk \citep{adrian_brunnermeier_2016covar,acharya_pedersen_philippon_richardson_2017mes}.

In much of the literature on production networks a shock strikes a producer, either as a change in the demand for its output or as a change in its productivity, and the question is how the shock travels to other producers. In the benchmark of \citet{acemoglu_akcigit_kerr_2016} a demand shock travels upstream, to the suppliers of the producer, and a productivity shock travels downstream, to its customers. A closure strikes a link instead, and so both of its ends at once: the seller as a loss of orders and the buyer as a loss of inputs. One event is therefore a shock to demand on one shore and a shock to production on the other. We divide every loss by which of the two it began with.

The short-run production-network model we solve is not itself new.
\citet{bhatt2026_sanctions} build it to study economic power under trade
sanctions, where a restriction severs every link between the two countries it sets apart. What we supply it with is economic geography, which fixes the severed links from outside the trading pair. Supplying the geography raises questions
the sanctions setting never puts. The sharpest is how to treat the several
gates that a single voyage may cross, since a shipment stopped at the first
gate cannot be stopped again at the second gate.

The geography we invoke has a literature of its own, though it is usually put
to a different question. Since \citet{krugman1991} the new economic geography
has explained where activity locates and why some places grow rich.
\citet{fujita_krugman_venables1999} set out the framework, and
\citet{redding_venables2004} find that access to markets and to suppliers
accounts for much of the cross-country variation in income per head. The cost of poor placement is counted there in growth forgone.
\citet{limao_venables2001} measure the penalty a landlocked country pays in
transport costs, and \citet{gallup_sachs_mellinger1999} tie coastal access and
the tropics to the level of development. We take the same geography and ask
what it does to macroeconomic risk rather than to growth. A country's sea routes fix not
only how rich it can become but how much of what it has can be removed at a
stroke, and that burden falls hardest where the growth literature already
locates the disadvantage. African economies, which sell one or two primary
commodities through a single corridor and buy their
downstream inputs back along the same few routes, are the plainest case.

\subsection{Organization of the paper}
The paper proceeds as follows. Section~\ref{sec:model} sets up the
short-run production-network model and the benchmark-selected equilibrium
that defines the post-closure counterfactual.
Section~\ref{sec:mapping} maps route geography onto the extended input--output matrix and builds the incidence matrix, through which every closure enters the model.
Section~\ref{sec:results} states the model calibration and reports the
master exposure matrix of losses from each single closure, gate by gate.
Section~\ref{sec:structure} examines the structure of that exposure, the
heavy tail across countries and the exporter--importer asymmetry, which we decompose by source and by channel. Section~\ref{sec:joint}
closes several chokepoints at once. The section sets out how the loss from a joint closure compares with the sum of the losses from its single closures, and reports the three joint scenarios.  Section~\ref{sec:conclusion} concludes the paper.
Online Appendix~A states and proves the formal results. Online
Appendix~B carries the data construction, the reroutability matrices,
the parameter estimation, the robustness checks, and the notation. The same appendix sets out what the closures on record, the 2026 closure of the Strait of Hormuz among them, tell us about the parameters of the model, and why they cannot test the losses it reports.\footnote{All code and results behind the paper are public.
The source repository is at
\url{https://bitbucket.org/VipinVeetil/chokepoints}. The calibrated model inputs are archived on Zenodo
(DOI:~\href{https://doi.org/10.5281/zenodo.21263266}{10.5281/zenodo.21263266}),
and \href{https://pypi.org/project/maritime-choke/}{\texttt{pip install
maritime-choke}} installs the solver together with an interactive interface.
The model can also be run in a web browser, with nothing to install, through a Streamlit app (\url{https://maritime-chokepoints.streamlit.app}).
Online Appendix~\ref{app:replication} sets out what each holds.}

\section{The Model}
\label{sec:model}

The model takes the world input--output table as a production network of
country-sectors and asks what output that network loses when a set of
maritime passages closes. A closure turns into a loss of output in stages, each of which we study in turn. A severed shipment may first reach the same
buyer by a longer voyage, at a cost recorded in the reroutability matrices of
Section~\ref{subsec:chokepoint_restrictions}, whose entries are the shares of a trade flow that survive the detour. The trade that rerouting
cannot save is then re-matched, buyers toward surviving same-sector
suppliers and sellers toward surviving same-sector buyers, through the rationed exchange at fixed prices of Section~\ref{subsec:rerouting_matrix}, in which what still crosses the closed passage is held fixed. And whatever the exchange cannot reconstitute reaches production as a distorted bundle of intermediate inputs, through the CES technology of
Section~\ref{subsec:input_bundle_distortion}.
Section~\ref{subsec:short_run_activity} assembles the three stages into a fixed point and states which equilibrium the model reports. Section~\ref{subsec:activity_decomposition} divides the loss of activity at that equilibrium by its source and by the way it reaches a producer, and Section~\ref{subsec:country_world_losses} defines the loss measures and divides them in the same way. The model is the fixed-price general equilibrium of \citet{bhatt2026_sanctions}, with a closure in place of a sanction, and the counterfactual is short run throughout.\footnote{As in their model, where a sanction bars entries of the input--output matrix and leaves final demand outside the exchange, a closure here severs intermediate flows alone, and final demand stays autonomous.} Prices are fixed, benchmark input
proportions are optimal at those prices, and no producer expands beyond its
benchmark sales, so a closure works through quantities alone.

\subsection{Notation and preliminaries}
Matrices are uppercase bold and vectors lowercase bold. Greek letters denote parameters, and calligraphic letters
denote sets and operators. Inequalities between vectors and matrices are
entrywise. Indices \(i,j\) range over the country-sectors, \(c,c'\) over the
\(C\) countries, \(r,r'\) over the \(S\) sectors, and \(k\) over the \(K\)
chokepoints. An ordered pair always reads in the direction the goods move: in \(cc'\) the first index is the exporting country and the second the
importing one, and in \(ij\) the first index is the selling country-sector
and the second the buying one. A double-struck one denotes an indicator, and its subscripts say what it
indicates, while the bold \(\mathbf 1\) is the vector of ones and never an indicator. Accents and superscripts carry a fixed meaning throughout. We give the meaning of each where it first appears, and Online Appendix~\ref{app:notation} catalogs them in
full.

Let \(\mathcal C\) denote the set of countries, with \(C:=|\mathcal C|\), and
\(\mathcal S\) the set of sectors, with \(S:=|\mathcal S|\). The units of the world trade network are the country-sectors
\(i\in\mathcal{CS}:=\mathcal C\times\mathcal S\), so
\(|\mathcal{CS}|=CS\). The projections \(c(i)\) and \(r(i)\) return the
country and the sector of \(i\), and \(\mathcal N_r:=\{i:\,r(i)=r\}\) collects the
country-sectors of sector \(r\).
Let \(Z_{ij}\geq 0\) be the benchmark intermediate flow from seller
country-sector \(i\) to buyer country-sector \(j\), collected in the \(CS\times CS\) matrix
\(\mathbf Z\). The matrix \(\mathbf Z\) has one row for each country-sector in its role as a seller and one column for each country-sector in its role as a buyer, and every matrix of flows or of input shares in the paper is laid out in the same way, with sellers in the rows and buyers in the columns. Whenever buyer \(j\)'s benchmark intermediate spending
\(\sum_i Z_{ij}\) is positive, the benchmark input-share matrix
\(\mathbf A=[a_{ij}]\) is defined by \(a_{ij}:=Z_{ij}/\sum_{i'}Z_{i'j}\). The matrix \(\mathbf A\) is again of dimension \(CS\times CS\), and
each of its columns describes the composition of a buyer's
intermediate-input bundle and sums to one.

Let \(s_i\) be the benchmark `gross activity' of country-sector \(i\), by which we mean its gross output, and let
\(f_i\geq 0\) be the autonomous non-intermediate demand for \(i\)'s output,
final consumption and every other use not generated by intermediate
purchases inside the network. Let \(\beta_j\in(0,1)\) be the leakage parameter, the fraction of country-sector \(j\)'s activity that leaves the intermediate-goods network as value added and net taxes instead of being spent on intermediate inputs. Its intermediate spending is then \((1-\beta_j)s_j=\sum_i Z_{ij}\). We collect the leakage parameters in the \(CS\times CS\) diagonal matrix
\(\boldsymbol\beta:=\operatorname{diag}(\beta_1,\ldots,\beta_{CS})\). The two primitives describe the two sides of a country-sector's
ledger, \(\beta_i\) governing its spending on inputs and \(f_i\) the
demand for its output, and at benchmark they are
tied by the world accounting identity \(\sum_i f_i=\sum_i\beta_i s_i\). The
benchmark activity vector then satisfies
\begin{equation}
\mathbf s
=
\mathbf f+\mathbf A(\mathbf I-\boldsymbol\beta)\mathbf s .
\label{eq:benchmark_activity_choke}
\end{equation}
Here \(\mathbf I\) is the \(CS\times CS\) identity matrix, and \((\mathbf I-\boldsymbol\beta)\mathbf s\) is the vector of the intermediate spending of the buyers. Naturally, the term \(a_{ij}(1-\beta_j)s_j\) is the
intermediate demand that buyer \(j\) directs to seller \(i\). The product \(\mathbf A(\mathbf I-\boldsymbol\beta)\mathbf s\) adds these terms up along the row of each seller, and so gives the intermediate demand that each seller receives from all of its buyers together. Since each column of \(\mathbf A(\mathbf I-\boldsymbol\beta)\) sums to \(1-\beta_j\), which is below one, a dollar of demand placed anywhere sets off a chain of upstream orders that fades geometrically rather than echoing without end. The geometric fading of those orders is why \eqref{eq:benchmark_activity_choke} has a unique nonnegative solution \citep[Lemma~A.1]{bhatt2026_sanctions}.

\subsection{Chokepoint closure and route reroutability}
\label{subsec:chokepoint_restrictions}

Let \(\mathcal K\) be the set of maritime chokepoints, with
\(K:=|\mathcal K|\). A chokepoint matters for the model only through the
benchmark flows that pass through it. For a chokepoint \(k\in\mathcal K\),
define the transit indicator \(\mathbbm 1^{(k)}_{ij}\in\{0,1\}\) to equal one if the benchmark flow from
seller country-sector \(i\) to buyer country-sector \(j\) transits chokepoint
\(k\), and zero otherwise. We construct these transit indicators, when we map the chokepoints onto the trade flows, from
route geography, the restriction to the sectors whose goods travel by sea, and the observed flows.

The salvage a longer voyage affords is not uniform across trading pairs. The reroutability matrix of Online Appendix~\ref{app:reroutability}
records it. For each
chokepoint \(k\) and ordered country pair \((c,c')\), let
\(e^{(k)}_{cc'}\in[0,1]\) denote the `reroutability' of the pair, by which we mean the share of the benchmark flow from
country \(c\) to country \(c'\) that survives when chokepoint \(k\) closes
and the flow takes the shortest detour around it. The complementary share
\(1-e^{(k)}_{cc'}\) is lost on that detour. We call it the `lost share' of the pair. A reroutability of one means that the
detour costs nothing, and a reroutability of zero means that no viable detour exists and the whole flow is lost.
Reroutability is the same for every sector, because it is a property of the sea route rather than of the commodity carried.

For each chokepoint \(k\), collect these shares in the \(C\times C\)
reroutability matrix \(\mathbf E_k=[e^{(k)}_{cc'}]\). We call the collection of these matrices, one for each of the \(K\) chokepoints, the `reroutability profile', and write it \(\mathbf E=(\mathbf E_k)_{k\in\mathcal K}\).
The reroutability profile \(\mathbf E\) is exogenous. We read it off sailing distances
and detour geometry, and hold it fixed across every scenario and every parameter setting. The one exception is the robustness check that varies the detour geometry (Online Appendix~\ref{app:robustness_detour}).

A closure set is a set \(\cset{}\subseteq\mathcal K\) of chokepoints shut
together. What a closure does depends on the detours around the closed gates as well as on the gates themselves, so
we write \(\omega=(\cset{},\mathbf E)\) for the pair and call \(\omega\) the `scenario'. Post-closure objects carry \(\omega\): a scalar as a parenthetical superscript and the corresponding bold object as a subscript.
Because the reroutability profile \(\mathbf E\) is exogenous, a scenario is determined by its
closure set. We fix one scenario throughout, and name it explicitly only where a
comparison across closure sets needs the distinction.

Gates that share a detour have to be told apart from gates that do not. When
two closed gates lie on one voyage and a single longer voyage avoids them
both, the flow pays for that detour once, however many of the two are shut.
Suez and Bab-el-Mandeb are two such gates, since a voyage around the Cape of Good Hope avoids them both. We say that two chokepoints belong to the same `detour class' when the closure of either forces the same physical detour on the flows that transit both, and a chokepoint that shares its detour with no other forms a detour class by itself. The detour classes \(\mathcal D_1,\ldots,\mathcal D_N\) partition \(\mathcal K\), and Section~\ref{sec:step1_def} assigns the twelve
chokepoints of the paper to their classes. The disrupted fraction of link \(i\to j\)
under the closure set \(\cset{}\) is
\begin{equation}
d_{ij}
:=
1-\prod_{n}
\Bigl(1-\max_{k\in \cset{}\cap\mathcal D_n}\,\bigl(1-e^{(k)}_{c(i)c(j)}\bigr)\,\mathbbm 1^{(k)}_{ij}\Bigr),
\label{eq:partial_choke_disruption}
\end{equation}
with the convention \(\max\emptyset=0\). Here \(\bigl(1-e^{(k)}_{c(i)c(j)}\bigr)\mathbbm 1^{(k)}_{ij}\) is the share of the flow on link \(i\to j\) that the closure of chokepoint \(k\) alone would cost, the lost share of the country pair where the link transits the chokepoint and zero where it does not. The disrupted fraction \(d_{ij}\in[0,1]\) is the share of the benchmark trade from \(i\) to \(j\) that the scenario puts beyond reach once the shipment has taken whatever detour it has. It depends on the sea routes alone, so it is settled before any economic response to the closure.

Note that a country-sector link \(i\to j\) inherits the reroutability
\(e^{(k)}_{c(i)c(j)}\) of the countries it runs between. A flow that transits no closed
chokepoint has \(d_{ij}=0\), and a flow that transits a single closed
chokepoint on an unreroutable country pair has \(d_{ij}=1\). The gates of a detour
class share one detour, which is why the closed gates of a class enter \eqref{eq:partial_choke_disruption} through the largest lost share among them. Across classes each factor is the share of the
flow that survives the gates of one class. A shipment must clear every closed gate on its
route, and only what survives one class of gates is exposed to the next class, so the
share surviving the scenario is the product of the per-class survivals and
\(d_{ij}\) is one minus that product. The disrupted fraction is therefore the union
of the loss events along the route rather than their sum, so a flow that transits two closed chokepoints is not removed twice. Gulf crude bound for
northwest Europe, for instance, clears Hormuz, then Bab-el-Mandeb, then
Suez, and then Gibraltar on a single voyage. Shut Hormuz and the cargo never sails. Shut Suez
as well and nothing further is lost, since there is nothing left at the
second gate to lose. When the closure set holds one gate, or at most one gate of each
class, every maximum is taken over a single gate and the disrupted fraction
is one minus the plain product of the per-gate survivals.

Closures on record have seldom stopped every ship. Blockades are imperfectly enforced,
and what keeps moving through a closed passage can differ by what is shipped
(Online Appendix~\ref{app:est_delta}). We call the share of the disrupted trade that still crosses the closed passage the `residual transit share'. We let the residual transit share differ across the seaborne sectors \(\mathcal M\subset\mathcal S\), those whose output physically travels
by ship. We take that share from the record of past closures and treat it as a primitive of the model. The residual transit is trade that still moves through the passage by sea. It should not be mistaken for a bypass. Two economies whose sea route is closed cannot move their displaced trade onto a pipeline, a railway, or a road between them, because the capacity of each is given in the short run.

\begin{assumption}[Residual transit share]
\label{ass:attenuation}
For each seaborne sector \(r\in\mathcal M\) there is a residual transit share
\(\delta_r\), bounded away from both ends uniformly in \(r\),
\(0<\underline\delta\le\delta_r\le\bar\delta<1\), with \(\underline\delta\)
and \(\bar\delta\) fixed. A
link \(i\to j\) whose disrupted fraction under the closure set \(\cset{}\) is
\(d_{ij}\) retains, after closure, the share
\(1-(1-\delta_{r(i)})\,d_{ij}\) of its benchmark weight.
\end{assumption}

Assumption~\ref{ass:attenuation} says that a closure stops some of the trade it reaches and never all of it. A fully exposed link
(\(d_{ij}=1\)) keeps the residual
\(\delta_{r(i)}\,a_{ij}\), a clear link (\(d_{ij}=0\)) is untouched, and a link between the two loses the share \(1-\delta_{r(i)}\) of its disrupted fraction.
The residual transit share is indexed by the seller's sector because it is a property of the
shipped good. The assumption is in part a statement of fact, since most closures on record have let some trade through, and in part what the formal results of the paper stand on. A positive residual transit share leaves some flow on every link that a closure reaches, so the closure itself removes no trading relationship of the benchmark. The existence and the stability of the post-closure equilibrium are proved on that property (Online Appendix~\ref{app:stability}).

One further part of a disrupted flow is held fixed with the residual transit. The trade on a link that moves by pipeline, road, rail, or air does not cross the passage at all, and it stays at its benchmark value, since the capacity of each of these modes is given in the short run. We hold that trade fixed together with the residual transit in the rationed exchange of Section~\ref{subsec:rerouting_matrix}, so it is neither displaced nor free to carry displaced cargo. Section~\ref{sec:seashare} measures it link by link.

The shipped good decides both which flows a closure reaches and how severely the closure affects the flows it does reach. Only the sales of the sectors in \(\mathcal M\) carry
chokepoint incidence, so services, utilities, and construction never lose
a sale directly, whereas the residual transit share \(\delta_r\) measures how much of a
reached flow still moves.

Under Assumption~\ref{ass:attenuation}, collect the post-closure weights in the
\(CS\times CS\) disrupted matrix \(\check{\mathbf A}=[\check a_{ij}]\),
\begin{equation}
\check a_{ij}
:=
\bigl(1-(1-\delta)\,d_{ij}\bigr)\,a_{ij}.
\label{eq:disrupted_matrix}
\end{equation}
The symbol \(\delta\) without a subscript abbreviates the residual transit share \(\delta_{r(i)}\) of the seller's sector in Assumption~\ref{ass:attenuation}, and a caron marks an object the closure has disrupted and the rationed exchange has not yet touched. The bracket separates
the part of the order the closure never reached from the part that leaks
through it, \(1-(1-\delta)\,d_{ij}=(1-d_{ij})+\delta\,d_{ij}\). Because
\(\delta>0\), the support of \(\check{\mathbf A}\) is exactly the benchmark support \(\operatorname{supp}(\mathbf A)\).

\subsection{Rationed exchange at fixed prices}
\label{subsec:rerouting_matrix}

A closure forces adjustment on the buying and the selling end of the market at once, so the re-matching is two-sided. Prices are fixed in the short run, so the re-matching cannot run through prices and takes the form of rationing, as in the fixed-price equilibria of \citet{dreze1975} and \citet{benassy1975}, in which agents trade under quantity constraints that are consistent with one another. A buyer that loses a supplier seeks the same input from the suppliers it can still reach, in proportion to what it already bought from each, and a seller that loses a customer seeks buyers among the customers it can still reach, in proportion to what it already sold to each. We take this exchange from \citet{bhatt2026_sanctions}. Not all displaced trade finds a new partner within the horizon of a closure. A single unrecovered share \(\tau\in[0,1]\) of the displaced trade is written off symmetrically on both sides, and the exchange places the rest. Because the country-sectors that supply one input are substitutes for one another and the suppliers of different inputs are not, the exchange re-matches one input sector at a time, and the unrecovered share \(\tau\) measures how far that substitution falls short. We call the flows of one input sector \(r\), from the country-sectors \(\mathcal N_r\) that supply the input to all of their buyers, the `sector-\(r\) block'. The sector-\(r\) block is a \(C\times CS\) matrix, with a row for each of the \(C\) country-sectors that supply the input, one in every country, and a column for each of the \(CS\) country-sectors that may buy it.

The disrupted flows are \(\check Z_{ij}:=Z_{ij}\bigl(1-(1-\delta)d_{ij}\bigr)
=\check a_{ij}(1-\beta_j)s_j\), the disrupted shares of \eqref{eq:disrupted_matrix}
scaled by benchmark spending. We collect them in the \(CS\times CS\) disrupted transactions matrix \(\check{\mathbf Z}\). Fix an input sector \(r\). Buyer \(j\)'s
displaced sector-\(r\) input share is
\begin{equation}
\ell_{rj}
:=
\sum_{i\in\mathcal N_r}(1-\delta)\,d_{ij}\,a_{ij},
\label{eq:partial_disrupted_mass}
\end{equation}
the recoverable part \(1-\tau\) of which the surviving suppliers are
intended to absorb. With the benchmark sector-\(r\) input share
\(\alpha_{rj}:=\sum_{i\in\mathcal N_r}a_{ij}\), the exchange restores buyer \(j\)'s purchases from the sector-\(r\) block to its surviving sector-\(r\) budget, the benchmark input share less the unrecovered part of the displaced one, as a share of its spending,
\begin{equation}
t^{(\omega)}_{rj}
:=
\alpha_{rj}-\tau\,\ell_{rj},
\label{eq:choke_coltarget}
\end{equation}
carrying the flow \(t^{(\omega)}_{rj}(1-\beta_j)s_j\), and it restores seller \(i\)'s sales to its surviving capacity
\begin{equation}
g^{(\omega)}_i
:=
(s_i-f_i)-\tau\,Z^{\rm disp}_i,
\qquad
Z^{\rm disp}_i:=\sum_j(1-\delta)\,d_{ij}\,Z_{ij},
\label{eq:choke_rowtarget}
\end{equation}
its benchmark intermediate sales \(s_i-f_i=\sum_j Z_{ij}\) of
\eqref{eq:benchmark_activity_choke} less the unrecovered share of its
displaced sales \(Z^{\rm disp}_i\). The capacities of a block's sellers and the budgets of its buyers are the row totals and the column totals that the re-matched flows of the block have to meet. A seller's row total is all that it sells to the buyers of the block, and a buyer's column total is all that it buys from the sellers of the block.

Both the budget \(t^{(\omega)}_{rj}\) and the capacity \(g^{(\omega)}_i\) sit at or below their benchmark values. The short-run bound on
sales is therefore built into the budgets and capacities themselves, since under them no buyer purchases more of an input, and no seller sells more, than it did before the closure. Since the unrecovered share strips the same fraction of the same displaced trade from the capacities of a block's sellers and from the budgets of its buyers, the row totals and the column totals of the block add up to the same amount (Online Appendix~\ref{app:proof_model}). Their agreement is
necessary for the exchange to have a solution, because the row totals of any matrix add up to the same amount as its column totals.

The agreement is not sufficient, because the exchange can re-place displaced
trade only along the trade links that the closure leaves open, which we call the block's `open links'. Suppose, for instance, that every supplier of an input to some buyer sits behind a closed
passage that has no detour. That buyer's budget still counts the share of its displaced
purchases that is not written off, but no open link reaches it from any seller, so that part of
its budget cannot be spent. Since the row and column totals of the block agree with that budget
counted, the block's sellers are left with the same amount of capacity that no buyer they can
reach will take. More generally, the open links of a block may be too few to carry all of its displaced trade, so the row and column totals can agree in sum while no allocation over
the open links meets them all.
A closure of the Strait of Hormuz leaves the crude oil and gas block with more displaced trade
than its open links can carry, since the Gulf producers lose the buyers that their exports
reached and those buyers find no other producer with capacity to spare. The part of a block's
capacities, and equally of its budgets, that no allocation over its open links can place is the block's placement deficit, which we write \(U_r\) for the sector-\(r\) block and which Online Appendix Lemma~\ref{prop:wellposed} computes as the shortfall of a maximum flow.

A block with a placement deficit leaves the exchange without a solution, since no allocation
meets all of its row and column totals, so the equilibrium built on the exchange cannot be computed until the shortfall is booked somewhere. We book it in two slack
accounts, which we add to every sector-\(r\) block that carries a placement deficit and to no
other. The unsold-output account is an extra column of the block, a buyer that belongs to no
country-sector and absorbs the displaced sales that no buyer reachable over an open link can
take. The unmet-demand account is an extra row, a seller that belongs to no country-sector and
absorbs the displaced purchases that no reachable seller can fill. The \(C\times CS\) block then
becomes a \((C+1)\times(CS+1)\) block, in which each account takes the placement deficit \(U_r\)
as its total and the exchange always has a solution. Booking the shortfall on both sides of the
market keeps the exchange even-handed, since the two sides already write off the same
unrecovered share \(\tau\) of their displaced trade and are re-matched by the same proportional
rule. In the Hormuz closure the same quantity is booked as oil and gas that the Gulf producers
could not sell and as oil and gas that their buyers could not obtain. The accounts are dropped
once the exchange is solved, so the transactions matrix after the exchange keeps the \(CS\times CS\) layout of
the benchmark. What they absorbed enters no other block and generates no downstream demand, and
it appears only as sales that the block's sellers do not make and inputs that its buyers do not
receive.

\begin{definition}[Two-sided effective demand matrix]
\label{def:choke_reroute}
Fix a scenario \(\omega=(\cset{},\mathbf E)\), an unrecovered share
\(\tau\in[0,1]\), and residual transit shares
\((\delta_r)_{r\in\mathcal M}\) as in Assumption~\ref{ass:attenuation}.
Divide each entry of the disrupted
transactions matrix \(\check{\mathbf Z}\) into an unexposed flow and a residual transit,
\begin{equation}
\check Z_{ij}=(1-d_{ij})\,Z_{ij}+Y^{(\omega)}_{ij},
\qquad
Y^{(\omega)}_{ij}:=\delta_{r(i)}\,d_{ij}\,Z_{ij}.
\label{eq:choke_split}
\end{equation}
Hold the residual transit \(Y^{(\omega)}_{ij}\) fixed. On each sector-\(r\) block the unexposed flows \(\mathring Z_{ij}:=(1-d_{ij})Z_{ij}\) are rationed at fixed prices to the capacities \eqref{eq:choke_rowtarget} and the budgets \eqref{eq:choke_coltarget} net of that transit. The allocation, which we call the `re-matched flows', is the generalized Kullback--Leibler projection of the unexposed flows onto those row and column totals. The links on which the unexposed flow is positive are the block's open links. Where the row and column totals cannot all be met on the open links, the exchange is taken on the block augmented by the two slack accounts. Write \(Z^{(\omega)}_{ij}\) for the residual transit plus the re-matched flow.
The effective demand matrix \(\mathbf A_\omega=[a^{(\omega)}_{ij}]\) is
\(a^{(\omega)}_{ij}:=Z^{(\omega)}_{ij}/\bigl((1-\beta_j)s_j\bigr)\), and a
block with no disrupted link is left at benchmark.
\end{definition}

Definition~\ref{def:choke_reroute} treats the two parts of a disrupted flow differently. The residual transit \(Y^{(\omega)}_{ij}\), the part that still crosses the closed passage, stays as the closure left it. The unexposed flow \(\mathring Z_{ij}\), the part of the benchmark flow that the closure does not reach, is the only part the exchange rescales. Column \(j\) of the effective demand matrix \(\mathbf A_\omega\) records, as shares of buyer \(j\)'s benchmark intermediate spending, what buyer \(j\) orders from each seller once the displaced trade has been re-matched.

The treatment of the residual transit is where our model departs from that of \citet{bhatt2026_sanctions}. In their exchange the residual that a sanction leaves on a barred link is scaled along with every other surviving flow until the totals of each seller and each buyer are met, so displaced trade can return across the barred link. Here the residual transit is held fixed, so displaced trade must be re-placed over an open route, and the part that finds none is booked in the slack accounts. The difference in who imposes the restriction is what justifies the departure. A sanction is imposed by one of the trading parties. Where the severance is costly enough, the pair has a Coasean incentive to bargain flow back onto the restricted link through exemptions, waivers, or tolerated evasion. An exchange that rescales such links therefore has an economic interpretation. A chokepoint closure is imposed by a third party or by nature, so the severed pair has no counterparty to bargain with. What still crosses the passage is then a datum of the closure rather than something the pair can adjust, which is why we hold it fixed.

Within each block the capacities and the budgets meet at fixed prices. A supplier whose remaining customers would overdraw its capacity rations them, and a supplier whom its customers would leave with spare capacity is rationed by them. The allocation is the one at which no seller and no buyer rescales further. Online Appendix Lemma~\ref{prop:wellposed} shows that this allocation exists and is unique, and that of all the allocations meeting the row and column totals it departs least from the unexposed flows. We call the allocation of a sector-\(r\) block `interior' when it places a positive flow on every one of the block's open links. Which is to say that every trading relationship the closure leaves open still carries some trade, however little, once the displaced trade has been re-matched. On such a block the allocation is a two-sided proportional rationing of the unexposed flows,
\begin{equation}
Z^{(\omega)}_{ij}
=
Y^{(\omega)}_{ij}+u_i\,(1-d_{ij})\,Z_{ij}\,v_j,
\qquad i\in\mathcal N_r,
\label{eq:choke_reroute}
\end{equation}
in which the unexposed flow of each link is multiplied by a positive scaling \(u_i\) of the link's seller and a positive scaling \(v_j\) of its buyer.
At these scalings the purchases of every buyer \(j\) add up to its budget,
\(\sum_{i\in\mathcal N_r}Z^{(\omega)}_{ij}=t^{(\omega)}_{rj}(1-\beta_j)s_j\), and the sales of every
sector-\(r\) seller \(i\) add up to its capacity, \(\sum_j Z^{(\omega)}_{ij}=g^{(\omega)}_i\). On a block that carries the slack accounts, each budget and each capacity is met less the part of it that the accounts have absorbed. The scalings are the multipliers on the row and column totals, \(u_i\) on seller \(i\)'s capacity and \(v_j\) on buyer \(j\)'s budget.

Since the residual transit carries no scaling, the scalings of a block's sellers and buyers are a complete record of how its unexposed flows adjusted (Online Appendix~\ref{app:proof_model} writes the exchange for a whole block in matrix form). What the exchange pins down at each link is the product \(u_iv_j\),
the joint adjustment of the link's seller and buyer. Multiplying every seller's scaling on a
block by one common factor and dividing every buyer's by it leaves every product, and with it
the allocation, unchanged, which is why the scalings carry meaning only relative to one another.
In the re-matched flows, the ratio of a buyer's purchases from two sellers \(i\) and \(i'\)
is their ratio in the unexposed flows multiplied by \(u_i/u_{i'}\). A seller whose scaling is high
relative to those of the other sellers of the block is one that leans recovered output onto the
buyers it can still reach. The ratio of a seller's sales to two buyers \(j\) and \(j'\) is likewise their ratio in the unexposed flows multiplied by \(v_j/v_{j'}\), and a buyer whose scaling
is high relative to those of the other buyers of the block is one that leans harder on its
surviving suppliers.

Where the allocation of a block is interior, its scalings can also be thought of as
pseudo-prices, the prices at which an unconstrained buyer would choose the purchases it is
rationed to \citep{rothbarth1941,nearyroberts1980}. The pseudo-price of seller \(i\)'s good is
the reciprocal of its scaling \(u_i\), so a seller with displaced output to re-place sells as
though it had cut its price relative to the other sellers of the block, by as much as the buyers
it can still reach need to absorb that output. The pseudo-prices record the work that a change
in relative prices would have done had prices been free to move. Online
Appendix~\ref{app:proof_model} takes up the blocks on which the allocation is not interior,
where no pseudo-prices exist.

Suppose that nothing were written off, \(\tau=0\). A seller's capacity
\eqref{eq:choke_rowtarget} and a buyer's budget \eqref{eq:choke_coltarget} fall short of their
values before the closure only by the unrecovered part of the seller's displaced sales and of
the buyer's displaced purchases. At \(\tau=0\) the capacity of every seller \(i\) and the
sector-\(r\) budget of every buyer \(j\) return to those values, \(g^{(\omega)}_i=s_i-f_i\) and
\(t^{(\omega)}_{rj}=\alpha_{rj}\). On a sector-\(r\) block that carries no placement deficit the exchange
meets every one of these capacities and budgets, which is to say that each buyer of the block
buys as much of input \(r\) as it did before the closure and each seller sells as much, though
from and to different partners. The exchange then leaves the sellers and the buyers of such a block whole, so what makes a closure costly on such blocks is the unrecovered share rather than the exchange itself. The argument is that of \citet[Lemma~A.6]{bhatt2026_sanctions}, which Online Appendix~\ref{app:stability} carries over to an exchange that holds the residual transit fixed. A placement deficit is the one
loss that no value of the unrecovered share removes. Where displaced demand has no
open link to travel along, the slack accounts absorb it even at
\(\tau=0\), so the unrecovered share measures the write-off conditional on a feasible exchange, and the deficit \(U_r\) measures the failure of the exchange itself.

\subsection{Realized retention, the intermediate input efficiency factor, and the output ratio}
\label{subsec:input_bundle_distortion}

The effective demand is not yet an equilibrium, because the
matrix \(\mathbf A_\omega\) only records what each buyer asks of each
surviving supplier once the closure and the re-matching have settled. The matrix records those orders on the presumption that every supplier meets them, which is to say that each supplier \(i\) ships to each of its buyers \(j\) all of the input that \(j\) orders from it, the share \(a^{(\omega)}_{ij}\) of \(j\)'s benchmark intermediate spending. A supplier short of its own inputs cannot meet its orders. We call the share of its orders that supplier \(i\) actually ships its `fulfillment rate'
\(h_i\in[0,1]\).

A buyer's output depends on what its suppliers deliver rather than on what
it ordered from them. The realized sectoral input-retention ratio \(\theta^{(\omega)}_{rj}\) measures, input sector by input sector, how much of its benchmark input a buyer receives. The intermediate input efficiency factor \(\kappa^{(\omega)}_j\) aggregates these ratios over the buyer's input sectors, and the output ratio \(G^{(\omega)}_j\) says how much of its benchmark output the buyer can still produce. All three of these objects come from \citet{bhatt2026_sanctions}, their equations~10 to~12. Here each is evaluated at the effective demand \(\mathbf A_\omega\) of Definition~\ref{def:choke_reroute}, which holds the residual transit fixed, rather than at the effective demand a sanction leaves. The output ratio is written \(G^{(\omega)}_j\) rather than
their \(\lambda^{(\omega)}_j\), since this paper reserves \(\lambda\) for
the proportional loss. Every supplier \(i\) fills the same share \(h_i\) of each order it receives, so its buyers are served pro rata. Once shipped, the deliveries of different sellers within one
input sector are interchangeable at the buyer. Collect the fulfillment rates in the fulfillment vector
\(\mathbf h\in[0,1]^{CS}\), one coordinate for each country-sector.
Buyer \(j\)'s realized sectoral input-retention
ratio is
\begin{equation}
\theta^{(\omega)}_{rj}(\mathbf h)
\;:=\;
\frac{\sum_{i\in\mathcal N_r}h_i\,a^{(\omega)}_{ij}}{\alpha_{rj}}
\;\in\;[0,1],
\qquad r\in\mathcal R_j,
\label{eq:choke_tilde_f}
\end{equation}
with \(\mathcal R_j:=\{r:\alpha_{rj}>0\}\) the support of \(j\)'s benchmark input bundle, and Online Appendix Lemma~\ref{lem:choke_kernel} keeps the ratio at or below one. At \(\mathbf h=\mathbf 1\) the ratio isolates the contraction the exchange produces on its own, the effective demand retention
\(\bar\theta^{(\omega)}_{rj}:=\theta^{(\omega)}_{rj}(\mathbf 1)
=\sum_{i\in\mathcal N_r}a^{(\omega)}_{ij}\big/\alpha_{rj}\).
Fulfillment cuts the realized ratio further wherever a supplier is itself short. In a closure of the Strait of Hormuz, for instance, a Korean refinery that runs on Gulf crude cannot replace all of the crude it loses, so it fills only part of every order it receives. A Korean chemical works that buys all of its naphtha from that refinery receives none of it across a closed passage, so its effective demand retention of naphtha stays at one, and yet its realized retention falls to the share of each order that the refinery can fill.
We keep the effective demand retention and the realized retention apart throughout, because the residual transit, the part of a disrupted flow that still crosses the closed passage, floors the effective demand retention whatever the equilibrium. Realized retention
is that floor scaled by the order-weighted fulfillment of the block's own
suppliers, and so has a positive lower bound only in an equilibrium in which every supplier ships a positive share of its orders
(Online Appendix~\ref{app:stability}).

Buyer \(j\)'s intermediate input efficiency factor is
\begin{equation}
\kappa^{(\omega)}_j(\mathbf h;\rho)
\;:=\;
\left(
\sum_{r\in\mathcal R_j}\alpha_{rj}\,
\theta^{(\omega)}_{rj}(\mathbf h)^{\rho}
\right)^{\!1/\rho},
\label{eq:choke_ces_kappa}
\end{equation}
the \(\boldsymbol\alpha\)-weighted \(\rho\)-power mean of the retention
ratios, with substitution parameter
\(\rho\in[\underline\rho,\overline\rho]\subset(-\infty,0)\) bounded away
from the Cobb--Douglas and Leontief poles. Since \(\rho\) is negative, the inputs of different sectors are complements, so the power mean lies below the weighted average of the retention ratios unless they are all equal. A buyer that loses the same share of every input sector loses that share of its bundle of intermediate inputs and no more, whereas a buyer whose shortfall is concentrated in one input sector loses more than its average shortfall, because the scarce input holds back the use of the others. Complementarity also makes every input sector essential, so a buyer that receives nothing of one of them has \(\kappa^{(\omega)}_j=0\).

The intermediate input efficiency factor reaches output through the outer nest, which combines
the intermediate composite with the non-intermediate block. The non-intermediate block consists of the producer's primary inputs, which stay at their benchmark level. The outer nest weights the two blocks by the input--output table's own cost shares, which are the leakage parameter \(\beta_j\) for the non-intermediate block and \(1-\beta_j\) for the intermediate composite,
\begin{equation}
G_j(\kappa)
:=
\bigl[\beta_j+(1-\beta_j)\,\kappa^{\varrho}\bigr]^{1/\varrho},
\qquad
\varrho\in[\underline\varrho,\overline\varrho]\subset(-\infty,0).
\label{eq:choke_outer}
\end{equation}
Here the outer-nest curvature \(\varrho\) is negative, so the intermediate composite and the primary inputs are complements, and it is bounded away from the Cobb--Douglas and Leontief poles as the substitution parameter \(\rho\) is. The output ratio of producer \(j\) is the outer nest evaluated at its intermediate input efficiency factor, and the \(CS\times CS\) diagonal matrix \(\mathbf G_\omega\) carries the output ratio of country-sector \(j\) in row \(j\) and column \(j\),
\begin{equation}
G^{(\omega)}_j(\mathbf h)
:=
G_j\bigl(\kappa^{(\omega)}_j(\mathbf h;\rho)\bigr),
\label{eq:choke_lambda}
\end{equation}
\begin{equation}
\mathbf G_\omega(\mathbf h)
:=
\diag\bigl(G^{(\omega)}_1,\ldots,G^{(\omega)}_{CS}\bigr).
\label{eq:choke_Kmatrix}
\end{equation} At the boundary \(G_j(0)=0\), the limit that a negative \(\varrho\) delivers, so a producer whose input bundle collapses
delivers nothing. The log-log slope of \(G_j\) with respect to the intermediate input efficiency factor, the output elasticity of intermediates
\begin{equation}
\widetilde e_j(\kappa)
:=
\frac{(1-\beta_j)\kappa^{\varrho}}{\beta_j+(1-\beta_j)\kappa^{\varrho}}
\;\in\;[1-\beta_j,1),
\label{eq:choke_shadow}
\end{equation}
is the share of a producer's output that a marginal degradation of
its bundle of intermediate inputs takes away. The tilde keeps the output elasticity of intermediates apart from the reroutability \(e^{(k)}_{cc'}\).

\subsection{Order fulfillment and the post-closure equilibrium}
\label{subsec:short_run_activity}

The effective demand \(\mathbf A_\omega\) of
Section~\ref{subsec:rerouting_matrix} says what is ordered, and the output ratio
\(G^{(\omega)}_j\) of Section~\ref{subsec:input_bundle_distortion} says what a producer can deliver given what it receives, so deliveries and output ratios determine each other. The post-closure equilibrium is
a fixed point of that system, a set of deliveries and output ratios that reproduce one another.

Recall that a supplier's fulfillment rate \(h_i\) is the share of its orders that it actually ships. We assume that each country-sector fills its orders according to its own output ratio \(G^{(\omega)}_i\). Which is to say that a supplier whose inputs allow it to make only a share of its benchmark output fills that same share of every order it receives, from other producers and from final users alike. We call this assumption the `fulfillment rule', which is the counterpart, for a chokepoint closure, of the fulfillment condition in Definition~1 of \citet{bhatt2026_sanctions}.

\begin{definition}[Fulfillment rule]
\label{def:fulfillment}
At a fulfillment vector \(\mathbf h\in[0,1]^{CS}\), supplier \(i\) delivers
the share
\begin{equation}
h_i'
\;=\;
G^{(\omega)}_i(\mathbf h)
\;=\;
G_i\bigl(\kappa^{(\omega)}_i(\mathbf h;\rho)\bigr)
\;\in\;[0,1]
\label{eq:choke_fulfillment}
\end{equation}
of the effective demand it faces. The fulfillment rule
\(\mathcal G_\omega\colon[0,1]^{CS}\to[0,1]^{CS}\) stacks these
coordinates, \(\mathcal G_\omega(\mathbf h)
:=\bigl(G^{(\omega)}_i(\mathbf h)\bigr)_i\), and a fulfillment vector
is consistent when it is a fixed point of \(\mathcal G_\omega\).
\end{definition}

Since the capacities of the exchange keep the effective demand on every supplier within its benchmark sales (Online Appendix Lemma~\ref{lem:choke_kernel}), no supplier is asked for more than it sold before the closure, so a supplier that still received all of its inputs could meet all of its orders. A supplier under-delivers only when its own inputs fall short and its output ratio falls below one with them. Under-delivery in equilibrium reflects degraded productivity and never physical scarcity. Online Appendix~\ref{app:modelrobust} reports how the losses change under an alternative rule that treats what a supplier can still make as a capacity: the supplier fills its orders in full whenever that capacity covers them.

Write \(\mathbf M_0:=\mathbf A(\mathbf I-\boldsymbol\beta)\) and \(\mathbf M_\omega:=\mathbf A_\omega(\mathbf I-\boldsymbol\beta)\) for the propagation matrices of the benchmark and of the scenario, whose entry in row \(i\) and column \(j\) is the intermediate demand that a unit of buyer \(j\)'s activity places on seller \(i\).
For a closure set \(\cset{}\), a reroutability profile \(\mathbf E\), admissible \((\tau,\rho,\varrho)\), and residual transit shares \((\delta_r)_{r\in\mathcal M}\) as in Assumption~\ref{ass:attenuation}, the post-closure map on
\([\mathbf 0,\mathbf s]\times[0,1]^{CS}\) is
\begin{equation}
\mathcal T_\omega(\mathbf x,\mathbf h)
\;:=\;
\Bigl(
\underbrace{\mathbf G_\omega(\mathbf h)\,\mathbf f}_{\substack{\text{delivered}\\\text{autonomous demand}}}
\;+\;
\underbrace{\mathbf G_\omega(\mathbf h)\,\mathbf M_\omega\,\mathbf x}_{\substack{\text{delivered}\\\text{intermediate demand}}},\;\;
\underbrace{\mathcal G_\omega(\mathbf h)}_{\substack{\text{updated}\\\text{fulfillment rates}}}
\Bigr).
\label{eq:choke_operator}
\end{equation}
The product \(\mathbf M_\omega\mathbf x\) adds up, along the row of each seller, the intermediate orders that the buyers place with it at the activity \(\mathbf x\). We write \(D^{(\omega)}_i(\mathbf x):=[\mathbf M_\omega\mathbf x]_i\) for the sum along the row of seller \(i\), its total effective intermediate demand. The output ratio matrix \(\mathbf G_\omega(\mathbf h)\) of \eqref{eq:choke_Kmatrix} in front of the product multiplies the row of each seller by its output ratio.

The post-closure equilibrium is a pair fixed by the map,
\((\mathbf s_\omega,\mathbf h_\omega)=\mathcal T_\omega(\mathbf s_\omega,\mathbf h_\omega)\).
Its fulfillment rates satisfy \(\mathbf h_\omega=\mathcal G_\omega(\mathbf h_\omega)\). Since the fulfillment rule sets each supplier's fulfillment rate equal to its output ratio, the output ratio matrix at that fixed point is the diagonal matrix of fulfillment rates,
\(\mathbf G_\omega(\mathbf h_\omega)=\operatorname{diag}(\mathbf h_\omega)\).
The activity component then satisfies
\begin{equation}
\mathbf s_\omega
\;=\;
\operatorname{diag}(\mathbf h_\omega)\,\mathbf f
\;+\;
\operatorname{diag}(\mathbf h_\omega)\,\mathbf M_\omega\,\mathbf s_\omega.
\label{eq:choke_activity}
\end{equation}
Coordinatewise, \eqref{eq:choke_activity} reads
\(s^{(\omega)}_{i}=h^{(\omega)}_{i}\bigl(f_i+D^{(\omega)}_i(\mathbf s_\omega)\bigr)\), the share \(h^{(\omega)}_i\) of everything asked of producer \(i\).

The deliveries a producer receives fix its bundle of intermediate inputs,
the input bundle fixes its output ratio, and that output ratio is the share of its orders that it fills in the next round. We call this repeated application of the fulfillment rule the `delivery recursion'. The delivery recursion
closes in \(\mathbf h\) alone, so the post-closure fixed point is triangular. The fulfillment rates are determined first, and activity then follows from a single linear system solved at those rates. Since each column of the propagation matrix \(\mathbf M_\omega\) sums to at most \(1-\beta_j\) (Online Appendix Lemma~\ref{lem:choke_kernel}) and no supplier ships more than it is asked for, upstream orders fade geometrically as they do at the benchmark. That fading is why the linear system has exactly one solution at any given fulfillment rates. The routing and the rationed exchange are computed once per scenario, before the fixed
point is solved, so \(\mathbf A_\omega\) is a fixed input to the iteration. The closure's geography still acts
at every round, because each round propagates through \(\mathbf A_\omega\)
rather than through the benchmark matrix, but the severed links are cut
once, in the construction of \(\mathbf A_\omega\).

The post-closure map \(\mathcal T_\omega\) can have more than one fixed point, even though
the benchmark economy has a single equilibrium. The benchmark equilibrium is unique because it
takes every supplier to meet its orders, so activity solves a linear system with a single
solution. Once the model lets a supplier fill less than its orders, whether it fills them depends on what its own suppliers deliver, and that dependence runs in a circle through the network. Since
every input sector that a producer uses is essential, a group of producers that depend only on
one another for some input they use can all stop together, each for want of what the others no
longer make. The extreme case is the complete collapse \((\mathbf 0,\mathbf 0)\), which is a fixed point on any economy whose active producers all use intermediate inputs, and at which no producer delivers anything because none receives anything. We therefore select the fixed point that the delivery recursion reaches from the benchmark, at which every supplier
fills its orders in full, when the shortfalls of each round lower the fulfillment rates of the
next.

\begin{definition}[Benchmark-selected equilibrium]
\label{def:selection}
For \(\mathbf h\in[0,1]^{CS}\), let
\begin{equation}
\mathbf x(\mathbf h)
:=
\bigl[\mathbf I-\mathbf G_\omega(\mathbf h)\,\mathbf M_\omega\bigr]^{-1}
\mathbf G_\omega(\mathbf h)\,\mathbf f.
\label{eq:choke_selected}
\end{equation}
Set \(\mathbf h^{(0)}=\mathbf 1\) and iterate
\(\mathbf h^{(u+1)}=\mathcal G_\omega(\mathbf h^{(u)})\). The sequence
is entrywise nonincreasing and converges to a limit \(\mathbf h_\omega\). The pair
\((\mathbf s_\omega,\mathbf h_\omega)\) with \(\mathbf s_\omega:=\mathbf x(\mathbf h_\omega)\) is a fixed point of
\(\mathcal T_\omega\), called the `benchmark-selected equilibrium'.
\end{definition}

The sequence of fulfillment vectors that Definition~\ref{def:selection} generates, which we call the `benchmark path', decreases to the greatest fixed point of the fulfillment rule \(\mathcal G_\omega\). The rule is monotone,
since a producer that receives more from its suppliers can make more and so fills more of every order it receives. A path that starts with every supplier filling its orders in full, and that lowers each rate only as far as the shortfalls of the previous round require, never falls below any consistent fulfillment vector, so it comes to rest on the consistent vector that lies above all the others. The comparison is country-sector by country-sector and involves no trade-off among them, since at the selected fixed point every supplier fills at least as large a share of its orders as at any other fixed point. No producer's activity falls when a
fulfillment rate rises, so the selected equilibrium also carries the smallest loss of activity
of any fixed point for every country-sector at once, and with it the smallest GDP loss for every
economy and for the world (Online Appendix Proposition~\ref{prop:equilibrium}(ii)). Every loss
in the paper is in that sense the smallest that the model admits at its parameters.

Existence and selection of the benchmark-selected equilibrium follow by adapting Proposition~1 of \citet{bhatt2026_sanctions}. We call an equilibrium `interior' when every fulfillment rate is positive, \(\mathbf h_\omega>\mathbf 0\). Which is to say that every producer still fills some share of the orders it receives, however small, and no producer has stopped altogether. Interior uniqueness follows by adapting Proposition~1.2 of their
Online Appendix, in that an interior selected equilibrium is the only fixed point at which no producer has stopped. Every other fixed point is a collapse of some group of producers that depend only on one another for an input. Online Appendix Proposition~\ref{prop:equilibrium} states and proves the result, and its part (iii) gives a condition under which the selected equilibrium is interior. The condition is sufficient and not necessary. Every equilibrium we report is interior by direct check (Online Appendix~\ref{app:stability_numerics}).\footnote{The selected equilibrium is also determinate, and the primary inputs of the producer are what
make it so. The primary inputs stay at their benchmark level and cushion a shortfall of
intermediate inputs, so output falls by less than the input bundle does. Were the intermediate
composite and the primary inputs perfect complements, at the excluded pole \(\varrho\to-\infty\)
of the outer nest, a uniform cut in the share of orders that every supplier fills would cut every output ratio in the same proportion and so reproduce itself. Every uniformly scaled copy of a consistent fulfillment vector would then be consistent as well, as Lemma~2.1 of the Online Appendix of \citet{bhatt2026_sanctions} shows for the benchmark of an economy without a sanction.}

\subsection{Decomposition of the loss of economic activity due to a chokepoint closure}
\label{subsec:activity_decomposition}

A detour saves part of the trade that a closure reaches, part still crosses the passage, and the rationed exchange re-matches part of the rest. The remainder is an order that its seller loses and an input that its buyer loses. A closure is therefore two shocks at once: a loss of orders on the demand side of the producers that sell through the chokepoint and a shortfall of inputs on the production side of the producers that buy through it. We call the losses that begin on the demand side `lost orders' and the losses that begin on the production side `input shortfalls'. Each of the two strikes some producers directly and reaches others through the network, from their buyers or from their suppliers. We now divide the loss of activity by its source and by its channel, i.e., the way in which it reaches a producer.

We take lost orders first. Consider the model with the output ratio matrix \(\mathbf G_\omega\) held at the identity, in which every producer goes on filling all the orders it receives, whatever inputs reach it. This is an economy that suffers from lost orders but not from input shortfalls. Its equilibrium condition \eqref{eq:choke_activity} reduces to the benchmark system \eqref{eq:benchmark_activity_choke} with the propagation matrix \(\mathbf M_\omega\) of the scenario in place of the benchmark propagation matrix \(\mathbf M_0\). Since \(\mathbf I-\mathbf M_\omega\) is invertible (Online Appendix Lemma~\ref{lem:choke_kernel}), that system has a single solution, which is the activity with lost orders alone,
\begin{equation}
\mathbf s^{\ordonly}_\omega
:=(\mathbf I-\mathbf M_\omega)^{-1}\,\mathbf f.
\label{eq:orders_alone}
\end{equation}
The economy with lost orders alone still loses activity, because what a
producer makes follows the orders it receives and the closure removes orders. A Gulf crude producer whose Asian buyers the closure cuts off, and who can place only part of its crude elsewhere, makes less because it is asked for less, so it buys less from its own suppliers, who
in turn make less. The loss of activity in this economy, \(\mathbf s-\mathbf s^{\ordonly}_\omega\), is the one that a demand-driven
input--output calculation on the post-closure network would generate.

Naturally, no producer that is short of inputs could in fact keep filling its orders. This shortfall of inputs is the second source of loss. A refinery that has
lost its crude still holds orders for its fuel, but it fills only the share of them that its
input bundle allows. That shortfall is a failure of supply rather than a fall in demand.
Recall that the output ratio \(G^{(\omega)}_i\) is the share of its benchmark output that
producer \(i\) can still make with the inputs it receives, so \(1-G^{(\omega)}_i\) is the
share it cannot make for want of inputs. A producer goes short of inputs because the closure has cut off part of its own purchases, and also because its suppliers, themselves short of inputs, fill only part of what it still orders from them. The two are told apart by the output ratio \(G^{(\omega)}_i(\mathbf 1)\) that producer \(i\) would have if every one of its suppliers filled its orders in full, which is its fulfillment rate after the first step of the benchmark path of Definition~\ref{def:selection}. We accordingly divide the supply shortfall of producer \(i\) at the equilibrium
\((\mathbf s_\omega,\mathbf h_\omega)\), i.e., the part of everything asked of it that it fails to deliver, into what the closure causes and what its suppliers cause,
\begin{align}
\zeta^{(\omega)}_i
&:=\bigl(1-G^{(\omega)}_i(\mathbf 1)\bigr)\,
\bigl(f_i+D^{(\omega)}_i(\mathbf s_\omega)\bigr)
& &\text{(direct input shortfall),}\label{eq:choke_zeta}\\
\xi^{(\omega)}_i
&:=\bigl(G^{(\omega)}_i(\mathbf 1)-G^{(\omega)}_i(\mathbf h_\omega)\bigr)\,
\bigl(f_i+D^{(\omega)}_i(\mathbf s_\omega)\bigr)
& &\text{(downstream propagation).}\label{eq:choke_xi}
\end{align}

The loss of activity \(\mathbf s-\mathbf s_\omega\) is the sum of the loss through lost orders, \(\mathbf s-\mathbf s^{\ordonly}_\omega\), and the loss through input shortfalls, \(\mathbf s^{\ordonly}_\omega-\mathbf s_\omega\), which is what the shortage of inputs adds to it. Subtracting the system that \(\mathbf s^{\ordonly}_\omega\) solves from the benchmark system \eqref{eq:benchmark_activity_choke}, and the equilibrium condition \eqref{eq:choke_activity} from the system that \(\mathbf s^{\ordonly}_\omega\) solves, yields
\begin{equation}
\begin{aligned}
\mathbf s-\mathbf s^{\ordonly}_\omega
&=(\mathbf M_0-\mathbf M_\omega)\,\mathbf s
+\mathbf M_\omega\bigl(\mathbf s-\mathbf s^{\ordonly}_\omega\bigr),\\
\mathbf s^{\ordonly}_\omega-\mathbf s_\omega
&=\boldsymbol\zeta^{(\omega)}
+\mathbf M_\omega\bigl(\mathbf s^{\ordonly}_\omega-\mathbf s_\omega\bigr)
+\boldsymbol\xi^{(\omega)}.
\end{aligned}
\label{eq:choke_three_term}
\end{equation}
The second line uses the fact that the orders which a producer receives and does not fill are its supply shortfall. Table~\ref{tab:loss_decomposition} sets out the terms of \eqref{eq:choke_three_term} by their channel, and so gives every part of a producer's loss a source and a channel.

\begin{table}[H]
\centering
\caption{The loss of activity from a closure, by its source and by the channel through which it reaches a producer.}
\label{tab:loss_decomposition}
\small
\begin{tabular}{l c c c c}
\toprule
 & & \multicolumn{2}{c}{Network propagation} & \\
\cmidrule(lr){3-4}
 & Direct & Upstream & Downstream & Total \\
 & & (from buyers) & (from suppliers) & \\
\midrule
Lost orders & $(\mathbf M_0-\mathbf M_\omega)\,\mathbf s$ & $\mathbf M_\omega\bigl(\mathbf s-\mathbf s^{\ordonly}_\omega\bigr)$ & $\mathbf 0$ & $\mathbf s-\mathbf s^{\ordonly}_\omega$ \\
\quad(demand side) & & & & \\
\addlinespace
Input shortfalls & $\boldsymbol\zeta^{(\omega)}$ & $\mathbf M_\omega\bigl(\mathbf s^{\ordonly}_\omega-\mathbf s_\omega\bigr)$ & $\boldsymbol\xi^{(\omega)}$ & $\mathbf s^{\ordonly}_\omega-\mathbf s_\omega$ \\
\quad(production side) & & & & \\
\midrule
Total & $(\mathbf M_0-\mathbf M_\omega)\,\mathbf s+\boldsymbol\zeta^{(\omega)}$ & $\mathbf M_\omega(\mathbf s-\mathbf s_\omega)$ & $\boldsymbol\xi^{(\omega)}$ & $\mathbf s-\mathbf s_\omega$ \\
\bottomrule
\end{tabular}
\par\smallskip
{\footnotesize\textit{Note:} Every entry is a vector with one coordinate for each country-sector, at the scenario $\omega$. Each row sums to its last entry and each column to its bottom entry. The activity with lost orders alone $\mathbf s^{\ordonly}_\omega$ is defined in \eqref{eq:orders_alone}, and the direct input shortfall $\boldsymbol\zeta^{(\omega)}$ and downstream propagation $\boldsymbol\xi^{(\omega)}$ in \eqref{eq:choke_zeta} and \eqref{eq:choke_xi}.\par}
\end{table}

The columns of Table~\ref{tab:loss_decomposition} divide a producer's loss by its channel. The first column holds what the closure itself does to the producer, which loses the orders whose shipments crossed the chokepoint and the inputs that it bought through it. The direct demand shortfall \((\mathbf M_0-\mathbf M_\omega)\mathbf s\) is the change in the network valued at benchmark
activity, i.e., the shortfall in the demand for a producer's output due to the closure of the chokepoint. The shortfall is net of what the exchange re-matches. We call the sum of the direct demand shortfall and the direct input shortfall the `direct loss' of the producer. The second and third columns reach the producer through the network, from its two sides. Upstream propagation reaches it from its buyers, who order less because they have contracted, and so travels from a buyer to its suppliers, i.e., up the supply chain. Downstream propagation reaches it from its suppliers, who deliver less because they are short of inputs, and so travels from a supplier to its buyers, i.e., down the supply chain. We call the sum of upstream and downstream propagation `network propagation'.

The rows of the table divide the same loss by where it begins, i.e., by whether it begins with the orders that the closure removes or with a shortage of inputs. The columns cannot make that division, because the two sources travel differently. Lost orders travel upstream alone, whereas input shortfalls travel both ways, so the downstream column of the table holds input shortfalls alone and the upstream column holds both. Note that what separates the two sources in the upstream column is the activity with lost orders alone \(\mathbf s^{\ordonly}_\omega\).

What makes the two sources travel differently is the fulfillment rate of a producer, which depends on the inputs that it receives and not on its orders. A producer that loses orders makes less and orders less from its suppliers, but it still fills in full every order that remains, so its customers lose nothing by it. A producer that is short of inputs orders less from its suppliers because it can make less, and for the same reason it delivers less to its customers.\footnote{The division of a loss into the direct effect of a shock and its propagation to suppliers and customers is that of the literature on production networks \citep{acemoglu_akcigit_kerr_2016,carvalho_etal_2021_covid}. In the competitive economy of \citet{acemoglu_akcigit_kerr_2016} a demand shock travels upstream alone and a productivity shock downstream alone. Input shortfalls travel both ways here because prices are fixed, so a producer which makes less buys less.} The Gulf crude producer that loses its Asian buyers and the refinery that has lost its crude both order less from their suppliers. Both cuts reach the suppliers as upstream propagation, although the first began with the orders that the closure removed and the second with a shortage of inputs.

In the economy with lost orders alone no producer goes short of inputs, so every order withdrawn there traces back to an order that the closure removed. Whatever the full model loses beyond that economy, \(\mathbf s^{\ordonly}_\omega-\mathbf s_\omega\), began with a shortage of inputs. Solving each line of \eqref{eq:choke_three_term} for its left side gives the loss through lost orders as \((\mathbf I-\mathbf M_\omega)^{-1}(\mathbf M_0-\mathbf M_\omega)\mathbf s\) and the loss through input shortfalls as \((\mathbf I-\mathbf M_\omega)^{-1}(\boldsymbol\zeta^{(\omega)}+\boldsymbol\xi^{(\omega)})\). The inverse \((\mathbf I-\mathbf M_\omega)^{-1}\), i.e., the Leontief inverse of the post-closure network, carries the orders that the closure removes and the supply shortfalls through every round of orders that they set off.\footnote{No entry of Table~\ref{tab:loss_decomposition} is negative, so every entry is a part of the loss and never an offset against it. This is because the exchange adds no flow to the benchmark sales of any seller, the output ratio of a producer never exceeds one and does not fall when its suppliers deliver more, and the inverse \((\mathbf I-\mathbf M_\omega)^{-1}\) has no negative entry (Online Appendix Lemma~\ref{lem:choke_kernel}). The direct input shortfall and downstream propagation are valued at the post-closure demand \(f_i+D^{(\omega)}_i(\mathbf s_\omega)\) that a producer actually faces and not at its benchmark demand. An order withdrawn from a producer which could not have filled it is therefore counted once, in the direct demand shortfall or in upstream propagation.}

\subsection{Measure of country and world GDP losses due to maritime chokepoint closure}
\label{subsec:country_world_losses}

We now index activity by the closure set to compare losses across maritime chokepoint closures. We read these losses through the decomposition of Section~\ref{subsec:activity_decomposition}. The closure set \(\cset{}\) is the set of chokepoints shut together. We write \(\Delta s^{(\cset{})}_{i}:=s_i-s^{(\cset{})}_{i}\) for the activity loss of country-sector \(i\) under \(\cset{}\), which is never negative (Online Appendix Proposition~\ref{prop:equilibrium}). Write \(\gamma_i\) for the benchmark GDP
content of a unit of country-sector \(i\)'s activity, the value added that the input--output table records per unit of its gross output,\footnote{The GDP content \(\gamma_i\) differs from the leakage parameter \(\beta_i\) only by net taxes on products. Where a formal statement measures GDP by the accounting share, \(\gamma_i=\beta_i\), it says so.} and \(\mathcal{CS}(c):=\{i\in\mathcal{CS}:c(i)=c\}\) for country
\(c\)'s set of country-sectors. The GDP loss\footnote{We measure the GDP loss by value-added shares rather than by final expenditure because the model is written in changes against the benchmark. The loss counts the orders that a closure removes and the output that producers short of an input can no longer make. Final expenditure values levels, and in levels at fixed prices every lost intermediate sale is offset by a purchase that is no longer made, so it records only the output that fails to reach final users.  A producer short of an input is short of it in every unit that it still makes, so in levels the input that it does without is recorded as value added of its own. The losses that we report are gross GDP losses and not welfare losses, since a welfare measure would credit the expenditure switching, inventory rundown, and price adjustment that the short-run quantity model excludes.} of country \(c\) under closure set \(\cset{}\) is
\begin{equation}
\Delta\mathrm{GDP}_c(\cset{}) := \sum_{i\in\mathcal{CS}(c)} \gamma_i\,\Delta s^{(\cset{})}_{i}.
\label{eq:country_loss_choke}
\end{equation}
The notation leaves implicit the dependence of the GDP loss \(\Delta\mathrm{GDP}_c\) on the reroutability profile \(\mathbf E\), the unrecovered share \(\tau\), the residual transit share \(\delta\), the substitution parameter \(\rho\), and the outer-nest curvature \(\varrho\). 

The proportional loss of country \(c\), its GDP loss as a share of its benchmark GDP, is
\begin{equation}
\lambda_c(\cset{})
:=
\frac{\Delta\mathrm{GDP}_c(\cset{})}
{\sum_{i\in\mathcal{CS}(c)}\gamma_i s_i}.
\label{eq:country_loss_share_choke}
\end{equation}
The world loss is the share of benchmark world GDP that the closure removes,
\begin{equation}
L_W(\cset{})
:=
\frac{\sum_{c\in\mathcal C} \Delta\mathrm{GDP}_c(\cset{})}
{\sum_{i}\gamma_i s_i}.
\label{eq:world_loss_choke}
\end{equation}

Table~\ref{tab:loss_decomposition}, taken at the scenario \(\omega\) of the closure set \(\cset{}\), holds country-sector by country-sector, so the GDP loss of a country divides in the same way once every entry is weighted by the GDP content \(\gamma_i\). We measure how much of a country's loss reaches it through the network by the upstream and the downstream network-propagation share of the country, which are the totals of the upstream and of the downstream column of the table as shares of its GDP loss,
\begin{equation}
\eta^{\mathrm{up}}_c(\cset{})
:=\frac{\sum_{i\in\mathcal{CS}(c)}\gamma_i\,\bigl[\mathbf M_\omega(\mathbf s-\mathbf s_\omega)\bigr]_i}{\Delta\mathrm{GDP}_c(\cset{})},
\qquad
\eta^{\mathrm{dn}}_c(\cset{})
:=\frac{\sum_{i\in\mathcal{CS}(c)}\gamma_i\,\xi^{(\omega)}_i}{\Delta\mathrm{GDP}_c(\cset{})}.
\label{eq:network_shares}
\end{equation}
Their sum, \(\eta_c(\cset{}):=\eta^{\mathrm{up}}_c(\cset{})+\eta^{\mathrm{dn}}_c(\cset{})\), is the network-propagation share of country \(c\). The rest of the loss, \(1-\eta_c(\cset{})\), is the direct-loss share of the country, i.e., the part of its GDP loss that arises as the direct loss of its producer country-sectors. A producer country-sector is reached through the network by its buyers and suppliers at home as well as abroad, so the network-propagation share counts the propagation of a loss among the producer country-sectors of a country together with the propagation that arrives from other countries. The lost-orders share of country \(c\) is the total of the first row of the table as a share of its GDP loss (\(\sum_{i\in\mathcal{CS}(c)}\gamma_i\,[\mathbf s-\mathbf s^{\ordonly}_\omega]_i/\Delta\mathrm{GDP}_c(\cset{})\)). Its input-shortfall share is the same share of the second row (\(\sum_{i\in\mathcal{CS}(c)}\gamma_i\,[\mathbf s^{\ordonly}_\omega-\mathbf s_\omega]_i/\Delta\mathrm{GDP}_c(\cset{})\)).

For the world loss the sums run over all country-sectors. We write \(\eta^{\mathrm{up}}_W\), \(\eta^{\mathrm{dn}}_W\), and \(\eta_W\) for its network-propagation shares. The direct loss of the closure set \(\cset{}\) is \(\bigl(1-\eta_W(\cset{})\bigr)L_W(\cset{})\), and its `network multiplier' is
\begin{equation}
\nu_W(\cset{}):=\frac{1}{1-\eta_W(\cset{})}.
\label{eq:network_multiplier}
\end{equation}
The network multiplier is the GDP that the world loses for each dollar of direct loss, once the producer country-sectors which the closure reaches have ordered less from their suppliers and delivered less to their buyers.

% ================================================================

\section{Mapping Chokepoints onto Trade Flows}
\label{sec:mapping}
% ================================================================

A chokepoint enters the model only through the country-sector links that its
closure severs. We now identify those links and collect them in the chokepoint incidence matrix $\boldsymbol{\Gamma}$, which has a row for each of the \(K\) chokepoints and
a column for each bilateral intermediate flow of the extended input--output table, and whose entry records
whether the flow physically transits the passage. We first fix the set of passages that qualify as
chokepoints, then restrict attention to the sectors whose output
travels by sea, and finally assign every bilateral flow to the chokepoints
its least-cost route crosses. Note that the same good may cross no
chokepoint, one chokepoint, or several on the voyage from its seller's
economy to its buyer's. Which chokepoints a good crosses, and so how many, is a property of the route between the two economies rather than of the good, so the incidence matrix records the chokepoints crossed by each bilateral intermediate flow, i.e., by the sales of each seller country-sector to each buyer country-sector. Alongside the incidence matrix we measure the seaborne share of each such flow, i.e., the share of a seller country-sector's sales to a buyer country-sector that moves by ship, since a closure reaches only that part.

The flows themselves come from the extended matrix \(\mathbf Z^*\), the empirical counterpart of the benchmark matrix \(\mathbf Z\) of Section~\ref{sec:model}, whose base is the OECD Inter-Country Input--Output table at reference year 2019.\footnote{We
anchor on 2019, the last pre-pandemic reference year, because the
benchmark should describe the world economy in its ordinary state
(Online Appendix~\ref{sec:data}). On the reference years from 2011 to 2022 the world loss of a Hormuz closure runs from $2.0$ to $5.2$ percent of world GDP, moving with the energy trade that the strait carries, and the Strait of Hormuz is the costliest gate in every year (Online Appendix~\ref{app:yearrobust}).} That table identifies 80 economies individually and pools all others into a single Rest-of-World composite. The composite conceals the economies whose trade is most chokepoint-dependent: the Gulf states apart from Saudi Arabia and the UAE (Qatar, Kuwait, Oman, Bahrain, Iran, Iraq), the Red Sea rim, and much of Africa. We resolve these economies by combining two datasets. The OECD table supplies the levels, and the GTAP~10 database, whose 141 regions identify most of these economies individually, supplies the bilateral shares by which each Rest-of-World cell is divided into seven maritime-relevant blocks.\footnote{Because GTAP contributes only the within-composite shares, re-aggregating the seven blocks recovers the OECD
table exactly. Online Appendix~\ref{sec:data} sets out the construction, the
sector concordance, and the consistency checks.} Gulf-state exporters (OtherGulf), North African LNG suppliers (OtherNorthAfrica), and East African transit economies (OtherEastAfrica) thus appear as distinct rows and columns of \(\mathbf Z^*\) rather than disappearing into an undifferentiated rest of the world. The 80 identified economies and the seven blocks make up the extended economy set \(\mathcal C^*=\{1,\ldots,C^*\}\), \(C^*=87\). We work with that set in the rest of the paper. With the \(S=50\) sectors of the OECD table, \(\mathbf Z^*\) is therefore of dimension \(C^*S\times C^*S\), which is \(4{,}350\times4{,}350\), and it is laid out like \(\mathbf Z\). An entry \(Z^*_{(c,r),(c',r')}\) is the benchmark intermediate flow from seller country-sector \((c,r)\) to buyer country-sector \((c',r')\), the country and sector indices written out explicitly.

% ================================================================
\subsection{Chokepoints and detour routes}
\label{sec:step1_def}
% ================================================================

Not every narrow passage is a chokepoint. Which passages qualify depends on the routes that ships follow. We begin by specifying the routes. Let $\mathcal{P}(c, c')$ denote the set of maritime routes from the seaboard of country $c$ to the seaboard of country $c'$. Let
$p^*(c,c') \in \mathcal{P}(c,c')$ be the least-cost route, the one that minimizes sailing distance, and with it transit time. Routes and distances are taken at the resolution of fourteen maritime regions, with each economy assigned to the region of its principal seaboard (Online Appendix~\ref{app:reroutability}).\footnote{The region of Sub-Saharan Africa pools the Atlantic and the Indian Ocean seaboards of the continent, so its route to the Mediterranean runs through the Red Sea. The routing therefore charges the trade of the Atlantic seaboard, from Senegal to Angola, with the Mediterranean and the Black Sea to Suez and Bab-el-Mandeb, although such a cargo would in fact pass Gibraltar. Without that trade the losses of these economies under a closed Suez Canal fall by half to four fifths, and the world loss by about \(6\%\).} We assume that every ship follows the least-cost route at the benchmark and keeps to it until a
closure occurs. A maritime passage enters
$\mathcal{K}$ as a chokepoint only if it carries positive benchmark
trade mass for which no within-route alternative exists. Formally, passage $k$ is a chokepoint when there is a set of country
pairs $(c,c')$ carrying positive benchmark mass such that $k$ lies on
the least-cost route $p^*(c,c')$, and every alternative route from $c$ to $c'$ that avoids
$k$ carries a strictly positive detour cost. On that set of country pairs the reroutability
$e^{(k)}_{cc'}$ of Section~\ref{subsec:chokepoint_restrictions} is
then strictly below one.

A passage that lies on no least-cost route for any significant flow in
$\mathbf Z^*$ is instead a detour route, the longer way around a gate that a
corridor falls back on when the gate shuts. That such a passage carries
no benchmark mass is not an empirical finding but a consequence of least-cost
routing.\footnote{The small normal-times traffic that does use the detour
passages, such as the deep-draft tankers that exceed Malacca's limit and take Lombok, is counted with the flows of the primary corridor.} Economically, a
detour route is an outside option rather than a conduit, so it enters the model not as a row of
$\boldsymbol\Gamma$ but through the detour distance $\Delta_k$ that sets
the reroutability $e^{(k)}_{cc'}$ (Online Appendix~\ref{app:reroutability}). In normal times almost no Asia--Europe
cargo goes around the Cape of Good Hope, for instance, because the route through Suez and Bab-el-Mandeb is shorter. When war-risk premia closed the Red Sea
in 2023--24, the diverted traffic went around the Cape, though nothing about the
Cape itself had changed. What changed was the cost of the passage it stands in
for. The longer way around a gate can also run through chokepoints that carry regular traffic. When Gibraltar is closed, a ship bound from the Mediterranean to the Atlantic leaves through Suez and Bab-el-Mandeb before it rounds the Cape of Good Hope.

The reroutability rule converts the detour distance into the share of a flow that survives the
longer voyage,
\begin{equation}
e^{(k)}_{cc'}
\;=\;
\frac{D_{cc'}}{D_{cc'}+\Delta_k}.
\label{eq:reroutability_rule}
\end{equation}
Here $D_{cc'}$ is the benchmark voyage distance from $c$ to $c'$ and
$\Delta_k$ the extra distance the detour adds, both in nautical miles. The denominator
is the detoured voyage, so the lost share $1-e^{(k)}_{cc'}=\Delta_k/(D_{cc'}+\Delta_k)$ is the fraction of that voyage
which is pure detour. The model treats the same fraction of the flow as lost, so a long voyage absorbs a given detour more easily than a short one.

Applied to the extended matrix, the chokepoint criterion yields $K = 12$ chokepoints and six
detour routes. Four of the twelve are terminal gates, the sole sea
exits of enclosed basins, namely the Strait of Hormuz for the Persian Gulf, the
Turkish Straits for the Black Sea, the Danish Straits (including the
\O resund) for the Baltic, and the Kerch Strait for the Sea of Azov. A terminal gate has no detour, so the reroutability $e^{(k)}_{cc'}$ of every country pair whose trade transits
the gate is zero. The other eight are bypassable gates, which lie on through-routes with a
longer way around. On the
Asia--Europe corridor these are the Suez Canal, the Bab-el-Mandeb / Gulf of Aden
passage, and the Strait of Gibraltar. On the Pacific arc they are the Strait of Malacca and the Taiwan--Luzon Corridor, in which we pool the Taiwan and
Luzon straits because the two carry the same East Asian outbound flows at the regional resolution of
$\mathbf Z^*$. The Panama Canal, the Strait of Dover, and the Korea
Strait complete the eight bypassable gates.

Each of the six detour routes is the longer way around one or more of the eight bypassable gates. The Cape of Good Hope stands
in for Suez, Bab-el-Mandeb, and Gibraltar together, and Cape Horn for
Panama. The Lombok and Sunda Straits stand in for Malacca, the
east-of-Taiwan passage for the Taiwan--Luzon Corridor, the
around-Scotland route for Dover, and the around-Japan route for the
Korea Strait.\footnote{Taiwan with mainland China, at the Taiwan--Luzon Corridor, and Japan with Korea, at the Korea Strait, enter at \(e=0.4\), an override of the rule set for the pair. Voyage distances are taken between regions, so the rule \eqref{eq:reroutability_rule} has no distance \(D_{cc'}\) for two economies of the same region, and both pairs lie within East Asia. For Taiwan and mainland China the override is the rule applied to the pair itself, and Japan and Korea take the same value (Online Appendix~\ref{app:reroutability}).}

This assignment of detour routes to gates fixes the detour classes of
Section~\ref{subsec:chokepoint_restrictions}. Suez, Bab-el-Mandeb, and
Gibraltar share the Cape of Good Hope detour and form one class, and each
of the other nine gates, whether terminal or bypassable, forms a class of
its own,
\begin{equation}
\begin{aligned}
\mathcal D \;=\; \bigl\{\;
& \{\text{Hormuz}\},\
  \{\text{Turkish}\},\
  \{\text{Danish}\},\
  \{\text{Kerch}\},\
  \{\text{Dover}\},\\
& \{\text{Suez},\,\text{Bab-el-Mandeb},\,\text{Gibraltar}\},\
  \{\text{Malacca}\},\\
& \{\text{Panama}\},\
  \{\text{Taiwan--Luzon}\},\
  \{\text{Korea}\}
\;\bigr\}.
\end{aligned}
\label{eq:detour_classes}
\end{equation}

The distinction between chokepoints and detours also separates geographically the
two ways in which the model lets a severed trade be saved: the re-matching of the trade with another partner and the rerouting of the shipment itself. The unrecovered share $\tau$ governs the re-matching, and the reroutability $e^{(k)}_{cc'}$ governs the rerouting. After a closure a buyer may turn to suppliers reachable by
unaffected routes, though only to suppliers of the same input sector, which is the substitution the unrecovered share $\tau$
governs. A
Japanese refiner that has lost Saudi crude through Hormuz turns
to Russian ESPO, US Light Sweet, Brazilian, and West African crude,
all of them crude, and all of them reaching Japan by routes that bypass
Hormuz. That substitution is the re-matching of
Definition~\ref{def:choke_reroute}, which recovers up to $(1-\tau)$ of the displaced
mass. Rerouting, by contrast, is not a choice of the buyer at all. Whether the
severed shipment from a given supplier still reaches its buyer by a longer
voyage around the closed passage is a fact of the sea route. The reroutability $e^{(k)}_{c(i)c(j)}$ records the share of the flow
that survives once that voyage is taken. The Japanese refiner's case separates re-matching from rerouting cleanly. Hormuz is the sole exit of the Persian Gulf, so the Saudi barrel has no detour and its reroutability is zero. The refiner's whole recovery therefore runs through re-matching. Blocked at Malacca instead, the same cargo would
have reached Japan by the Lombok and Sunda detour, so its reroutability would be close to one.

% ================================================================
\subsection{The seaborne sectors}
\label{sec:seaborne}

Not all output travels by sea, so the second step is to fix the sectors a
closure can reach. We restrict the chokepoint analysis to the seaborne sectors $\mathcal M \subset \mathcal{S}$, which are the
$27$ goods-producing sectors of the OECD table,\footnote{Sector codes follow ICIO 2025. Of the mining sectors the text names \texttt{B05} (coal), \texttt{B06} (crude petroleum and natural gas), \texttt{B07} (metal ores), and \texttt{B08} (other mining and quarrying). Of the manufacturing sectors it names \texttt{C10T12} (food, beverages, and tobacco), \texttt{C19} (refined petroleum), \texttt{C20} (chemicals), \texttt{C24A} (iron and steel), \texttt{C24B} (non-ferrous metals), \texttt{C26} (electronics), and \texttt{C29} (motor vehicles). Outside the goods sectors it uses \texttt{D} (electricity and gas supply) and \texttt{F} (construction).}
the agriculture, mining, and manufacturing sectors whose output is
physically transportable by ship,
\begin{equation}
\begin{aligned}
\mathcal M \;=\; \bigl\{\;
& \text{A01},\;\text{A02},\;\text{A03},\;
  \text{B05},\;\text{B06},\;\text{B07},\;\text{B08},\;\text{B09},\\
& \text{C10T12},\;\text{C13T15},\;\text{C16},\;\text{C17\_18},\;
  \text{C19},\;\text{C20},\;\text{C21},\;\text{C22},\\
& \text{C23},\;\text{C24A},\;\text{C24B},\;\text{C25},\;
  \text{C26},\;\text{C27},\;\text{C28},\;\text{C29},\\
& \text{C301},\;\text{C302T309},\;\text{C31T33}\;\bigr\}.
\end{aligned}
\label{eq:SM}
\end{equation}

The other 23 ICIO sectors, comprising utilities (D, E), construction
(F), and all services (G through T), produce output that does not physically
travel by ship and so cannot be severed at a passage. All flows
$Z^*_{(c,r),(c',r')}$ with $r \notin \mathcal M$ accordingly carry
chokepoint incidence zero.

The sector restriction applies to the seller's sector only, since it is the shipped object, the seller's output, that
transits the strait. Whether a tanker of Gulf LNG passes Hormuz does not
depend on whether its cargo is bought by a chemical works (\texttt{C20}, a goods sector) or by a power utility (\texttt{D}, a non-goods sector). A closure stops ships, not transactions, so what it severs is fixed by the
cargo and the route. A buyer outside the seaborne sectors therefore loses its seaborne inputs as directly as any other buyer. At each of the six gates of Online Appendix Table~\ref{tab:seller_buyer}, construction (\texttt{F}) is among the sectors that buy most of the trade the closure reaches, and at Hormuz and Malacca power utilities are second only to refining (\texttt{C19}).

\subsection{The routing rule}
\label{sec:step2}

We now assign country pairs to the gates their trade crosses. Let $\mathcal{G}_k \subset \mathbb{R}^2$ be the geographic boundary of
chokepoint $k$, the polygon of latitude--longitude coordinates that
demarcates the passage on the map. The routing indicator
\begin{equation}
\mathbbm 1^{(k)}_{cc'} := \mathbbm 1\!\left[
p^*(c,c') \cap \mathcal{G}_k \neq \emptyset
\right]
\label{eq:routing}
\end{equation}
equals one when the least-cost route $p^*(c,c')$ of Section~\ref{sec:step1_def} crosses the
passage. For energy commodities (\texttt{B06}, \texttt{C19}) and bulk goods
(\texttt{B07}, \texttt{B08}, \texttt{C24A}, \texttt{C24B}, \texttt{C10T12}) a unique
primary route exists, so the geometry of the route determines
$\mathbbm 1^{(k)}_{cc'}$. For manufactured goods with
multiple potential ports (\texttt{C26}, \texttt{C29}), by contrast, we assign
$\mathbbm 1^{(k)}_{cc'} = 1$ only when the majority of bilateral trade value uses the route
crossing chokepoint~$k$, based on shipping-lane data and port geography. Online Appendix Table~\ref{tab:routing} records the resulting assignments for the economy-pair groups that carry the most trade, such as Gulf exports to East Asia and Europe, East Asian exports to Europe and North America, and Russia's exports from its Baltic, Black Sea, and Pacific ports.

\subsection{The flow-level assignment and the incidence matrix}
\label{sec:step3}

We now combine the sector restriction with the routing rule. A
bilateral flow
$Z^*_{(c,r),(c',r')}$ is assigned to chokepoint~$k$ if and only if it passes each of three
filters,
\begin{equation}
\mathbbm 1^{(k)}_{(c,r,c',r')} = 1 \iff
\underbrace{\mathbbm 1^{(k)}_{cc'} = 1}_{\text{(i) route crosses }k}
\;\wedge\;
\underbrace{r \in \mathcal M}_{\text{(ii) shipped good travels by sea}}
\;\wedge\;
\underbrace{Z^*_{(c,r),(c',r')} > 0}_{\text{(iii) flow exists}}.
\label{eq:three_filters}
\end{equation}

Saudi crude sold to Chinese refineries,
$Z^*_{\texttt{SAU\_B06},\,\texttt{CHN\_C19}}$, illustrates the three
filters at work. The cargo leaves Ras Tanura on the Persian Gulf, and the
least-cost route to China's refining centers transits
Hormuz, Malacca, and the Taiwan--Luzon Corridor on its final northbound
leg, so $\mathbbm 1^{(\text{Hormuz})}_{\text{SAU},\text{CHN}} = 1$. The route crosses the gate,
the shipped good travels by sea, and the flow exists, so
$\mathbbm 1^{(\text{Hormuz})}_{(\text{SAU},\texttt{B06},\text{CHN},\texttt{C19})} = 1$.\footnote{The extension of the matrix is what
makes many assignments possible at all. Kuwaiti crude to China appears as
$Z^*_{\texttt{OtherGulf\_B06},\,\texttt{CHN\_C19}}$ rather than dissolving
into an undifferentiated ROW row, and the same holds for Qatari LNG bound
for Europe and for East African flows through the Indian Ocean.}
Another illustrative flow is that of Taiwanese semiconductors sold to
mainland electronics assemblers,
$Z^*_{\texttt{TWN\_C26},\,\texttt{CHN\_C26}}$. The voyage from Kaohsiung to the mainland ports runs the length
of the Taiwan Strait, so
$\mathbbm 1^{(\text{TL})}_{(\text{TWN},\texttt{C26},\text{CHN},\texttt{C26})} = 1$.

Taiwan and mainland China, at the Taiwan--Luzon Corridor, and Japan and Korea, at the Korea Strait, are the only two pairs of East Asian economies whose trade the routing sends through a gate. A north-China shipment of electronics to Incheon illustrates the routing filter (\(\mathbbm 1^{(k)}_{cc'}=1\)) refusing what the sector filter (\(r\in\mathcal M\)) and the flow filter (\(Z^*_{(c,r),(c',r')}>0\)) pass. The shipment is a
seaborne good and a positive flow, so it passes filters (ii) and (iii) of
\eqref{eq:three_filters}. Its voyage,
though, is a Yellow Sea crossing, and at the region resolution of the
routing it meets no gate's boundary \(\mathcal G_k\), so filter (i)
fails and the flow carries incidence zero at all twelve gates. Apart from the trade between Taiwan and mainland China, the incidence of the Taiwan--Luzon Corridor therefore
falls on East Asia's long-haul exchanges with the rest of the world
rather than on the region's internal trade.

Collecting the answers of the three filters of
equation~\eqref{eq:three_filters} yields the chokepoint incidence matrix,
\begin{equation}
\boldsymbol{\Gamma}
=
\bigl[\mathbbm 1^{(k)}_{m}\bigr]_{k\in\mathcal K,\;m}
=
\begin{array}{c@{\ }c}
 & \text{\scriptsize flows, country-sector to country-sector} \\
\text{\scriptsize\shortstack{chokepoint\\ gates}} &
\begin{pmatrix}
\mathbbm 1^{(1)}_{m_1} & \mathbbm 1^{(1)}_{m_2} & \cdots & \mathbbm 1^{(1)}_{m_M}\\
\vdots & \vdots & & \vdots\\
\mathbbm 1^{(K)}_{m_1} & \mathbbm 1^{(K)}_{m_2} & \cdots & \mathbbm 1^{(K)}_{m_M}
\end{pmatrix}
\end{array}
\;\in\;\{0,1\}^{K \times M}.
\label{eq:incidence_def}
\end{equation}
Here a flow $m=\langle i,j\rangle$ is an ordered seller--buyer pair of country-sectors, and $M$ is the number of such flows with positive mass
whose seller sector lies in \(\mathcal M\). Row \(k\) lists every flow whose voyage crosses gate \(k\), and column \(m\) lists every gate standing on flow \(m\)'s route. The entry $\mathbbm 1^{(k)}_{m}=\mathbbm 1^{(k)}_{ij}$ in row \(k\) and column \(m\), one when the flow passes all three filters at gate \(k\) and zero otherwise, is the transit indicator of Section~\ref{subsec:chokepoint_restrictions}. Recall that the transit indicators enter the model through the
disrupted fraction \(d_{ij}\) of equation~\eqref{eq:partial_choke_disruption}, the share of the benchmark trade from \(i\) to \(j\) that a closure puts beyond reach.
The columns of the incidence matrix \(\boldsymbol\Gamma\) fall into one block for each of the \(C^*\times C^*\) economy pairs, and the block \(\boldsymbol\Gamma^{(c,c')}\), with \(K\) rows, collects the columns of the flows from economy \(c\) to economy \(c'\). A block is identically zero whenever no route between
the pair crosses any chokepoint (Germany to France, for instance).

The rows of \(\boldsymbol\Gamma\) overlap, since a flow whose route crosses
several gates is included in the row of each. On the long corridors such overlap is the usual case. An
illustrative flow is Saudi crude bound for Germany,
$Z^*_{\texttt{SAU\_B06},\texttt{DEU\_C19}}$, which crosses
Hormuz, Bab-el-Mandeb, Suez, and Gibraltar on one voyage, so
$\mathbbm 1^{(\text{Hormuz})}_{m} =
\mathbbm 1^{(\text{Bab-el-Mandeb})}_{m} = \mathbbm 1^{(\text{Suez})}_{m} =
\mathbbm 1^{(\text{Gibraltar})}_{m} = 1$.
East Asian electronics bound for Northwest Europe cross five gates on one voyage: the Taiwan--Luzon
Corridor, Malacca, Bab-el-Mandeb, Suez, and Gibraltar (Online
Appendix Table~\ref{tab:routes}). When several of the gates on such a voyage close together, the disrupted fraction \(d_{ij}\) of
equation~\eqref{eq:partial_choke_disruption} counts the flow as lost only once.

\subsection{The seaborne share of a trade flow}
\label{sec:seashare}

Recall that a seaborne sector is one of the goods-producing sectors \(\mathcal M\) whose output can travel by ship (Section~\ref{sec:seaborne}). Such a sector does not ship all of its output by sea. Natural gas and a
good part of the crude oil that crosses a land border move through
pipelines, neighboring economies trade by road and rail, and goods that
are light for their value, electronics and pharmaceuticals among them,
travel by air. We therefore measure, for each link \(i\to j\) whose seller
sector lies in \(\mathcal M\), the seaborne share \(\mu_{ij}\in[0,1]\),
which is the share of the benchmark flow \(Z_{ij}\) that moves by ship. The
seaborne share \(\mu_{ij}\) depends on the seller's sector \(r(i)\) and on
the two economies \(c(i)\) and \(c(j)\) that the link joins.

The measurement rests on customs records that report the value of trade by
mode of transport. The European Union and the United Kingdom keep such
records for their trade with every outside partner, product by product
\citep{eurostat_comext_modes}, so a pair of economies in which one side reports takes its seaborne share from them directly. A pair in which
neither side reports takes the share that the exporter's goods of the same
sector show on arrival in Europe. Trade inside the continental network of
Europe and Central Asia, where road, rail, and pipeline carry most of the
cargo and the external records are silent, takes its share from the
international transport margins of GTAP \citep{Aguiar2019}. Outside Europe and Central Asia the
customs records tell sea from air but not from land, so trade which crosses a land border
there, such as that of Jordan with its Gulf neighbors, is still charged to the sea route between
the two economies.\footnote{Online Appendix~\ref{app:land_neighbors} measures what this charge does to the losses. Taking all trade between land neighbors there off the sea routes lowers the world loss of a Hormuz closure from $2.83$ to $2.79$ percent of world GDP, and the world loss of no closure by more than a tenth. Among the economies the charge matters most for Jordan. The routing sends the crude oil that Jordan buys from Saudi Arabia through the Strait of Hormuz, whereas it is shipped from Yanbu to Aqaba inside the Red Sea. With that trade taken off the gates, Jordan's loss under a Hormuz closure falls from $7.4$ to $1.7$ percent of its GDP.} Online
Appendix~\ref{app:seashare} sets out the construction and reports how much
of the trade on each gate's routes is seaborne.

A closure acts only on the part of a flow that moves by sea. On link \(i\to j\) that part is \(\mu_{ij}Z_{ij}\), of which the closure puts the disrupted fraction \(d_{ij}\) of equation~\eqref{eq:partial_choke_disruption} beyond reach. Of the seaborne trade put beyond reach, the residual transit share \(\delta\) of Assumption~\ref{ass:attenuation} still crosses the closed passage, so the closure displaces the flow \((1-\delta)\,\mu_{ij}\,d_{ij}\,Z_{ij}\). The trade which moves by pipeline, road, rail,
or air between the two economies is held fixed with the residual transit of
equation~\eqref{eq:choke_split}, as Section~\ref{subsec:chokepoint_restrictions} sets out, so the cargo which the sea no longer
carries must be re-placed over an open route.\footnote{With the seaborne share, the residual transit of the link \(i\to j\) is \(Y^{(\omega)}_{ij}=\bigl[\delta\,\mu_{ij}+(1-\mu_{ij})\bigr]d_{ij}Z_{ij}\) and its unexposed flow stays \((1-d_{ij})Z_{ij}\). The computation is thus the model of Section~\ref{sec:model} with the residual transit share of the link raised from \(\delta\) to \(\delta\mu_{ij}+(1-\mu_{ij})\), a number that lies between \(\delta\) and one and reaches one only on a link that carries nothing by sea, which no closure touches. Assumption~\ref{ass:attenuation} indexes the residual transit share by the seller's sector, and the seaborne share lets it vary also with the pair of economies, above the same floor \(\underline\delta\). The formal results of Section~\ref{sec:model} use the residual transit share through its positivity and that floor alone, so they hold with the residual transit share of the link in the place of \(\delta_{r(i)}\). Were the non-seaborne flow left with the unexposed flow instead, where the exchange rescales it, the loss at Hormuz would fall to the value that it takes when the exchange rescales the residual transit, whereas the loss at every other gate would be all but unchanged (Online Appendix~\ref{app:seashare}).} Overall, the incidence matrix \(\boldsymbol\Gamma\) and the seaborne share \(\mu_{ij}\) together say how much of the world's intermediate trade a closure can reach.\footnote{The flows whose sea route crosses at least one of the twelve gates carry about two fifths of the world's cross-border intermediate trade by value. Their seaborne part is a little over a quarter of that trade and under \(4\%\) of all intermediate purchases, domestic ones included.}

\section{Single Chokepoint Closures}
\label{sec:results}
% ================================================================

We now close each of the twelve chokepoints in turn on the extended
intermediate-flow matrix $\mathbf Z^*$, under the incidence specification of
Section~\ref{sec:mapping}. We collect the results in the master exposure matrix
$\boldsymbol\Psi = [\lambda_c(\cset{}_k)]_{c\in\mathcal C^*,\, k\in\mathcal K}$,
which holds the proportional GDP loss of each of the \(87\) economies
under each of the twelve single closures. Here \(\cset{}_k:=\{k\}\) is
the closure set that shuts gate \(k\) alone, and the matrix is of dimension \(C^*\times K\), with a row for each economy and a column for each single closure. We report the loss to a country as a proportion of its GDP.\footnote{Every loss the paper reports is measured against the solved no-closure
equilibrium of the model rather than against the row totals of the assembled table. Online Appendix~\ref{oa:data_integrity} sizes the gap between the two.}

% ----------------------------------------------------------------
\subsection{Calibration}
\label{subsec:calibration}
% ----------------------------------------------------------------

Beyond the data and the reroutability profile, the model carries the substitution parameter
$\rho\in[\underline\rho,\overline\rho]\subset(-\infty,0)$ and the outer-nest
curvature $\varrho\in[\underline\varrho,\overline\varrho]\subset(-\infty,0)$. The model also carries the
unrecovered share $\tau\in[0,1]$ and the
residual transit share $\delta$ of Assumption~\ref{ass:attenuation}. We calibrate each of the four on evidence that predates the closures we study. The substitution parameter and the outer-nest curvature come
from the structural production-function
literature (Online Appendix~\ref{app:est_rho}). The unrecovered share is set against the firm-level evidence on how buyers replace a lost supplier (Online Appendix~\ref{app:est_tau}), and the residual transit share from the transit record
of nine historical disruptions (Online Appendix~\ref{app:est_delta}).
The baseline calibration takes the two curvatures and the residual transit share from the intervals this
evidence supports. The evidence does not pin down the unrecovered share, which we set at \(0.3\). The baseline is therefore \((\tau,\rho,\varrho)=(0.3,-1,-1)\) with a uniform
residual transit share \(\delta=0.10\) on all seaborne goods
(Section~\ref{sec:seaborne}).\footnote{In computing the intermediate input efficiency factor \(\kappa^{(\omega)}_j\) of \eqref{eq:choke_ces_kappa}, we treat as fully delivered any input sector that makes up less than \(0.05\%\) of the buyer's benchmark input bundle, so that a negligible input that a closure cuts off cannot by itself stop its buyer. Treating every input sector alike raises the world loss of each closure at the baseline calibration by between \(0.3\) and \(0.7\) percent of the loss itself. The loss of a closed Strait of Hormuz rises from \(2.825\%\) to \(2.833\%\) of world GDP, and no loss rises by as much as a hundredth of a percentage point of world GDP (Online Appendix~\ref{app:stability_numerics}).} Online
Appendix Table~\ref{tab:calibration} summarizes the calibration parameter
by parameter, with the interpretation, the evidence, and the supported
range of each, and Online Appendix~\ref{app:robustness} varies the parameters across those ranges.

We estimate the residual transit share on the record of past disruptions and then apply it at every gate.\footnote{A ``closed'' strait is not an empty one. When war-risk premia drove the container trade from the Red Sea in 2023--24, box transits fell by roughly $90\%$ at the trough, while the tankers kept nearly half of the corridor's oil moving \citep{unctad_2024_troubled_waters,imf_portwatch,eia_redsea_2024}. The uniform baseline holds every seaborne good to the residual of the containers, a tenth, so it is the more severe of the calibrations that the record admits (Online Appendix~\ref{app:est_delta}). The closure of the Strait of Hormuz in 2026 came nearer to a tenth than any closure before it, the strait carrying about a quarter of its benchmark oil over the first five and a half months (Online Appendix~\ref{app:hormuz_record}). The milder commodity-differentiated residual transit share, with \(\delta_r=0.45\) on the energy sectors and \(\delta=0.10\) on the other sectors, enters the robustness exercises of Online Appendix~\ref{app:modelrobust}.} The model allows this because it keeps apart the two things that decide what a closure does. The geography of a gate decides how much of each flow the closure reaches and how much of it a detour can save. The intensity of the closure decides how much of the reached trade still passes through. A single severity number per closure would merge the two. Such a number could always be chosen to fit the one episode at its gate, so the fit would establish nothing beyond itself. Geography differs across gates but not over time, so we measure the reroutability $e^{(k)}_{cc'}$ from sailing distances alone. The residual transit share $\delta_r$ depends on the cargo and on how hard the closure is enforced, not on the passage, so we measure it from the transit record of realized disruptions wherever they occurred. A gate that has never been closed, such as Malacca, then takes its geography from the map and its residual transit share from closures elsewhere.

The master exposure matrix $\boldsymbol\Psi$ measures unbuffered steady-state exposure, the loss of each economy from a sustained closure of each chokepoint once the post-closure equilibrium has settled. The flows of the benchmark are those of a year, so each loss is the share of a year's GDP that a closure sustained through the year would cost. The unrecovered share is accordingly the share of the displaced trade that finds no new partner within the year. The exposure is unbuffered because the model has no inventory state, no strategic-reserve drawdown, no emergency mobilization of bypass pipelines, and no price-induced demand substitution. The overland trade that the customs records cannot tell
from sea trade outside Europe and Central Asia is charged to the sea route
(Section~\ref{sec:seashare}), which raises the measured loss as well. In these respects the reported losses $\lambda_c(\cset{}_k)$ are upper bounds, the values that the losses would take if the topology of supply were the only thing standing between the closure and output.\footnote{The buffers are substantial. Each member of the IEA that imports oil must hold stocks, public and industry together, equal to at least $90$ days of its net imports \citep{iea_oil_stocks_2024}. The Saudi and Emirati crude pipelines that end outside the strait carried more once the 2026 closure mobilized them, but the chokepoint incidence does not credit them (Online Appendix~\ref{app:hormuz_pipeline}). A
buffered short-run calculation would lower the effective $\tau$ of the
crude-oil sector for as long as the reserves last. } In two other respects they are lower bounds, since the model leaves out the final goods that a closure stops (Online Appendix~\ref{app:final_goods}) and reports, of the equilibria that it admits, the one with the smallest loss (Section~\ref{subsec:short_run_activity}).

% ----------------------------------------------------------------
\subsection{The master exposure matrix}
\label{subsec:master_matrix_results}
% ----------------------------------------------------------------

A row of the master exposure matrix $\boldsymbol\Psi$ is an economy's exposure profile across
the world's chokepoints, and a column is a gate's reach across the world's economies. Ranked by
total exposure $\sum_{k\in\mathcal K} \lambda_c(\cset{}_k)$, the nine most exposed of the $87$ economies are
the United Arab Emirates, OtherGulf, Saudi Arabia, Brunei, Korea, Kazakhstan, Taiwan, Jordan,
and Thailand.\footnote{The total counts the corridor of Suez and Bab-el-Mandeb twice, since every flow that the routing sends through one of the two gates also passes the other. Counted once, the same nine economies lead in a slightly
different order. Jordan owes its place among the nine to the trade with Saudi Arabia that the routing charges to the Strait of Hormuz. With that trade taken off the gates, Turkey takes Jordan's place (Online Appendix~\ref{app:land_neighbors}).} Kazakhstan apart, the Hormuz cell leads each of their rows. Large exposure is
rare, with only about one in ten of the $1{,}044$ matrix entries exceeding $1\%$ of GDP.
Of the fifteen cells above $5\%$, fourteen lie in the Hormuz column and the fifteenth is Turkey's
cell on the Turkish Straits. The cells above $10\%$ are five country--chokepoint pairs of the
Hormuz column.

\subsubsection{The world loss of each closure}
Table~\ref{tab:share_world_loss} reports the world GDP loss under
each single closure, together with the top-five contributing economies as
shares of that loss. Hormuz dominates at about $2.8\%$ of world GDP. The strait is twenty-one
nautical miles wide at its narrowest point, and the intermediate goods that transit it
in a benchmark year amount to about $1\%$ of world GDP. Writing off every transiting
dollar as lost would put the cost at that $1\%$, and letting the
displaced trade re-source without friction would put it near
zero. The equilibrium loss lies above both of those naive estimates, even though the model lets the displaced trade reroute and re-source.
No detour leads out of the Gulf, so a longer voyage salvages nothing. Re-sourcing recovers less than a third of the displaced trade, since the sellers of the Gulf reach few buyers except through the strait. At every other gate it recovers seven tenths. The shipments that the re-sourcing does not recover cost more than their own value. The seller loses an outlet, and the buyer loses an input that is complementary to the rest of what it uses. Both losses then travel on through the world trade network. The $2.8\%$ is the loss that remains once the re-sourcing has done its work. That residual, not the trade share, is what the model measures. The number depends on holding fixed what still crosses the closed strait. The exchange of \citet{bhatt2026_sanctions} rescales what still crosses a severed link, so it lets re-placed trade pass back through the strait, and under that exchange the same closure costs $1.01\%$ of world GDP (Online Appendix~\ref{app:modelrobust}).

Hormuz dominates because Gulf exporters of crude oil, refined products, and petrochemical feedstocks supply the world's refining, chemical, and metals industries. When the strait closes, their buyers find few other sellers of the same inputs with output to spare in the short run. These buyers' intermediate input efficiency factors \(\kappa^{(\omega)}_j\) fall, and so do those of the buyers further along the world refining and petrochemical chain.

Behind Hormuz the gates rank in the order of the seaborne intermediate trade that their closures
reach (Online Appendix Table~\ref{tab:seashare_gate}). The Panama Canal comes first at about
$0.32\%$, since its closure reaches more seaborne intermediate trade than that of any gate but Hormuz.
Five gates follow close behind: Gibraltar ($0.28\%$), the Suez Canal and Bab-el-Mandeb ($0.26\%$), the Turkish Straits ($0.26\%$), and the Danish Straits ($0.24\%$). The European gates among them would stand higher if all the trade on their routes moved by sea. More than half of the trade on the routes through the Danish Straits and Gibraltar, and
two-fifths of the trade on the routes through the Turkish Straits, moves by road, rail, pipeline,
or air, where a closure does not reach it.

The other five gates each cost less than $0.2\%$ of world GDP. Short detours hold down the two Asian gates, the Strait of Malacca at $0.17\%$ and the Taiwan--Luzon Corridor at $0.14\%$. The Lombok--Sunda and east-of-Taiwan passages add little to the benchmark voyage, so the reroutability $e^{(k)}_{cc'}$ is close to one. The Korea Strait costs $0.05\%$, and Dover and Kerch $0.01\%$ or less. The routing charges Dover only with the Baltic trade that has already cleared the Danish Straits, and Kerch only with Ukraine's seaborne exports (Online Appendix~\ref{app:reroutability}). Each of these three gates reaches trade concentrated on a few economies, so its world loss is small even though it leaves a distinct mark on those economies.

\begin{table}[H]
\centering
\caption{World GDP loss and top-five contributing economies under
each single-chokepoint closure.}
\label{tab:share_world_loss}
\small
% >>> inlined from 04_outputs/tex_tables/tab_share_world_loss_K12.tex
\begin{tabular}{l r p{7cm}}
\toprule
Chokepoint & $L_W$ (\%) & Top-5 contributing economies (share of $L_W$) \\
\midrule
Strait of Hormuz & 2.83 & CHN (19\%), JPN (11\%), OtherGulf (9\%), SAU (8\%), IND (7\%) \\
Panama Canal & 0.32 & USA (26\%), CHN (25\%), BRA (6\%), JPN (6\%), KOR (4\%) \\
Strait of Gibraltar & 0.28 & CHN (11\%), USA (11\%), DEU (8\%), FRA (7\%), GBR (5\%) \\
Suez Canal & 0.26 & CHN (14\%), USA (7\%), DEU (6\%), OtherGulf (5\%), ITA (5\%) \\
Bab-el-Mandeb & 0.26 & CHN (14\%), USA (7\%), DEU (6\%), OtherGulf (5\%), ITA (5\%) \\
Turkish Straits & 0.26 & TUR (17\%), RUS (8\%), CHN (8\%), USA (4\%), DEU (4\%) \\
Danish Straits & 0.24 & RUS (12\%), SWE (9\%), USA (8\%), CHN (8\%), DEU (7\%) \\
Strait of Malacca & 0.17 & CHN (27\%), OtherGulf (7\%), JPN (7\%), IND (5\%), KOR (5\%) \\
Taiwan--Luzon Corridor & 0.14 & CHN (37\%), TWN (9\%), JPN (7\%), USA (6\%), KOR (5\%) \\
Korea Strait & 0.05 & CHN (24\%), JPN (19\%), RUS (18\%), KOR (17\%), USA (3\%) \\
Strait of Dover & 0.01 & USA (18\%), RUS (11\%), POL (7\%), SWE (7\%), OtherRestOfWorld (4\%) \\
Kerch Strait & 0.01 & UKR (24\%), ITA (8\%), ESP (7\%), USA (7\%), DEU (6\%) \\
\bottomrule
\end{tabular}
% <<< end 04_outputs/tex_tables/tab_share_world_loss_K12.tex
\par\smallskip
\footnotesize\textit{Note:} the world loss is
$\sum_c (\mathrm{GDP}_c / \text{world GDP})\cdot \lambda_c(\cset{}_k)$ at the
baseline calibration $(\tau,\rho,\varrho)=(0.3,-1,-1)$, in \% of world
GDP. Magnitudes are unbuffered steady-state losses, before
strategic-reserve drawdown and pipeline-bypass substitution
(Section~\ref{subsec:calibration}). Parenthetical entries are each
economy's share of the world loss, in \%.
\end{table}

\begin{figure}[H]
\centering
\includegraphics[width=0.92\linewidth]{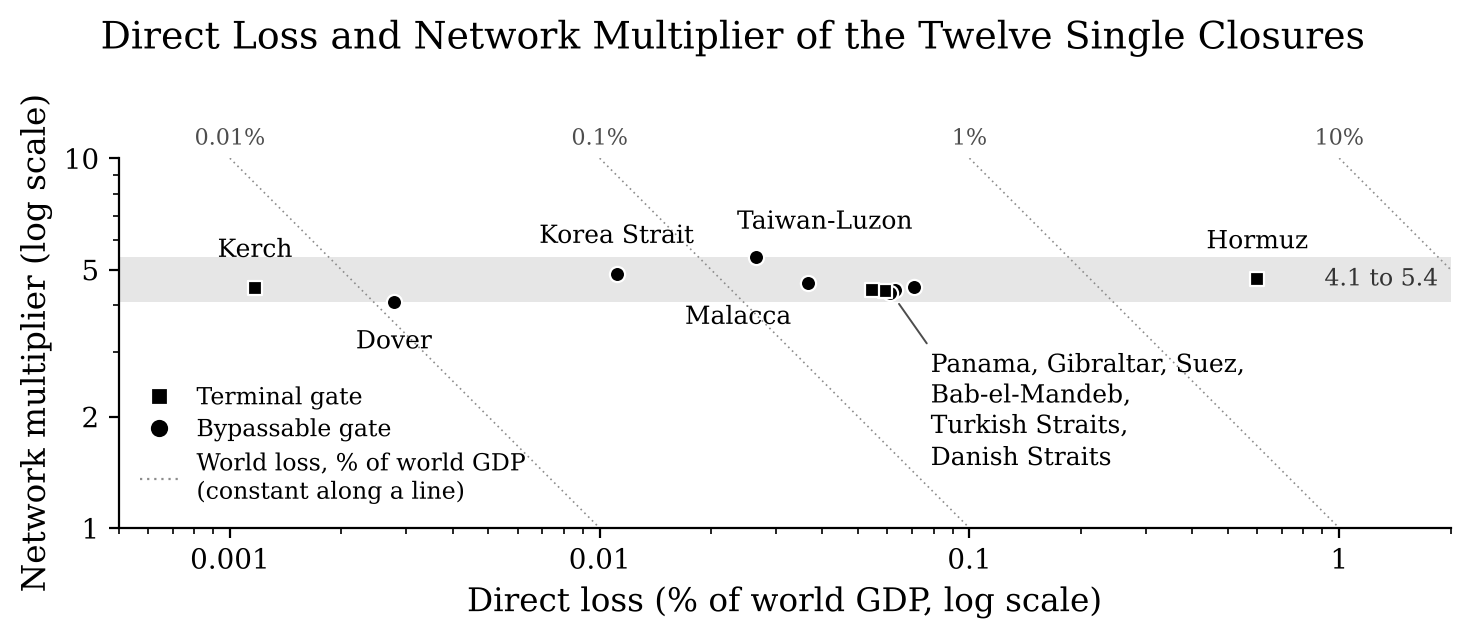}
\caption{The direct loss \((1-\eta_W)\,L_W\), in \% of world GDP, and the network multiplier \(\nu_W=1/(1-\eta_W)\) of the twelve single-chokepoint closures, on logarithmic axes. The world loss, the product of the two, is constant along each dotted line. The gray band spans the network multipliers of the twelve closures. Squares are terminal gates and circles bypassable gates. Suez and Bab-el-Mandeb coincide.}
\label{fig:network_multiplier}
\end{figure}

What sets these differences between the gates is the direct loss of each closure, not the network. Network propagation carries between $76$ and $82\%$ of the world loss of each of the twelve single closures, and input shortfalls account for between $66$ and $79\%$ of it (Online Appendix Table~\ref{tab:decomposition_world}). Since the network carries much the same share at every gate, the network multiplier \(\nu_W\) of \eqref{eq:network_multiplier}, the world loss per dollar of direct loss, varies little, between \(4.1\) at the Strait of Dover and \(5.4\) at the Taiwan--Luzon Corridor. The direct loss, by contrast, differs by a factor of about five hundred, from \(0.001\%\) of world GDP at Kerch to \(0.60\%\) at Hormuz (Figure~\ref{fig:network_multiplier}). The gates therefore rank by their direct loss exactly as they rank by their world loss, and across the eleven distinct corridors, Suez and Bab-el-Mandeb counting once, the direct loss accounts for \(99\%\) of the variance of the logarithm of the world loss. The gap of nearly nine to one between Hormuz and the Panama Canal, the next gate, is a factor of \(8.5\) in the direct loss and of only \(1.05\) in the network multiplier. Which is why the network sets the scale of what a closure costs, and economic geography sets the differences between the closures.

\subsubsection{The map of each closure}
\begin{figure}[H]
\centering
\includegraphics[width=\linewidth]{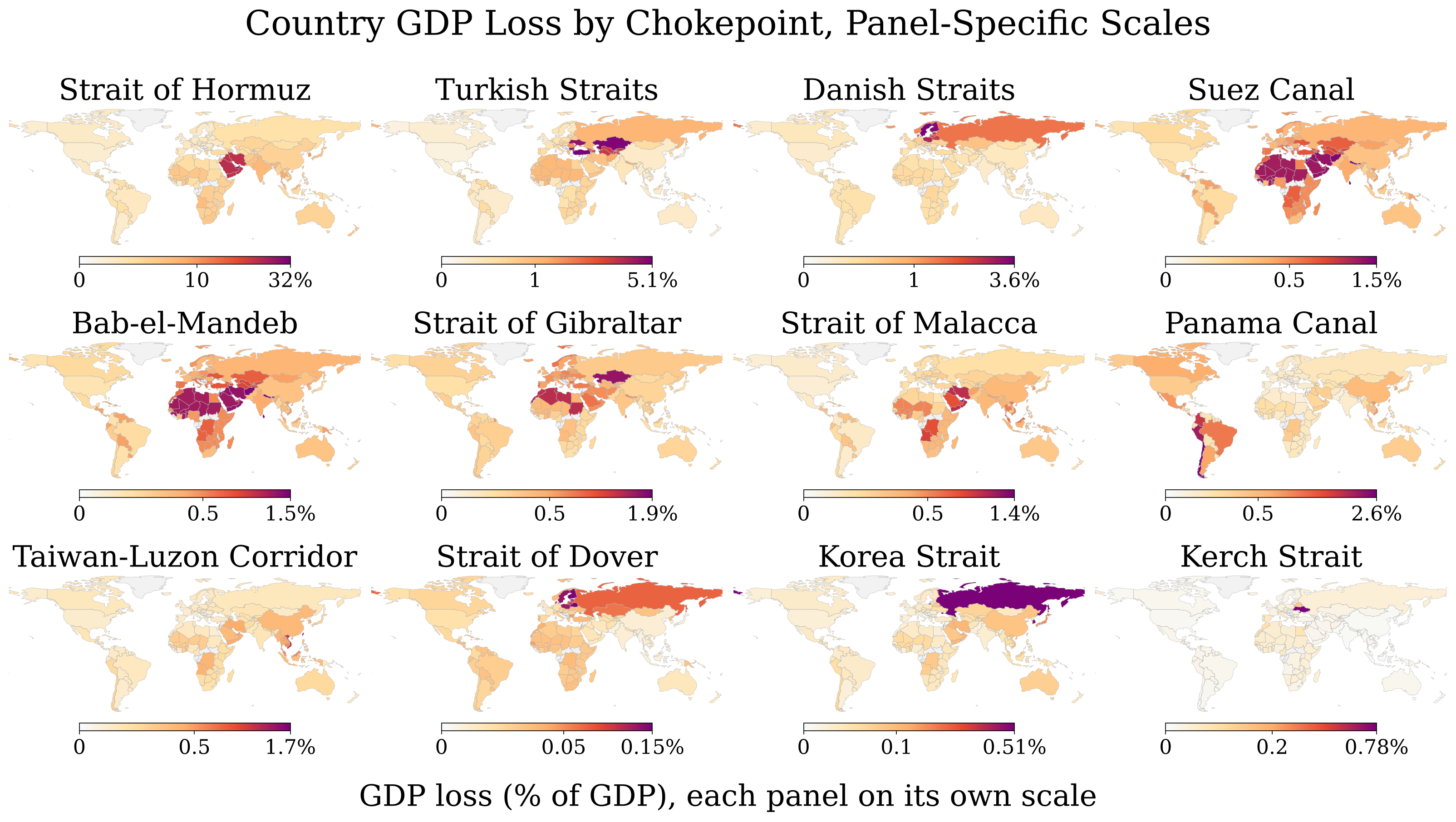}
\caption{Country-level GDP loss $\lambda_c(\cset{}_k)$ under each of the twelve single-chokepoint closures. Each panel has its own square-root color scale, from zero to the largest loss under that closure, in \% of GDP. The scale tops out at $32\%$ in the Hormuz panel and at $0.15\%$ in the Dover panel. ROW-block aggregates are expanded onto the territories of Online Appendix Table~\ref{tab:row_blocks}. Gray territories are outside the extended ICIO.}
\label{fig:heatmap_perpanel}
\end{figure}

Figure~\ref{fig:heatmap_perpanel} paints $\lambda_c(\cset{}_k)$ on a world
map for each closure $k$. Each of the twelve panels has a color scale of its own, drawn beneath it, which runs from zero to the largest loss of any economy under that closure. A color therefore compares economies within one panel, not across panels.

The geography of exposure differs closure by closure, and the
country-level losses painted in Figure~\ref{fig:heatmap_perpanel} fall
into regional patterns. The Strait of Hormuz panel shows a sharp Persian Gulf concentration, with the United Arab Emirates at
$32\%$, and Saudi Arabia and the OtherGulf block, which is chiefly Iran, Iraq, Kuwait, Qatar, Oman, and Bahrain, at $23\%$ each.\footnote{The block also holds Lebanon, Palestine, Syria, and Yemen. The regional data do not separate them from Iraq, so the routing sends their trade through the Strait of Hormuz with that of the Gulf (Online Appendix~\ref{sec:data}).} A second tier runs through the Asian buyers of Gulf energy,
Korea at about $10\%$, Thailand and Taiwan at about $8\%$, India at about $6\%$, and Japan at about $5.5\%$, whose refineries and chemical plants
draw most of their crude oil and gas through the strait. The Turkish Straits
panel concentrates in the Caspian and Black Sea rim, led by Turkey and Kazakhstan
at about $5\%$, with Bulgaria and Ukraine at about $4.5\%$ and the Central Asian block at about $3.3\%$. The Danish Straits panel paints the Baltic littoral, with Sweden and Finland at about $3.5\%$, Estonia and Poland at just under
$3\%$, and Lithuania at about $2\%$. The Suez and Bab-el-Mandeb panels are identical, since the routing
sends the same Asia--Europe flows through both gates and their incidence
rows coincide. Their common pattern is broad and shallow, reaching Jordan, South Asia, the Gulf exporters, and North and West Africa, with no cell above about $1.5\%$. The routing at the resolution of regions overstates the share of West Africa (Section~\ref{sec:step1_def}).
The Panama Canal panel, finally, produces a diffuse pattern, with the
Pacific facade of Latin America (Chile, Peru) as its named-economy cases.

What generates this geographic pattern of losses is the distinction between terminal and bypassable gates. Each closure's map has a depth, how much the worst-hit
economies lose, and a breadth, how many economies register a visible
loss. The depth of a map is the number at the upper end of its color scale, and its breadth is how far the color spreads beyond the worst-hit economies. A terminal gate carries $e^{(k)}_{cc'}=0$ for every country pair whose trade transits it, since
no detour salvages any part of the trade behind it, so all of that trade is disrupted, save the residual transit share $\delta_r$. Its panel is deep, and the deepest losses on the map fall on the rim of the enclosed basin. At the Turkish Straits and the Danish Straits the median economy loses about $0.2\%$ of GDP,
a small fraction of what the rim loses, and at Kerch it loses next to nothing, since the buyers at the far end re-source within sector and escape. 

A bypassable gate with a short detour sits at the other extreme. A large share of the world's long-haul routes touches the Taiwan--Luzon Corridor or Malacca, but the detour around either adds only a few hundred to a thousand nautical miles to voyages an order of magnitude longer. With reroutabilities near one, even a fully exposed flow loses only a small share of its value, so each panel is shallow. The median economy loses about $0.1\%$ of GDP at Malacca and half of that at the Taiwan--Luzon Corridor. The deepest
cell on the shallow panels is Taiwan's $1.7\%$ on the Taiwan--Luzon
Corridor. The corridor runs along Taiwan's own
coast, so nearly everything Taiwan ships must cross it. For Taiwan's trade beyond East Asia the east-of-Taiwan detour adds little to a long voyage, so its reroutability is near one. Taiwan's trade with mainland China is the exception, since the detour is longer than the crossing itself. The reroutability of the pair is $0.4$ (Section~\ref{sec:step1_def}), so a closure puts three fifths of the seaborne cross-Strait trade beyond reach. That trade is what deepens Taiwan's cell. The intermediate gates, Suez, Bab-el-Mandeb, and Gibraltar on the Cape detour and Panama on Cape Horn, sit between the extremes, with $e$ between $\nicefrac{1}{3}$ and $\nicefrac{4}{5}$. Their panels paint the corridor itself at moderate depth.

Depth and breadth are set by different pieces of geography, and neither
implies the other. Depth follows the detour distance $\Delta_k$ through
the reroutability rule \eqref{eq:reroutability_rule}, and breadth follows the number of economies whose
least-cost routes cross the gate. Kerch, a terminal gate on a very small basin, sits in the bottom band of
Table~\ref{tab:share_world_loss} by world loss, yet Ukraine loses about $0.8\%$ of GDP to it
and no other economy much more than a tenth of a percent, a deep, narrow map attached to
the smallest world loss of the twelve. The size of a closure's world loss and the
shape of its map are therefore different properties. In fact, the shape of what a closure costs can largely be read off the detour map before the model is solved. But the size of what it costs scales with the trade mass at stake, and only the solved model delivers it.

The detour map alone cannot explain the whole of that shape. Our model's post-closure equilibrium carries within it the re-matching of displaced trade and the complementarity of inputs, neither of which a map of routes records. The off-corridor world escapes through within-sector re-matching, but distant, exposed buyers are captured through the complementarity of their inputs. And it is this capture that generates the broadest map of all, at Hormuz. Hormuz is a terminal gate like Kerch, with the deepest rim of the twelve, but the cargo behind it is crude oil and gas, which its buyers cannot replace within the sector. The same capture is how Japan, an economy nowhere near the Gulf, comes to carry a visible cell on the Hormuz panel despite selling almost nothing through the strait. The capture also travels upstream, to economies that hardly trade through the strait. Brunei sells and buys through it intermediate inputs worth less than $1\%$ of its GDP, as do $58$ of the other $86$ economies. Yet Brunei loses $16\%$ of its GDP, $99\%$ of it through the network, since the Japanese, Indian, and Korean refineries and the Japanese utilities that buy most of its crude oil and gas are among the buyers that the closure leaves short of Gulf supply. Under the Hormuz closure the median economy takes $95\%$ of its loss through the network (Online Appendix~\ref{app:propccdf}).

\subsubsection{Exposure economy by economy}
\begin{figure}[H]
\centering
\includegraphics[width=0.85\linewidth]{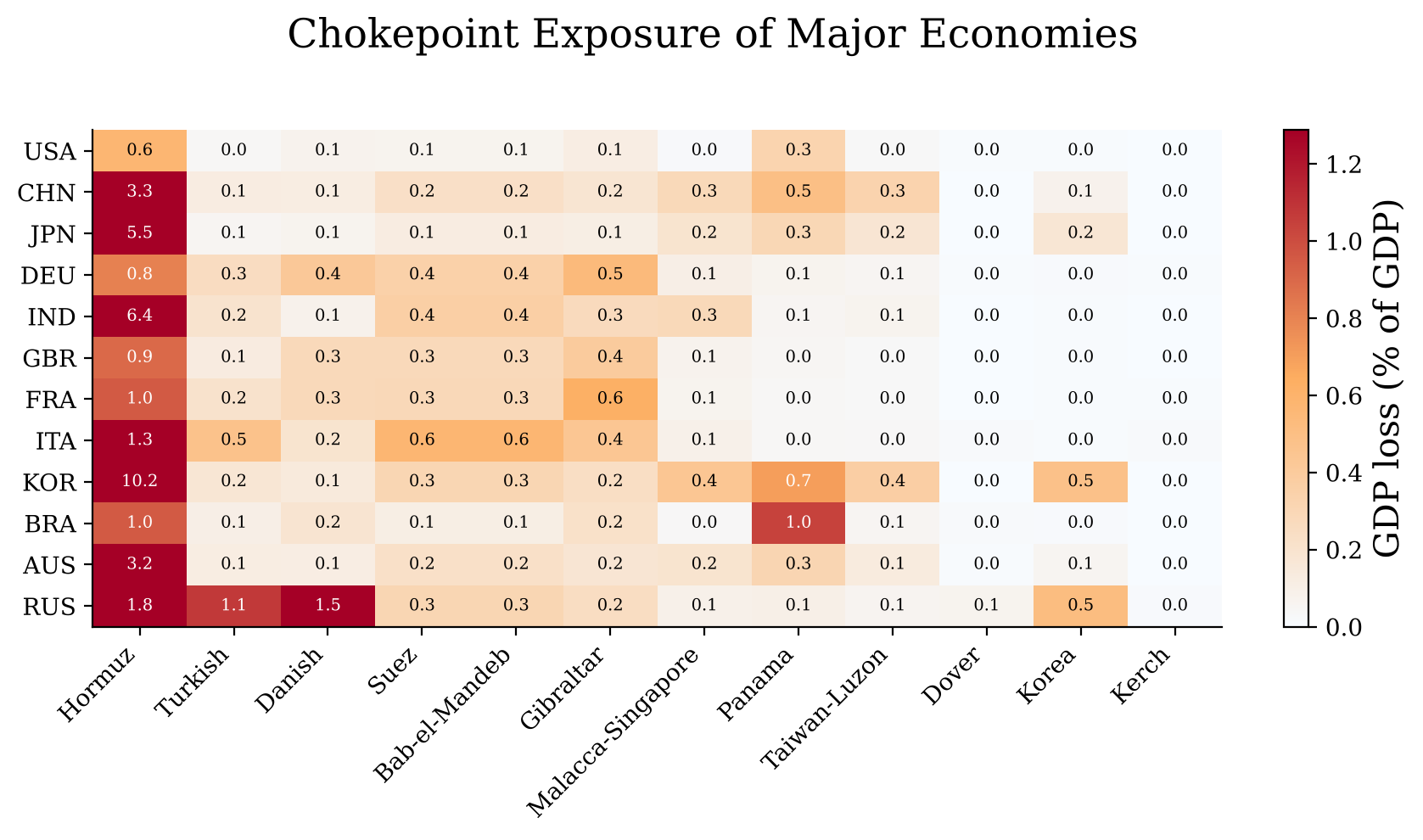}
\caption{Cross-chokepoint exposure profile of twelve major economies: the GDP loss $\lambda_c$ of each economy under each single closure, in percent of GDP.}
\label{fig:distribution_majors}
\end{figure}

Figure~\ref{fig:distribution_majors} shows the same geography one economy at a time, for twelve major economies. The panel divides into the Hormuz column and everything outside it.
Along the Hormuz column the energy importers India, Korea, and Japan
carry their largest cells, at $6.4$, $10.2$, and $5.5\%$ respectively,
and Australia and China carry theirs there too, at $3.2$ and $3.3\%$.
The deepest single cells outside that column belong to Russia, in the
Danish-Straits column ($1.5\%$) and the Turkish-Straits column ($1.1\%$), and to Brazil in the Panama column
($1.0\%$). The United States, Germany, France, the United Kingdom, and
Brazil have low profiles across all chokepoints, with no
cell above about $1\%$.

At the insulated end of the range, where no closure of any of the
twelve gates costs an economy much, the United States is the least exposed
major economy. Its largest single-chokepoint loss
is about $0.6\%$ of GDP (the Strait of Hormuz), with the Panama Canal behind at about $0.3\%$, and every other closure costs it materially
less. The United States is a large and sectorally diversified economy, self-sufficient in primary energy since the shale revolution, and it trades largely with its continental neighbors and across the open North Atlantic. It carries little intermediate-input mass through the twelve gates as a result, and re-matching at $\tau=0.3$ recovers most of what it does carry.

Western Europe and Russia sit in the middle. About half of a Western
European economy's foreign purchases of seaborne goods cross no gate at all
in the routing, since they come from within its own maritime region,
Norway's energy among them, or across the open North Atlantic. Germany, the United Kingdom, France, and Switzerland stay below $1\%$
under every closure, each peaking on Hormuz. The Netherlands, Spain, and
Italy lose between $1.1$ and $1.3\%$ there. The gates on Europe's own coast cost its large economies less,
Gibraltar taking between $0.5$ and $0.6\%$ each from France, the Netherlands, Spain, and Germany,
because much of what they trade with one another and with their neighbors moves over land.
Russia stays in low single digits on every closure, at about $1.8\%$ on Hormuz and $1.5\%$ on the Danish
Straits, which its Baltic-port exports at Primorsk and Ust-Luga cross. Its seaborne exports are
divided among the Baltic, the Black Sea, and the Pacific, so no single closure reaches more than about a third of them.

Africa sits toward the exposed end, below the Gulf and the Asian buyers of Gulf energy and about level with the Black Sea and Baltic rims of the Turkish and Danish panels. Under a closed Hormuz, Angola loses about $5.9\%$ of GDP, South Africa about $4.4\%$, the Democratic Republic of the Congo about $3.8\%$, and the African ROW aggregates between $2.2$ and $4.5\%$. These losses come through the buyers of African commodities, which are among the economies that a Hormuz closure contracts. North Africa spreads its exposure across gates. Tunisia, for instance, loses about $1.9\%$ on Gibraltar and a little over $1\%$ on each of Hormuz and the Turkish Straits. Across the twelve closures the African economies take an unusually large part of their loss directly, between a third and a half of it on the gates that their own trade crosses, because their trade is narrow on both sides (Online Appendix~\ref{app:africa}).

% ================================================================
\section{The Structure of Exposure: Tails, Asymmetry, and Channels}
\label{sec:structure}
% ================================================================

Section~\ref{sec:results} reported what each closure costs the world and where on the map its losses fall. We now study how those losses are distributed across economies and country-sectors, and which mechanisms of the model shape that distribution. The losses are heavy-tailed. The Gulf exporters lose between a fifth and a third of their GDP behind a closed Hormuz, and ten of the $4{,}299$ country-sectors lose more than half of their output (Sections~\ref{subsec:heavy_tail} to~\ref{subsec:mean_tail}). At every gate the exporting end of the severed corridor loses more, in proportion to its GDP, than the importing end. The asymmetry arises from the orders that a closure takes directly from the exporters, which network propagation, weighing more on the importers, offsets only in part (Section~\ref{subsec:asymmetry}). 

% ----------------------------------------------------------------
\subsection{The heavy tail of cross-country exposure}
\label{subsec:heavy_tail}
% ----------------------------------------------------------------

The cross-country distribution of chokepoint exposure is heavy-tailed.
We measure the tail through a country's worst-case exposure
$\lambda_c^{\max} := \max_{k\in\mathcal K}\lambda_c(\cset{}_k)$, the
proportional GDP loss it would face under the worst single-chokepoint
closure. The top five values, the United Arab Emirates ($32\%$), Saudi Arabia ($23\%$),
OtherGulf ($23\%$), Brunei ($16\%$), and Korea ($10\%$), account for a third of the cross-country sum
$\sum_c \lambda_c^{\max}$, and the top ten for about $45\%$. Four of the top five are named economies rather than constructed
ROW blocks, so this concentration of exposure is not an artifact of pooling many small countries into one row.

\begin{figure}[H]
\centering
\includegraphics[width=\linewidth]{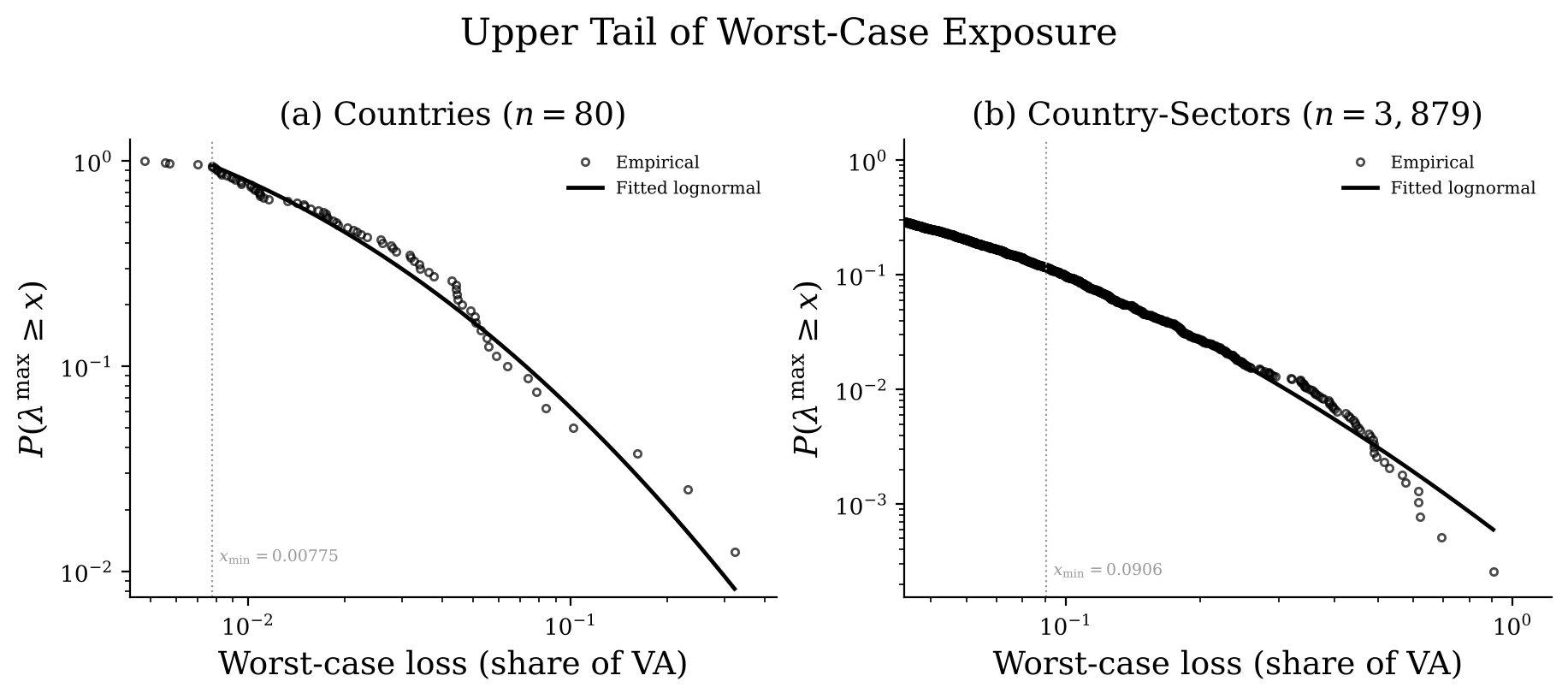}
\caption{Upper tail of worst-case exposure at the baseline calibration, for the $80$ named economies (a) and their $3{,}879$ loss-carrying country-sectors (b). Points are the empirical CCDF, the fraction of units with loss above $x$. The line is a lognormal fitted above the Kolmogorov--Smirnov-minimizing threshold, marked dotted. The power-law and exponential fits reported in the text are not drawn.}
\label{fig:tail_fits}
\end{figure}

We examine the distribution of losses at two levels of aggregation: across
the $80$ named economies and across their $3{,}879$ loss-carrying
country-sectors, those with positive benchmark activity and a positive
worst-case loss.\footnote{The seven
constructed Other* blocks are excluded from the figure and the estimates.
Each pools many small economies, so its position in the size distribution
is an artifact of our aggregation rather than a measurement of any
economy's exposure.} At the country-sector level we use each country-sector's worst-case exposure
$\lambda_i^{\max}:=\max_{k\in\mathcal K}\lambda_i(\cset{}_k)$, where
$\lambda_i(\cset{}_k):=\Delta s^{(\cset{}_k)}_{i}/s_i$ is its proportional activity
decline and, since country-sector $i$'s GDP is $\gamma_i s_i$, also its
proportional GDP loss. Figure~\ref{fig:tail_fits} plots the empirical complementary cumulative distribution function (CCDF) at each level against a lognormal fitted above a
threshold chosen to minimize the Kolmogorov--Smirnov distance, in the manner
of \citet{clauset2009powerlaw}.

The lognormal describes both distributions well. At the country level that threshold is a loss of $0.8\%$ of GDP, below which lie only four of the $80$ named economies. The lognormal is therefore fitted to $76$ economies and describes nearly the whole of the cross-country distribution, not its upper end alone. At the country-sector level the threshold is a loss of $9\%$ of activity, which $452$ of the $3{,}879$ loss-carrying country-sectors exceed. Worst-case exposure carries a coefficient of variation of $1.39$, against the $1$ that an exponential would deliver, and a likelihood-ratio test rejects the exponential at the country-sector level but not on the $76$ observations of the country level. A power law fitted above the same thresholds is rejected by its goodness-of-fit test at the country level but not at the country-sector level, where the likelihood ratio nonetheless favors the lognormal (Online Appendix~\ref{app:zipfdetail}). Whatever parametric family a reader prefers, aggregate
chokepoint exposure is concentrated in a small number of catastrophically exposed economies,
with a second tier behind them, most of it the Asian buyers of Gulf energy.

% ----------------------------------------------------------------
\subsection{The fall in output across country-sectors, gate by gate}
\label{subsec:sector_declines}
% ----------------------------------------------------------------

\begin{figure}[H]
\centering
\includegraphics[width=\linewidth]{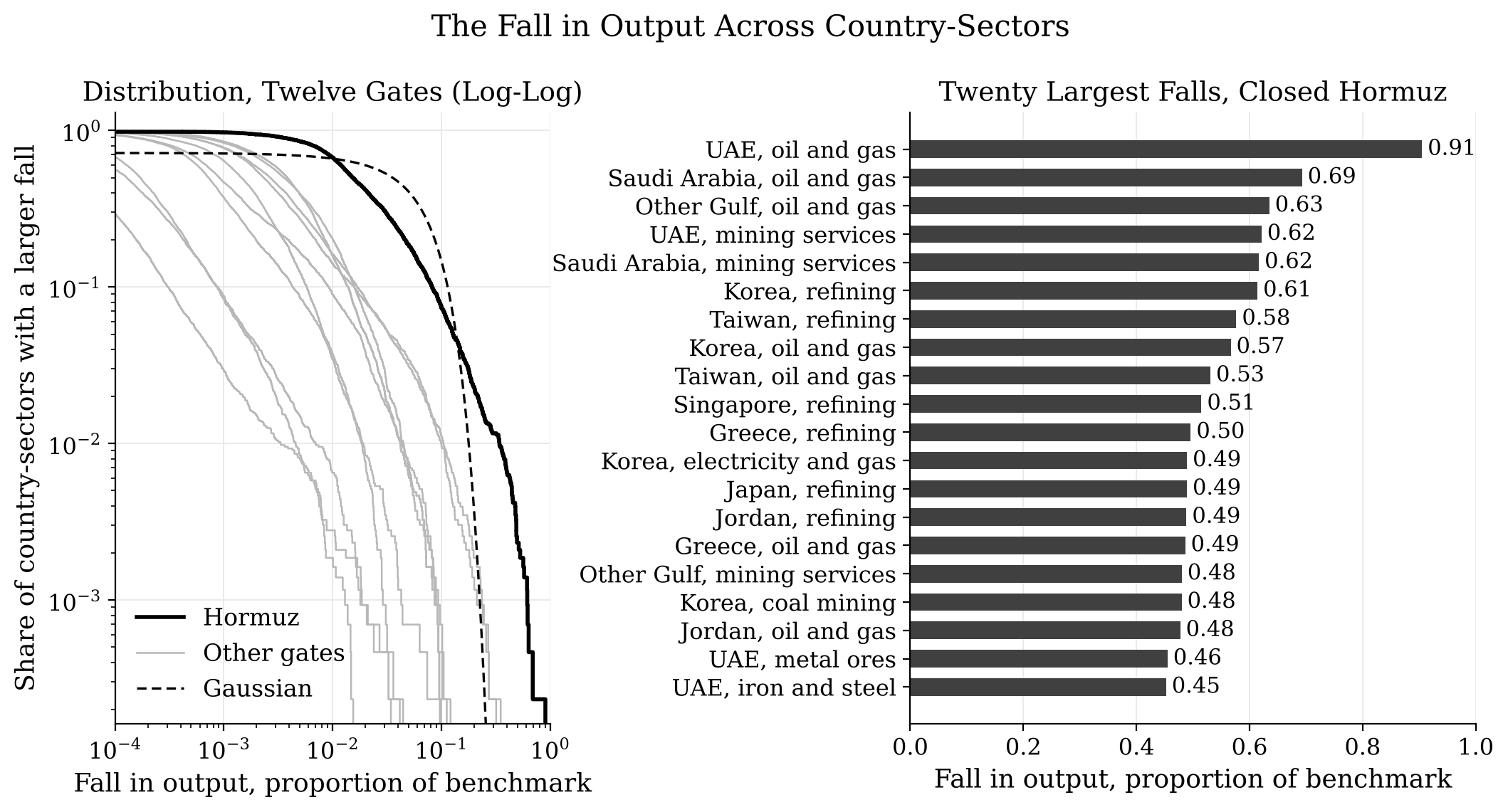}
\caption{The fall in output across country-sectors at the baseline calibration. The left panel gives the CCDF of the proportional activity decline across the $4{,}299$ country-sectors with positive benchmark activity, unweighted, under a closed Hormuz (black) and under each of the eleven other gates (grey), on logarithmic axes with declines and shares as proportions. The dashed curve is the Gaussian with the mean and the standard deviation of the declines under Hormuz. It begins at $0.72$ because $28\%$ of it lies below zero. The right panel names the twenty largest declines under a closed Hormuz.}
\label{fig:sector_declines}
\end{figure}

The worst-case exposure of a country-sector takes the largest of its losses over the twelve gates. We now look instead at the losses under each gate separately, for their distribution under one gate may be very different from that under another. Figure~\ref{fig:sector_declines} plots the CCDF of the proportional activity decline \(\lambda_i(\cset{}_k)\) across country-sectors under each closure. Its right panel names the twenty largest declines under a closed Hormuz. The GDP loss of a country is the GDP-weighted mean of the declines of its country-sectors, but the mean hides how unevenly they are spread. Under a closed Hormuz two thirds of the world's country-sectors lose more than \(1\%\) of their output, the median loses \(1.5\%\), and nearly one in five loses more than \(5\%\). Output falls by more than a tenth at \(321\) of the \(4{,}299\) country-sectors, which make \(5\%\) of world output, and by more than half at ten, among them the oil and gas extraction of the three Gulf economies and the refineries of Korea, Taiwan, and Singapore. The twenty largest declines all lie at the two ends of the severed corridor: at the Gulf sellers and at the refining, power, and energy sectors of their buyers in East Asia and the Mediterranean. No other gate has a tail of that length. Under the Turkish Straits, which come next, \(48\) country-sectors lose more than a tenth of their output and none loses half, and under the Panama Canal only two lose more than a tenth. 

On the logarithmic axes of the left panel of Figure~\ref{fig:sector_declines}, the CCDF of Hormuz lies above that of every other gate over its whole length. Which is to say that, at every threshold from \(0.01\%\) of benchmark output to the largest decline under Hormuz, more country-sectors exceed the threshold under a closed Hormuz than under any other closure. The dashed curve of that panel is the Gaussian distribution with the mean and the standard deviation of the declines under Hormuz, which are \(3.6\%\) and \(6.1\%\) of benchmark output. Between declines of \(1\%\) and \(14\%\) of output the Gaussian curve lies above the curve of Hormuz, but beyond \(14\%\) it falls away below it. Under that Gaussian a single country-sector would lose more than a quarter of its output and none would lose half, but under a closed Hormuz \(63\) lose more than a quarter and ten lose more than half. The largest decline lies fourteen standard deviations above the mean. Which is to say that the tail of the declines is far longer than their mean and their standard deviation would suggest, so a summary of the declines by those two numbers would hide the country-sectors that a closure harms most. Online Appendix Table~\ref{tab:sector_declines_top} lists the five largest declines under every closure, and the replication package holds the decline of every country-sector under every scenario.

% ----------------------------------------------------------------
\subsection{Mean, tail, and concentration}
\label{subsec:mean_tail}
% ----------------------------------------------------------------

Two economies that lose the same amount on average can face very different worst cases, depending on whether their exposure is concentrated on one gate or spread across many. We therefore compare the worst-case exposure $\lambda_c^{\max}$ of each economy, its largest loss over the twelve single closures, with its mean exposure $\bar\lambda_c$. The mean is the average loss over the eleven distinct corridors, which we collect in \(\mathcal K^{\mathrm d}\), rather than over the twelve gates, because Suez and Bab-el-Mandeb carry identical incidence, so closing either severs the same flows and costs every economy the same.\footnote{We keep the two as separate gates because they are separate passages that have closed separately. The Suez Canal closed from 1967 to 1975, and again for six days in 2021, while Bab-el-Mandeb stayed open, and the Houthi campaign of 2023--24 drove the container lines from Bab-el-Mandeb while the canal stayed open. Their incidence coincides only because the routing works between the fourteen maritime regions of Online Appendix Table~\ref{tab:regions}, and none of these regions lies on the Red Sea between the two gates. With ports resolved, the trade of the Red Sea coast would cross one gate and not the other, as Saudi crude loaded at Yanbu crosses Bab-el-Mandeb on its way to Asia and never reaches Suez.} An average over the twelve gates would count that corridor twice, whereas the worst case is the same whether the corridor is counted once or twice.

Mean and worst-case exposure rank the economies much alike. Their Spearman rank correlation across the $87$ economies is $0.94$, and eight of the ten economies with the highest mean exposure are among the ten with the highest worst-case exposure. What separates economies is how their exposure is distributed across the gates: whether it is concentrated on one gate or spread across the distinct corridors. The ratio $\lambda_c^{\max}/\bar\lambda_c$ measures that concentration. A
country with a high mean but a low tail-to-mean ratio is broadly exposed, with many chokepoints
hurting it moderately. A country with a high tail and a low mean has, instead, a single Achilles
heel, one chokepoint that hurts it catastrophically while the others barely register.

\begin{figure}[H]
\centering
\includegraphics[width=0.72\linewidth]{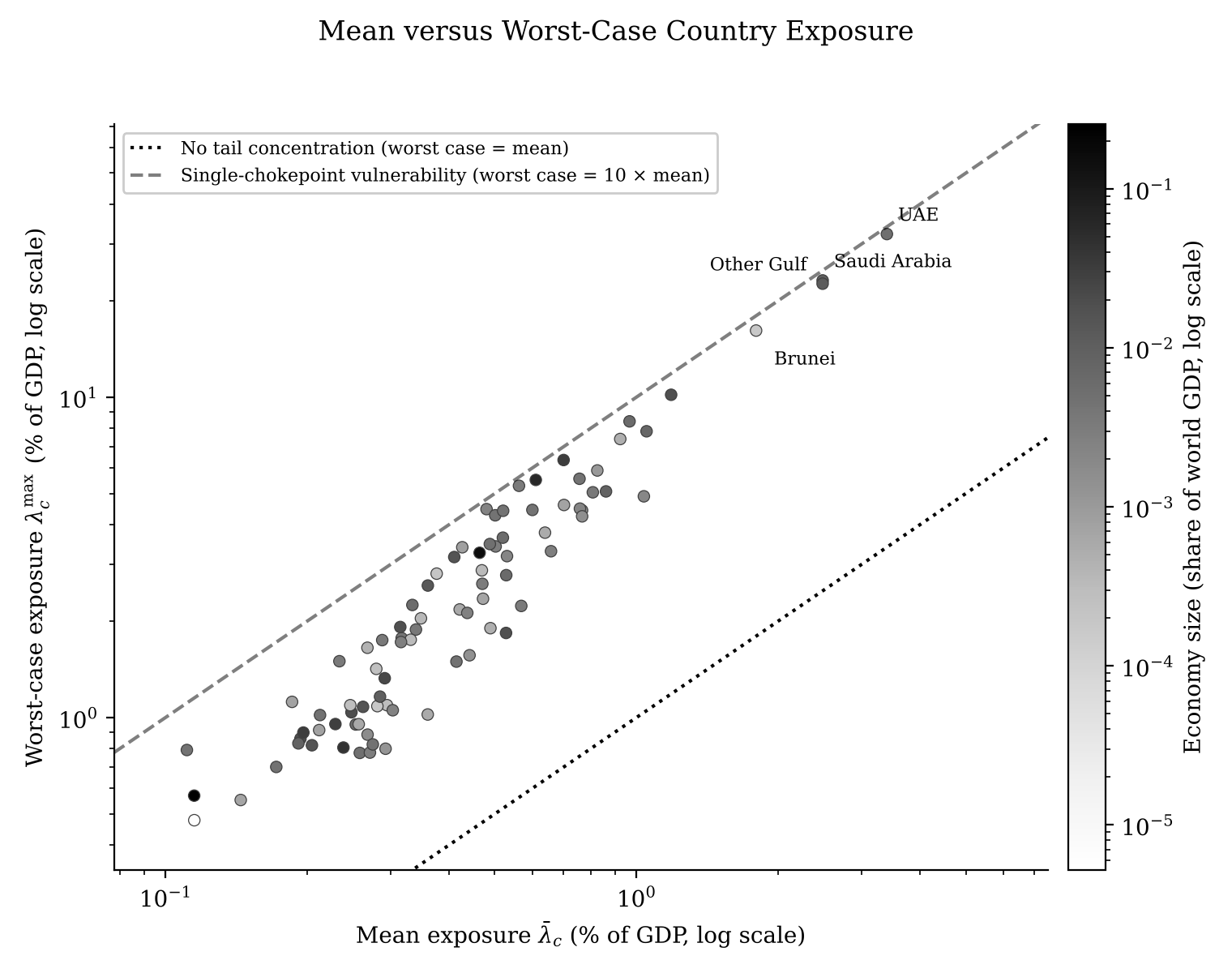}
\caption{Mean exposure \(\bar\lambda_c\) against worst-case exposure \(\lambda_c^{\max}\), on logarithmic axes, each point shaded by the economy's share of world GDP. The dashed ray $\lambda^{\max}=10\,\bar\lambda$ marks worst cases an order of magnitude above the mean, and the single-corridor envelope $\lambda^{\max}=11\,\bar\lambda$ lies just above it. Only the four worst-hit economies are labeled.}
\label{fig:mean_vs_tail}
\end{figure}

Figure~\ref{fig:mean_vs_tail} scatters
$(\bar\lambda_c, \lambda_c^{\max})$ across the extended economy set, with
each economy shaded by its share of world GDP. The diagonal
$\lambda^{\max} = \bar\lambda$ identifies uniform exposure, but no country sits on it. The upper ray $\lambda^{\max} = 10\,\bar\lambda$ marks worst
cases an order of magnitude above the mean, the signature of a country
whose chokepoint risk is concentrated on one gate. No economy reaches the ray. The shading makes the
composition of its neighborhood plain, since the economies nearest to the ray are
those whose chokepoint risk is Hormuz and little else, large economies among them. India, Japan, and Korea sit
just below the ray beside New Zealand, Pakistan, and the Philippines, with worst cases between eight and a half and
nine and a half times their means. China lies further below, at seven times, while the United
States and the large European economies sit well below, at between three and five times. The United Arab
Emirates, Saudi Arabia, and OtherGulf occupy the high end of the neighborhood, each losing
substantially from one closure and little from any other.
Table~\ref{tab:tail_mean_ratio} ranks economies by the ratio
$\lambda_c^{\max}/\bar\lambda_c$. The ten highest ratios run from $8.6$
to $9.5$ against a ceiling of eleven, the value the ratio takes when a
country's entire exposure flows through a single one of the eleven
distinct corridors. The other ROW-block aggregates
(OtherNorthAfrica, OtherSouthAsia, OtherCentralAsia, OtherWestAfrica) sit further from the ray, because their exposure is spread across several gates rather than concentrated on one. A normalized Herfindahl index of the corridor shares gives much the same
ordering as the ratio, with the United Arab Emirates and Pakistan at its head. Online Appendix~\ref{app:tailcomp} reports it economy by economy.

\begin{table}[H]
\centering
\caption{Top-10 economies by the ratio of worst-case to mean exposure,
$\lambda_c^{\max}/\bar\lambda_c$. The mean is taken over the eleven
distinct corridors, so the ratio runs
from $1$, uniform exposure, to $11$, exposure through a single corridor.
The binding chokepoint is the gate attaining $\lambda_c^{\max}$.}
\label{tab:tail_mean_ratio}
\small
% >>> inlined from 04_outputs/tex_tables/tab_tail_mean_ratio.tex
\begin{tabular}{cllccc}
\toprule
Rank & Economy & Binding chokepoint & $\lambda_c^{\max}$ (\%) & $\bar\lambda_c$ (\%) & $\lambda_c^{\max}/\bar\lambda_c$ \\
\midrule
1 & ARE & Strait of Hormuz & 32.3 & 3.4 & 9.5 \\
2 & PAK & Strait of Hormuz & 5.3 & 0.6 & 9.4 \\
3 & SAU & Strait of Hormuz & 23.1 & 2.5 & 9.3 \\
4 & NZL & Strait of Hormuz & 4.5 & 0.5 & 9.3 \\
5 & OtherGulf & Strait of Hormuz & 22.6 & 2.5 & 9.1 \\
6 & IND & Strait of Hormuz & 6.4 & 0.7 & 9.1 \\
7 & JPN & Strait of Hormuz & 5.5 & 0.6 & 9.0 \\
8 & BRN & Strait of Hormuz & 16.1 & 1.8 & 9.0 \\
9 & THA & Strait of Hormuz & 8.4 & 1.0 & 8.7 \\
10 & KOR & Strait of Hormuz & 10.2 & 1.2 & 8.6 \\
\bottomrule
\end{tabular}
% <<< end 04_outputs/tex_tables/tab_tail_mean_ratio.tex
\end{table}

% ----------------------------------------------------------------
\subsection{Exporter--importer asymmetry}
\label{subsec:asymmetry}
% ----------------------------------------------------------------

Recall from Section~\ref{subsec:activity_decomposition} that a severed trade is two losses at once, since the buyer at one end loses an input and the seller at the other loses an outlet. The rationed exchange offers buyers and sellers the identical formal remedy. Buyers re-match displaced demand onto surviving same-sector
suppliers, and sellers re-match displaced output onto surviving same-sector
buyers. The common unrecovered share \(\tau\) strips the sellers' capacities and the buyers' budgets alike, and the residual transit enters the buying side and the selling side symmetrically.\footnote{The symmetry is exact. Transposing the disrupted transactions matrix, which puts every seller in the place of a buyer and every buyer in the place of a seller, and exchanging the capacities with the budgets returns the transpose of the exchange (Online Appendix~\ref{app:proof_model}).} Put simply, the fixed-price equilibrium re-matches displaced trade without favoring the exporter or the importer, since a corridor whose exporting and importing sides were mirror images in the benchmark
data would see the same share of its displaced trade re-placed on either side.

Table~\ref{tab:asymmetry_by_chokepoint} sets the losses of the exporting and the importing side against each other at all twelve closures. Every economy both sells and buys through a gate, so for each closure we assign it to a side by its net position. An economy whose disrupted sales through the gate exceed its disrupted purchases belongs to the exporting side of the corridor, and an economy whose disrupted purchases are the larger belongs to the importing side. We then compare the mean proportional GDP loss of the two sides, \(\lambda_X\) for the exporting side and \(\lambda_M\) for the importing side, each mean weighted by disrupted trade mass.\footnote{Weighting by trade mass rather than GDP makes the statistic the typical proportional loss of the trade at stake on each side.} The
exporting side loses more than the importing side at every one of the twelve gates. The gap in levels is by far the largest at Hormuz. Its exporters lose about $25\%$ of GDP against the importers' roughly $5.3\%$, a ratio of about $4.6\times$. The ratios fan out from Kerch, at about $35\times$, to the Danish Straits, at $1.3\times$.\footnote{The comparison is made on trade in intermediate inputs. On the rough count of Online Appendix~\ref{app:final_goods}, which adds the final goods that each closure stops as sales lost to the producers that shipped them, the exporting side still loses more at every one of the twelve gates. The ratio is wider at Hormuz, Panama, the Danish Straits, Dover, and Kerch and narrower at the other seven. Online Appendix~\ref{app:asym_three} names the leading net exporters and net importers at Hormuz, the Panama Canal, and the Turkish Straits, and shows that the comparison does not turn on which economies are named.} Kerch's importers re-source away from the Azov almost freely while its severed exporters cannot, whereas the exporting and importing sides of the Danish Straits come closest to symmetry in the benchmark trade data.

\begin{table}[H]
\centering
\caption{Exporter--importer asymmetry under each single closure, at the baseline calibration. Each side's loss is the mean proportional GDP loss of its economies, weighted by disrupted trade mass, in \% of GDP. The ratio is the exporting side's loss over the importing side's, computed on unrounded values. Suez and Bab-el-Mandeb share a row, since their incidence rows are identical.}
\label{tab:asymmetry_by_chokepoint}
\small
\begin{tabular}{l r r r}
\toprule
Chokepoint & Exporting side (\%) & Importing side (\%) & Ratio \\
\midrule
Kerch Strait & 0.78 & 0.02 & 35.4 \\
Strait of Dover & 0.08 & 0.02 & 5.2 \\
Strait of Hormuz & 24.68 & 5.31 & 4.6 \\
Strait of Malacca & 0.95 & 0.30 & 3.2 \\
Taiwan--Luzon Corridor & 0.89 & 0.37 & 2.4 \\
Korea Strait & 0.51 & 0.21 & 2.4 \\
Panama Canal & 1.16 & 0.48 & 2.4 \\
Strait of Gibraltar & 0.81 & 0.50 & 1.6 \\
Suez Canal and Bab-el-Mandeb & 0.74 & 0.48 & 1.5 \\
Turkish Straits & 2.33 & 1.60 & 1.5 \\
Danish Straits & 1.49 & 1.11 & 1.3 \\
\bottomrule
\end{tabular}
\end{table}

There are two distinct forces at work, whose sum determines whether exporters lose more than importers. The first is network propagation, which hurts both ends of a severed trade but hurts the importers more. A seller that loses a sale buys less from its own suppliers, who then sell less in turn. These upstream producers lose output but not productivity, because a producer that is only asked for less still receives all the inputs it orders and fills every order that remains. A buyer that loses an input loses productivity as well. Its inputs are complements in production, so whatever share of its input mass the exchange cannot reconstitute degrades its whole bundle of intermediate inputs through the CES factor \(\kappa^{(\omega)}_j\). That loss of productivity then travels downstream to the buyer's customers as well as upstream to its suppliers (Section~\ref{subsec:activity_decomposition}).

The second force is economic geography, which by itself favors neither the exporter nor the importer. The exchange writes off the same displaced trade at both ends of a corridor, so whichever end has fewer alternatives to the corridor bears the larger loss in proportion to its size. Empirically, economic geography makes the exporters the more dependent end.\footnote{The dependence belongs to the route geography and not to the economies. \citet{bhatt2026_sanctions} run the same short-run model under trade sanctions, where the sales of the Gulf exporters spread across many buyers. There Saudi Arabia holds the favorable side of $65$ of its $79$ bilateral relationships, and its mean vulnerability of $0.15\%$ lies in the lower half of the twenty-five largest economies. Behind a closed Hormuz, by contrast, Saudi Arabia loses about $23\%$ of its GDP and the United Arab Emirates about a third. Diversified across counterparties and concentrated behind one strait are the same economy seen from two sides, and only the route geography tells them apart.} A commodity exporter often ships nearly all of its exports through a single corridor, whereas its buyers draw on suppliers spread across many routes, so the trade that a closure severs is a large part of what the exporter sells and a small part of what the importer buys. How far the second force outweighs the first is settled corridor by corridor by the bipartite geometry of the trade through each closed gate.

We find that all of the exporter--importer asymmetry arises from the direct loss of the exporters, most of it the orders that the closure takes from them. In fact, network propagation works in the opposite direction, i.e., dollar for dollar of direct loss it costs the importing side more than the exporting side, as the first force predicts. It does so at every one of the twelve gates, and so narrows the asymmetry everywhere without reversing it anywhere.

To see precisely how economic geography overwhelms network propagation to generate the exporter--importer asymmetry, begin by defining the direct loss and the network multiplier of each side. The direct loss of an economy is the part of its loss that the closure inflicts on its own producers, through the orders they lose and the inputs they go without. It comes to \((1-\eta_c)\lambda_c\) of the economy's GDP, with \(\eta_c\) the network-propagation share of Section~\ref{subsec:country_world_losses}. Write \(\lambda^{\mathrm{dir}}_X\) and \(\lambda^{\mathrm{dir}}_M\) for the means of these direct losses over the economies of the exporting and of the importing side, weighted as in Table~\ref{tab:asymmetry_by_chokepoint}. The direct loss of a side reflects economic geography, since it is large when the closure cuts a large part of what the side sells or buys and the exchange can re-place little of it. The network multiplier of a side is its loss per dollar of its direct loss, \(\nu_X:=\lambda_X/\lambda^{\mathrm{dir}}_X\) for the exporting side and \(\nu_M:=\lambda_M/\lambda^{\mathrm{dir}}_M\) for the importing side. It reflects network propagation, since it counts what the network adds, through the buyers and suppliers of the side's producers, to each dollar that the closure takes from the side directly. A network multiplier of one would mean that the side loses only what the closure takes from it directly. The ratio of the losses of the two sides is then the product of the ratio of their direct losses and the ratio of their network multipliers,
\begin{equation}
\frac{\lambda_X}{\lambda_M}
=\frac{\lambda^{\mathrm{dir}}_X}{\lambda^{\mathrm{dir}}_M}\cdot\frac{\nu_X}{\nu_M}.
\label{eq:asymmetry_split}
\end{equation}

Hormuz makes the decomposition concrete. Its importing side loses \(0.56\%\) of GDP directly, and the network multiplies this by \(9.5\), to \(5.3\%\). Its exporting side loses \(15.9\%\) directly, \(28.5\) times as much, and the network multiplies this by only \(1.6\), to \(24.7\%\). The ratio of network multipliers of \(0.16\) thus pulls the ratio of direct losses of \(28.5\) down to a ratio of losses of \(4.6\). The same holds at every gate (Table~\ref{tab:asymmetry_decomposition}). The ratio of network multipliers lies between \(0.16\), at Hormuz, and \(0.91\), at Panama, and so is below one everywhere, whereas the ratio of direct losses lies between \(1.6\), at the Danish Straits, and \(54\), at Kerch. What sets a gate's position on the scale of Table~\ref{tab:asymmetry_by_chokepoint} is therefore how unequally the surviving markets fall on its exporting and its importing side. This gradient is a property of commodity-concentrated export geography rather than of any single passage.

\begin{table}[H]
\centering
\caption{Decomposition of the exporter--importer asymmetry under each single closure, at the baseline calibration. The ratio of direct losses and the ratio of network multipliers are the two factors of \eqref{eq:asymmetry_split}, whose product is the ratio of Table~\ref{tab:asymmetry_by_chokepoint}. The input-shortfall share is the part of each side's loss that input shortfalls account for, in percent, with each economy weighted as in Table~\ref{tab:asymmetry_by_chokepoint}. Suez and Bab-el-Mandeb share a row.}
\label{tab:asymmetry_decomposition}
\small
\begin{tabular}{l r r r r}
\toprule
 & \multicolumn{2}{c}{Ratio of} & \multicolumn{2}{c}{Input-shortfall share (\%)} \\
\cmidrule(lr){2-3}\cmidrule(lr){4-5}
Chokepoint & direct losses & network multipliers & exporting side & importing side \\
\midrule
Kerch Strait & 54.3 & 0.65 & 0 & 94 \\
Strait of Dover & 6.4 & 0.81 & 43 & 81 \\
Strait of Hormuz & 28.5 & 0.16 & 18 & 98 \\
Strait of Malacca & 8.0 & 0.40 & 37 & 84 \\
Taiwan--Luzon Corridor & 4.0 & 0.60 & 52 & 80 \\
Korea Strait & 5.5 & 0.44 & 30 & 82 \\
Panama Canal & 2.6 & 0.91 & 52 & 75 \\
Strait of Gibraltar & 2.7 & 0.60 & 42 & 79 \\
Suez Canal and Bab-el-Mandeb & 2.6 & 0.58 & 52 & 81 \\
Turkish Straits & 2.5 & 0.59 & 29 & 77 \\
Danish Straits & 1.6 & 0.83 & 45 & 76 \\
\bottomrule
\end{tabular}
\end{table}

The decomposition by source of Section~\ref{subsec:activity_decomposition} gives the asymmetry its economic content. The importing side owes more of its loss to input shortfalls than the exporting side at every gate, between \(75\) and \(98\%\) of it against between \(0\) and \(52\%\) (Table~\ref{tab:asymmetry_decomposition}). Under a closed Hormuz, lost orders account for \(81\) to \(84\%\) of the loss of the three Gulf exporters, and input shortfalls for \(98\) to \(99\%\) of the loss of China, Japan, India, and Korea, which sell almost nothing through the strait (Online Appendix Table~\ref{tab:decomposition_hormuz}). The exporter thus loses demand that the closure physically cuts off, whereas the importer loses productivity in an input bundle that it can no longer assemble in benchmark proportions.

What a Gulf exporter ``redirects'' when its only sea exit is shut invites
the objection that there is nowhere for the output to go. The exchange never invents a trading partner, since its support is contained in the benchmark's, so a severed supplier can lean only on
relationships already present there. Behind a closed Hormuz only two outlets survive: the exporter's
own domestic sectors and its intra-Gulf customers. Every other Gulf sale transits the strait, but the two outlets are far too small to absorb those sales.
Gulf buyers absorb about $37\%$ of Gulf energy output in the benchmark. The two outlets take just $24\%$ of the UAE's
displaced sales and $15\%$ of Saudi Arabia's, so the exchange writes the rest off. What cushions the exporter is accordingly not the re-placement but the sales that never crossed the strait. The Strait of Hormuz
accounts for a minority of the UAE's benchmark intermediate sales, roughly
two-fifths. Most of the rest, $49\%$ of the total, is domestic and crosses no water. Intra-Gulf trade and the sales of sectors whose output does not travel by sea carry what is left.\footnote{The cushion is not an artifact of the residual transit through the closed passage. At the baseline \(\delta=0.10\), with only a tenth of a severed flow still transiting, the UAE's Hormuz loss is about \(32\%\) of GDP, severe but well short of the near-total loss that losing its sole export outlet would imply. Lowering the residual transit share to \(0.01\) raises the loss only to about \(35\%\), and raising it to \(0.20\) lowers it to about \(29\%\) (Online Appendix~\ref{app:robustness_delta}), so the residual transit is the smaller part of the cushion.}

The asymmetry is also why the head of the heavy tail of
Section~\ref{subsec:heavy_tail} is populated by exporters, since the two
findings describe the same geography. A chokepoint's incidence falls on the exporters behind it: the Gulf behind Hormuz, the Caspian producers behind the Turkish Straits, and the Baltic exporters behind the Danish Straits. At most gates the deepest losses therefore land on a small set of
exporting economies, and the union over the twelve gates is still a
small set against the $87$ economies of the master matrix. The buyers at the far end of a corridor re-source within sector, so at most gates they escape with small losses, which is the
asymmetry seen from the importing side. Hormuz is the gate at which the buyers do not escape. The crude oil and gas that leave the Gulf are more than the surviving producers can replace, so the crude oil and gas block carries a placement deficit (Section~\ref{subsec:rerouting_matrix}). The Asian buyers of Gulf energy form a second tier of the tail, at $5$ to $10\%$ of GDP against the $23$ to $32\%$ of the Gulf exporters.
The concentration of exports behind one gate is compounded by a
concentration of imports along the same routes. An exporter whose output
leaves by one lane usually buys along the same few routes, so the
economies deepest on the exporting side of a gate are deep on its importing side too. Their worst cases therefore concentrate on a single corridor rather than spreading across
several. Online Appendix Table~\ref{tab:herfindahl} lists the economies whose exposure is most concentrated, each with its binding gate.

% ================================================================
\section{Joint Chokepoint Closures}
\label{sec:joint}
% ================================================================

Geopolitical risk rarely arrives one chokepoint at a time, since a regional
escalation or a sharp rise in war-risk premia typically impairs several corridors at once. The single-closure exercises of Section~\ref{sec:results} are silent on such episodes. This section
 closes chokepoints jointly, in three scenarios drawn from
recognizable international-political fault lines: a Middle East
escalation, an East Asian security crisis, and a Russia--Europe
confrontation. The question a joint closure poses is how the sum of the losses from its single closures, each gate closed alone, compares with the loss from closing the same gates together. The answer turns on the economic geography of the closed gates: on whether they sit
in series or in parallel on the routes they cut, on how much cargo two of them would both take, and on whether gates that take different cargo carry the same inputs to common buyers.
Section~\ref{subsec:superadditivity} sets out the argument, Section~\ref{subsec:joint_scenarios} builds the three
scenarios, and Section~\ref{subsec:joint_empirics} reports the
results. Online Appendix~\ref{app:robustness} examines the robustness of these results and of the findings of Sections~\ref{sec:results} and~\ref{sec:structure}.

% ----------------------------------------------------------------
\subsection{A joint closure and the sum of its single closures}
\label{subsec:superadditivity}
% ----------------------------------------------------------------

Define the `additivity index' of a closure set \(\cset{}\) as
\begin{equation}
S(\cset{})
\;:=\;
\frac{L_W(\cset{})}
{\displaystyle\sum_{k\in \cset{}} L_W(\cset{}_k)}.
\label{eq:superadd_index}
\end{equation}
The joint closure costs less than the sum of its single closures when \(S(\cset{})<1\), in which case we call the closure set `sub-additive', and more when \(S(\cset{})>1\). The index is defined only
where its denominator is strictly positive. Each constituent loss is
nonnegative, so what that asks is that some single closure of
\(\cset{}\) already cost the world something.

Whether the index lies below one or above it depends first on what a single closure leaves in place. A closed gate does not take from the world all the trade that crossed it. A shipment that has a detour takes it and loses only the lost share of its country pair, the residual transit share \(\delta\) of what is disrupted still crosses the passage, and the rationed exchange re-matches all of the displaced trade but the unrecovered share \(\tau\). What a second closure adds to the first therefore depends on where its gate stands in relation to the first gate.

Once its flows are routed, the trade table says where one gate stands in relation to another. The routing of Section~\ref{sec:mapping} sends each flow of the table along a route that crosses a known set of gates. A buyer whose supplies arrive through two passages has divided its purchases between the suppliers behind each, so the two passages carry different flows of the table, each a part of what the buyer buys. A flow that must clear two gates one after the other is one and the same flow at both. The table thus records, flow by flow, whether two gates carry the same trade or different trade, and whether the different trade is the same input of a common buyer. The table does not record how much of a flow a longer voyage would save, which the reroutability profile \(\mathbf E\) takes from sailing distances.

Two gates can sit in series or in parallel. We say that two gates sit `in series' when they belong to one detour class, i.e., when a shipment which must clear both to reach
its destination avoids both by a single detour. Saudi crude bound for a German refinery, for
instance, must clear Bab-el-Mandeb and then Suez, but it avoids both by rounding the Cape. Closing
either gate already forces the detour, so closing the other takes nothing more from the shipment.
The disrupted fraction of such a flow under the joint closure is the larger of its two
single-closure fractions and not their sum, which is what the maximum within a detour class in
\eqref{eq:partial_choke_disruption} records. Gates in series duplicate one another, so their joint loss falls below the sum of their single losses.

We say that two gates sit `in parallel' when they belong to different detour classes. The pure
case is that of two gates which stop different shipments. The same German
refinery buys Caspian crude through the Turkish Straits and Russian crude through the Danish
Straits, and none of these cargoes crosses both gates. Nothing is duplicated in this case, so the joint closure takes from the refinery the sum of what the two single closures take. In reality no set of gates is purely in series or purely in parallel. The nearest to the pure series case in our routing are Bab-el-Mandeb and Suez, which it charges with the same flows. Between the two pure cases lie gates of different classes that some shipments cross in turn, as Gulf crude bound for Europe crosses Hormuz and then the Red Sea gates. Such gates duplicate one another on the shipments that they share and on no others, and on a shared shipment only in the part that each of them would take, since only what survives the gates of one class is exposed to those of the next.

The cost of two closures need not add up the way the trade that they displace does, because inputs are complements in production. Online Appendix Proposition~\ref{prop:joint_pair} takes up the question in the simplest configuration that poses it, i.e., two gates that reach one buyer and no other, where the buyer sells to final users alone and so passes its shortfall on to no other producer. The proposition shows that a joint closure can cost the world less than the sum of its single closures, and that it can cost more.

What decides the sign of the difference is the geography that the trade table records. Two gates in series that stop the same shipments cost together what either costs alone, whatever the parameters of the model. Two gates in parallel that take different inputs from the buyer cost less together than the sum of what each costs alone. The reason is that the scarcest input holds back a bundle of complementary inputs, so a buyer which is already short of one input loses less by going short of another. A car plant that is already short of chips cannot use all the steel it has, so losing part of its steel as well costs it little more. Two gates in parallel that take the same input from the buyer, each from suppliers of its own, cost more together than the sum of what each costs alone, unless that input makes up most of the buyer's cost. Two shortfalls of one input are one larger shortfall of that input. Because the input is a complement to the buyer's other inputs, each further unit that goes missing costs the buyer more than the last. The German refinery that has lost its Caspian crude runs on the crude it has left, so the loss of its Russian crude as well costs it more than the same loss would cost a fully supplied refinery.\footnote{The statement on different inputs holds whenever the outer nest is no more complementary than the inner one, \(\varrho\ge\rho\), as at the baseline. The statement on a shared input holds at the baseline curvatures when the input makes up less than half of the buyer's cost. At other curvatures it holds for a small enough unrecovered share when the input makes up no more than a third of the buyer's input bundle.} 

But how far a joint loss departs from the sum of the single losses is what the parameters of the model govern. And the reason is rather simple. A closure leaves a buyer short by the unrecovered share \(\tau\) of the purchases that it displaces, since the exchange re-matches the rest. The loss from a single closure is proportional to this shortfall. The interaction of two closures, the amount by which their joint loss departs from the sum of their single losses, is proportional to the product of their two shortfalls. Relative to the sum of the single losses, the departure is therefore of the size of one shortfall, i.e., proportional to the unrecovered share \(\tau\) times the share of the buyer's purchases that a closure displaces. Which is why two closures in parallel that each displace a small share of a buyer's purchases are nearly additive, whichever way they interact. The substitution parameter \(\rho\) scales the interaction itself, since the more complementary the inputs are, the more two shortfalls of one input compound and the more two shortfalls of different inputs offset one another. The outer-nest curvature \(\varrho\) adds to the compounding, because the cushion that primary inputs provide against a shortfall of inputs thins as the input bundle shrinks. The write-off of the unrecovered share works the other way, since the write-off is a share of what a buyer purchases, and a buyer that the other closure has already made smaller purchases less. 

Online Appendix Proposition~\ref{prop:joint_pair} settles the sign of the difference for a single buyer that sells to final users alone. A closure of the world's gates differs from that case in two ways. It reaches many buyers through many inputs at once, each of them passing its shortfall on to its own customers. Two closures can also shut, between them, every route over which the trade that either displaces alone would have been re-placed, so this trade finds no partner and is lost whatever the unrecovered share. The index \(S(\cset{})\) of the solved model takes all of this into account, so we measure it for three closure sets that a single political escalation could bring about.

% ----------------------------------------------------------------
\subsection{The three scenarios}
\label{subsec:joint_scenarios}
% ----------------------------------------------------------------

The three fault lines are recognizable because the states that would
close the gates have said so on the record, and in the Middle East have
already acted (Online Appendix~\ref{app:scenariopol}). For each scenario we name the gates and the cargo they carry, say why a single escalation would close them together, and note which of them sit in series, in one detour class of \eqref{eq:detour_classes}, and which in parallel.

\subsubsection{Middle East regional escalation}
The first scenario closes the Strait of Hormuz together with Bab-el-Mandeb and the Suez Canal,
\(\cset{}^{ME}=\{\text{Hormuz},\text{Bab-el-Mandeb},\text{Suez}\}\). Hormuz is the sole sea exit of Persian Gulf energy exports, so closing it removes the bulk of the crude, refined products, and petrochemical feedstocks that supply Asian and European refining and chemical chains. Bab-el-Mandeb and Suez govern the Asia--Europe corridor through the Red Sea, so closing them severs the route on which Asian electronic components and machinery parts reach European producers and European chemicals reach Asian producers. A regional escalation can join the Hormuz and Red Sea theaters through proxy conflict or retaliation against shipping, as the war of 2026 has begun to do (Online Appendix~\ref{app:scenariopol}). Hormuz is terminal, whereas Bab-el-Mandeb and Suez are bypassable by the voyage around the Cape. The two Red Sea gates share that detour and so lie in one detour class, in series, while Hormuz forms a class of its own. Consider a shipment from Asia to Europe, which crosses both gates. The detour around the Cape salvages about \(73\%\) of it, so closing either gate disrupts about \(27\%\) of it. Closing both gates disrupts no more, since the shipment still needs only the one detour. This is what the maximum within a detour class in \eqref{eq:partial_choke_disruption} records, so the joint closure disrupts \(1-\min(e_{\text{Bab-el-Mandeb}},e_{\text{Suez}})\approx 0.27\) of the shipment. Treating the two gates as independent would instead let the shipment keep \(73\%\) at each gate, or \(0.73\times0.73\approx 53\%\) in all, and so put the disrupted share at about \(47\%\), counting the Cape detour twice.

\subsubsection{East Asian security crisis}
The second scenario closes three East Asian chokepoints at once,
\(\cset{}^{EA}=\{\text{Taiwan--Luzon},\allowbreak\,\text{Malacca},\allowbreak\,
\text{Korea}\}\). The Taiwan--Luzon Corridor is East Asia's Pacific gateway, carrying the region's long-haul exchanges with the rest of the world, the transpacific deep-water route to the Americas and the outbound legs of the Asia--Europe trades. The Strait of Malacca carries the long-distance Asia--Indian Ocean flow, including the Gulf--Asia oil corridor and the Asia--Europe container route. The Korea Strait carries Japan--Korea bilateral trade together with the Russian Far-East maritime exit toward its East Asian buyers. The three gates belong to one strategic and economic system, so a Taiwan contingency would draw the wider Indo-Pacific into the closure (Online Appendix~\ref{app:scenariopol}). All three gates are bypassable: the corridor by the passage east of Taiwan, Malacca by the Lombok and Sunda Straits, and the Korea Strait by the route around Japan. Each lies in a detour class of its own, so the three sit in parallel and each enters \eqref{eq:partial_choke_disruption} through a factor of its own.

\subsubsection{Russia--Europe maritime escalation}
The third scenario closes three chokepoints on Russia's European sea routes, \(\cset{}^{RE}=\{\text{Turkish Straits},\text{Danish Straits},
\text{Kerch Strait}\}\). The Turkish Straits connect the Black Sea to the Mediterranean, and the Danish Straits connect the Baltic Sea to the North Sea. The Kerch Strait is the sole maritime exit of the Sea of Azov, the outlet of the ports of southern Russia (Rostov-on-Don, Taganrog) and eastern Ukraine (Mariupol, Berdyansk).\footnote{The routing has no port resolution, so it charges the gate with Ukraine's seaborne exports to Europe, the Mediterranean, Africa, and the Americas but with no Russian flow. These exports are chiefly grain, basic metals, ores, and food products. The routing therefore overstates what the gate takes from Ukraine, whose Black Sea ports it cannot tell from its Azov ports.} Unlike the Middle East and East Asian scenarios, which disrupt container shipping the world over, the Russia--Europe scenario is a Europe-centered shock to energy, food, fertilizers, metals, and industrial inputs. Since the three gates govern Russia's western and southern seaborne trade, a Russia--Europe escalation can impair access at all three at once (Online Appendix~\ref{app:scenariopol}). Each would close through a mechanism of its own: the Montreux Convention's wartime regime at the Turkish Straits, sabotage and naval inspection in the Baltic approaches, and Russian physical control at Kerch. All three gates are terminal, so no detour salvages any part of the benchmark flows that transit them. As in the East Asian scenario, each lies in a detour class of its own and the three sit in parallel.

% ----------------------------------------------------------------
\subsection{Empirical results for the three joint scenarios}
\label{subsec:joint_empirics}
% ----------------------------------------------------------------

Each of the three joint closures costs the world less than the sum of its single closures, and it spreads that loss around the world as its single closures do. Table~\ref{tab:joint_world_loss} reports the world losses at the baseline calibration \((\tau,\rho,\varrho)=(0.3,-1,-1)\) of Section~\ref{subsec:calibration}. The Middle East scenario costs \(3.00\%\) of world GDP against \(3.35\%\) for the sum of its three single closures, a tenth less, so its index is \(S(\cset{}^{ME})=0.89\). The East Asian scenario costs \(0.36\%\) against \(0.37\%\), and the Russia--Europe scenario \(0.49\%\) against \(0.50\%\), so their indices, \(S(\cset{}^{EA})=0.98\) and \(S(\cset{}^{RE})=0.97\), put both within \(3\%\) of that sum. The same holds economy by economy. Every one of the \(87\) economies loses less under each joint closure than the sum of its losses under the single closures, save Angola under the Russia--Europe scenario, which loses \(0.3\%\) more.

\begin{table}[H]
\centering
\caption{World GDP loss \(L_W(\cset{})\) of each joint closure against the sum of the world losses of its single closures, at the baseline calibration \((\tau,\rho,\varrho)=(0.3,-1,-1)\). The index \(S(\cset{})\) of \eqref{eq:superadd_index} is computed on unrounded losses, and Table~\ref{tab:share_world_loss} gives the losses of the single closures.}
\label{tab:joint_world_loss}
\small
\begin{tabular}{l l r r r}
\toprule
Scenario & Gates & $L_W(\cset{})$ & $\sum_{k\in\cset{}} L_W(\cset{}_k)$ & $S(\cset{})$ \\
& & (\% GDP) & (\% GDP) & \\
\midrule
$\cset{}^{ME}$ & Hormuz, Suez, Bab-el-Mandeb & 3.00 & 3.35 & 0.89 \\
$\cset{}^{EA}$ & Malacca, Taiwan--Luzon, Korea & 0.36 & 0.37 & 0.98 \\
$\cset{}^{RE}$ & Turkish, Danish, Kerch & 0.49 & 0.50 & 0.97 \\
\bottomrule
\end{tabular}
\end{table}

The Middle East scenario falls short of the sum of its single closures chiefly because of its serial pair. Bab-el-Mandeb and Suez carry the same flows, so closing both is closing one, and the sum of the three single closures counts the loss from that one closure twice, \(0.26\%\) of world GDP each time. Set against the losses from Hormuz and a single Red Sea gate, the two corridors that the scenario in fact closes, the joint loss gives an index of \(0.97\). The rest of the shortfall is the Gulf cargo bound for Europe. About a quarter of the trade that crosses Hormuz goes on through the Red Sea, and what the closure of Hormuz has taken from such a shipment the closure of the Red Sea gates cannot take again.

The East Asian and Russia--Europe scenarios are nearly additive because their gates take little of the same trade, for a different reason in each. In the East Asian scenario the Korea Strait shares no shipment with the other two gates. Malacca and the Taiwan--Luzon Corridor share many, since about three quarters of the trade that crosses Malacca goes on through the corridor, but each has a detour that saves most of a shipment. The closure of Malacca takes on average about \(12\%\) of the trade that crosses it and the closure of the corridor about \(7\%\), so the part of a shared shipment that both would take, the product of the two, is small. In the Russia--Europe scenario the gates are terminal, so a shipment that crosses two of them is lost once but counted twice in the sum of the single closures. Such shipments are few. Every shipment through Kerch also clears the Turkish Straits, but Kerch carries little, and only about \(3\%\) of the seaborne trade that crosses the Turkish or the Danish Straits also crosses the other. The trade that the joint closure displaces is accordingly \(99\%\) of the sum of what its single closures displace in the East Asian scenario and \(97\%\) in the Russia--Europe scenario. Each ratio is within half a hundredth of the index of its scenario, although it involves only the routes, the detours, and the trade table and none of the parameters \(\tau\), \(\rho\), and \(\varrho\) (Online Appendix~\ref{app:joint_displaced}).\footnote{The three scenarios are not special. Across the \(66\) pairwise closures of the twelve gates, the index of a pair is, to within \(0.013\), the trade that its joint closure displaces over the sum of what its two single closures displace, unless Hormuz is one of the two gates. A dollar of trade displaced at Hormuz costs the world about three times as much as one displaced at any other gate, which can lift the index of a pair with Hormuz above that ratio, by as much as \(0.036\). Only two pairs cost more than the sum of their two single closures, Hormuz with Panama and Hormuz with the Korea Strait, and by less than \(1\%\) (Online Appendix~\ref{app:pairwise_grid}).}

Nor do the closures of either scenario compound at their buyers, although they take the same input from common buyers over different routes. Crude oil and gas is the largest seller sector at the Turkish and the Danish Straits and among the three largest at Malacca and at the Taiwan--Luzon Corridor, and two shortfalls of one input are one larger shortfall of it (Online Appendix Proposition~\ref{prop:joint_pair}). But compounding is proportional to the share of the same input that two closures take from a buyer, and that share is small. Where one closure of the East Asian scenario displaces part of a buyer's purchases of an input, another closure of the scenario displaces on average less than \(2\%\) of the buyer's purchases of that input, and in the Russia--Europe scenario less than \(3\%\). Stronger complementarity moves both scenarios toward the sum of their single closures without carrying them across. At the harshest corner of the parameter ranges, where both curvatures stand at \(-4\) and the unrecovered share at \(0.5\), the two indices rise only to \(0.99\) and \(0.98\) (Online Appendix~\ref{app:modelrobust}).

\begin{figure}[H]
\centering
\includegraphics[width=0.85\linewidth]{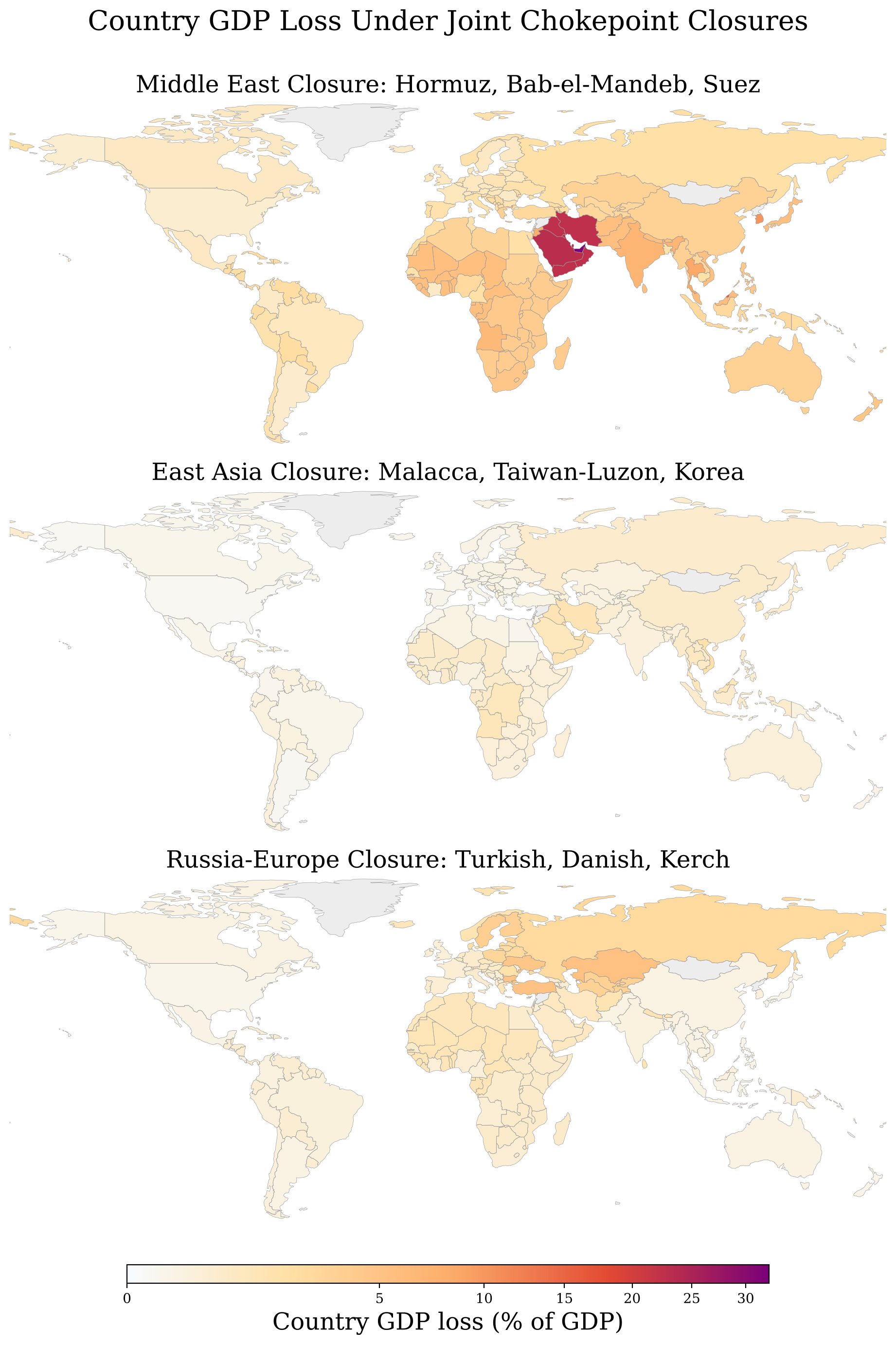}
\caption{Country-level GDP loss \(\lambda_c(\cset{})\) under the three
joint-closure scenarios, at the baseline calibration
\((\tau,\rho,\varrho)=(0.3,-1,-1)\). The color scale is shared
across panels and square-root transformed. Gray territories are outside
the extended ICIO universe, and the rest-of-world aggregates are spread over their territories by the mapping of Online Appendix Table~\ref{tab:row_blocks}.}
\label{fig:joint_choropleth}
\end{figure}

The joint losses fall where the single losses fall. Across the \(87\) economies, the correlation of an economy's loss under a joint closure with the sum of its losses under the single closures is at least \(0.997\) in each scenario. The fifteen economies that lose the most under each joint closure, listed in Table~\ref{tab:joint_country_top15}, are the fifteen that lose the most under its single closures together, with a single exception in the Middle East scenario. Figure~\ref{fig:joint_choropleth} maps the losses of the three scenarios on one color scale.

\begin{table}[H]
\centering
\caption{The fifteen economies that lose the most under each joint closure, with their GDP loss \(\lambda_c(\cset{})\), at the baseline calibration \((\tau,\rho,\varrho)=(0.3,-1,-1)\).}
\label{tab:joint_country_top15}
\small
\begin{tabular}{rlr@{\quad}lr@{\quad}lr}
\toprule
  & \multicolumn{2}{c}{$\cset{}^{ME}$} & \multicolumn{2}{c}{$\cset{}^{EA}$} & \multicolumn{2}{c}{$\cset{}^{RE}$} \\
  \cmidrule(lr){2-3}\cmidrule(lr){4-5}\cmidrule(lr){6-7}
Rank & Country & $\lambda_c$ (\%) & Country & $\lambda_c$ (\%) & Country & $\lambda_c$ (\%) \\
\midrule
1 & ARE & 32.3 & BRN & 2.3 & KAZ & 5.2 \\
2 & SAU & 23.2 & TWN & 2.2 & TUR & 5.2 \\
3 & OtherGulf & 22.7 & ARE & 2.0 & BGR & 4.8 \\
4 & BRN & 16.4 & VNM & 1.8 & UKR & 4.7 \\
5 & KOR & 10.3 & OtherGulf & 1.5 & SWE & 3.7 \\
6 & THA & 8.5 & KOR & 1.3 & FIN & 3.5 \\
7 & TWN & 7.9 & AGO & 1.3 & OtherCentralAsia & 3.4 \\
8 & JOR & 7.8 & MYS & 1.3 & EST & 3.0 \\
9 & IND & 6.9 & COD & 1.2 & POL & 3.0 \\
10 & AGO & 6.5 & SAU & 1.1 & BLR & 2.7 \\
11 & MYS & 5.9 & THA & 1.0 & RUS & 2.5 \\
12 & OtherSouthAsia & 5.6 & KHM & 0.8 & LTU & 2.3 \\
13 & OtherWestAfrica & 5.6 & SGP & 0.8 & LVA & 2.2 \\
14 & PAK & 5.6 & OtherWestAfrica & 0.7 & DNK & 2.0 \\
15 & JPN & 5.5 & MMR & 0.7 & ROU & 1.9 \\
\bottomrule
\end{tabular}
\end{table}

The Middle East panel is the Hormuz panel of Figure~\ref{fig:heatmap_perpanel} with the Red Sea corridor added, and the addition is small where the Hormuz loss is deep. The United Arab Emirates loses \(32.3\%\) of GDP under the joint closure, as under Hormuz alone, since the Gulf cargo that Hormuz has stopped cannot be stopped again at Suez. Korea and Japan buy little through the Red Sea, so their losses barely move, to \(10.3\) and \(5.5\%\) from \(10.2\) and \(5.5\%\) under Hormuz alone. The Red Sea gates add most to the losses of South Asia and of West and North Africa, whose exports and imported inputs both run through the Mediterranean--Suez--Hormuz corridor, and to those of Croatia, Portugal, and Spain. The South Asian and West African aggregates lose \(5.6\%\) each against \(4.2\) and \(4.5\%\) under Hormuz alone.\footnote{The two West African figures rise in part because the routing sends the trade of the Atlantic seaboard with the Mediterranean through the Red Sea (Section~\ref{sec:step1_def}).}

The East Asian panel is the quietest of the three, with most economies at the low end of the scale. The largest losses are those of Brunei (\(2.3\%\) of GDP), Taiwan (\(2.2\%\)), the United Arab Emirates (\(2.0\%\)), and Vietnam (\(1.8\%\)). Taiwan loses \(1.7\%\) under the Taiwan--Luzon Corridor alone, since the closure severs not only its long-haul trade beyond East Asia but also three fifths of its seaborne trade with mainland China (Section~\ref{subsec:master_matrix_results}). The largest part of that trade is the \texttt{C26} electronics shipped to mainland assemblers. The Emirates and Brunei lose energy exports to North-East Asian buyers, the Emirati ones across Malacca and the corridor and Brunei's across the corridor alone, and Angola and the Democratic Republic of the Congo lose \(1.3\) and \(1.2\%\) on the crude, cobalt, and copper that they ship to Chinese buyers. China itself loses only \(0.7\%\). The Lombok and Sunda detours save most of the energy and Asian inputs that it buys across Malacca, and its trade with Europe runs over many overlapping routes. The panel is quiet because of these detours and diversified routes, and says nothing about the stakes of the crisis.\footnote{The detours matter for the result, since re-solving the East Asian
scenario with the Lombok--Sunda and east-of-Taiwan detours closed raises
its world loss many times over, to about \(3.5\%\), with
Vietnam's loss reaching \(19.2\%\) of GDP and Taiwan's \(16.9\%\) (Online
Appendix~\ref{app:robustness_detour}).}

The Russia--Europe panel sets the Turkish and Danish panels of Figure~\ref{fig:heatmap_perpanel} side by side, a tight band along the Caspian, the Black Sea, and the Baltic. The Turkish Straits shut the main sea outlet of Caspian and Ukrainian grain, fertilizer, metals, and oil, and the Danish Straits the principal sea outlet of Russian energy. Along each rim the joint closure adds little to the closure of the rim's own strait. Kazakhstan and Turkey lose \(5.2\%\) each against \(4.9\) and \(5.1\%\) under the Turkish Straits alone, and Finland and Poland lose \(3.5\) and \(3.0\%\) against \(3.4\) and \(2.8\%\) under the Danish Straits alone. The joint closure adds most to the loss of Russia, which loses \(2.5\%\) of GDP, a little less than the sum of the \(1.5\%\) that it loses under the Danish Straits alone and the \(1.1\%\) under the Turkish Straits alone, since the scenario severs both parts of its maritime export base. Russia still loses less than the Caspian and Black Sea rim economies, because about half of its seaborne exports, those to East Asia and to its neighbors on the Baltic and the Black Sea, cross none of the three gates. Buyers further west lose far less, Germany and the Netherlands \(0.7\%\) and the United Kingdom \(0.4\%\), since their alternative supplies partly bypass the closed gates: Norwegian gas, North American LNG, Middle Eastern crude through Suez, and Mediterranean pipelines. The economies most damaged by a Russia--Europe maritime escalation are thus not Russia or its principal adversaries, but the transit and rim economies that depend on either side of the Russian maritime trade for their own intermediate inputs.

\section{Concluding Remarks}
\label{sec:conclusion}
What does it cost the world economy when one of its critical maritime
chokepoints closes? The simple answer is the value of tonnage passing through
the strait. That answer, however, is far from the truth. Much of maritime
trade involves intermediate inputs, and their disruption can generate
significant contraction in downstream production. We have, in this paper,
developed a model to assess the size of this contraction. It treats the world
input--output table as a production network of country-sectors. That table already carries the maritime geography inside it. India buys as much Emirati crude as it does because a particular passage lets the cargo through, and every other coefficient in the table is likewise the residue of the routes that were open when it was measured. The model not
only allows rerouting intermediate inputs around closed straits, but also
permits buyers to find alternate sellers and sellers to find alternate buyers.
Put simply, it lets economic actors circumvent the closure and, even
more so, the problems that arise from it.

This flexibility, though, is
bounded by the realities of economic geography
and by how little capacity can expand in the short run. There is no way
around the only exit from a bay. Alternate suppliers cannot raise
production overnight, however much they might wish to. A factory short
of one input cannot make do with more of another when the two are
complements. Interacting with the world trade network, these three
rigidities break the proportion between the cost of closing a passage and the tonnage that transits it. A passage
that carries little can still impose a large loss on a productive economy,
when what little it carries has no near substitute and feeds much
downstream. A passage that carries a great deal, by contrast, can close at
little cost, when the same goods arrive by other routes and sellers.

The cost of a closure must therefore be computed, because no reading of the
tonnage that passes through the chokepoint will give it. What we do is impose a different maritime geography on the same
economy, close a passage, let the network settle, and read the difference.
We do this for twelve critical chokepoints, one at a time and then several at
once.

Our counterfactual experiments generate three regularities. The first of which is that chokepoint exposure is heavy-tailed across
countries. A closed Strait of Hormuz takes production worth double-digit shares of GDP from the Gulf exporters behind it. It takes production worth about a tenth of Korea's GDP as well, an economy that sells almost nothing through the strait.
Yet most of the world
loses a few percent or less, even under its worst single closure. The deepest losses come from closing the terminal gates, the sole sea exits of enclosed basins such as the Persian Gulf and the Black Sea, since no detour can salvage any of the trade that must pass through them.

The second regularity is that the exporting and the importing end of a severed corridor do not
share its loss evenly, and the imbalance is large. At every one of
the twelve gates, the exporting side loses more than the importing
side. Along the major corridors the exporter--importer loss ratio runs from less than twofold at the gates of Europe and the Red Sea, through about two and a half to three at the gates of East Asia and at Panama, to between four and five at Hormuz, the largest of the energy gates. What gives
the finding its force is that the asymmetry is nowhere written into the rationed exchange, which
is neutral between buyers and sellers. It grants buyers
and severed suppliers the identical formal opportunity to re-match, so the tilt in the division of
the loss is not manufactured by the exchange. What supplies the tilt
is economic geography. Ours is a world where the economies selling through a chokepoint tend
to be commodity exporters with a narrow basket, often a single major good such
as energy. The economies buying through it are usually more diversified,
drawing the same input from several sources. When the corridor closes, that
asymmetry decides who pays. The severed exporter has few off-corridor buyers for its stranded output, so whatever it cannot place is a market simply lost.
The importer, meeting the same closure from the other side, re-sources across
the suppliers it keeps and carries only the residue, a productivity loss in input bundles it can no longer assemble in benchmark proportions.

The third regularity concerns chokepoints forced shut together, as political
conflict rarely strikes a single corridor alone. Economic geography decides how the loss from closing the gates together compares with the sum of the losses from closing each alone. A flow lost to one closure cannot be lost
again to another, so the joint closure disrupts the union of what its single closures disrupt, not their sum, and two gates on a shared corridor largely duplicate one
another. Gates with detours of their own take different cargo for the most part, so their losses add. Inputs are complements in production, so closures that leave the same producers short of inputs twice over could cost more together than the sum of what each costs alone.

Our three scenarios show gates that duplicate one another and gates whose losses add, but none of them costs more than the sum of its single closures. A Middle East escalation would close Hormuz with
Bab-el-Mandeb and Suez, severing Gulf energy and the Asia--Europe corridor at
once. But Suez and Bab-el-Mandeb sit in series on one corridor and share the
Cape-of-Good-Hope detour, so the second gate duplicates the first. The joint closure therefore costs about a tenth less than the sum of its single closures. An East Asian crisis, a Taiwan contingency, would close the Taiwan--Luzon Corridor, the Strait of Malacca, and the Korea Strait. These gates have detours of their own, and the joint closure costs within two percent of what its single closures cost together. A Russia--Europe
confrontation would shut the Turkish, Danish, and Kerch Straits, the terminal
gates of Russia's seaborne trade. Starving Central and Eastern Europe of fuel,
food, and materials, it too costs about the sum of its single closures, three percent less.

The principal quantitative findings of our paper run against the grain of how
chokepoint risk is ordinarily assessed. What matters is not how much trade crosses a passage, but how badly the
world re-sources and re-sells around it once it is gone. The irreplaceability of what a passage carries, not its tonnage, is what chokepoint exposure measures.
Which is why the deepest losses settle on economies with narrow export
baskets and import-dependent downstream sectors. It is also why an economy
can be devastated by a strait through which it itself sells almost
nothing. For the economy that would insure against such a shock, the
counsel is not to take itself off the map, since geography is seldom so
obliging, but to hold several routes and several suppliers, and to
keep its distance from single-gate dependence.\footnote{Risk is hedged most where it is most salient, since those most exposed to a risk tend to be most aware of it. India and China, whose import corridors pass Hormuz and Malacca, spent the 2010s building strategic petroleum reserves, and China's crude inventories stood near 1.4 billion barrels in December 2025, the world's largest \citep{eia_tie_apr2026}. But a hedge is incomplete wherever the risk is too new to have been priced, too costly to cover, or simply unseen, so no landscape of risk is hedged flat. Our model measures the exposure whether or not it is hedged.}

\newpage
\bibliography{ref}
\bibliographystyle{aea}

% ============================================================================
% Online Appendices A and B. Each file holds the body of the appendix as it
% stands in the working paper, and opens by restarting the counters and
% setting the A. or B. numbering that the appendix carries.
% ============================================================================
% ============================================================================
% Online Appendix A, input by v17_choke_points.tex after the paper.
% From the first line after \maketitle on, this file is the body of the
% appendix as it stands in the working paper. The lines before it do what the
% appendix's own preamble does when it is compiled as a separate document:
% they restart the counters and set the A. numbering, the line spacing and
% the title block.
% ============================================================================
\clearpage
\setcounter{part}{0}\setcounter{section}{0}\setcounter{subsection}{0}\setcounter{subsubsection}{0}\setcounter{equation}{0}\setcounter{table}{0}
\setcounter{figure}{0}\setcounter{footnote}{0}\setcounter{theorem}{0}\setcounter{proposition}{0}\setcounter{lemma}{0}\setcounter{corollary}{0}
\setcounter{claim}{0}\setcounter{assumption}{0}\setcounter{remark}{0}\setcounter{definition}{0}\setcounter{example}{0}\setcounter{result}{0}
\renewcommand{\thepart}{A.\Roman{part}}
\renewcommand{\thesection}{A.\arabic{section}}
\renewcommand{\theHsection}{A.\arabic{section}}
\renewcommand{\thesubsection}{A.\arabic{section}.\arabic{subsection}}
\renewcommand{\theHsubsection}{A.\arabic{section}.\arabic{subsection}}
\renewcommand{\theequation}{A.\arabic{equation}}
\renewcommand{\thetable}{A.\arabic{table}}
\renewcommand{\thefigure}{A.\arabic{figure}}
\renewcommand{\thetheorem}{A.\arabic{theorem}}
\renewcommand{\theproposition}{A.\arabic{proposition}}
\renewcommand{\thelemma}{A.\arabic{lemma}}
\renewcommand{\thecorollary}{A.\arabic{corollary}}
\renewcommand{\theassumption}{A.\arabic{assumption}}
\renewcommand{\thedefinition}{A.\arabic{definition}}
% the same prefix for the link anchors, so that no two anchors in the document share a name
\def\theHequation{A.\arabic{equation}}\def\theHtable{A.\arabic{table}}\def\theHfigure{A.\arabic{figure}}
\def\theHtheorem{A.\arabic{theorem}}\def\theHproposition{A.\arabic{proposition}}\def\theHlemma{A.\arabic{lemma}}
\def\theHcorollary{A.\arabic{corollary}}\def\theHclaim{A.\arabic{claim}}\def\theHassumption{A.\arabic{assumption}}
\def\theHremark{A.\arabic{remark}}\def\theHdefinition{A.\arabic{definition}}\def\theHexample{A.\arabic{example}}
\def\theHresult{A.\arabic{result}}
\onehalfspacing
\setlength{\emergencystretch}{0pt}
\OArestoretitle
\title{\textbf{Online Appendix~A}\\[0.4ex]
to\\[0.4ex]
{\large\textbf{``Macroeconomic Risks from Maritime Trade Disruptions''}}\\[4ex]
\textbf{Statements and Proofs}}

\author{Vipin P. Veetil\thanks{Economics Area, Indian Institute of Management Kozhikode.} \and Fathimath S. Vemmarath\thanks{Indian Statistical Institute, Kolkata.}}

\date{\today}
\maketitle
\pdfbookmark[0]{Online Appendix A: Statements and Proofs}{onlineappendixA}

This online appendix states and proves the paper's formal results. It divides into two sets of claims. Part~A.I concerns the base model, the economy a closure acts on before any
question of where the closure falls. We establish the well-posedness of the rationed exchange with residual transit held fixed, and then give the existence and selection of the reported equilibrium. These results are adaptations of certain claims and proofs of \citet{bhatt2026_sanctions}, which they had made for a setting in which the exchange is constrained by the support of flows but not by the disruptive cut.

Part~A.II concerns the size of the loss from a closure. We show that, with GDP measured by its accounting share, the world loss through lost orders is the value of the orders that buyers no longer place with any seller, an identity that rests on the unit GDP multiplier of the benchmark network. We then compare the world loss of a joint closure with the sum of the world losses of its single closures at an isolated pair of gates, i.e., two gates that reach one buyer which sells to final users alone. The joint closure costs what the closure of one gate costs when the two gates stop the same shipments. Otherwise it costs less or more than the sum of its two single closures according to whether the two gates take different inputs or the same input from the buyer.

Unless stated otherwise, every result in this appendix is proved for the
exact theoretical model. The maintained domain has \(\beta_j\in(0,1)\),
each active benchmark input-coefficient column normalized to sum to one,
and exact benchmark accounting identities.
Online Appendix~\ref{oa:data_integrity} reconciles the exact model with the
assembled table. It documents the single
\(\beta_j^{\rm spend}\ne\beta_j^{\rm tech}\) outlier, the spectral-radius
bound used when the column-sum norm exceeds one, and the benchmark activity
vector, which is the solution of the benchmark system and not the row totals of the table. Results that require a
positive fulfillment floor or strict subhomogeneity are not asserted for
the coordinate with \(\beta_j^{\rm tech}=0\).

As in the paper, every matrix of flows, of input shares, or of propagation is of dimension \(CS\times CS\), with a row for each country-sector as a seller and a column for each country-sector as a buyer, and a sector-\(r\) block is the \(C\times CS\) submatrix on the rows of the \(C\) country-sectors that supply input \(r\). A matrix of any other dimension is given its dimension where it first appears. Sections, equations, and results cited without the prefix A or B are those of the paper.

% ================================================================
\part{The Base Model: The Rationed Exchange and the Equilibrium}
% ================================================================

% ----------------------------------------------------------------
\section{Well-Posedness of the Rationed Exchange with Residual Transit Held Fixed}
\label{app:proof_model}
% ----------------------------------------------------------------

The rationed exchange of Section~\ref{subsec:rerouting_matrix} of the paper differs from the exchange of \citet{bhatt2026_sanctions} in one respect, which is that it holds the residual transit fixed. Their exchange is that of their Section~2.2, in the biproportional form of their equation~9, which their Appendix~A.3 restates as a block-wise balancing program. Their Proposition~A.1 shows that its allocation is the biproportional adjustment of the disrupted flows to
the totals that the sales of each seller and the purchases of each buyer have to meet
\citep{sinkhorn1964,bacharach1970}.

The closure divides each entry of the disrupted transactions matrix \(\check{\mathbf Z}\) into an unexposed flow and
a residual transit,
\begin{equation*}\tag{\ref{eq:choke_split}}
\check Z_{ij}=(1-d_{ij})\,Z_{ij}+Y^{(\omega)}_{ij},
\qquad
Y^{(\omega)}_{ij}:=\delta_{r(i)}\,d_{ij}\,Z_{ij}.
\end{equation*}
Here \(d_{ij}\in[0,1]\) is the disrupted fraction of the link \(i\to j\),
a datum of the scenario's route geography, and \(\delta_{r(i)}\) the residual transit share of seller \(i\)'s sector, the only one this appendix uses. Wherever the seller's sector is fixed by the context, as in a display
summing over links that share a seller or over a sector block, we write
\(\delta\) for \(\delta_{r(i)}\). The exchange holds the residual transit fixed and allocates the unexposed flows to the row totals and the column totals it has to meet, the capacities of the sellers and the budgets of the buyers. Seller \(i\)'s row is restored to its capacity
\begin{equation*}\tag{\ref{eq:choke_rowtarget}}
g^{(\omega)}_i
:=
(s_i-f_i)-\tau\,Z^{\rm disp}_i,
\qquad
Z^{\rm disp}_i:=\sum_j(1-\delta)\,d_{ij}\,Z_{ij},
\end{equation*}
and buyer \(j\)'s sector-\(r\) block to its budget, as a share of its spending,
\begin{equation*}\tag{\ref{eq:choke_coltarget}}
t^{(\omega)}_{rj}
:=
\alpha_{rj}-\tau\,\ell_{rj},
\end{equation*}
with \(\ell_{rj}:=\sum_{i\in\mathcal N_r}(1-\delta)\,d_{ij}\,a_{ij}\)
buyer \(j\)'s displaced sector-\(r\) input share and
\(\alpha_{rj}:=\sum_{i\in\mathcal N_r}a_{ij}\) its benchmark counterpart. Collecting terms link by link,
\begin{equation}
t^{(\omega)}_{rj}
=
\sum_{i\in\mathcal N_r}\bigl(1-\tau(1-\delta)\,d_{ij}\bigr)\,a_{ij},
\label{eq:choke_target_link}
\end{equation}
and \(g^{(\omega)}_i=\sum_j\bigl(1-\tau(1-\delta)d_{ij}\bigr)Z_{ij}\) the
same way in flow units. A link \(i\to j\) that is fully exposed (\(d_{ij}=1\)) carries the bracket \(1-\tau(1-\delta)=\delta+(1-\tau)(1-\delta)\), in which the closure leaves the residual transit share \(\delta\) standing, the exchange restores the share \(1-\tau\) of the displaced remainder, and the rest is written off
as unrecovered. The retention factor \(1-\tau(1-\delta)\,d_{ij}\) of
\eqref{eq:choke_target_link} is one on every clear link. The retention factor is what the exchange aims at rather than what it achieves, but the allocation attains it wherever the capacities and the budgets can be met jointly on the open links. Summing
\eqref{eq:choke_target_link} over the block's sellers and buyers shows that the block totals of the capacities and the budgets agree,
\begin{align}
\sum_{i\in\mathcal N_r}g^{(\omega)}_i
&=
\sum_{i\in\mathcal N_r}\sum_j Z_{ij}
-\tau(1-\delta)\sum_{i\in\mathcal N_r}\sum_j d_{ij}\,Z_{ij} \notag\\
&=
\sum_j t^{(\omega)}_{rj}\,(1-\beta_j)s_j,
\label{eq:choke_margins_agree}
\end{align}
the counterpart of equation~A.2 of \citet{bhatt2026_sanctions} for links that a closure disrupts only in part.
On a block whose residual capacities and budgets, by which we mean the capacities and the budgets net of the residual transit, admit a plan positive on every open
link, the transactions after the exchange take the biproportional form
\begin{equation*}\tag{\ref{eq:choke_reroute}}
Z^{(\omega)}_{ij}
=
Y^{(\omega)}_{ij}+u_i\,(1-d_{ij})\,Z_{ij}\,v_j,
\qquad i\in\mathcal N_r,
\end{equation*}
with positive scalings \(u_i\) on the sellers and \(v_j\) on the buyers. Equation
\eqref{eq:choke_reroute} states the rationing one link at a time. For all the links of a block
at once, arrange the flows on the block's links in matrices with the \(C\) rows of the block's
sellers and the \(CS\) columns of its buyers, and write \(\mathbf Y_\omega=[Y^{(\omega)}_{ij}]\)
for the residual-transit matrix. The block's flows after the exchange are
\begin{equation}
\mathbf Z_\omega
=
\mathbf Y_\omega
+\mathbf D_1\bigl(\check{\mathbf Z}-\mathbf Y_\omega\bigr)\mathbf D_2,
\qquad
\mathbf D_1=\diag(\mathbf u),\quad\mathbf D_2=\diag(\mathbf v).
\label{eq:choke_diag}
\end{equation}
Here \(\mathbf D_1\) is the \(C\times C\) diagonal matrix of the sellers' scalings and
\(\mathbf D_2\) the \(CS\times CS\) diagonal matrix of the buyers' scalings, a buyer carrying a
separate scaling in each of its input sectors. Multiplying the matrix
\(\check{\mathbf Z}-\mathbf Y_\omega\) of the block's unexposed flows on the left by
\(\mathbf D_1\) multiplies the row of each seller by that seller's scaling, and multiplying the
same matrix on the right by \(\mathbf D_2\) multiplies the column of each buyer by that buyer's
scaling. The residual-transit matrix \(\mathbf Y_\omega\) stands outside both products, so the share of an exposed flow that crosses the closed passage is the residual transit share \(\delta\) itself. In the exchange of \citet{bhatt2026_sanctions} the residual on a barred link is scaled along with every other surviving flow, so the share of an exposed flow that crosses a barred link is the residual transit share multiplied by the scalings that the exchange assigns to the link's seller and buyer. Dividing each column \(j\) by buyer \(j\)'s benchmark intermediate spending
\((1-\beta_j)s_j\) turns the flows of \eqref{eq:choke_diag} into the input shares of the
effective demand matrix \(\mathbf A_\omega\). The Sinkhorn iteration \citep{demingstephan1940,sinkhorn1964} computes the two diagonal matrices by rescaling in turn the columns of the unexposed flows to the column totals and their rows to the row totals. One such pair of updates is a sweep.

Proportional rationing on both sides of a block is one of several rationing schemes of fixed-price equilibrium, but the allocation it delivers has properties that recommend the scheme
when the data say no more about the re-matching than the benchmark flows and the new totals. The
allocation opens no link that carried no trade at the benchmark, so a closure creates no trading
relationship that did not exist before it. Each seller scales the unexposed flow on all of its
links by one factor, and each buyer by one factor of its own. The ratio of a buyer's purchases
from two suppliers in the re-matched flows changes by a factor that depends on the two suppliers
alone, the same for every buyer. The pattern of association between sellers and buyers in the
unexposed flows, measured by their cross-product ratios, is left as it was
\citep{mosteller1968}. Among all allocations that meet the new totals, this allocation also
departs least from the unexposed flows in generalized Kullback--Leibler divergence
\citep{ireland_kullback_1968,bacharach1970,csiszar1975}. Minimum divergence is the precise sense in
which the exchange adds no structure beyond what the row and column totals force. Lemma~\ref{prop:wellposed} computes the allocation as that minimizer.
\citet{balinski_demange_1989} show that this two-sided proportional form is the only revision of
a matrix to new row and column totals that meets a short list of requirements of
proportionality. One of those requirements is that a block whose capacities and budgets all fall
by one common share sees every flow fall by that share.

The lemma concerns the block problem net of the residual transit. Write
\(\mathring Z_{ij}:=(1-d_{ij})Z_{ij}\),
\(\mathring g_i:=g_i^{(\omega)}-\sum_jY^{(\omega)}_{ij}\), and
\(\mathring t_{rj}:=t^{(\omega)}_{rj}(1-\beta_j)s_j
-\sum_{i\in\mathcal N_r}Y^{(\omega)}_{ij}\) for the unexposed flows and
the residual capacities and budgets.
The unexposed flows already deliver
$\sum_j\mathring Z_{ij}$ from seller $i$ and
$\sum_{i\in\mathcal N_r}\mathring Z_{ij}$ into buyer $j$, and the residual
capacities and budgets exceed these sums by exactly the nonnegative recoveries,
\begin{equation}
\mathring g_i=\sum_j\mathring Z_{ij}+(1-\tau)Z^{\rm disp}_i,
\qquad
\mathring t_{rj}=\sum_{i\in\mathcal N_r}\mathring Z_{ij}
+(1-\tau)(1-\beta_j)s_j\,\ell_{rj}.
\label{eq:wellposed_split}
\end{equation}
Section~\ref{subsec:rerouting_matrix} of the paper defines the placement deficit \(U_r\ge0\) of the block as the part of its capacities, and equally of its budgets, that no allocation over its open links can place, and Lemma~\ref{prop:wellposed} computes it as the shortfall of a maximum flow.

We make two reductions to the block problem before stating the lemma. Write \(\mathbf B^{*}\) for the re-matched flows the projection returns. A block whose residual capacities and budgets all
vanish carries \(\mathbf B^{*}=\mathbf 0\) and calls for no scaling at
all. On any other block, delete every row whose residual capacity is zero and every column whose residual
budget is zero before the first update. Every cell incident to such a node
is forced to zero, so the update would divide by a vanishing sum.
Let \(I_r\) and \(J_r\) denote the real seller
and buyer indices that survive the deletion, and write
\(\mathring{\mathbf K}:=[\mathring Z_{ij}]_{I_r\times J_r}\) for the
matrix of surviving unexposed flows, with a row for each surviving seller in \(I_r\) and a column for each surviving buyer in \(J_r\).
Write \(\widehat{\mathbf K}\), \(\widehat{\mathbf g}\), and
\(\widehat{\mathbf t}\) for the kernel of the scaling problem, the matrix the iteration rescales, and for the row totals and the column
totals that survive the deletion, and
\(E:=\operatorname{supp}(\widehat{\mathbf K})\) for the surviving
support.
The surviving triple consists of the unexposed flows with their residual capacities and
budgets on a block carrying no placement deficit, and of their slack
augmentation on a block carrying one.
On a block carrying no placement deficit,
\(U_r=0\), the surviving triple is
\((\widehat{\mathbf K},\widehat{\mathbf g},\widehat{\mathbf t})
=(\mathring{\mathbf K},\mathring{\mathbf g},\mathring{\mathbf t})\).
On a
block carrying the deficit \(U_r>0\), the augmentation appends an
unmet-demand seller \(\perp_D\) and an unsold-output buyer \(\perp_S\),
and the augmented kernel \(\widehat{\mathbf K}\) and augmented totals
\(\widehat{\mathbf g},\widehat{\mathbf t}\) are
\begin{equation}
\widehat{\mathbf K}
:=
\begin{bmatrix}
\mathring{\mathbf K} & \mathbf z^{\rm disp}\\
(\mathbf p^{\rm disp})^{\!\top} & 0
\end{bmatrix},
\qquad
\widehat{\mathbf g}:=(\mathring{\mathbf g},U_r),
\qquad
\widehat{\mathbf t}:=(\mathring{\mathbf t},U_r),
\label{eq:slack_augmented_problem}
\end{equation}
with \(z^{\rm disp}_i:=Z^{\rm disp}_i\) and
\(p^{\rm disp}_j:=(1-\beta_j)s_j\ell_{rj}\). The augmented kernel has one row more than the matrix \(\mathring{\mathbf K}\) of surviving unexposed flows, the row of the unmet-demand seller, and one column more, the column of the unsold-output buyer. The re-matched flows \(\mathbf B^{*}\) have the dimension of the kernel \(\widehat{\mathbf K}\) from which they are projected. In the support graph, a
real seller has an edge to \(\perp_S\) exactly when it has positive
displaced sales, and \(\perp_D\) has an edge to a real buyer exactly
when that buyer has positive displaced purchases. The slack--slack cell
is structurally zero. The slack flows serve only to make the augmented
problem feasible and are discarded before the real-to-real
transaction matrix \(\mathbf Z_\omega\), the effective demand coefficients, and the downstream demands are computed.

Positivity of the surviving kernel is a property of that triple and not
of the unexposed flows on their own. On a deficit-free block it follows from the absence of the deficit, since a surviving row with no open link would carry
its whole residual capacity unplaced, putting the block's maximum flow
below the common total \eqref{eq:choke_margins_agree} of the residual capacities and of the residual budgets, and the
same argument runs down the columns. The one case the argument excludes is a seller
whose sales the closure severs in full. That seller keeps the strictly
positive residual capacity \((1-\tau)Z^{\rm disp}_i\) of
\eqref{eq:wellposed_split} against an empty row of unexposed flows, so its block
carries a placement deficit and the unsold-output column supplies its
kernel entry. The unmet-demand row does the same for a buyer whose
sector-\(r\) purchases are severed in full. Every surviving row and every
surviving column of \(\widehat{\mathbf K}\) therefore holds a positive entry, so the scaling divisions meet no zero denominator at any sweep.
Call the solved problem strictly feasible when its transportation
polytope holds a plan positive on every edge of \(E\).

\begin{lemma}[Well-posedness of the rationed exchange with residual transit held fixed]
\label{prop:wellposed}
Fix \(\omega\), \(\tau\in[0,1]\), and residual transit shares
\((\delta_r)_{r\in\mathcal M}\) as in Assumption~\ref{ass:attenuation}. For a sector-\(r\) block, define the
placement deficit \(U_r\ge0\) as the shortfall of the maximum
bipartite flow, carried from sellers at the residual capacities to buyers at the residual budgets over the block's open links, below the common
total \eqref{eq:choke_margins_agree} of those capacities and budgets. The exchange is then well posed on every block, in that it is feasible and uniquely solved, it takes the biproportional form exactly when it is strictly feasible,
and the Sinkhorn iteration converges to that solution.
\begin{enumerate}
\item[\textnormal{(i)}] Sizing both
slack accounts at \(U_r\) makes the exchange of Definition~\ref{def:choke_reroute} feasible on every block. On the solved
problem the generalized Kullback--Leibler divergence from
\(\widehat{\mathbf K}\) has a unique minimizer \(\mathbf B^{*}\) over the
constraints on the row and column totals.
\item[\textnormal{(ii)}] Three statements
about that minimizer are equivalent, namely that the solved
problem is strictly feasible, that \(\operatorname{supp}(\mathbf B^{*})=E\), and that
\(\mathbf B^{*}=\diag(\mathbf u)\,\widehat{\mathbf K}\,\diag(\mathbf v)\)
for finite positive scalings. The scalings are then unique up to one
reciprocal gauge per connected component of the support graph, and any
feasible matrix of that form is the minimizer.
\item[\textnormal{(iii)}] The Sinkhorn
iteration converges in matrix entries to \(\mathbf B^{*}\) on every
block. Under strict feasibility the scaling iterates converge at a
local geometric rate once one gauge per support component is fixed.
Where strict feasibility fails no geometric rate holds, and the rate can be as slow as \(1/n\) in the number \(n\) of sweeps.
\end{enumerate}
\end{lemma}

\begin{proof}[Proof of Lemma~\ref{prop:wellposed}]
Fix a disrupted sector-$r$ block, with its unexposed flows
$\mathring{\mathbf Z}$ and its residual capacities and budgets
$(\mathring{\mathbf g},\mathring{\mathbf t})$ as defined before the lemma.
The unexposed flows are the disrupted matrix less the residual transit,
\(\mathring{\mathbf Z}=\check{\mathbf Z}-\mathbf Y_\omega\) by
\eqref{eq:choke_split}, so the caron carries the whole disrupted flow and the ring only the part the exchange may rescale.
The residual totals agree,
$\sum_{i\in\mathcal N_r}\mathring g_i=\sum_j\mathring t_{rj}$, because the
aggregate identity \eqref{eq:choke_margins_agree} nets the same transit
total from both sides. By \eqref{eq:wellposed_split}, the residual capacities and budgets exceed what the unexposed flows already deliver by exactly the nonnegative recoveries.
Parts (i) and (ii) run in four steps. We first give the placement deficit
a closed form, then show that the slack pair sized at that deficit leaves
the augmented polytope nonempty, then show that the divergence attains
its minimum there at exactly one point, which proves (i). We then show
that strict feasibility is exactly what puts that minimum in the interior
of the support, which proves (ii). Part (iii) reads the iteration as
cyclic information projection and closes with the two regimes of its convergence rate.

We first identify the placement deficit with the largest amount by which a set of sellers holds more residual capacity than the buyers it can reach hold in residual budgets.
Build the bipartite transportation network of the block. The network consists of a source, an arc
of capacity $\mathring g_i$ from the source to each seller $i$, an
uncapacitated arc from $i$ to $j$ for each
$(i,j)\in\operatorname{supp}(\mathring{\mathbf Z})$, an arc of capacity
$\mathring t_{rj}$ from each buyer $j$ to a sink, and the sink. A cut of
finite capacity severs no uncapacitated arc, so it is indexed by the
sellers it leaves on the source side, a subset
$\mathscr R\subseteq\mathcal N_r$ of the block's sellers. Such a cut
severs the source arcs of the sellers in $\mathcal N_r\setminus\mathscr R$
and the sink arcs of every buyer with an open link to a seller in $\mathscr R$,
the columns $\mathring{\mathcal N}(\mathscr R)$. Its capacity is
$\sum_{i\in\mathcal N_r\setminus\mathscr R}\mathring g_i
+\sum_{j\in\mathring{\mathcal N}(\mathscr R)}\mathring t_{rj}$. By the
max-flow--min-cut theorem the maximum flow is the least such capacity, so
its shortfall below the common total of the residual capacities and of the residual budgets is
\begin{equation}
U_r
\;=\;
\max_{\mathscr R\subseteq\mathcal N_r}\;\Bigl[\sum_{i\in\mathscr R}\mathring g_i
-\sum_{j\in \mathring{\mathcal N}(\mathscr R)}\mathring t_{rj}\Bigr],
\label{eq:wellposed_deficit}
\end{equation}
the largest Gale--Hoffman violation the block admits. The maximum is
nonnegative because $\mathscr R=\emptyset$ is admissible, and it vanishes
exactly when the residual capacities and budgets are jointly attainable on the block's open links \citep{gale1957flows,hoffman1960}.

We next show that the slack pair restores feasibility. By
\eqref{eq:wellposed_split} the unexposed flows are themselves a flow in that
network, weakly below the residual capacities and budgets, and they
saturate the source arc of every seller with no displaced sales and the
sink arc of every buyer with no displaced purchases. Augment it to a
maximum flow $F^{*}$ along simple augmenting paths in the residual network, since any augmenting walk can be shortened to a path by removing its cycles. Such a path
leaves the source once and enters the sink once, so it
traverses one source arc and one sink arc, each in the forward
direction, and a saturated arc admits no further forward flow. The saturated arcs therefore stay saturated, so under $F^{*}$ the unplaced
row residuals sit only on sellers with displaced sales and the unmet
column residuals only on buyers with displaced purchases. By
\eqref{eq:wellposed_deficit} each set of residuals totals $U_r$, and the
unsold-output column and the unmet-demand row are supported on exactly
these sellers and buyers, each carrying the total $U_r$. Routing the
unplaced row residuals into that column and covering the unmet column
residuals from that row therefore extends $F^{*}$ to a point of the
augmented polytope that meets every row and column total. Bounding each seller's
residual by \eqref{eq:wellposed_split} gives
$\mathring g_i-\sum_j F^{*}_{ij}\le\mathring g_i-\sum_j\mathring Z_{ij}
=(1-\tau)Z^{\rm disp}_i$, and summing over the block gives
$U_r\le(1-\tau)\sum_{i\in\mathcal N_r}Z^{\rm disp}_i$, so the placement deficit never exceeds the recoverable mass. On a deficit-free block $U_r=0$, no
slack is appended, and the same construction lands in the unaugmented
polytope.

We now prove existence and uniqueness of the minimizer, on the solved
problem the two reductions have left, which is the kernel \(\widehat{\mathbf K}\)
with the row totals \(\widehat{\mathbf g}\) and the column totals \(\widehat{\mathbf t}\). The
generalized Kullback--Leibler divergence is $+\infty$ at any matrix that
puts mass where its reference kernel has none. That problem is therefore
the minimization of
$\sum_{(i,j)\in E}\bigl(b_{ij}\log(b_{ij}/\widehat K_{ij})-b_{ij}
+\widehat K_{ij}\bigr)$
over the nonnegative matrices supported in \(E\) that meet the row and column totals of the solved
problem. On an augmented block the slack cells enter that sum with their
own kernel entries, the displaced sales and the displaced purchases. The
polytope is convex, closed and bounded, hence compact by the
Heine--Borel theorem, and it is nonempty by
the preceding step, which exhibited a feasible point of exactly this
augmented polytope. On it the objective is continuous under the
convention $0\log 0=0$, and it is strictly convex because $b\mapsto b\log b$
is strictly convex on $[0,\infty)$. Weierstrass' theorem therefore gives
a minimizer, and strict convexity puts it at a single point, so the re-matched flows exist and are unique \citep{csiszar1975}. This proves
(i).

We next prove part (ii), the strict-feasibility equivalence. Suppose the solved
polytope holds a plan \(\mathbf B^{\circ}\) positive on every edge of
\(E\). If the minimizer zeroed some edge of \(E\), the segment
\(\mathbf B_\lambda:=(1-\lambda)\mathbf B^{*}+\lambda\mathbf B^{\circ}\)
would stay feasible for \(\lambda\in[0,1]\). The one-sided derivative of
the objective at \(\lambda=0\) would then be \(-\infty\), because
\(x\mapsto x\log(x/\widehat K_{ij})-x+\widehat K_{ij}\) has derivative
\(\log(x/\widehat K_{ij})\), which diverges at the origin. A feasible
\(\mathbf B_\lambda\) at small \(\lambda>0\) would therefore carry a
strictly lower objective value than \(\mathbf B^{*}\), contradicting its
minimality, so \(\operatorname{supp}(\mathbf B^{*})=E\). Since the minimizer is positive on \(E\), the ordinary Lagrange conditions read
\(\log(B^{*}_{ij}/\widehat K_{ij})=\log u_i+\log v_j\) on \(E\), the row
and column multipliers written in the scalings' own units. Those
conditions are the factorization
\(\mathbf B^{*}=\diag(\mathbf u)\,\widehat{\mathbf K}\,\diag(\mathbf v)\)
with \(\mathbf u\) and \(\mathbf v\) finite and positive
\citep{sinkhorn1964,bacharach1970}. That factorization is in turn positive
on every edge of \(E\) and feasible, hence itself a strictly feasible
plan, and a minimizer supported on all of \(E\) is one as well. The three
statements are therefore equivalent. Since the program is convex with linear constraints, those conditions are sufficient as well as necessary, so any
feasible matrix of that form is the minimizer. The conditions
leave exactly one gauge freedom per connected component of the support graph, namely the scaling of that
component's \(u\) entries by a positive constant and of its \(v\) entries by the reciprocal. Where strict feasibility fails, the minimizer zeros an edge of \(E\) and \eqref{eq:choke_reroute} survives only as a limit of positive scalings. This
proves (ii).

Adding back the residual-transit matrix $\mathbf Y_\omega$, whose entries
are data, and recovering
\(a^{(\omega)}_{ij}=Z^{(\omega)}_{ij}/\bigl((1-\beta_j)s_j\bigr)\) assembles \(\mathbf A_\omega\), in which disrupted cells carry their transit, open links
their re-matched flows, and untouched blocks their benchmark values. Its
support is the benchmark one on a deficit-free block that is strictly
feasible, and contained in the benchmark one otherwise, which is the statement that
Lemma~\ref{lem:choke_kernel} makes about the support.

It remains to prove part (iii), the convergence of the Sinkhorn iteration.
Each row rescaling replaces the current iterate by the matrix closest to it in the divergence
among those with the required row totals, and each column
rescaling does the same among those with the required column totals. A half-sweep is therefore
the information projection of the current iterate onto one of two affine
families. Their intersection is the solved polytope, nonempty by the
preceding step, and it holds \(\mathbf B^{*}\) at finite divergence from the starting kernel, since \(\mathbf B^{*}\) is supported inside \(E\). Those three facts are
the hypotheses of the cyclic-projection theorem of \citet[Theorem~3.2]{csiszar1975}, which carries the iterates to the unique
minimizer on every block, augmented or not
\citep[see also][]{bacharach1970}.\footnote{What the projections drive
down is the divergence \(D(\mathbf B^{*}\Vert\mathbf B^{(n)})\) from the
minimizer to the iterate, not the objective
\(D(\mathbf B\Vert\widehat{\mathbf K})\), which does not descend. The
iteration starts at
\(\mathbf B^{(0)}=\widehat{\mathbf K}\), where the objective is zero, so
any rescaling that moves the iterate raises it. It rises to the
constrained minimum \(D(\mathbf B^{*}\Vert\widehat{\mathbf K})\) without
being monotone on the way, since each half-sweep leaves the family the
previous half-sweep had reached.}

Strict feasibility delivers a local geometric rate on any support. Linearize
one full sweep in the logarithms of the column factors at the fixed point.
Its Jacobian is
\(\mathbf P:=\diag(\widehat{\mathbf t})^{-1}(\mathbf B^{*})^{\!\top}
\diag(\widehat{\mathbf g})^{-1}\mathbf B^{*}\), a square matrix with a row and a column for each column of the solved problem, well defined because
\(\mathbf B^{*}\) is positive on \(E\). On each connected component of the
support graph \(\mathbf P\) is row-stochastic with positive diagonal,
irreducible, and reversible with stationary weights
\(\widehat{\mathbf t}\), and it is positive semidefinite because it is
similar to a Gram matrix. Its eigenvalue one is simple and carries that
component's gauge, and every eigenvalue on the gauge quotient is strictly
below one. The sweep is accordingly a local contraction there, so the iterates converge at a geometric rate once one gauge per
component is fixed. The coefficient is read off \(\mathbf B^{*}\) and so
depends on the row and column totals, which carry \(\delta\).\footnote{Parts (i) and (ii) apply the information projection of \citet{csiszar1975}, and part (iii) his convergence theorem for cyclic projections \citep[Theorem~3.2]{csiszar1975}. The lemma is the counterpart, for an exchange that holds the residual transit fixed, of Propositions~A.1 and~A.2 of \citet{bhatt2026_sanctions}, which their Appendix~A proves for the exchange that rescales the residual transit.}

Where strict feasibility fails no geometric rate holds. Finite positive
scalings do not exist in the limit there, so the iterates can approach the minimizer at a polynomial rate instead. Proposition~3 of \citet{qu2025sinkhorn} characterizes Sinkhorn convergence in that case as sublinear, and their Section~5.3 shows that some entry of the iterate then approaches its limit no faster than order \(1/n\). By the same proposition, finite positive scalings on a connected support are equivalent to linear convergence. The failure is not a property of the placement deficit, since a block with none can exhibit it, as blocks of the extended ICIO do (Online Appendix~\ref{app:stability_numerics}). This
proves (iii).
\end{proof}

The pseudo-prices of Section~\ref{subsec:rerouting_matrix} of the paper are the reciprocals of
the seller scalings of Lemma~\ref{prop:wellposed}(ii), so the lemma settles when they exist.
Like the prices of a general equilibrium, the pseudo-prices are pinned down only relative to one another. One common scalar remains free on each group of sellers and buyers that the block's
links, slack accounts included, tie together, which is the reciprocal gauge of part (ii). The pseudo-prices need an interior allocation, but a closure that severs many links can deny a block
one by leaving a group of buyers able to reach only sellers whose capacities their budgets use
up exactly. Those sellers then have nothing left for the other buyers their open links reach, so
every allocation that meets the row and column totals leaves those links empty. The rescaling
does not create these zeros but converges to them, and since a positive scaling times a positive
flow is never zero, it reaches them only in the limit, with some scalings running to zero or
growing without bound. No finite price makes a buyer choose to buy nothing on a link it is free
to use, so on such a block the scalings are no longer pseudo-prices. Online
Appendix~\ref{app:stability_numerics} reports how often, on the data, the open links of a block
fall into several groups and the allocation leaves open links empty.

The exchange treats its two sides alike. Write
\(\mathsf B(\check{\mathbf Z};\mathbf g,\mathbf t)\) for the rationed exchange of Definition~\ref{def:choke_reroute} with capacities \(\mathbf g\) and budgets \(\mathbf t\). Then
\(\mathsf B(\check{\mathbf Z}^{\!\top};\mathbf t,\mathbf g)
=\mathsf B(\check{\mathbf Z};\mathbf g,\mathbf t)^{\!\top}\), with the
scalings \(\mathbf u\) and \(\mathbf v\) exchanged. Recall that the rows of the disrupted transactions matrix \(\check{\mathbf Z}\) are sellers and its columns are buyers, so its transpose describes the economy in which every seller has taken the place of a buyer and every buyer that of a seller. Transposition carries every ingredient of the exchange onto its mirror image. The
residual transit transposes entrywise, the capacities and the budgets net of that transit swap roles,
and the slack pair is exchanged. The placement deficit is invariant
because the maximum bipartite flow of the transposed network equals
that of the original. The generalized Kullback--Leibler divergence is
invariant under joint transposition of its argument and its reference,
so the uniqueness of Lemma~\ref{prop:wellposed}(i)
forces the transpose of the exchange to be the exchange of the transpose.

% ================================================================
\section{The Effective Demand and the Equilibrium It Supports}
\label{app:stability}

% ================================================================

\subsection{The exchange as an information projection, and the effective demand it leaves}
The effective demand matrix \(\mathbf A_\omega\) of Definition~\ref{def:choke_reroute} is the
result of running the iteration of Lemma~\ref{prop:wellposed} on the unexposed flows \(\mathring{\mathbf Z}\) and the residual capacities and budgets \((\mathring{\mathbf g},\mathring{\mathbf t})\) of Online Appendix~\ref{app:proof_model}, and adding the residual transit back. The iteration runs sector block by sector block, on the block augmented by the slack pair wherever
the placement deficit is positive.
Part (iii) of that lemma governs the convergence and the rate regimes.

On each disrupted sector block, the re-matched flows of Definition~\ref{def:choke_reroute} are, by Lemma~\ref{prop:wellposed}(i), the unique solution of
\begin{equation*}
\min_{\mathbf B\ge 0}\;
\sum_{i,j}\Bigl( b_{ij}\log\frac{b_{ij}}{\mathring Z_{ij}}-b_{ij}+\mathring Z_{ij}\Bigr)
\qquad\text{s.t.}\qquad
\sum_j b_{ij}=\mathring g_i,\quad
\sum_{i\in\mathcal N_r} b_{ij}=\mathring t_{rj},
\end{equation*}
the projection of the unexposed flows onto the
constraints of the residual capacities and budgets in generalized Kullback--Leibler divergence
\citep{ireland_kullback_1968,bacharach1970,csiszar1975}.  On a placement-deficit
block the same program runs on the augmented kernel and totals, the slack
pair entering as one additional row and column. Where the projection is positive on every open link, the first-order conditions give
\(b_{ij}=\mathring Z_{ij}\,u_i\,v_j\) with \(\log u_i\) and
\(\log v_j\) the multipliers on the row and column constraints. Where it
sits on the boundary the multipliers diverge, so the same product form survives only as a limit. In either case the transactions after the exchange are \(\mathbf Z_\omega=\mathbf Y_\omega+[b_{ij}]\)
and the effective demand coefficients are
\(a^{(\omega)}_{ij}=Z^{(\omega)}_{ij}/\bigl((1-\beta_j)s_j\bigr)\). On a
placement-deficit block the bracket holds the real-to-real submatrix of
the augmented minimizer, its slack row and column dropped.

We next record the bookkeeping that the effective demand inherits from the capacities and
the budgets, namely its support, its column sums, the seller-side bound, and dissipativity.

\begin{lemma}[Post-closure propagation]
\label{lem:choke_kernel}
For every scenario \(\omega\) and every \((\tau,\delta)\),
\(\mathbf A_\omega\) is nonnegative with
\(\operatorname{supp}(\mathbf A_\omega)\subseteq\operatorname{supp}(\mathbf A)\).
Every disrupted cell of the benchmark support is strictly positive,
and every used sector block of every buyer's column is strictly
positive in total. On placement-deficit-free blocks that are strictly feasible in the sense
of Lemma~\ref{prop:wellposed}(ii), the support equals
\(\operatorname{supp}(\mathbf A)\). The sector-\(r\) block of an affected buyer \(j\)'s column of \(\mathbf A_\omega\)
sums to the budget \(t^{(\omega)}_{rj}=\alpha_{rj}-\tau\ell_{rj}\) on a block whose
residual capacities and budgets are jointly attainable, and to
\(t^{(\omega)}_{rj}-U_{rj}/\bigl((1-\beta_j)s_j\bigr)\) on a placement-deficit block. Here
\(U_{rj}\ge0\) is the unmet demand absorbed at \((r,j)\), with
\(\sum_j U_{rj}=U_r\). In either case every column of \(\mathbf A_\omega\)
sums to at most one. On the seller side
\(\sum_j Z^{(\omega)}_{ij}\le g^{(\omega)}_i\le s_i-f_i\), with equality in
the first inequality away from deficit blocks, so the exchange adds no flow to any seller's benchmark sales. Consequently
\(\mathbf A_\omega(\mathbf I-\boldsymbol\beta)\) is column-substochastic
under the maintained \(\beta_j\in(0,1)\), with
\(\|\mathbf A_\omega(\mathbf I-\boldsymbol\beta)\|_1\le
\|\mathbf A(\mathbf I-\boldsymbol\beta)\|_1<1\).\footnote{The appendix
draws dissipativity from the spectral radius
\(r_{\mathrm{sp}}(\mathbf A_\omega(\mathbf I-\boldsymbol\beta))<1\)
rather than from the norm bound, which one accounting column of the
assembled table fails (Online Appendix~\ref{oa:data_integrity}). Under
either route the resolvent
\((\mathbf I-\mathbf A_\omega(\mathbf I-\boldsymbol\beta))^{-1}\) exists,
equals its Neumann sum, and is nonnegative.}
\end{lemma}

\begin{proof}[Proof of Lemma~\ref{lem:choke_kernel}]
Write \(\mathbf Z_\omega=\mathbf Y_\omega+[b_{ij}]\) for the decomposition
\eqref{eq:choke_split} into residual transit and re-matched flows, and fix
a disrupted sector-\(r\) block.

We first establish nonnegativity and support. The transit entries
\(Y^{(\omega)}_{ij}=\delta_{r(i)}d_{ij}Z_{ij}\) are data, strictly
positive wherever \(d_{ij}Z_{ij}>0\) by
Assumption~\ref{ass:attenuation}. Where the block is strictly feasible,
Lemma~\ref{prop:wellposed}(ii) gives \(b_{ij}=\mathring Z_{ij}u_iv_j\)
with \(u_i,v_j>0\), and \(\mathring Z_{ij}\) vanishes only where the
benchmark flow does, so the support is exactly the benchmark one. Where
strict feasibility fails the minimizer empties an open link and only
the containment survives, every disrupted cell still held positive by its
transit. Untouched blocks stay at benchmark, so
\(\mathbf A_\omega\ge\mathbf 0\) with
\(\operatorname{supp}(\mathbf A_\omega)\subseteq\operatorname{supp}(\mathbf A)\).
Block totals stay positive throughout. The unmet-demand account is supported only on buyers with displaced purchases, so a column with no displaced sector-\(r\) purchase keeps its full target \(\alpha_{rj}\), and
a disrupted one keeps at least its transit.

We next bound the column block sums. The column constraint of
Definition~\ref{def:choke_reroute} is met exactly where the residual
capacities and budgets are jointly attainable, so the block sums to
\(t^{(\omega)}_{rj}\), and on a placement-deficit block the unmet-demand
account absorbs \(U_{rj}\ge0\) of that target, with
\(\sum_jU_{rj}=U_r\) by Lemma~\ref{prop:wellposed}(i). Since
\(\ell_{rj}\ge0\) and \(U_{rj}\ge0\), each block sum is at most
\(\alpha_{rj}\), and summing over \(r\in\mathcal R_j\) with
\(\sum_r\alpha_{rj}\le1\) bounds every column of \(\mathbf A_\omega\) by
one.

On the seller side the row constraint gives
\(\sum_jZ^{(\omega)}_{ij}=g^{(\omega)}_i\) away from deficit blocks and
\(\sum_jZ^{(\omega)}_{ij}\le g^{(\omega)}_i\) on them, the unsold-output
account carrying the difference, and \(g^{(\omega)}_i\le s_i-f_i\) because
\(\tau Z^{\rm disp}_i\ge0\) in \eqref{eq:choke_rowtarget}.

Substochasticity and the norm bound close the argument. Lemma~A.2 of
\citet{bhatt2026_sanctions} derives both from three inputs alone, namely nonnegativity, block sums at or below their benchmark values, and
\(\beta_j\in(0,1)\). The preceding steps supply the first two for the rationed exchange, the third is maintained throughout, and their
argument then applies unchanged.
\end{proof}

Lemma~\ref{lem:choke_kernel} also carries the full-recovery limit of \citet[Lemma~A.6]{bhatt2026_sanctions} over to an exchange that holds the residual transit fixed. Their argument needs only that
both targets sit at their benchmark values when \(\tau=0\) and that each
buyer's block sums then return to the benchmark sector-\(r\) input share
\(\alpha_{rj}\). The budget \eqref{eq:choke_coltarget} and the capacity \eqref{eq:choke_rowtarget} sit at their benchmark values at
\(\tau=0\), and the lemma returns the
block sums to \(\alpha_{rj}\) on every placement-deficit-free block with the residual transit held fixed, so the
argument runs unchanged.

\subsection{Existence, selection, and interior uniqueness}
The no-closure economy solves the benchmark activity equation
\begin{equation*}\tag{\ref{eq:benchmark_activity_choke}}
\mathbf s
=
\mathbf f+\mathbf A(\mathbf I-\boldsymbol\beta)\mathbf s,
\end{equation*}
which Proposition~\ref{prop:equilibrium}(i) nests. Its unique nonnegative solution is Lemma~A.1 of \citet{bhatt2026_sanctions}, whose argument needs only a strictly positive leakage \(\beta_j\) at every buyer. The post-closure map whose fixed points the proposition studies is
\begin{equation*}\tag{\ref{eq:choke_operator}}
\mathcal T_\omega(\mathbf x,\mathbf h)
\;:=\;
\Bigl(
\mathbf G_\omega(\mathbf h)\,\mathbf f
\;+\;
\mathbf G_\omega(\mathbf h)\,\mathbf M_\omega\,\mathbf x,\;\;
\mathcal G_\omega(\mathbf h)
\Bigr).
\end{equation*}
Here \(\mathbf M_\omega\) is the propagation matrix of the scenario, \(\mathbf G_\omega\) the diagonal output ratio matrix, and \(\mathcal G_\omega\) the fulfillment rule. A buyer with no benchmark intermediate use takes \(\kappa^{(\omega)}_j=1\).

\begin{proposition}[The reported equilibrium]
\label{prop:equilibrium}
Fix \(\omega\), admissible \((\tau,\rho,\varrho)\), and residual transit shares \((\delta_r)_{r\in\mathcal M}\) as in
Assumption~\ref{ass:attenuation}. The post-closure map then has a
fixed point, the benchmark path selects the greatest one, and under the
interiority condition that selection is interior and the only strictly
positive fixed point.
\begin{enumerate}
\item[\textnormal{(i)}] The map
\(\mathcal T_\omega\) has at least one fixed point
\((\mathbf s_\omega,\mathbf h_\omega)\in[\mathbf 0,\mathbf s]\times[0,1]^{CS}\),
with \(\mathbf 0\le\mathbf s_\omega\le\mathbf s\), so the loss vector
\begin{equation*}
\boldsymbol\Delta\mathbf s
\;:=\;
\mathbf s-\mathbf s_\omega
\end{equation*}
is nonnegative entry by entry. When \(\omega\) severs no link, the
benchmark \((\mathbf s,\mathbf 1)\) is a fixed point, so the no-closure
economy of \eqref{eq:benchmark_activity_choke} is nested.
\item[\textnormal{(ii)}] The benchmark
path of Definition~\ref{def:selection} decreases to \(\mathbf h_\omega\),
the greatest fixed point of \(\mathcal G_\omega\) on \([0,1]^{CS}\),
and every fixed point of \(\mathcal T_\omega\) satisfies
\(\mathbf h^*\le\mathbf h_\omega\) and \(\mathbf x^*\le\mathbf s_\omega\)
entrywise.
\item[\textnormal{(iii)}] Suppose the scenario
satisfies the interiority condition
\begin{equation}
\kappa^{(\omega)}_i(\mathbf 1)
\;>\;
(1-\beta_i)^{-1/\varrho}
\qquad\text{for every active }i.
\label{eq:choke_cushion}
\end{equation}
Then \(\mathbf h_\omega\ge c_\omega\mathbf 1\) for a floor
\(c_\omega>0\) computable before any equilibrium is solved.
\item[\textnormal{(iv)}] Suppose the selected equilibrium is interior,
\(\mathbf h_\omega>\mathbf 0\). Then it is the only strictly positive
fixed point of \(\mathcal G_\omega\) in \([0,1]^{CS}\), and the reduced
iteration converges to it from every strictly positive start. Every other
fixed point has some producer at exactly zero fulfillment, with one of
that producer's used input sectors supplied entirely from within the zero
set.
\end{enumerate}
\end{proposition}

\begin{proof}[Proof of Proposition~\ref{prop:equilibrium}]
We first prove (ii), and with it the existence claim of (i). Both nests are read at the boundary by their
continuous extensions, \(\kappa^{(\omega)}_j:=0\) wherever
\(\theta^{(\omega)}_{rj}=0\) at some \(r\in\mathcal R_j\) and
\(G_j(0):=0\), the limits negative \(\rho\) and negative \(\varrho\)
deliver, so \(\mathcal G_\omega\) is continuous on \([0,1]^{CS}\). The map is entrywise
nondecreasing, and it maps \(\mathbf 1\) weakly below \(\mathbf 1\), so
the benchmark path of Definition~\ref{def:selection} is nonincreasing and its limit \(\mathbf h_\omega\) is a fixed point of \(\mathcal G_\omega\). This is the
monotone argument of Proposition~1 of
\citet{bhatt2026_sanctions}, which needs only those two properties. Every fixed point \(\mathbf h^*\) of \(\mathcal G_\omega\) lies below \(\mathbf 1\), so monotonicity keeps it below every iterate of the path and \(\mathbf h^*\le\mathbf h_\omega\) in the limit, which makes \(\mathbf h_\omega\) the greatest fixed point. The activity
map \eqref{eq:choke_selected} is entrywise nondecreasing in \(\mathbf h\)
through its nonnegative Neumann series, bounded uniformly by
Lemma~\ref{lem:choke_kernel}. The vector \(\mathbf s_\omega:=\mathbf x(\mathbf h_\omega)\) is therefore nonnegative, the pair \((\mathbf s_\omega,\mathbf h_\omega)\) is a fixed point of \(\mathcal T_\omega\), and every other fixed point has \(\mathbf x^*\le\mathbf s_\omega\). This proves (ii).

We next prove the rest of part (i), the bound \(\mathbf s_\omega\le\mathbf s\) and benchmark nesting. The capacities of the exchange cap the activity coordinate coordinatewise, in that for every \(\mathbf x\in[\mathbf 0,\mathbf s]\) and every supplier
\(i\), \(D^{(\omega)}_i(\mathbf x)\le g^{(\omega)}_i\le s_i-f_i\), so
the activity map returns at most
\(f_i+(s_i-f_i)=s_i\).\footnote{Effective demand
\(D^{(\omega)}_i(\mathbf x)=[\mathbf A_\omega(\mathbf I-\boldsymbol\beta)\mathbf x]_i\)
is nondecreasing in \(\mathbf x\), so on \([\mathbf 0,\mathbf s]\) it is
largest at \(\mathbf x=\mathbf s\), where it equals the delivered row sum
\(\sum_j Z^{(\omega)}_{ij}\). The seller-side step of
Lemma~\ref{lem:choke_kernel} bounds that sum by the capacity \(g^{(\omega)}_i\), itself at most \(s_i-f_i\) because
\(\tau Z^{\rm disp}_i\ge0\) in \eqref{eq:choke_rowtarget}. The first
inequality binds at \(\mathbf x=\mathbf s\) on placement-deficit-free
blocks, and the second exactly at sellers with no exposed sales or at
\(\tau=0\). The bound is the counterpart, for links that a closure disrupts only in part, of Lemma~A.4 of
\citet{bhatt2026_sanctions}, whose argument needs the row constraint to
bind at every seller, relaxed here to an inequality by the slack
accounts.} At the fulfillment vector \(\mathbf h_\omega\) the activity map is entrywise nondecreasing in \(\mathbf x\) and carries \([\mathbf 0,\mathbf s]\) into itself, so its iterates from \(\mathbf s\) decrease to a fixed point in that interval, and the fixed point is \(\mathbf s_\omega\) because the resolvent of Lemma~\ref{lem:choke_kernel} makes it unique. When \(\omega\) severs no
link, \(\theta^{(\omega)}_{rj}(\mathbf 1)=1\) on \(\mathcal R_j\), so
\(\mathcal G_\omega(\mathbf 1)=\mathbf 1\) and
\eqref{eq:choke_activity} reduces to the benchmark equation
\eqref{eq:benchmark_activity_choke}. This proves (i).

We next prove part (iii), the interiority floor. Define that floor as
\begin{equation}
c_i
:=
\Biggl[\frac{\beta_i}
{1-(1-\beta_i)\bigl(\kappa^{(\omega)}_i(\mathbf 1)\bigr)^{\varrho}}\Biggr]^{1/\varrho}
\in(0,1],
\qquad
c_\omega:=\min_i c_i.
\label{eq:choke_floor}
\end{equation}
The interiority condition \eqref{eq:choke_cushion} is what makes \(c_i\) well
defined. We show that the order interval \([c_\omega\mathbf 1,\mathbf 1]\) is invariant under
\(\mathcal G_\omega\). Let \(\mathbf h\ge c_\omega\mathbf 1\). By
homogeneity and monotonicity of the power mean,
\(\kappa^{(\omega)}_i(\mathbf h)\ge
\kappa^{(\omega)}_i(c_\omega\mathbf 1)
=c_\omega\,\kappa^{(\omega)}_i(\mathbf 1)\), so
\(G^{(\omega)}_i(\mathbf h)\ge
G_i\bigl(c_\omega\,\kappa^{(\omega)}_i(\mathbf 1)\bigr)\).
Writing \(\kappa:=\kappa^{(\omega)}_i(\mathbf 1)\) and
\(c\le c_i\), the inequality \(G_i(c\,\kappa)\ge c\) rearranges, using
\(\varrho<0\), to
\(c^{\varrho}\ge\beta_i/\bigl(1-(1-\beta_i)\kappa^{\varrho}\bigr)\), whose
right side is positive exactly under the interiority condition
\eqref{eq:choke_cushion} and which holds with equality at \(c=c_i\) of
\eqref{eq:choke_floor}. Since \(c\mapsto c^{\varrho}\) is decreasing, it
holds for every \(c\le c_i\), in particular \(c=c_\omega\). The benchmark
path starts at \(\mathbf 1\) and therefore never leaves
\([c_\omega\mathbf 1,\mathbf 1]\), so its limit obeys
\(\mathbf h_\omega\ge c_\omega\mathbf 1\). This proves (iii).

We next prove part (iv), interior uniqueness and convergence. Both are the envelope argument of Proposition~1.2 of the Online Appendix of \citet{bhatt2026_sanctions}. That argument
uses only entrywise monotonicity, the degree-one homogeneity of
\(\boldsymbol\kappa^{(\omega)}\), and the strict subhomogeneity of the
outer aggregator, all of which the rationed exchange preserves. The proof of their Proposition~1 establishes the first property and their Lemma~A.3 the third, and the second holds for any power mean.
The one ingredient to re-verify is that \(\mathbf h>\mathbf 0\) implies
\(\kappa^{(\omega)}_i(\mathbf h)>0\) for every active \(i\). Under the rationed exchange this holds because every used sector block of
every buyer's column keeps strictly positive delivered mass
(Lemma~\ref{lem:choke_kernel}), so no used input sector of an active
producer loses its entire effective demand mass. With that, two strictly
positive fixed points \(\mathbf g,\mathbf h\) with
\(t:=\min_i g_i/h_i<1\) would satisfy
\(g_i\ge G_i(t\,\kappa^{(\omega)}_i(\mathbf h))>t\,h_i\) for every
\(i\), contradicting attainment of the minimum. For convergence, write
\(t_0:=\min\{1,\min_i h^{(0)}_i/h^{(\omega)}_{i}\}\) and
\(T_0:=\max\{1,\max_i h^{(0)}_i/h^{(\omega)}_{i}\}\). Then
\begin{equation}
t_0^{\,\bar e_*^{\,u}}\,\mathbf h_\omega
\;\le\;
\mathbf h^{(u)}
\;\le\;
T_0^{\,\bar e_*^{\,u}}\,\mathbf h_\omega,
\qquad
\bar e_*
:=
\max_i \widetilde e_i\bigl(t_0\,
\kappa^{(\omega)}_i(\mathbf h_\omega)\bigr)
\;<\;1.
\label{eq:choke_envelope}
\end{equation}
Here \(\widetilde e_i\) is the output elasticity of intermediates
\eqref{eq:choke_shadow}, and the envelope follows from the log-log
concavity bound
\(G_i(t\kappa)\ge t^{\,\widetilde e_i(t\kappa)}G_i(\kappa)\) exactly as
in the argument of \citet{bhatt2026_sanctions}. Since \(\bar e_*<1\), both exponents tend to
zero and \(\mathbf h^{(u)}\to\mathbf h_\omega\).\footnote{The modulus
\(\bar e_*\) is relative rather than unweighted, and the two are
different objects. The benchmark modulus \(\max_j(1-\beta_j)\) bounds the delivery Jacobian's row
sums in the unweighted sup norm. No modulus below one is available in the
\(\ell_1\) norm on levels, where any bound carries the largest row sum of
\(\mathbf A_\omega\), which Lemma~\ref{lem:choke_kernel} leaves free and
which exceeds one wherever a seller supplies many buyers.} This proves
convergence and interior uniqueness.

It remains to verify the boundary statement in part~\textnormal{(iv)}.
Let \(\mathbf h^*\) be a fixed point with \(h^*_i=0\) at an active
producer \(i\). Since \(\beta_i\in(0,1)\) and \(\varrho<0\), the continuous
extension of the outer aggregator satisfies \(G_i(\kappa)=0\) if and only
if \(\kappa=0\). Hence \(\kappa^{(\omega)}_i(\mathbf h^*)=0\). Because
\(\rho<0\), the power mean \(\kappa^{(\omega)}_i\) can vanish only if
\(\theta^{(\omega)}_{ri}(\mathbf h^*)=0\) for at least one used sector
\(r\in\mathcal R_i\). But
\[
\theta^{(\omega)}_{ri}(\mathbf h^*)
=\frac{1}{\alpha_{ri}}
\sum_{k\in\mathcal N_r}h^*_k\,a^{(\omega)}_{ki}
\]
is a sum of nonnegative terms. Thus every effective-demand-positive
supplier \(k\) in that sector, \(a^{(\omega)}_{ki}>0\), also has
\(h^*_k=0\). One used input sector of \(i\) is therefore supplied
entirely from within the zero set, as asserted. This proves (iv).
\end{proof}

Parts (iii) and (iv) of Proposition~\ref{prop:equilibrium} need \(\beta_j\in(0,1)\) strictly,
since at \(\beta_i=0\) the outer aggregator is the identity and the strict subhomogeneity the
uniqueness argument turns on degrades to an equality. The interiority condition then reads
\(\kappa^{(\omega)}_i(\mathbf 1)>1\) and cannot hold, so the floor \eqref{eq:choke_floor}
collapses to zero. Parts (i) and (ii) use neither subhomogeneity nor the interiority condition.

The interiority condition \eqref{eq:choke_cushion} asks that enough of the damping the leakage
provides remain once input bundles have degraded, since the output elasticity of intermediates \eqref{eq:choke_shadow} rises toward one as they do. Online Appendix~\ref{app:stability_numerics} reports where the condition, which is sufficient and not necessary, fails on the data
and verifies interiority directly.

\subsection{The two retention objects}
Two retention objects have to be kept apart here, and the residual transit floors only the first of them. Write
\(e^{(\omega)}_{rj}:=\bigl(\sum_{i\in\mathcal N_r}d_{ij}a_{ij}\bigr)
/\alpha_{rj}\in[0,1]\) for the exposed share of buyer \(j\)'s benchmark
sector-\(r\) block, and
\(\bar\theta^{(\omega)}_{rj}:=\theta^{(\omega)}_{rj}(\mathbf 1)
=\bigl(\sum_{i\in\mathcal N_r}a^{(\omega)}_{ij}\bigr)/\alpha_{rj}\) for
its effective demand retention. Recall from Lemma~\ref{lem:choke_kernel} that \(U_{rj}\ge0\) is the unmet demand the block absorbs at that buyer, and write \(\upsilon^{(\omega)}_{rj}:=U_{rj}/\bigl(\alpha_{rj}(1-\beta_j)s_j\bigr)\) for its share of the block. Dividing the delivered column total of
Lemma~\ref{lem:choke_kernel} by \(\alpha_{rj}\) gives the exact identity
\begin{equation}
\bar\theta^{(\omega)}_{rj}
=1-\tau(1-\delta_r)\,e^{(\omega)}_{rj}-\upsilon^{(\omega)}_{rj}.
\label{eq:standing_retention}
\end{equation}
The first two terms of \eqref{eq:standing_retention} are the intended retention ratio
\(t^{(\omega)}_{rj}/\alpha_{rj}=1-\tau\ell_{rj}/\alpha_{rj}\), the share of buyer \(j\)'s
benchmark sector-\(r\) input that its budget keeps, which is the \(a_{ij}\)-weighted mean of the
retention factors of \eqref{eq:choke_target_link}. The effective demand retention falls short of
the intended retention ratio by the unmet-demand share \(\upsilon^{(\omega)}_{rj}\) alone.
A deficit-free block carries \(\upsilon^{(\omega)}_{rj}=0\) and therefore
retains \(1-\tau(1-\delta_r)e^{(\omega)}_{rj}\) exactly, which the
exposed share bounds below by \(1-\tau(1-\delta_r)\). The residual transit
holds \(a^{(\omega)}_{ij}\ge\delta_rd_{ij}a_{ij}\) cell by cell whatever
the deficit, so \(\bar\theta^{(\omega)}_{rj}\ge\delta_re^{(\omega)}_{rj}\)
in general, and every used block keeps \(\bar\theta^{(\omega)}_{rj}>0\).
An unexposed block sits at its benchmark value of one, and an exposed
one keeps its transit.

A fully exposed block admits exact formulas. There \(e^{(\omega)}_{rj}=1\)
forces \(d_{ij}=1\) at every positive cell, the unexposed flows vanish in
that column, and
\(a^{(\omega)}_{ij}=\delta_ra_{ij}\),
\(\bar\theta^{(\omega)}_{rj}=\delta_r\), and
\(\upsilon^{(\omega)}_{rj}=(1-\tau)(1-\delta_r)\) by
\eqref{eq:standing_retention}. The residual transit share is thus exactly the
effective demand retention of a fully exposed block, and such a block
carries a placement deficit at every \(\tau<1\), so the deficit-free
case and the fully exposed case never coincide away from that boundary.

Realized retention is the second object, and it factors. Write
\(\mathcal S^{(\omega)}_{rj}:=\{i\in\mathcal N_r:a^{(\omega)}_{ij}>0\}\)
for the block's effective demand support and
\(\pi^{(\omega)}_{ij}:=a^{(\omega)}_{ij}\big/
\sum_{k\in\mathcal N_r}a^{(\omega)}_{kj}\) for the weights it induces,
positive on that support and summing to one. Then
\begin{equation}
\theta^{(\omega)}_{rj}(\mathbf h)
=\bar\theta^{(\omega)}_{rj}
\sum_{i\in\mathcal S^{(\omega)}_{rj}}\pi^{(\omega)}_{ij}\,h_i,
\label{eq:realized_retention}
\end{equation}
so realized retention is effective demand retention times the
order-weighted mean fulfillment of the block's own suppliers. Two
consequences of \eqref{eq:realized_retention} follow at once. Realized retention is strictly positive exactly when at least one supplier of \(\mathcal S^{(\omega)}_{rj}\) has strictly
positive fulfillment, and at the benchmark-selected equilibrium, with
\(\underline h_\omega:=\min_ih^{(\omega)}_i\),
\begin{equation}
\theta^{(\omega)}_{rj}(\mathbf h_\omega)
\;\ge\;\underline h_\omega\,\bar\theta^{(\omega)}_{rj}
\;\ge\;\underline h_\omega\,\delta_r\,e^{(\omega)}_{rj}
\;\ge\;0,
\label{eq:transit_floor}
\end{equation}
which reads \(\underline h_\omega\delta_r\) on a fully exposed block.

The last inequality of \eqref{eq:transit_floor} is not strict without a
positive fulfillment bound, and the residual transit share alone supplies none. Under
the interiority condition \eqref{eq:choke_cushion},
Proposition~\ref{prop:equilibrium}(iii) delivers
\(\underline h_\omega\ge c_\omega>0\) and with it
\(\theta^{(\omega)}_{rj}(\mathbf h_\omega)\ge c_\omega\delta_r
\ge c_\omega\underline\delta>0\) on every fully exposed block. The
same conclusion holds with \(\underline h_\omega\) in place of
\(c_\omega\) wherever interiority is verified after solving. Without one of the two, realized retention can vanish at a strictly positive residual transit share, as it does in a cycle of two producers whose leakage parameters and residual transit shares violate the interiority condition \eqref{eq:choke_cushion}.
With \(\rho<0\) the collapse of
any used block, not only a fully exposed one, sends the intermediate input efficiency factor to zero, and an interior fulfillment vector together
with \eqref{eq:transit_floor} is what keeps the feedback of the
fulfillment fixed point away from that boundary.

Existence of the post-closure equilibrium asks only that each \(\delta_r\) be positive. Its
stability asks for the uniform floor \(\underline\delta\), which costs nothing with finitely many seaborne sectors, \(\underline\delta=\min_r\delta_r\). A complete closure, \(\delta_r=0\),
violates Assumption~\ref{ass:attenuation}. Fully severed links then drop out of the trade
network and the formal results no longer cover the case, although the equilibrium can still be
computed, and the loss it returns is larger (Online Appendix~\ref{app:stability_numerics}).

\newpage
% ================================================================
\part{The World Loss through Lost Orders and Joint Closures}
% ================================================================

This part holds an accounting identity for the world loss through lost orders and the comparison of a joint closure with the sum of its single closures on which Section~\ref{subsec:superadditivity} of the paper draws.

% ----------------------------------------------------------------
\section{The World Loss through Lost Orders}
\label{app:loss_identity}
% ----------------------------------------------------------------

GDP is measured by its accounting share throughout this section, \(\gamma_j=\beta_j\), so the GDP of country-sector \(j\) is \(\beta_js_j\), its gross output less its intermediate purchases, and benchmark world GDP is \(\boldsymbol\gamma^{\!\top}\mathbf s>0\). Every column of \(\mathbf A\) sums to one, \(\beta_j\in(0,1)\), and \(s_j>0\) at every country-sector. We write \(\mathbf M_0:=\mathbf A(\mathbf I-\boldsymbol\beta)\) and \(\mathbf M_\omega:=\mathbf A_\omega(\mathbf I-\boldsymbol\beta)\) for the propagation matrices of the benchmark and of a scenario \(\omega\), \(\mathbf s^{\ordonly}_\omega:=(\mathbf I-\mathbf M_\omega)^{-1}\mathbf f\) for the activity with lost orders alone, \(L^{\ordonly}_W:=\boldsymbol\gamma^{\!\top}(\mathbf s-\mathbf s^{\ordonly}_\omega)/\boldsymbol\gamma^{\!\top}\mathbf s\) for the world loss through lost orders, i.e., the world loss of the economy with lost orders alone, and
\begin{equation}
q_j(\omega)
:=(1-\beta_j)-\sum_i\bigl[\mathbf M_\omega\bigr]_{ij}
\label{eq:dissipative_shortfall}
\end{equation}
for the fall in the column of buyer \(j\) in the propagation matrix, i.e., the intermediate orders per unit of its activity that buyer \(j\) no longer places with any seller. By Lemma~\ref{lem:choke_kernel} the sector-\(r\) block of column \(j\) of \(\mathbf A_\omega\) sums to \(\alpha_{rj}-\tau\ell_{rj}-U_{rj}/((1-\beta_j)s_j)\), and hence
\begin{equation}
q_j(\omega)
=(1-\beta_j)\Bigl(\tau\sum_{r}\ell_{rj}+\bar U_j\Bigr)\;\ge\;0,
\label{eq:colsum_loss}
\end{equation}
where \(\bar U_j:=\sum_rU_{rj}/((1-\beta_j)s_j)\) is the unmet demand that the column absorbs on placement-deficit blocks. The fall in the column is therefore the unrecovered share of the buyer's displaced purchases plus what the exchange cannot place for it. The identity of this section rests on one accounting property of the benchmark network, which we state first.

\begin{lemma}[Unit benchmark GDP multipliers]
\label{lem:unit_multiplier}
Suppose that GDP is measured by its accounting share \(\gamma_j=\beta_j\). The benchmark propagation matrix
\(\mathbf M_0\), of dimension \(CS\times CS\), then
satisfies
\[
\boldsymbol\gamma^{\!\top}(\mathbf I-\mathbf M_0)^{-1}\;=\;\mathbf 1^{\!\top}.
\]
\end{lemma}

\begin{proof}
Because each column of \(\mathbf A\) sums to one, the column sums of
\(\mathbf M_0\) are
\(\mathbf 1^{\!\top}\mathbf M_0=(1-\beta_j)_j\), so
\(\mathbf 1^{\!\top}(\mathbf I-\mathbf M_0)=(\beta_j)_j=\boldsymbol\gamma^{\!\top}\).
Multiply on the right by \((\mathbf I-\mathbf M_0)^{-1}\). The entry in row \(i\) and column \(j\) of that inverse is the activity of country-sector \(i\) that a unit of autonomous demand at country-sector \(j\) calls forth, so the identity says that each column of the inverse sums to one once every row is weighted by its GDP content.
\end{proof}

A dollar of demand placed anywhere therefore generates a dollar of GDP along the orders that it sets off. Lemma~\ref{lem:loss_identity} applies the same accounting to the orders that a closure removes.

\begin{lemma}[The world loss through lost orders]
\label{lem:loss_identity}
Suppose that GDP is measured by its accounting share. For every scenario \(\omega\) the world loss through lost orders satisfies
\begin{equation}
(\boldsymbol\gamma^{\!\top}\mathbf s)\,L^{\ordonly}_W
=\sum_j q_j(\omega)\,\bigl[\mathbf s^{\ordonly}_\omega\bigr]_j.
\label{eq:world_loss_colsum}
\end{equation}
\end{lemma}

\begin{proof}
By the proof of Lemma~\ref{lem:unit_multiplier}, \(\boldsymbol\gamma^{\!\top}=\mathbf 1^{\!\top}(\mathbf I-\mathbf M_0)\), and the benchmark activity solves \((\mathbf I-\mathbf M_0)\mathbf s=\mathbf f\), so \(\boldsymbol\gamma^{\!\top}\mathbf s=\mathbf 1^{\!\top}\mathbf f\). The activity with lost orders alone solves \(\mathbf s^{\ordonly}_\omega=\mathbf f+\mathbf M_\omega\mathbf s^{\ordonly}_\omega\), so \(\mathbf 1^{\!\top}\mathbf f=\mathbf 1^{\!\top}(\mathbf I-\mathbf M_\omega)\,\mathbf s^{\ordonly}_\omega\). Hence
\begin{equation*}
\boldsymbol\gamma^{\!\top}\bigl(\mathbf s-\mathbf s^{\ordonly}_\omega\bigr)
=\mathbf 1^{\!\top}(\mathbf I-\mathbf M_\omega)\,\mathbf s^{\ordonly}_\omega
-\mathbf 1^{\!\top}(\mathbf I-\mathbf M_0)\,\mathbf s^{\ordonly}_\omega
=\mathbf 1^{\!\top}(\mathbf M_0-\mathbf M_\omega)\,\mathbf s^{\ordonly}_\omega.
\end{equation*}
Entry \(j\) of the row vector \(\mathbf 1^{\!\top}(\mathbf M_0-\mathbf M_\omega)\) is \(q_j(\omega)\) by \eqref{eq:dissipative_shortfall}, because column \(j\) of \(\mathbf M_0\) sums to \(1-\beta_j\). This proves \eqref{eq:world_loss_colsum}.
\end{proof}

The right side of \eqref{eq:world_loss_colsum} is the value of the orders that buyers no longer place with any seller, each buyer taken at its activity with lost orders alone. The identity values every such dollar at one, wherever in the network the order is removed and however many rounds the loss travels upstream. The distribution of the surviving orders of a column among the sellers of a block does not enter the right side, so it acts on the world loss through lost orders only through the activity vector.

% ----------------------------------------------------------------
\section{A Joint Closure against the Sum of Its Single Closures}
\label{app:joint_pair}
% ----------------------------------------------------------------

Section~\ref{subsec:superadditivity} of the paper compares the world loss of a joint closure with the sum of the world losses of its single closures. We make the comparison exact in the simplest configuration that poses it, in which two gates reach one buyer and no other and the buyer passes its shortfall on to no other producer. Write \(p^{(k)}_{ij}:=\bigl(1-e^{(k)}_{c(i)c(j)}\bigr)\mathbbm 1^{(k)}_{ij}\) for the share of the flow on link \(i\to j\) that the closure of gate \(k\) alone would cost. We say that gate \(k\) exposes the link \(i\to j\) when \(a_{ij}\,p^{(k)}_{ij}>0\), i.e., when the link carries benchmark trade that transits the gate and that the detour around the gate does not save in full.

\begin{definition}[Isolated pair of gates]
\label{def:isolated_pair}
Two gates form an isolated pair at a buyer \(j^*\), at the unrecovered share \(\tau\), when the following hold.
\begin{enumerate}
\item[\textnormal{(i)}] The buyer sells to final users alone, \(a_{j^*j}=0\) for every country-sector \(j\).
\item[\textnormal{(ii)}] Every link that either gate exposes is a link into the buyer \(j^*\).
\item[\textnormal{(iii)}] Every sector block is placement-deficit-free under the closure of either gate and under the closure of both.
\end{enumerate}
\end{definition}

The configuration is a device for setting apart what two closures do to one another at a single buyer, but it describes no pair of the world's gates. Condition~(ii) leaves the purchases of every other buyer in place. Condition~(i) keeps the one producer that goes short out of every input bundle. Condition~(iii) makes the shortfall of the buyer the unrecovered share of its displaced purchases and nothing more.

Fix an isolated pair of gates at \(j^*\), and label the two gates \(1\) and \(2\). Write \(\cset{}:=\{1,2\}\) for the joint closure and \(\cset{}_1\) and \(\cset{}_2\) for the two single closures. The buyer is the same throughout. We drop its index from the objects that belong to it. For each of the three closure sets \(\cset{}'\) we write \(\ell_r(\cset{}')\) for the displaced sector-\(r\) input share \eqref{eq:partial_disrupted_mass} of the buyer under \(\cset{}'\), and \(\ell(\cset{}'):=\sum_r\ell_r(\cset{}')\) for the share of all its intermediate purchases that \(\cset{}'\) displaces. We write \(\alpha_r:=\alpha_{rj^*}\) for the benchmark input shares of the buyer and \(\beta:=\beta_{j^*}\) for its leakage parameter, and
\begin{equation}
\kappa(\cset{}')
:=\biggl(\sum_{r\in\mathcal R_{j^*}}\alpha_r\Bigl(1-\frac{\tau\,\ell_r(\cset{}')}{\alpha_r}\Bigr)^{\!\rho}\biggr)^{1/\rho}
\label{eq:pair_kappa}
\end{equation}
for the intermediate input efficiency factor \eqref{eq:choke_ces_kappa} of a buyer that is short of the share \(\tau\ell_r(\cset{}')/\alpha_r\) of each input sector \(r\). Whether a block is placement-deficit-free can depend on the unrecovered share \(\tau\), so a statement that holds at several values of \(\tau\) concerns the values at which the gates form an isolated pair. The proposition gives the two world losses in closed form, compares the trade that the three closures displace, and signs the departure of the joint loss from the sum of the two single losses.

\begin{proposition}[A joint closure at an isolated pair of gates]
\label{prop:joint_pair}
Suppose that GDP is measured by its accounting share, \(\gamma_j=\beta_j\). Fix an isolated pair of gates at a buyer \(j^*\).
\begin{enumerate}
\item[\textnormal{(i)}] For each of the three closure sets \(\cset{}'\),
\begin{equation}
\begin{aligned}
(\boldsymbol\gamma^{\!\top}\mathbf s)\,L^{\ordonly}_W(\cset{}')
&=s_{j^*}\,(1-\beta)\,\tau\,\ell(\cset{}'),\\
(\boldsymbol\gamma^{\!\top}\mathbf s)\,L_W(\cset{}')
&=s_{j^*}\Bigl[1-\bigl(1-(1-\beta)\,\tau\,\ell(\cset{}')\bigr)\,G_{j^*}\bigl(\kappa(\cset{}')\bigr)\Bigr].
\end{aligned}
\label{eq:pair_loss}
\end{equation}
\item[\textnormal{(ii)}] For every input sector \(r\), \(\ell_r(\cset{})\le\ell_r(\cset{}_1)+\ell_r(\cset{}_2)\), and equality holds at every \(r\) if and only if the two gates expose no common link. If the two gates belong to one detour class and \(p^{(1)}_{ij^*}=p^{(2)}_{ij^*}\) on every link into the buyer, then \(L_W(\cset{})=L_W(\cset{}_1)=L_W(\cset{}_2)\).
\item[\textnormal{(iii)}] Suppose that the two gates expose no common link. Then \(L^{\ordonly}_W(\cset{})=L^{\ordonly}_W(\cset{}_1)+L^{\ordonly}_W(\cset{}_2)\), and
\begin{equation}
\begin{aligned}
(\boldsymbol\gamma^{\!\top}\mathbf s)\Bigl[L_W(\cset{})-L_W(\cset{}_1)-L_W(\cset{}_2)\Bigr]
={}&s_{j^*}\,(1-\beta)\,\tau^{2}\,\biggl[(1-\rho)\sum_{r\in\mathcal R_{j^*}}\frac{\ell_r(\cset{}_1)\,\ell_r(\cset{}_2)}{\alpha_r}\\
&-\bigl((3-\rho)-\beta\,(3-\varrho)\bigr)\,\ell(\cset{}_1)\,\ell(\cset{}_2)\biggr]+O\bigl(\tau^{3}\bigr),
\end{aligned}
\label{eq:pair_second_order}
\end{equation}
while \((\boldsymbol\gamma^{\!\top}\mathbf s)\,L_W(\cset{}_k)=2\,s_{j^*}(1-\beta)\,\tau\,\ell(\cset{}_k)+O(\tau^{2})\) for \(k=1,2\). Each remainder is at most a multiple of the stated power of \(\tau\), and the multiple does not depend on \(\tau\).
\item[\textnormal{(iv)}] Suppose that the two gates expose no common link and that each gate exposes some link. If no input sector holds a link that gate \(1\) exposes and a link that gate \(2\) exposes, and \(\varrho\ge\rho\), then \(L_W(\cset{})<L_W(\cset{}_1)+L_W(\cset{}_2)\). If instead every exposed link belongs to one input sector \(r\), the bracket of \eqref{eq:pair_second_order} has the sign of \((1-\rho)+\alpha_r\bigl(\beta(3-\varrho)-(3-\rho)\bigr)\), which is positive whenever \(\alpha_r\le\tfrac13\), and \(L_W(\cset{})>L_W(\cset{}_1)+L_W(\cset{}_2)\) then holds at every small enough \(\tau\). At \(\rho=\varrho=-1\) that inequality holds at every \(\tau\) when \(\alpha_r(1-\beta)<\tfrac12\), and the reverse strict inequality holds at every \(\tau\) when \(\alpha_r(1-\beta)>\tfrac12\).
\end{enumerate}
\end{proposition}

\begin{proof}
Fix one of the three closure sets, write \(\omega\) for its scenario, and write \(\ell_r\), \(\ell\), and \(\kappa\) without their argument. By condition~(ii) of Definition~\ref{def:isolated_pair} the disrupted fraction \eqref{eq:partial_choke_disruption} is zero on every benchmark link that does not lead to \(j^*\), so \(\ell_{rj}=0\) for every other buyer \(j\) and every \(r\). By condition~(iii) the exchange meets every budget \eqref{eq:choke_coltarget} (Lemma~\ref{lem:choke_kernel}), so the sector-\(r\) block of column \(j\) of \(\mathbf A_\omega\) sums to \(\alpha_{rj}\) at every \(j\ne j^*\) and to \(\alpha_r-\tau\ell_r\) at \(j^*\). Row \(j^*\) of \(\mathbf A_\omega\) is zero, because row \(j^*\) of \(\mathbf A\) is zero by condition~(i) and \(\operatorname{supp}(\mathbf A_\omega)\subseteq\operatorname{supp}(\mathbf A)\).

We first prove (i), and begin with the fulfillment vector. Start the benchmark path of Definition~\ref{def:selection} at \(\mathbf h=\mathbf 1\). The retention ratio \eqref{eq:choke_tilde_f} of a block is then its total over its benchmark share, which is one at every \(j\ne j^*\) and \(1-\tau\ell_r/\alpha_r\) at \(j^*\). The first step of the path therefore leaves every coordinate other than \(j^*\) at one and takes coordinate \(j^*\) to \(G_{j^*}(\kappa)\). The one supplier whose rate is now below one is \(j^*\), which has zero weight in every column, so the second step changes no retention ratio. The path has reached its limit, with \(h^{(\omega)}_i=1\) for \(i\ne j^*\) and \(h^{(\omega)}_{j^*}=G_{j^*}(\kappa)\).

We next compute the two losses. Row \(j^*\) of \(\mathbf M_\omega\) is zero and every other producer fills its orders in full, so the equilibrium condition \eqref{eq:choke_activity} reads \(s^{(\omega)}_i=f_i+[\mathbf M_\omega\mathbf s_\omega]_i\) at \(i\ne j^*\) and \(s^{(\omega)}_{j^*}=h^{(\omega)}_{j^*}f_{j^*}\), and the benchmark system gives \(s_{j^*}=f_{j^*}\) by the same zero row. Summing the coordinates,
\[
\mathbf 1^{\!\top}(\mathbf I-\mathbf M_\omega)\,\mathbf s_\omega
=\mathbf 1^{\!\top}\mathbf f-\bigl(1-h^{(\omega)}_{j^*}\bigr)s_{j^*}.
\]
By the proof of Lemma~\ref{lem:unit_multiplier}, \(\boldsymbol\gamma^{\!\top}=\mathbf 1^{\!\top}(\mathbf I-\mathbf M_0)\) and \(\boldsymbol\gamma^{\!\top}\mathbf s=\mathbf 1^{\!\top}\mathbf f\), and entry \(j\) of \(\mathbf 1^{\!\top}(\mathbf M_0-\mathbf M_\omega)\) is the fall \(q_j(\omega)\) of \eqref{eq:dissipative_shortfall}. Hence
\[
\boldsymbol\gamma^{\!\top}(\mathbf s-\mathbf s_\omega)
=\mathbf 1^{\!\top}\mathbf f-\mathbf 1^{\!\top}(\mathbf I-\mathbf M_\omega)\,\mathbf s_\omega+\mathbf 1^{\!\top}(\mathbf M_0-\mathbf M_\omega)\,\mathbf s_\omega
=\bigl(1-h^{(\omega)}_{j^*}\bigr)s_{j^*}+\sum_jq_j(\omega)\,s^{(\omega)}_j.
\]
By \eqref{eq:colsum_loss} and condition~(iii), \(q_j(\omega)=0\) for \(j\ne j^*\) and \(q_{j^*}(\omega)=(1-\beta)\tau\ell\). Substituting \(s^{(\omega)}_{j^*}=h^{(\omega)}_{j^*}s_{j^*}\) and \(h^{(\omega)}_{j^*}=G_{j^*}(\kappa)\) gives the second line of \eqref{eq:pair_loss}. For the first line, Lemma~\ref{lem:loss_identity} gives \((\boldsymbol\gamma^{\!\top}\mathbf s)L^{\ordonly}_W=q_{j^*}(\omega)\,[\mathbf s^{\ordonly}_\omega]_{j^*}\), and \([\mathbf s^{\ordonly}_\omega]_{j^*}=f_{j^*}=s_{j^*}\) because row \(j^*\) of \(\mathbf M_\omega\) is zero. This proves (i).

We next prove (ii). Fix a link \(i\to j^*\), write \(p_k:=p^{(k)}_{ij^*}\), and write \(d(\cset{}')\) for the disrupted fraction \eqref{eq:partial_choke_disruption} of the link under \(\cset{}'\), so \(d(\cset{}_k)=p_k\). If the two gates belong to one detour class, then \(d(\cset{})=\max\{p_1,p_2\}\), and if they belong to different classes, then \(d(\cset{})=1-(1-p_1)(1-p_2)=p_1+p_2-p_1p_2\). In both cases \(d(\cset{})\le p_1+p_2\), with equality exactly when \(p_1p_2=0\). The displaced input share \eqref{eq:partial_disrupted_mass} adds the disrupted fractions of the links of a sector with the weights \((1-\delta)\,a_{ij^*}\), which are positive on the benchmark links. The inequality therefore passes to \(\ell_r\), and equality holds at every \(r\) exactly when \(a_{ij^*}p_1p_2=0\) on every link, i.e., when the two gates expose no common link. If the gates belong to one class and \(p_1=p_2\) on every link, then \(d(\cset{})=d(\cset{}_1)=d(\cset{}_2)\) on every link, so the three closure sets have the same displaced input shares and, by \eqref{eq:pair_loss}, the same world loss. This proves (ii).

We next prove (iii). For a vector \(\boldsymbol\epsilon=(\epsilon_r)_{r\in\mathcal R_{j^*}}\) with entries in \([0,1)\), write \(\bar\epsilon:=\sum_r\alpha_r\epsilon_r\) and \(|\boldsymbol\epsilon|:=\max_r\epsilon_r\), and
\[
\kappa(\boldsymbol\epsilon):=\Bigl(\sum_r\alpha_r(1-\epsilon_r)^{\rho}\Bigr)^{1/\rho},
\qquad
\Phi(\boldsymbol\epsilon):=1-\bigl(1-(1-\beta)\,\bar\epsilon\bigr)\,G_{j^*}\bigl(\kappa(\boldsymbol\epsilon)\bigr).
\]
By \eqref{eq:pair_loss}, \((\boldsymbol\gamma^{\!\top}\mathbf s)L_W(\cset{}')=s_{j^*}\,\Phi(\boldsymbol\epsilon(\cset{}'))\), where \(\epsilon_r(\cset{}'):=\tau\ell_r(\cset{}')/\alpha_r\) is the shortfall of the buyer in input sector \(r\). When the two gates expose no common link, part~(ii) gives \(\boldsymbol\epsilon(\cset{})=\boldsymbol\epsilon(\cset{}_1)+\boldsymbol\epsilon(\cset{}_2)\) and \(\ell(\cset{})=\ell(\cset{}_1)+\ell(\cset{}_2)\), and the first line of \eqref{eq:pair_loss} then gives the additivity of the world loss through lost orders.

Expanding each power, \(\sum_r\alpha_r(1-\epsilon_r)^{\rho}=1-\rho\,\bar\epsilon+\tfrac12\rho(\rho-1)\sum_r\alpha_r\epsilon_r^2+O(|\boldsymbol\epsilon|^3)\), and raising the sum to the power \(1/\rho\) gives
\[
1-\kappa(\boldsymbol\epsilon)
=\bar\epsilon+\tfrac12(1-\rho)\Bigl(\sum_r\alpha_r\epsilon_r^2-\bar\epsilon^{\,2}\Bigr)+O\bigl(|\boldsymbol\epsilon|^3\bigr).
\]
The outer nest has \(G_{j^*}(1)=1\), slope \(1-\beta\), and second derivative \(-\beta(1-\beta)(1-\varrho)\) at \(\kappa=1\), so \(1-G_{j^*}(\kappa)=(1-\beta)(1-\kappa)+\tfrac12\beta(1-\beta)(1-\varrho)(1-\kappa)^2+O((1-\kappa)^3)\). Since \(\Phi=(1-G_{j^*})+(1-\beta)\,\bar\epsilon-(1-\beta)\,\bar\epsilon\,(1-G_{j^*})\), the two expansions give
\begin{equation}
\Phi(\boldsymbol\epsilon)
=2(1-\beta)\,\bar\epsilon
+\tfrac12(1-\beta)\Bigl[(1-\rho)\sum_r\alpha_r\epsilon_r^2-\bigl((3-\rho)-\beta(3-\varrho)\bigr)\bar\epsilon^{\,2}\Bigr]
+O\bigl(|\boldsymbol\epsilon|^3\bigr).
\label{eq:pair_expansion}
\end{equation}
The first term is linear in \(\boldsymbol\epsilon\) and the second is a quadratic form, so for two such vectors \(\boldsymbol\epsilon\) and \(\boldsymbol\epsilon'\)
\[
\Phi(\boldsymbol\epsilon+\boldsymbol\epsilon')-\Phi(\boldsymbol\epsilon)-\Phi(\boldsymbol\epsilon')
=(1-\beta)\Bigl[(1-\rho)\sum_r\alpha_r\epsilon_r\epsilon'_r-\bigl((3-\rho)-\beta(3-\varrho)\bigr)\bar\epsilon\,\bar\epsilon'\Bigr]+O\bigl((|\boldsymbol\epsilon|+|\boldsymbol\epsilon'|)^3\bigr).
\]
Set \(\boldsymbol\epsilon=\boldsymbol\epsilon(\cset{}_1)\) and \(\boldsymbol\epsilon'=\boldsymbol\epsilon(\cset{}_2)\), for which \(\sum_r\alpha_r\epsilon_r\epsilon'_r=\tau^2\sum_r\ell_r(\cset{}_1)\ell_r(\cset{}_2)/\alpha_r\) and \(\bar\epsilon\,\bar\epsilon'=\tau^2\ell(\cset{}_1)\ell(\cset{}_2)\). The last display is then \eqref{eq:pair_second_order}, and \eqref{eq:pair_expansion} at \(\boldsymbol\epsilon(\cset{}_k)\) is the expansion of the single losses. A displaced input share is at most \((1-\underline\delta)\,\alpha_r\), so \(\epsilon_r(\cset{}')\le\tau(1-\underline\delta)\). The function \(\Phi\) has bounded derivatives of every order on the cube of vectors with entries in \([0,1-\underline\delta]\), so each remainder is at most a multiple of the stated power of \(\tau\) with a multiple that does not depend on \(\tau\). This proves (iii).

We last prove (iv). Write \(\boldsymbol\epsilon:=\boldsymbol\epsilon(\cset{}_1)\) and \(\boldsymbol\epsilon':=\boldsymbol\epsilon(\cset{}_2)\), neither of which is zero because each gate exposes a link. Write \(G_1\), \(G_2\), and \(G_{12}\) for the output ratio \(G_{j^*}(\kappa(\cdot))\) of the buyer at \(\boldsymbol\epsilon\), at \(\boldsymbol\epsilon'\), and at \(\boldsymbol\epsilon+\boldsymbol\epsilon'\), and \(q_1:=(1-\beta)\,\bar\epsilon\) and \(q_2:=(1-\beta)\,\bar\epsilon'\) for the fall \eqref{eq:dissipative_shortfall} in the column of the buyer under each single closure. The definition of \(\Phi\) gives
\begin{equation}
\Phi(\boldsymbol\epsilon+\boldsymbol\epsilon')-\Phi(\boldsymbol\epsilon)-\Phi(\boldsymbol\epsilon')
=-\bigl[(1-G_1)+(1-G_2)-(1-G_{12})\bigr]-q_1\,(G_1-G_{12})-q_2\,(G_2-G_{12}).
\label{eq:pair_identity}
\end{equation}
Suppose first that no input sector holds a link that gate \(1\) exposes and a link that gate \(2\) exposes, and that \(\varrho\ge\rho\). Then \(\epsilon_r\epsilon'_r=0\) for every \(r\). Write \(Y(\boldsymbol\epsilon):=\sum_r\alpha_r\bigl((1-\epsilon_r)^{\rho}-1\bigr)\). Then \(\kappa(\boldsymbol\epsilon)^{\rho}=1+Y(\boldsymbol\epsilon)\) and \(1-G_{j^*}(\kappa(\boldsymbol\epsilon))=\psi(Y(\boldsymbol\epsilon))\) with
\[
\psi(t):=1-\bigl[\beta+(1-\beta)(1+t)^{\varrho/\rho}\bigr]^{1/\varrho},
\qquad
\psi'(t)=\frac{1-\beta}{-\rho}\,(1+t)^{\varrho/\rho-1}\,\bigl[\beta+(1-\beta)(1+t)^{\varrho/\rho}\bigr]^{1/\varrho-1}.
\]
The function \(Y\) is positive at \(\boldsymbol\epsilon\) and at \(\boldsymbol\epsilon'\), and \(Y(\boldsymbol\epsilon+\boldsymbol\epsilon')=Y(\boldsymbol\epsilon)+Y(\boldsymbol\epsilon')\) because no input sector is short under both closures. The function \(\psi\) vanishes at zero, and its derivative is positive and strictly decreasing on \(t\ge0\), since \(0<\varrho/\rho\le1\) and \(1/\varrho-1<0\). A strictly concave function that vanishes at zero is strictly sub-additive, so \(1-G_{12}<(1-G_1)+(1-G_2)\), and \(G_{12}\le\min\{G_1,G_2\}\) because \(\psi\) is increasing. The right side of \eqref{eq:pair_identity} is therefore negative, which is the first claim.

Suppose next that every exposed link belongs to one input sector \(r\). Then \(\boldsymbol\epsilon\) and \(\boldsymbol\epsilon'\) vanish outside coordinate \(r\). The bracket of \eqref{eq:pair_second_order} is then
\[
\frac{\ell(\cset{}_1)\,\ell(\cset{}_2)}{\alpha_r}\,\Bigl[(1-\rho)+\alpha_r\bigl(\beta(3-\varrho)-(3-\rho)\bigr)\Bigr].
\]
For \(\alpha_r\le\tfrac13\) the term in square brackets is at least \((1-\rho)-\tfrac13(3-\rho)=-\tfrac23\rho>0\). The leading term of \eqref{eq:pair_second_order} is then positive and of the order of \(\tau^2\), so it exceeds the remainder at every small enough \(\tau\). At \(\rho=\varrho=-1\), write \(\varsigma:=\alpha_r(1-\beta)\) for the share of the input sector in the cost of the buyer and \(a:=\epsilon_r\) for a shortfall of that sector. Then \(G_{j^*}(\kappa(\boldsymbol\epsilon))=(1-a)/(1-(1-\varsigma)a)\), and
\[
\Phi(\boldsymbol\epsilon)=\phi(a):=\frac{\varsigma\,a\,(2-a)}{1-(1-\varsigma)\,a},
\qquad
\phi''(a)=\frac{2\,\varsigma\,(1-2\varsigma)}{\bigl(1-(1-\varsigma)\,a\bigr)^{3}}.
\]
The function \(\phi\) vanishes at zero, and on \([0,1)\) it is strictly convex when \(\varsigma<\tfrac12\) and strictly concave when \(\varsigma>\tfrac12\). The function \(\phi\) is therefore strictly super-additive in the first case and strictly sub-additive in the second, which is the last claim. This proves (iv).
\end{proof}

The bracket of \eqref{eq:pair_second_order} adds up three interactions of the two closures at the buyer. Since \((3-\rho)-\beta(3-\varrho)=(1-\rho)+2(1-\beta)-\beta(1-\varrho)\), the bracket equals
\begin{equation}
(1-\rho)\biggl[\sum_{r\in\mathcal R_{j^*}}\frac{\ell_r(\cset{}_1)\,\ell_r(\cset{}_2)}{\alpha_r}-\ell(\cset{}_1)\,\ell(\cset{}_2)\biggr]
+\beta\,(1-\varrho)\,\ell(\cset{}_1)\,\ell(\cset{}_2)
-2\,(1-\beta)\,\ell(\cset{}_1)\,\ell(\cset{}_2).
\label{eq:pair_three_terms}
\end{equation}
The first term is the complementarity of inputs, the second belongs to the outer nest, and the third is the write-off of the unrecovered share, the three interactions that Section~\ref{subsec:superadditivity} of the paper describes. The square bracket of the first term is \(\tau^{-2}\) times the covariance of the two shortfalls across the input sectors of the buyer, which is positive when the two closures take the same inputs and negative when they take different ones. Dividing \eqref{eq:pair_second_order} by the sum of the two single losses gives the index \eqref{eq:superadd_index} of two gates that expose no common link,
\begin{equation}
S(\cset{})-1
=\tau\,\frac{(1-\rho)\sum_r\ell_r(\cset{}_1)\,\ell_r(\cset{}_2)/\alpha_r-\bigl((3-\rho)-\beta(3-\varrho)\bigr)\,\ell(\cset{}_1)\,\ell(\cset{}_2)}{2\,\bigl(\ell(\cset{}_1)+\ell(\cset{}_2)\bigr)}
+O\bigl(\tau^{2}\bigr).
\label{eq:pair_index}
\end{equation}
The departure of the index from one is thus proportional to the unrecovered share to the first order, and it is small when each closure displaces a small share of the purchases of the buyer. The index of two gates in one detour class that expose the same links to the same degree is \(\tfrac12\) at every value of the parameters, by part~(ii).

The proposition holds the placement deficits at zero and the network at a single buyer. Where a block carries a placement deficit the shortfall of a buyer exceeds the unrecovered share of its displaced purchases. The deficit of a joint closure can exceed the sum of the deficits of its single closures, since two closures can between them shut every open link that either alone leaves. Where the buyer supplies other producers its shortfall becomes theirs, and the closure sets of the paper reach many buyers through many inputs at once. We therefore compute the indices of Section~\ref{subsec:joint_empirics} of the paper on the solved model. We compare them with the proposition and do not derive them from it. Part~(iv) also leaves out an outer nest that is more complementary than the inner one, \(\varrho<\rho\). The bracket of \eqref{eq:pair_second_order} is then positive for two gates that take different inputs whenever \(\beta>(3-\rho)/(3-\varrho)\), so such a pair can compound as well.

% ============================================================================
% Online Appendix B, input by v17_choke_points.tex after the paper.
% From the first line after \maketitle on, this file is the body of the
% appendix as it stands in the working paper. The lines before it do what the
% appendix's own preamble does when it is compiled as a separate document:
% they restart the counters and set the B. numbering, the line spacing and
% the title block.
% ============================================================================
\clearpage
\setcounter{part}{0}\setcounter{section}{0}\setcounter{subsection}{0}\setcounter{subsubsection}{0}\setcounter{equation}{0}\setcounter{table}{0}
\setcounter{figure}{0}\setcounter{footnote}{0}\setcounter{theorem}{0}\setcounter{proposition}{0}\setcounter{lemma}{0}\setcounter{corollary}{0}
\setcounter{claim}{0}\setcounter{assumption}{0}\setcounter{remark}{0}\setcounter{definition}{0}\setcounter{example}{0}\setcounter{result}{0}
\renewcommand{\thesection}{B.\arabic{section}}
\renewcommand{\theHsection}{B.\arabic{section}}
\renewcommand{\thesubsection}{B.\arabic{section}.\arabic{subsection}}
\renewcommand{\theHsubsection}{B.\arabic{section}.\arabic{subsection}}
\renewcommand{\theequation}{B.\arabic{equation}}
\renewcommand{\thetable}{B.\arabic{table}}
\renewcommand{\thefigure}{B.\arabic{figure}}
\renewcommand{\thetheorem}{B.\arabic{theorem}}
\renewcommand{\theproposition}{B.\arabic{proposition}}
\renewcommand{\thelemma}{B.\arabic{lemma}}
\renewcommand{\thecorollary}{B.\arabic{corollary}}
\renewcommand{\theassumption}{B.\arabic{assumption}}
\renewcommand{\thedefinition}{B.\arabic{definition}}
\renewcommand{\theremark}{B.\arabic{remark}}
% the same prefix for the link anchors, so that no two anchors in the document share a name
\def\theHequation{B.\arabic{equation}}\def\theHtable{B.\arabic{table}}\def\theHfigure{B.\arabic{figure}}
\def\theHtheorem{B.\arabic{theorem}}\def\theHproposition{B.\arabic{proposition}}\def\theHlemma{B.\arabic{lemma}}
\def\theHcorollary{B.\arabic{corollary}}\def\theHclaim{B.\arabic{claim}}\def\theHassumption{B.\arabic{assumption}}
\def\theHremark{B.\arabic{remark}}\def\theHdefinition{B.\arabic{definition}}\def\theHexample{B.\arabic{example}}
\def\theHresult{B.\arabic{result}}
\onehalfspacing
\setlength{\emergencystretch}{0pt}
\OArestoretitle
\title{\textbf{Online Appendix~B}\\[0.4ex]
to\\[0.4ex]
{\large\textbf{``Macroeconomic Risks from Maritime Trade Disruptions''}}\\[4ex]
\textbf{Data, Calibration, and Additional Results}}

\author{Vipin P. Veetil\thanks{Economics Area, Indian Institute of Management Kozhikode.} \and Fathimath S. Vemmarath\thanks{Indian Statistical Institute, Kolkata.}}

\date{\today}
\maketitle
\pdfbookmark[0]{Online Appendix B: Data, Calibration, and Additional Results}{onlineappendixB}

This online appendix carries the data, the calibration, and the
additional results behind the paper. Section~\ref{sec:data} constructs
the extended 87-economy intermediate-flow matrix $\mathbf Z^*$ from OECD
ICIO levels and GTAP~10 within-composite shares.
Section~\ref{app:reroutability} builds the reroutability matrices $\mathbf E_k$ from route
geography, records the routing assignments, and measures the seaborne share of each flow.
Section~\ref{app:estimation} calibrates the residual transit share $\delta$, the substitution parameter $\rho$, the outer-nest curvature $\varrho$, and the unrecovered share $\tau$ on evidence that predates the closures the paper studies, and Section~\ref{app:stability_numerics} reports how the rationed
exchange and the fulfillment fixed point behave on the data. Sections~\ref{app:addl_decomp}, \ref{app:addl_structure}, and~\ref{app:addl_joint} collect the
additional results behind Sections~\ref{sec:results}, \ref{sec:structure},
and~\ref{sec:joint} of the paper, and Section~\ref{app:robustness} checks the paper's four headline findings across the calibration
rectangle, the detour distances, the input--output vintages, and the maintained structure of the
exchange and the fulfillment rule. Section~\ref{app:replication}
documents the code, data, and replication package, and
Section~\ref{app:notation} catalogs the notation. Its companion, Online
Appendix~A, states and proves the formal results. Sections, equations, tables, figures, and results cited without the prefix A or B are those of the paper.

% ================================================================
\section{The Extended Input--Output Matrix}
\label{sec:data}
% ================================================================

The bilateral intermediate-flow matrix $\mathbf{Z}^*$ that drives the
model of Section~\ref{sec:model} must resolve trade at the
country-sector level with enough geographic granularity to assign flows
to physical chokepoints. The OECD Inter-Country
Input--Output table \citep{OECD2025} does not, since it pools every economy that it does not identify individually into a single Rest of World (ROW) composite. For most input--output applications the pooling is innocuous, but for chokepoint analysis it conceals the economies whose trade is most route-sensitive, among them Qatar, Kuwait, Oman, Bahrain, Iran, and Iraq behind the Strait of Hormuz, Yemen at Bab-el-Mandeb, and Libya and Algeria on the Mediterranean. Under the standard table the Strait of Hormuz appears to involve only Saudi Arabia and the United Arab Emirates on the exporting side, since every other Gulf supplier is part of an undifferentiated ROW seller whose exposure cannot be mapped to a transit route. We therefore construct an extended matrix
$\mathbf{Z}^* \in \mathbb{R}_+^{C^*S \times C^*S}$ with
$C^* = 87$ economies at the same
$S = 50$ sectors, the OECD table supplying the levels and the GTAP~10 database \citep{Aguiar2019} the within-ROW shares that divide each ROW cell into seven maritime-relevant blocks. The extended matrix has \(4{,}350\) rows and \(4{,}350\) columns, a row for each country-sector as a seller and a column for each country-sector as a buyer.

% ================================================================
\subsection{The OECD ICIO 2025 baseline}
\label{sec:Zmatrix}
% ================================================================

The OECD Inter-Country
Input--Output database provides the natural anchor. Its
2025 edition covers 81 economies and 50 ISIC Revision~4 sectors for
reference year 2019, giving a 4{,}050$\times$4{,}050 flow matrix with
well-documented accounting consistency. We anchor on 2019 even though the 2025 edition carries reference years through 2022,
because every later year is contaminated for our purpose. The benchmark should
describe the world economy in its normal state. But 2020 is the pandemic
cross-section, 2021 the rebound and container-market dislocation, and 2022 carries the
full-scale war in Ukraine, which rearranged exactly the energy flows whose severance
the counterfactuals study. The closures are therefore measured against the last normal-times network.

Let $\mathcal{C} = \{1, 2, \ldots, C\}$ be the set of $C = 81$
economies in the OECD ICIO 2025 baseline (80 individually
identified countries plus the ROW composite) and let
$\mathcal{S} = \{1, 2, \ldots, S\}$ be the set of $S = 50$ ISIC
Revision~4 sectors. The baseline intermediate-flow matrix is
$\mathbf{Z} \in \mathbb{R}_+^{CS \times CS}$, with element
$Z_{(c,r),(c',r')}$ the value (current USD millions, year 2019) of
output of sector $r$ in country $c$ used as intermediate input by
sector $r'$ in country $c'$. Its rows are the supplying country-sectors and its columns the using ones, and both carry labels of the
form \texttt{ISO3\_SECTORCODE}. For example,
$Z_{\texttt{SAU\_B06},\,\texttt{CHN\_C19}} \approx 19{,}000$ million
USD records Saudi Arabia's crude-petroleum sector supplying roughly
USD~19~bn to China's petroleum-refining sector in 2019.

We write $\mathbf{Z}^{cc'} \in \mathbb{R}_+^{S \times S}$ for the block of sector flows from country $c$ to
country $c'$. Sector codes follow ISIC Rev.~4 at ICIO 2025
resolution, with \texttt{B06} crude petroleum and gas extraction,
\texttt{C19} petroleum refining, \texttt{C26} electronics, and
\texttt{C29} motor vehicles.

The OECD 2025 edition is the extended edition, which
disaggregates China into a processing-trade component
(\texttt{CN1}) and a non-processing component (\texttt{CN2}), and
Mexico into a maquila component (\texttt{MX1}) and non-maquila
(\texttt{MX2}). We aggregate these back to standard \texttt{CHN}
and \texttt{MEX} entries before constructing $\mathbf{Z}^*$, so
that the accounting is identical to the standard 81-economy ICIO
table. Gross output $s_{(c,r)} = \sum_{c',r'} Z_{(c,r),(c',r')} +
f_{(c,r)}$ is formed by adding final-demand row sums $f_{(c,r)}$
(household consumption, government, investment, and exports to
final demand), which are available as additional columns in the ICIO
file.

% ================================================================
\subsection{Using GTAP to disaggregate the rest-of-world block}
\label{sec:gtap_satellite}
% ================================================================

We use the GTAP 10 Data Base \citep{Aguiar2019} to construct
within-ROW bilateral shares.\footnote{We use GTAP~10 because it is the most recent release freely available to academic users. GTAP supplies only the division of the OECD rest-of-world composite into our seven blocks, the levels being fixed by the OECD table at its 2019 reference year, so we expect the shares to differ little across releases, though we have not checked it.} GTAP provides a complete global
social accounting matrix with 141 regions and 65 sectors for base
year 2014. Its geographic scope individually resolves all the
Gulf states, the main African economies, and most of the
rest-of-world regions that OECD ICIO aggregates into ROW.

The share construction draws on three arrays of GTAP, all in USD millions of 2014. The first is the imports $\text{VIWS}(r, q, q')$ of commodity $r$ from source region $q$ by destination region $q'$, at world prices. The other two are the demand $\text{VIFM}(r, r', q')$ of industry $r'$ in region $q'$ for imported commodity $r$ and the demand $\text{VDFM}(r, r', q)$ of industry $r'$ in region $q$ for domestic commodity $r$.

GTAP records VIFM at the destination level, so it tells us how much imported oil industry $r'$ in China uses, but not which fraction
came from Saudi Arabia and which from Kuwait. The bilateral dimension
is captured by VIWS. We combine VIFM and VIWS via the import-proportionality assumption standard in the
input--output literature \citep{Tukker2016}. Within each
destination-commodity cell, the share of imports sourced from
region $q$ is the same across all using industries.

The bilateral intermediate flow of commodity $r$ from region $q$ to
industry $r'$ in region $q'$ is then the destination's demand for the
imported commodity, divided across sources by their shares in VIWS,
\begin{equation}
  Z_{\text{GTAP}}(q,r;\,q',r')
  = \frac{\text{VIWS}(r,q,q')}
         {\displaystyle\sum_{q''=1}^{141} \text{VIWS}(r,q'',q')}
    \times \text{VIFM}(r,r',q')
  \qquad (q \neq q'),
\label{eq:bilateral_import}
\end{equation}

with the convention that the flow is zero when the denominator is
zero.

Domestic flows (same origin and destination region) are taken
directly from VDFM:
\begin{equation}
  Z_{\text{GTAP}}(q,r;\,q,r') = \text{VDFM}(r,r',q)
\label{eq:bilateral_domestic}
\end{equation}

The result is a bilateral intermediate-flow tensor $Z_{\text{GTAP}}
\in \mathbb{R}_+^{141 \times 65 \times 141 \times 65}$. Summing it over sources $q$ recovers VIFM exactly, and only the shares of VIWS enter the construction.

% ================================================================
\subsection{The seven-block partition and sector concordance}
\label{sec:partition}
% ================================================================

We partition the 141 GTAP regions into 87 extended economies. The
75 GTAP regions that correspond to economies individually identified
in OECD ICIO 2025 are mapped one-to-one to their ISO3 codes. Three composite regions, \texttt{xac} (Angola and the Democratic Republic of the Congo), \texttt{xse} (Myanmar and Timor-Leste), and \texttt{xef} (Iceland and Liechtenstein), consist almost wholly of economies that the ICIO identifies individually. We leave them out so that their trade does not enter the shares of the ROW composite. The
remaining 63 GTAP regions, those that fall inside the OECD ROW
composite, are aggregated into seven geographic blocks chosen to
isolate the economically and geographically distinct parts of ROW
that are relevant to maritime chokepoints
(Table~\ref{tab:row_blocks}).

\begin{table}[H]
\centering
\caption{The seven rest-of-world blocks and their GTAP-region
constituents. Representative members are shown, but the complete 63-region concordance ships with the replication package (Online
Appendix~\ref{app:replication}).}
\label{tab:row_blocks}
\small
\begin{tabular}{ll}
\toprule
Block & GTAP regions (representative members) \\
\midrule
\textbf{OtherGulf} &
  \texttt{bhr}, \texttt{irn}, \texttt{kwt}, \texttt{omn},
  \texttt{qat}, \texttt{xws} \\
& (Bahrain, Iran, Kuwait, Oman, Qatar; xws covers Iraq, \\
& Lebanon, Palestine, Syria, and Yemen) \\[2pt]
\textbf{OtherNorthAfrica} &
  \texttt{xnf} (Algeria, Libya, Western Sahara) \\[2pt]
\textbf{OtherWestAfrica} &
  \texttt{ben}, \texttt{bfa}, \texttt{gha}, \texttt{gin},
  \texttt{tgo}, \texttt{xwf}, \texttt{xcf} \\[2pt]
\textbf{OtherEastAfrica} &
  \texttt{eth}, \texttt{ken}, \texttt{mdg}, \texttt{moz},
  \texttt{rwa}, \texttt{tza}, \texttt{uga}, \texttt{xec},
  and others \\[2pt]
\textbf{OtherSouthAsia} &
  \texttt{npl}, \texttt{lka}, \texttt{xsa}
  (Nepal, Sri Lanka, rest of South Asia) \\[2pt]
\textbf{OtherCentralAsia} &
  \texttt{kgz}, \texttt{tjk}, \texttt{xsu}
  (Kyrgyzstan, Tajikistan, rest of former Soviet CA) \\[2pt]
\textbf{OtherRestOfWorld} &
    all remaining GTAP regions not covered above \\
  & (Americas ex-ICIO, Pacific islands, Caucasus, etc.) \\
\bottomrule
\end{tabular}
\end{table}

\noindent
The GTAP aggregate region \texttt{xws} (Rest of Western
Asia) covers Iraq, Lebanon, Palestine, Syria, and Yemen. Yemen is a Red Sea
transit state that would ideally sit in a block of its own, but it
cannot be separated from the Gulf
aggregate at GTAP's geographic resolution, and the Levantine members of the
aggregate are routed with the Gulf for the same reason. Djibouti, Eritrea,
and Sudan belong to the aggregate \texttt{xec} (Rest of Eastern Africa) and
enter OtherEastAfrica. We therefore form no OtherRedSea block.

GTAP's 65 sectors must be mapped to OECD ICIO's 50 ISIC Rev.~4
sectors before the bilateral flows can be used as share donors.
The concordance assigns each GTAP sector to one or two ICIO
sectors with non-negative weights that sum to one. Most mappings
are one-to-one (e.g.\ GTAP \texttt{oil} $\to$ ICIO \texttt{B06},
\texttt{i\_s} $\to$ \texttt{C24A}, and \texttt{wtp} $\to$
\texttt{H50}). One-to-many mappings arise where GTAP aggregates sectors that ICIO 2025 disaggregates. For
example, \texttt{oxt} (other mining) is divided equally between \texttt{B07} (metal ores)
and \texttt{B08} (other quarrying), and \texttt{otn} (other transport equipment)
is divided 70/30 between \texttt{C302T309} and the newly disaggregated \texttt{C301}
(shipbuilding).

% ================================================================
\subsection{Constructing the extended intermediate-flow matrix by disaggregating the ROW cells}
\label{sec:Zstar}
% ================================================================

We assemble the extended matrix $\mathbf{Z}^*$ in three steps, each
preserving exact aggregation consistency with the OECD baseline.

In the first step, OECD-to-OECD flows pass through unchanged, in that for all pairs $(c, c')$ where both countries are individually
identified in OECD ICIO ($c, c' \neq \text{ROW}$), the flows are
copied without modification,
\begin{equation}
  Z^*_{(c,r),(c',r')} = Z_{(c,r),(c',r')}
  \qquad \forall\; c,c' \neq \text{ROW}
\label{eq:oecd_block}
\end{equation}

In the second step, for cells where one party is ROW and the
other is an individually identified OECD economy, the OECD cell value
is distributed across the seven new blocks using GTAP-derived shares.
For ROW-outbound flows, in which ROW supplies commodity $r$ to OECD
country $c'$ for use in sector $r'$, the OECD cell is divided across the
blocks $b \in \mathcal{B}$ in proportion to their GTAP-concorded flows,
\begin{equation}
  Z^*_{(b,r),(c',r')}
  = \frac{\tilde{Z}_{\text{GTAP}}(b,r;\,c',r')}
         {\displaystyle\sum_{b' \in \mathcal{B}}
          \tilde{Z}_{\text{GTAP}}(b',r;\,c',r')}
    \times Z_{(\text{ROW},r),(c',r')},
\label{eq:zstar_out}
\end{equation}

where $\tilde{Z}_{\text{GTAP}}$ denotes GTAP flows after sector concordance and
$\mathcal{B} = \{\text{OtherGulf}, \ldots, \text{OtherRestOfWorld}\}$ is the set of
seven blocks. The denominator sums only over the 63 GTAP regions that map into ROW
(not the 75 regions that map to individually identified OECD economies).

ROW-inbound flows are handled symmetrically, the shares taken over the
buying block's own index. In the third step, for cells where both parties are ROW, the
OECD cell value is distributed across all ordered block pairs
$(b, b') \in \mathcal{B} \times \mathcal{B}$ in proportion to the joint
GTAP flows,
\begin{equation}
  Z^*_{(b,r),(b',r')}
  = \frac{\tilde{Z}_{\text{GTAP}}(b,r;\,b',r')}
         {\displaystyle\sum_{\bar{b}, \bar{b}' \in \mathcal{B}}
          \tilde{Z}_{\text{GTAP}}(\bar{b},r;\,\bar{b}',r')}
    \times Z_{(\text{ROW},r),(\text{ROW},r')}.
\label{eq:zstar_rowrow}
\end{equation}

When the GTAP denominator in any of these shares is zero, because no
GTAP region in $\mathcal{B}$ records a flow for that
commodity-sector pair, flows are divided uniformly across all seven
blocks. The fallback applies to a small fraction of the value, $2.7$ percent of the sales of the ROW composite to the individually identified economies, $2.6$ percent of its purchases from them, and $5.0$ percent of the flows inside the composite,
concentrated in service sectors with sparse GTAP coverage.

By construction, equations \eqref{eq:zstar_out}--\eqref{eq:zstar_rowrow}
satisfy exact aggregation consistency. Re-aggregating all seven
blocks back into a single ROW row or column recovers the OECD
baseline exactly,
\begin{equation}
  \sum_{b \in \mathcal{B}} Z^*_{(b,r),(c',r')}
  = Z_{(\text{ROW},r),(c',r')}
  \qquad \forall\; c' \neq \text{ROW},\; \forall\; r, r'
\label{eq:agg_consistency}
\end{equation}

and analogously for ROW-inbound and ROW-to-ROW cells. We verify these identities numerically. The maximum absolute deviation across all cells is at or near machine zero.

The OECD ICIO reference year is 2019 and the GTAP 10 base year is 2014.
We use GTAP only for within-ROW shares, not for levels, so
the critical assumption is that the relative bilateral
distribution of flows across ROW blocks is sufficiently stable
between 2014 and 2019. That stability assumption is standard practice in the IOT extension
literature \citep{Tukker2016}. Production network shares
within a geographic sub-aggregate typically change slowly relative
to the aggregate level.
The resulting extended matrix $\mathbf{Z}^*$, together with the
derived input-coefficient matrix $\mathbf{A}^*$ of the same dimension and leakage
parameters $\boldsymbol{\beta}^*$, is the empirical object that
supplies the structural parameters to the model of
Section~\ref{sec:model}.

\subsection{Two integrity residuals of the assembled table}
\label{oa:data_integrity}

The assembled table carries two measurable departures from the exact accounting identities, so the results of the model are stated with both in view.
First, one active column has intermediate purchases above gross output, Cyprus's air-transport column
(\texttt{CYP\_H51}), with a recorded leakage of $-0.074$. We deliberately do not clip it, so the reported equilibria are solved with the column as recorded. The single share
$\beta_j$ of the formal model therefore divides in two at that column.
The spending share $\beta^{\rm spend}_j$ carries the recorded value and
enters the propagation matrix $\mathbf A(\mathbf I-\boldsymbol\beta)$,
so the accounting keeps what the table says. The technology share
$\beta^{\rm tech}_j$ is clipped into $[0,1]$ and enters the outer CES
aggregator alone, so the nonlinear map receives a valid share. The two
coincide at every other column of the table.

The first departure costs the column-sum bound and, at that one column, strict subhomogeneity, so the formal results are qualified accordingly. Because $\beta^{\rm spend}_j<0$, the column-sum
bound $\|\mathbf A(\mathbf I-\boldsymbol\beta)\|_1<1$ fails on the
data, at $1.074$, while the spectral radius of the benchmark
propagation matrix is $0.594$ and stays below one in every solved
scenario. Existence of the inverse $(\mathbf I-\mathbf M_0)^{-1}$, its
Neumann expansion, and its nonnegativity therefore rest on the spectral-radius
bound, which is the route recorded after
Lemma~\ref{lem:choke_kernel} for exactly this case. Because $\beta^{\rm tech}_j=0$ at that column,
its outer aggregator is the identity and its output elasticity of intermediates
is one, the unit gain of the excluded pole $\varrho\to-\infty$ of Section~\ref{subsec:short_run_activity}.
Strict subhomogeneity fails there, so neither part (iii) nor part (iv) of
Proposition~\ref{prop:equilibrium} reaches that coordinate, and neither
does the floor the interiority condition computes. What the paper reports is the benchmark-selected equilibrium,
which parts (i) and (ii) of that proposition deliver without either
property, so the reported object is well defined at the outlier and
only the uniqueness claim stops short of it. The formal results
maintain $\beta_j\in(0,1)$ throughout and do not claim otherwise.

The second departure is that the row totals of the assembled table do not solve the benchmark
system. The disaggregation of the rest-of-world composite takes the gross output of a block from its
purchases and its sales from the GTAP shares, and the two need not agree, so the row identity
$s_i=f_i+\sum_j Z_{ij}$ holds with slack. Across the $4{,}299$ active cells the absolute residual
averages $0.53$ percent of gross output, $0.20$ percent GDP-weighted, and about four fifths of it
sits in the seven blocks. The model therefore takes as its benchmark activity the solution of the
benchmark system and not the row totals. With $\mathbf s$ the vector of row totals,
\[
\breve{\mathbf s}_0=(\mathbf I-\mathbf M_0)^{-1}\mathbf f,
\qquad
\breve{\mathbf s}_0-\mathbf s
=-(\mathbf I-\mathbf M_0)^{-1}\mathbf r^{B},
\qquad
\mathbf r^{B}
:=\mathbf s-\mathbf f-\mathbf M_0\mathbf s.
\]
The input shares, the leakage shares, final demand, and the value-added shares stay those of the table, but the activity vector is replaced. The solution $\breve{\mathbf s}_0$ differs from the
row totals by $0.68$ percent on average across the active cells, $0.20$ percent in the named
economies and $6.1$ percent in the blocks, and by $0.27$ percent GDP-weighted. The solution lies
above the row total at $1{,}695$ cells and below it at $2{,}525$, but world GDP on the two vectors agrees to within $0.02$ percent.

All reported losses are computed against $\breve{\mathbf s}_0$, which is the equilibrium of the model
when no gate is closed, so the null scenario returns exactly zero and neither residual enters the
chokepoint-loss measurements. Because the benchmark solves the benchmark system, the solved
activity of every scenario lies at or below it, as part (i) of Proposition~\ref{prop:equilibrium}
states. The reported loss of a country-sector is its benchmark activity less its solved activity, with no projection between the two.\footnote{The code still clips each solved activity
vector to the benchmark box as a guard against solver noise. At the fifteen baseline scenarios the
clip moves no coordinate.} The decomposition of Section~\ref{subsec:activity_decomposition} of the paper is
computed on the same vectors, on which Table~\ref{tab:loss_decomposition} of the paper is exact. The
loss through lost orders is the loss of the economy with lost orders alone,
$\breve{\mathbf s}_0-(\mathbf I-\mathbf M_\omega)^{-1}\mathbf f$, and the loss through input shortfalls
is what the full model loses beyond it. No term of the decomposition is negative at any economy
under any of the fifteen scenarios. The formal results of Online Appendix~A are statements about the
exact model on its stated domain. This subsection is the numerical bridge between that model and the reported output.

% ================================================================
\section{Routes, Reroutability, and the Seaborne Share}
\label{app:reroutability}
% ================================================================

The reroutability matrices $\mathbf E_k$ of Section~\ref{subsec:chokepoint_restrictions} of the paper give, for each country pair, the share of a benchmark flow that survives when chokepoint $k$ closes and the cargo is left with the longer voyage around it. This section records what the paper's definition leaves open, namely the symmetry of the matrices, the maritime regions over which they are specified, the benchmark routes and detour distances, and the values the rule \eqref{eq:reroutability_rule} returns gate by gate. It then records where the incidence of each closure falls and measures the seaborne share of each flow.

\subsection{The reroutability matrix and its maritime regions}

For each maritime chokepoint $k$, the reroutability matrix
$\mathbf E_k=[e^{(k)}_{cc'}]$ is a $C^*\times C^*$ matrix over the
$C^*=87$ economies of the extended set of Online Appendix~\ref{sec:data}, namely the 80 individually identified OECD economies and the seven ROW blocks. The matrix has a row for each exporting economy and a column for each importing one.
Each block is assigned, like any economy, to the maritime region of its
principal seaboard. OtherGulf maps to the Persian Gulf, OtherNorthAfrica to
the Mediterranean, OtherWestAfrica and OtherEastAfrica to Sub-Saharan
Africa, OtherSouthAsia to South Asia, OtherCentralAsia to the Black Sea,
its seaborne outlet, and OtherRestOfWorld to the nominal ROW region. The matrices are
symmetric, $e^{(k)}_{cc'}=e^{(k)}_{c'c}$, because reroutability is a
property of the shared sea route, and homogeneous across sectors. The routings of OtherRestOfWorld are nominal, since that
aggregate is geographically heterogeneous.

The entries of the matrix follow the rule \eqref{eq:reroutability_rule} of the paper,
$e^{(k)}_{cc'}=D_{cc'}/(D_{cc'}+\Delta_k)$ for a flow whose benchmark route transits
chokepoint $k$, with $D_{cc'}$ the benchmark voyage distance in nautical miles and $\Delta_k$ the extra distance the
detour around $k$ adds. For a terminal
chokepoint, the sole gate of an enclosed basin, no detour
exists, $\Delta_k=\infty$, and $e^{(k)}_{cc'}=0$. For a
bypassable chokepoint a finite detour exists and
$e^{(k)}_{cc'}\in(0,1)$. A flow whose benchmark route does not
transit $k$ loses nothing to its closure and has $e^{(k)}_{cc'}=1$.

Voyage distances are taken at the resolution of fourteen
maritime regions. Each economy is assigned to the region of its
principal seaboard (Table~\ref{tab:regions}), and landlocked economies
are assigned to the region of the ports through which their
seaborne trade passes. Because the rule depends on $c$ and $c'$ only through their regions, the country-level
reroutability matrix is constant within each region block. Russia trades through several
basins, and its westbound flows are routed through the Baltic and the Danish Straits, its
southern flows through the Black Sea and the Turkish Straits, and its Pacific flows to East
Asia through the Korea Strait.

\begin{table}[H]
\centering
\caption{Maritime regions and economy assignment}
\label{tab:regions}
\small
\begin{tabular}{l l p{9.0cm}}
\toprule Code & Region & Economies \\ \midrule
EAS & East Asia        & CHN, HKG, JPN, KOR, TWN \\
SEA & Southeast Asia   & BRN, IDN, KHM, LAO, MMR, MYS, PHL, SGP, THA, VNM \\
SAS & South Asia       & BGD, IND, PAK, OtherSouthAsia \\
PGF & Persian Gulf     & ARE, SAU, OtherGulf \\
NWE & Northwest Europe & AUT, BEL, CHE, CZE, DEU, FRA, GBR, HUN, IRL, ISL, LUX, NLD, NOR, SVK \\
MED & Mediterranean    & CYP, EGY, ESP, GRC, HRV, ISR, ITA, JOR, MAR, MLT, PRT, SVN, TUN, OtherNorthAfrica \\
BLT & Baltic           & BLR, DNK, EST, FIN, LTU, LVA, POL, SWE \\
BLK & Black Sea        & BGR, KAZ, ROU, TUR, UKR, OtherCentralAsia \\
RUS & Russia           & RUS \\
NAM & North America    & CAN, MEX, USA \\
LAM & Latin America    & ARG, BRA, CHL, COL, CRI, PER \\
SSA & Sub-Saharan Africa & AGO, CIV, CMR, COD, NGA, SEN, STP, ZAF, OtherWestAfrica, OtherEastAfrica \\
OCE & Oceania          & AUS, NZL \\
ROW & Rest of the World & OtherRestOfWorld \\
\bottomrule
\end{tabular}
\end{table}

\subsection{From detour distance to the lost share}

A voyage detour increases the cost of moving cargo through
several channels. Fuel and crew time scale with distance and days
at sea, capital tied up in inventory at sea bears interest,
vessel asset utilization falls when the rotation is reshuffled,
and the cargo is exposed for longer to perishability and
contract-deadline penalties \citep{stopford2009,hummels2007}.
\citet{hummels_schaur2013} estimate that each day at sea is worth
a tariff of $0.6$--$2.1\%$ of cargo value, with higher figures
for time-sensitive manufactured goods.
\citet{djankov_freund_pham2010} find that each additional day a product is delayed before it is shipped reduces trade by more than one percent, and
\citet{brancaccio_kalouptsidi_papageorgiou2020} model the
endogenous response of freight rates to demand and routing
shocks. The functional form
$1-e^{(k)}_{cc'}=\Delta_k/(D_{cc'}+\Delta_k)$ of the lost share is the simplest one
consistent with these facts. The lost share is
zero when $\Delta_k=0$, tends to one as
$\Delta_k/D_{cc'}\to\infty$, and equals the share of the total
detoured voyage devoted to the detour, the ``wasted'' share of
transport effort. The implied magnitudes accord with the cost evidence. At a service speed of about $14$ knots, a Cape-of-Good-Hope detour of $\Delta\approx
3{,}500$ nautical miles adds about ten days at sea. Those ten added days imply a time cost of about
$14\%$ of cargo value at the central \citet{hummels_schaur2013} estimate, and roughly
double that once fuel, capital, and perishability components are added.
The reroutability rule returns a lost share $1-e\approx 0.22$--$0.27$ on the long Asia--Europe corridors, inside this band.

\subsection{Gates, detours, and benchmark routes}

The Persian Gulf, the Black Sea, the Baltic Sea, and the Sea of
Azov are fully enclosed basins, each with exactly one sea gate.
The Strait of Hormuz, the Turkish Straits, the Danish Straits,
and the Kerch Strait are therefore terminal. The Strait of Dover
is a bypassable gate, since a Channel transit can be salvaged by the
around-Scotland route, a detour of roughly $800$ nautical miles
that exists physically even though it carries no benchmark mass and
therefore does not itself enter $\mathcal K$. Being a detour
rather than a chokepoint is the operational distinction of
Section~\ref{sec:step1_def}. The remaining seven
chokepoints likewise lie on
through-routes with a navigable detour, listed in Table~\ref{tab:choke_class} together with the extra distance $\Delta_k$ the detour
adds and the range of the reroutability on the support of each gate. Bypass pipelines for Gulf
crude and the Kiel Canal for small Baltic vessels provide limited relief that the rule ignores, but Online Appendix~\ref{app:hormuz_pipeline} credits the Gulf crude pipelines separately.

\begin{table}[H]
\centering
\caption{Chokepoint classification, detour distances, and the range of the reroutability $e$ on the support of each gate. The ranges are those of the regional rule. The cross-Strait pair on the Taiwan--Luzon Corridor and the Japan--Korea pair at the Korea Strait enter through a country-pair override at $e=0.4$.}
\label{tab:choke_class}
\small
\begin{tabular}{l l r p{5.2cm} c}
\toprule
Chokepoint & Type & $\Delta_k$ (nm) & Detour & Range of $e$ \\ \midrule
Strait of Hormuz       & terminal & $\infty$ & none (sole exit of the Persian Gulf) & 0 \\
Turkish Straits        & terminal & $\infty$ & none (sole exit of the Black Sea) & 0 \\
Danish Straits         & terminal & $\infty$ & none (sole exit of the Baltic Sea) & 0 \\
Kerch Strait           & terminal & $\infty$ & none (sole exit of the Sea of Azov) & 0 \\
Suez Canal             & bypassable & 3{,}500  & round the Cape of Good Hope & 0.53--0.79 \\
Bab-el-Mandeb          & bypassable & 3{,}500  & round the Cape of Good Hope & 0.53--0.79 \\
Strait of Gibraltar    & bypassable & 4{,}000  & round Africa via Suez and the Cape & 0.33--0.77 \\
Strait of Malacca      & bypassable & 1{,}000  & Lombok / Sunda Strait & 0.71--0.93 \\
Panama Canal           & bypassable & 8{,}000  & round Cape Horn & 0.33--0.58 \\
Taiwan--Luzon Corridor & bypassable & 300      & east of Taiwan & 0.89--0.98 \\
Strait of Dover        & bypassable & 800      & around Scotland & 0.81--0.94 \\
Korea Strait           & bypassable & 1{,}500  & around the eastern coast of Japan & 0.40 \\
\bottomrule
\end{tabular}
\end{table}

Table~\ref{tab:routes} gives, for every pair of regions whose
least-cost maritime route transits at least one chokepoint, the
chokepoints crossed and the approximate voyage distance $D$.
Region pairs absent from the table transit none of the twelve chokepoints, so $e^{(k)}_{cc'}=1$
there for every $k$. Routes are symmetric. The East Asia--Latin America pair enters twice, with
one voyage distance for its transit of the Panama Canal and another for its transit of the
Taiwan--Luzon Corridor. The table does not list the Strait of Dover and the Kerch Strait, whose
supports lie inside the rows that cross the Danish Straits and the Turkish Straits. We describe those two supports once the rule has been applied.

\begin{longtable}{l l p{6.4cm} r}
\caption{Benchmark routes between maritime regions}
\label{tab:routes}\\
\toprule From & To & Chokepoints transited & $D$ (nm) \\ \midrule
\endfirsthead
\multicolumn{4}{l}{\small\itshape Table~\ref{tab:routes} (continued)}\\
\toprule From & To & Chokepoints transited & $D$ (nm) \\ \midrule
\endhead
\bottomrule \endfoot
Persian Gulf & East Asia & Hormuz, Malacca, Taiwan--Luzon & 6{,}500 \\
Persian Gulf & Southeast Asia & Hormuz, Malacca & 4{,}500 \\
Persian Gulf & South Asia & Hormuz & 1{,}500 \\
Persian Gulf & Northwest Europe & Hormuz, Bab-el-Mandeb, Suez, Gibraltar & 6{,}500 \\
Persian Gulf & Mediterranean & Hormuz, Bab-el-Mandeb, Suez & 5{,}000 \\
Persian Gulf & North America & Hormuz, Bab-el-Mandeb, Suez, Gibraltar & 9{,}500 \\
Persian Gulf & Black Sea & Hormuz, Bab-el-Mandeb, Suez, Turkish Straits & 6{,}000 \\
Persian Gulf & Baltic & Hormuz, Bab-el-Mandeb, Suez, Gibraltar, Danish Straits & 8{,}000 \\
Persian Gulf & Russia & Hormuz, Bab-el-Mandeb, Suez, Turkish Straits & 6{,}000 \\
Persian Gulf & Oceania & Hormuz & 6{,}000 \\
Persian Gulf & Sub-Saharan Africa & Hormuz & 4{,}500 \\
Persian Gulf & Latin America & Hormuz, Bab-el-Mandeb, Suez, Gibraltar & 11{,}000 \\
Persian Gulf & Rest of the World & Hormuz & 5{,}000 \\
East Asia & Northwest Europe & Taiwan--Luzon, Malacca, Bab-el-Mandeb, Suez, Gibraltar & 11{,}000 \\
East Asia & Mediterranean & Taiwan--Luzon, Malacca, Bab-el-Mandeb, Suez & 9{,}500 \\
East Asia & South Asia & Taiwan--Luzon, Malacca & 4{,}500 \\
East Asia & Southeast Asia & Taiwan--Luzon & 2{,}500 \\
East Asia & North America & Taiwan--Luzon, Panama & 9{,}000 \\
East Asia & Black Sea & Taiwan--Luzon, Malacca, Bab-el-Mandeb, Suez, Turkish Straits & 10{,}500 \\
East Asia & Baltic & Taiwan--Luzon, Malacca, Bab-el-Mandeb, Suez, Gibraltar, Danish Straits & 12{,}500 \\
East Asia & Russia & Korea Strait & 1{,}000 \\
East Asia & Sub-Saharan Africa & Taiwan--Luzon, Malacca & 8{,}000 \\
East Asia & Latin America & Panama & 10{,}000 \\
East Asia & Latin America & Taiwan--Luzon & 15{,}000 \\
East Asia & Rest of the World & Malacca & 6{,}000 \\
Southeast Asia & Northwest Europe & Malacca, Bab-el-Mandeb, Suez, Gibraltar & 9{,}500 \\
Southeast Asia & Mediterranean & Malacca, Bab-el-Mandeb, Suez & 8{,}000 \\
Southeast Asia & South Asia & Malacca & 2{,}500 \\
Southeast Asia & North America & Panama & 11{,}000 \\
Southeast Asia & Black Sea & Malacca, Bab-el-Mandeb, Suez, Turkish Straits & 9{,}000 \\
Southeast Asia & Baltic & Malacca, Bab-el-Mandeb, Suez, Gibraltar, Danish Straits & 11{,}000 \\
Southeast Asia & Sub-Saharan Africa & Malacca & 6{,}000 \\
Southeast Asia & Latin America & Panama & 11{,}000 \\
Southeast Asia & Rest of the World & Malacca & 5{,}000 \\
South Asia & Northwest Europe & Bab-el-Mandeb, Suez, Gibraltar & 7{,}000 \\
South Asia & Mediterranean & Bab-el-Mandeb, Suez & 5{,}500 \\
South Asia & North America & Bab-el-Mandeb, Suez, Gibraltar & 10{,}000 \\
South Asia & Black Sea & Bab-el-Mandeb, Suez, Turkish Straits & 6{,}500 \\
South Asia & Baltic & Bab-el-Mandeb, Suez, Gibraltar, Danish Straits & 8{,}500 \\
South Asia & Russia & Bab-el-Mandeb, Suez, Turkish Straits & 6{,}500 \\
Northwest Europe & Mediterranean & Gibraltar & 2{,}000 \\
Northwest Europe & Baltic & Danish Straits & 800 \\
Northwest Europe & Black Sea & Gibraltar, Turkish Straits & 3{,}500 \\
Northwest Europe & Russia & Danish Straits & 1{,}500 \\
Northwest Europe & Oceania & Gibraltar, Suez, Bab-el-Mandeb, Malacca & 12{,}000 \\
Mediterranean & North America & Gibraltar & 4{,}500 \\
Mediterranean & Baltic & Gibraltar, Danish Straits & 3{,}500 \\
Mediterranean & Black Sea & Turkish Straits & 1{,}200 \\
Mediterranean & Russia & Turkish Straits & 1{,}500 \\
Mediterranean & Oceania & Suez, Bab-el-Mandeb, Malacca & 10{,}000 \\
Mediterranean & Sub-Saharan Africa & Suez, Bab-el-Mandeb & 5{,}500 \\
Mediterranean & Latin America & Gibraltar & 6{,}000 \\
Mediterranean & Rest of the World & Suez, Bab-el-Mandeb & 4{,}000 \\
Baltic & North America & Danish Straits & 4{,}500 \\
Baltic & Black Sea & Danish Straits, Gibraltar, Turkish Straits & 4{,}500 \\
Baltic & Oceania & Danish Straits, Gibraltar, Suez, Bab-el-Mandeb, Malacca & 13{,}000 \\
Baltic & Latin America & Danish Straits & 6{,}500 \\
Baltic & Sub-Saharan Africa & Danish Straits & 6{,}000 \\
Baltic & Rest of the World & Danish Straits & 4{,}500 \\
Black Sea & North America & Turkish Straits, Gibraltar & 6{,}000 \\
Black Sea & Oceania & Turkish Straits, Suez, Bab-el-Mandeb, Malacca & 11{,}000 \\
Black Sea & Latin America & Turkish Straits, Gibraltar & 7{,}500 \\
Black Sea & Sub-Saharan Africa & Turkish Straits, Suez, Bab-el-Mandeb & 6{,}000 \\
Black Sea & Rest of the World & Turkish Straits & 4{,}000 \\
Russia & North America & Danish Straits & 5{,}000 \\
Russia & Latin America & Danish Straits & 7{,}000 \\
Russia & Sub-Saharan Africa & Danish Straits & 6{,}500 \\
North America & Oceania & Panama & 8{,}000 \\
North America & Latin America & Panama & 4{,}000 \\
\end{longtable}

Table~\ref{tab:routing} restates the assignments at the level of the
economy-pair groups that carry the most trade, the level at which
Section~\ref{sec:step2} of the paper applies them, together with the two
intra--East-Asian pairs the routing carries.

\begin{table}[H]
\centering
\caption{Geographic routing rules: chokepoints crossed by major trade routes}
\label{tab:routing}
\begin{tabular}{p{3.5cm}p{4cm}p{5cm}}
\toprule
Seller economy $c$ & Buyer economy $c'$ &
Chokepoints crossed ($\mathbbm 1^{(k)}_{cc'}=1$) \\
\midrule
SAU, ARE, OtherGulf &
East Asia (CHN, JPN, KOR, TWN) &
Hormuz, Malacca, Taiwan--Luzon \\
SAU, ARE, OtherGulf &
South Asia (IND, PAK, BGD) &
Hormuz \\
SAU, ARE, OtherGulf &
Europe (DEU, GBR, FRA, ITA, NLD) &
Hormuz, Bab-el-Mandeb, Suez (Gibraltar into Northwest Europe) \\
East Asia (CHN, JPN, KOR, TWN) &
Europe &
Taiwan--Luzon, Malacca, Bab-el-Mandeb, Suez (Gibraltar into Northwest Europe) \\
East Asia &
North America &
Taiwan--Luzon, Panama \\
TWN &
CHN (cross-Strait) &
Taiwan--Luzon \\
JPN &
KOR &
Korea Strait \\
OtherNorthAfrica &
Northwest Europe &
Gibraltar \\
RUS (Baltic ports) &
Global (westbound) &
Danish Straits \\
RUS, UKR (Black Sea) &
Mediterranean / Global &
Turkish Straits (Kerch in addition for Ukraine's exports to Europe, Africa, and the Americas) \\
RUS (Pacific ports) &
East Asia (CHN, JPN, KOR, TWN, HKG) &
Korea Strait \\
AUS &
East Asia (CHN, JPN, KOR) &
none (routes east of Malacca; Lombok--Sunda) \\
USA (Gulf Coast) &
East Asia &
Panama, Taiwan--Luzon \\
\bottomrule
\end{tabular}
\par\smallskip
\footnotesize\textit{Note:} Cape of Good Hope and Cape Horn are
rerouting alternatives, not chokepoints, and have transit-indicator
value $\mathbbm 1 = 0$ at the benchmark. The routing is regional, so
intra-regional East Asian flows and the Australia--East Asia routes cross
none of the twelve gates. The two country-pair exceptions are the ones the
corridor geometry assigns unambiguously. Cross-Strait TWN--CHN trade
transits the Taiwan--Luzon Corridor and Japan--Korea trade transits the
Korea Strait. 
\end{table}

Applying the rule to the routes of Table~\ref{tab:routes} and the detour distances of
Table~\ref{tab:choke_class} gives the reroutability of every region pair, and the last column of
Table~\ref{tab:choke_class} reports the range of the values on the support of each gate. The
terminal chokepoints (Hormuz, Turkish, Danish) return $e=0$ on their whole support, as does
Kerch on its support, Ukraine's seaborne exports to Europe, the Mediterranean, Africa, and the
Americas. The routing has no port resolution, so it cannot separate Ukraine's Azov ports from
its Black Sea ports, and it charges the Kerch Strait with those exports and with no Russian
flow. The remaining chokepoints return values between zero and one. Suez and Bab-el-Mandeb share
the Cape-of-Good-Hope detour and return identical values. Dover's support is the
Baltic--Atlantic corridor, the flows that have already exited the Danish Straits and transit the
Channel en route to the Mediterranean, the Americas, or beyond. At $\Delta=800$ nautical miles
around Scotland the rule returns $e$ between $0.81$ (Baltic--Mediterranean, $D=3{,}500$) and
$0.94$ (Baltic--Oceania, $D=13{,}000$) on that support.

The cross-Strait pair on the Taiwan--Luzon Corridor and the JPN--KOR pair at the Korea Strait
are the only two intra--East-Asian pairs the routing carries, and both enter through a country-pair override $e=0.4$, since the regional rule assigns no voyage within a region. For the cross-Strait pair the override is the rule applied at the country pair, with a crossing of about $200$ nautical miles against the $300$ of the passage east of Taiwan, so $e=200/(200+300)=0.4$. A crossing of $150$ or $250$ nautical miles would return $e=0.33$ or $0.45$ and put the world loss of a closed corridor at $0.148$ or $0.140\%$, against $0.144\%$. For the JPN--KOR pair the strict rule with $\Delta\approx1{,}500$ nautical miles and $D\approx700$ would return $e\approx0.32$. We set it to the cross-Strait value instead, so that the two pairs carry a uniform $e$. At the strict value the world loss of a closed Korea Strait would be $0.057\%$, against $0.054\%$. The rest of the support of the Korea Strait is the Pacific trade of Russia with China, Korea, Japan, Taiwan, and Hong Kong, which enters under the regional rule on the East
Asia--Russia row of Table~\ref{tab:routes}. At a benchmark voyage of $1{,}000$ nautical miles
and the detour of $1{,}500$ around the eastern coast of Japan the rule returns $e=0.4$, so the
whole support of the Korea Strait carries one value.

Distances and routings are specified between the fourteen regions and not between individual
ports, so the country-level reroutability matrix is constant within each region block, the
country-pair overrides excepted, and the remaining intra--East-Asian pairs stay outside the
chokepoint incidence at this resolution. Online Appendix~\ref{app:robustness_detour} assigns the
whole East Asian block at point resolution instead.

\subsection{Seller and buyer country-sector sets}
\label{app:sellerbuyer}

Table~\ref{tab:seller_buyer} records, for six major chokepoints, the
economies on each side of the passage and the sectors that carry most of the
seaborne value that the closure reaches, each list in descending order of that value. The sets are
descriptive, summarizing where each closure's incidence falls, and nothing
in them enters the construction of the incidence matrix.

\begin{table}[H]
\centering
\caption{Seller and buyer country-sector sets by chokepoint}
\label{tab:seller_buyer}
\small
\begin{tabular}{p{2.6cm}p{3.4cm}p{3.4cm}p{2cm}p{2cm}}
\toprule
Chokepoint &
$\mathcal{C}_k^{\text{sell}}$ &
$\mathcal{C}_k^{\text{buy}}$ &
$\mathcal{S}_k^{\text{sell}}$ &
$\mathcal{S}_k^{\text{buy}}$ \\
\midrule
Strait of Hormuz &
OtherGulf (QAT, KWT, OMN, BHR, IRN, IRQ), SAU, ARE &
CHN, IND, JPN, KOR, OtherGulf, ARE, SAU, USA, TWN, THA &
B06, C19, C20, C24B, C24A &
C19, D, C20, F, B06 \\
\midrule
Strait of Malacca &
CHN, OtherGulf, ARE, SAU, IND, KOR, DEU, IDN &
CHN, JPN, IND, KOR, THA, TWN, SGP, DEU &
B06, C20, C19, C24A, C28 &
C19, D, F, C20, C29 \\
\midrule
Suez Canal &
CHN, OtherGulf, SAU, DEU, IND, ITA, ARE, USA &
CHN, ITA, ESP, IND, DEU, USA, GBR, FRA &
B06, C20, C19, C28, C29 &
C19, F, C29, C20, C10T12 \\
\midrule
Bab-el-Mandeb &
CHN, OtherGulf, SAU, DEU, IND, ITA &
CHN, ITA, ESP, IND, DEU &
B06, C20, C19 &
C19, F, C29 \\
\midrule
Danish Straits &
RUS, SWE, FIN, POL, CHN, DEU, DNK, NOR &
SWE, POL, USA, DEU, FRA, CHN, FIN, DNK &
B06, C19, C20, C28, C24A &
C19, F, C29, C10T12, C20 \\
\midrule
Turkish Straits &
TUR, RUS, KAZ, OtherCentralAsia, CHN, UKR, ROU &
TUR, ITA, OtherCentralAsia, OtherGulf, DEU, IND, UKR &
B06, C20, C24A, C19, A01 &
C19, F, C24A, C10T12, B06 \\
\bottomrule
\end{tabular}
\end{table}

\subsection{The seaborne share of bilateral trade}
\label{app:seashare}

The incidence matrix assigns a flow to a gate by the sea route between the two economies, but it does not say how much of the flow uses that route.
We measure the seaborne share \(\mu_{ij}\) of each link, the share of the
benchmark trade from \(i\) to \(j\) that moves by ship, for every seller
sector in \(\mathcal M\) and every pair of economies.

The main source is the customs record of the external trade of the European
Union by mode of transport \citep{eurostat_comext_modes}. For 2019 it gives,
for each of the $27$ member states and the United Kingdom, the value of
imports from and of exports to every outside partner, by product and by the
mode of transport at the frontier of the Union (sea, rail, road, air, post,
fixed installations such as pipelines, inland waterway, and
self-propulsion). The records cover $1.9$ trillion euro of imports and as
much of exports. We concord the products to the $27$ seller sectors by
chapter of the Harmonized System, and by four-digit heading in the chapters that span several sectors. We leave out non-monetary gold. Among the
overseas partners of the reporters, $35$ percent of the value that moves by
sea or air moves by air. The air share is $84$ percent in other transport
equipment, $77$ percent in pharmaceuticals, $73$ percent in electronics, and
$59$ percent in other manufacturing, but it is below $1$ percent in coal, crude
oil and gas, and refined petroleum.

The seaborne share of a link is set by the kind of pair it joins. For a pair
that no land route joins and in which one economy is a reporter, the
seaborne share is the value moved by sea over the value moved by sea or
air, taken from the record of that reporter with that partner in that sector. Where the record holds less than $5$ million euro, the share is taken from the pooled record of all reporters with that partner, and failing
that from the pooled record of all reporters with their overseas partners.
For a pair in which neither economy is a reporter, the seaborne share is
the one that the exporter's goods of the same sector show on arrival in the
Union, under the same threshold, and failing that the share of the sector
in the overseas imports of the Union. A pair that joins Cyprus or Malta to another reporter, on which the external records are silent, takes that sector share as well. For a pair that joins a reporter to
one of its continental neighbors (Switzerland, Norway, Turkey, Ukraine,
Belarus, Russia, Kazakhstan, and the Central Asian block), the seaborne
share is the value moved by sea over the value moved by all recorded modes,
so road, rail, and pipeline count against it. A delivery recorded as
arriving by pipeline counts as non-seaborne only when a pipeline joins the
partner to the Union, since crude oil that lands at a port of the Union and
reaches an inland refinery by pipeline is recorded under that mode as well.

The external records are silent on trade inside the continental network of
Europe and Central Asia, which comprises the European members of the table other than Cyprus, Iceland, and Malta,
together with Belarus, Ukraine, Russia, Turkey, Kazakhstan, and the Central
Asian block. For the pairs inside this network, and for the trade between Russia and China, the seaborne share is the
water share of the international transport margins of GTAP
\citep{Aguiar2019}. We confine the margins to this network because outside
it they assign implausible modes to many importers, land transport
carrying a large part of the imports of Australia, for instance. We
take trade between China and Kazakhstan or
the Central Asian block to move overland in full, since Kazakhstan and the Central Asian economies are landlocked and reach China overland. We also take half of the
crude oil and gas that Russia sells to China to move by pipeline. About half of the crude oil that China bought from Russia in 2018 arrived through the Chinese branches of the East Siberia--Pacific Ocean pipeline \citep{bofit_russia_china_oil_2019}. The other land neighbors outside the network keep the share that their goods show on arrival in the Union, so their trade is charged to the sea route between them. Online Appendix~\ref{app:land_neighbors} corrects by hand the pairs on which geography leaves no doubt and bounds the error on the rest.

A closure displaces the seaborne share of a link, and the non-seaborne flow
of a link whose sea route is closed is held at its benchmark value in the
residual transit, as Section~\ref{sec:seashare} of the paper sets out. The
treatment of the non-seaborne flow matters at one gate. Left in the open trade, the non-seaborne
flow is rescaled by the exchange like any other open flow, and a link that
the closure severs in full regains a route on which displaced cargo can be
placed. At eleven of the twelve gates the two treatments return world
losses that agree to within $0.0002$ percentage points of world GDP. At
Hormuz, which has no detour, leaving the non-seaborne flow open lowers the
world loss from $2.83$ to $1.01$ percent of world GDP, the value the closure takes when the residual transit is rescaled (Online Appendix~\ref{app:modelrobust}). The loss falls that far although the non-seaborne flow is less than
$5$ percent of the trade on the gate's routes and a negligible part of its
crude oil. We hold the non-seaborne flow fixed for the reason that we hold
the transit fixed. What moves between two economies by pipeline, rail, or
air while their sea route is closed is a datum of the closure and not a
channel by which the cargo of the tankers can be re-placed.

Table~\ref{tab:seashare_gate} reports, gate by gate, the benchmark
intermediate trade that the closure would reach if all of it moved by ship,
and the part of it that is seaborne. The seaborne share is highest at
Hormuz, whose cargo is mostly crude oil and gas, and lowest at the Danish
Straits and Gibraltar, where more than half of the trade on the gate's
routes does not move by ship.

\begin{table}[H]
\centering
\caption{The seaborne share of the trade on each gate's routes. The first
column is the benchmark intermediate trade that a closure of the gate
reaches under the reroutability profile when every flow is taken to move by
ship, and the second is its seaborne part, both in billions of US dollars.}
\label{tab:seashare_gate}
\small
\begin{tabular}{lrrr}
\toprule
Gate & On the gate's routes & Seaborne & Seaborne share \\
\midrule
Strait of Hormuz & 837 & 797 & 0.95 \\
Turkish Straits & 396 & 239 & 0.60 \\
Danish Straits & 510 & 228 & 0.45 \\
Suez Canal & 336 & 247 & 0.73 \\
Bab-el-Mandeb & 336 & 247 & 0.73 \\
Strait of Gibraltar & 598 & 268 & 0.45 \\
Strait of Malacca & 179 & 144 & 0.80 \\
Panama Canal & 403 & 302 & 0.75 \\
Taiwan--Luzon Corridor & 195 & 126 & 0.65 \\
Strait of Dover & 21 & 11 & 0.52 \\
Korea Strait & 72 & 46 & 0.63 \\
Kerch Strait & 10 & 6 & 0.56 \\
\bottomrule
\end{tabular}
\end{table}

% ================================================================
\section{Parameter Estimation}
\label{app:estimation}
% ================================================================

The paper fixes the structural parameters within the ranges the evidence supports (Section~\ref{subsec:calibration}).
We calibrate here the three parameters that the evidence bears on separately, \((\delta,\rho,\tau)\), each on evidence from outside the model, and we set out how that evidence maps into the calibration. The outer-nest curvature \(\varrho\) is set with \(\rho\), on the elasticities between value added and intermediates of the production-function literature (Table~\ref{tab:calibration} and Online Appendix~\ref{app:est_rho}). Online Appendix~\ref{app:est_delta} estimates the residual transit share \(\delta\), the share of disrupted trade that still crosses a closed passage, from the transit and fleet record of nine historical
maritime-disruption episodes, where it is a reduced-form ratio inferred from measured transit counts, cargo volumes, and canal revenues. Online Appendix~\ref{app:hormuz_record} checks that calibration out of sample on the closure of Hormuz in 2026. Online Appendix~\ref{app:est_rho} takes the substitution parameter \(\rho\), the
curvature of the production technology across sectoral input bundles, from the
structural production-function literature that identifies it from natural experiments
and instrumented price variation. It then reconciles those estimates with our
short-run, country-sector setting. Online Appendix~\ref{app:est_tau} takes the unrecovered share \(\tau\), the share of severed same-sector trade not
re-sourced onto surviving suppliers, against the firm-level evidence on how buyers replace a lost supplier. Throughout we report ranges rather than point estimates,
and where the baseline calibration sits at an edge of the evidence we set it beside the robustness sweep of Online Appendix~\ref{app:robustness}.
Table~\ref{tab:calibration} collects the result, parameter by parameter.

\begin{table}[H]
\centering
\caption{The baseline calibration and the evidence behind it. Each parameter is set on evidence that predates the closures the paper studies. }
\label{tab:calibration}
\footnotesize
\renewcommand{\arraystretch}{1.35}
\begin{tabularx}{\textwidth}{@{}%
  >{\raggedright\arraybackslash}p{2.7cm}%
  >{\raggedright\arraybackslash}p{2.9cm}%
  >{\raggedright\arraybackslash}X%
  >{\raggedright\arraybackslash}p{2.9cm}@{}}
\toprule
Parameter (baseline) & Interpretation & Evidence & Supported range \\
\midrule
Unrecovered share, $\tau=0.3$ &
share of displaced same-sector trade not re-matched within the disruption
horizon &
output and sourcing responses of buyers to the 2011 T\=ohoku earthquake, to disasters and floods at suppliers, and to the 2020 lockdowns \citep{boehm_flaaen_2019,barrot_sauvagnat_2016,khanna_morales_pandalai_2022,castro_vincenzi_etal_2024,carvalho_etal_2021_covid,freund_etal_2022} & not pinned by the evidence; examined $[0.1,0.5]$ \\
\addlinespace
Substitution parameter, $\rho=-1$ &
curvature of technology across the sectoral input bundle &
structural production-function estimates from instrumented input prices and
supplier outages; short-run network-macro calibrations; micro-to-macro
aggregation results
\citep{atalay2017,boehm_flaaen_2019,baqaee_farhi_supply_2022,baqaee_farhi_2024,peter_ruane,oberfield_raval_2021} &
micro $\rho\lesssim-4$; macro $\rho\in[-4,-1]$; country-sector $\rho\in[-1,-0.25]$ \\
\addlinespace
Outer-nest curvature, $\varrho=-1$ &
substitution between the intermediate composite and the non-intermediate
(value-added) block &
value-added--vs--intermediates substitution elasticities from the production-function literature and network-macro calibrations \citep{atalay2017,oberfield_raval_2021,carvalho_etal_2021_covid,baqaee_farhi_supply_2022,baqaee_farhi_2024} &
estimates $\varrho\in[-1.5,-0.1]$; examined to $\varrho=-4$ \\
\addlinespace
Residual transit share, $\delta=0.10$ (uniform baseline) &
residual share of a fully exposed flow that still transits the closed
passage &
vessel-level transit record of nine maritime disruptions, 1967--2024:
AIS tracks, canal receipts, and tanker shares
\citep{unctad_2024_troubled_waters,eia_redsea_2024,imf_portwatch} &
boxes $[0.05,0.15]$; energy $[0.40,0.55]$ under attack \\
\bottomrule
\end{tabularx}
\end{table}

\subsection{Residual transit share \texorpdfstring{\(\delta\)}{delta} from historical episodes}
\label{app:est_delta}

By Assumption~\ref{ass:attenuation}, \(\delta\) is the through-passage
residual, the share of a fully exposed link's benchmark flow that still crosses
a closed chokepoint, the retained weight \(1-(1-\delta)d_{ij}\) evaluated at
\(d_{ij}=1\). The residual transit share is orthogonal to the reroutability \(e\) of
Section~\ref{subsec:chokepoint_restrictions}, the around-passage salvage a
longer voyage recovers. Rerouting sends a shipment around
the obstacle, and the residual transit share \(\delta\) lets it through despite the
closure. The empirical counterpart of \(\delta\) is
accordingly a ratio, the transit that continues through a passage
during a closure, divided by its pre-episode benchmark, measured at the steady-state trough of the episode and, wherever the data permit, separately by commodity. No
source reports \(\delta\) directly. Every value of the residual transit share in
Table~\ref{tab:delta_episodes} is accordingly an
inferred quantity,
constructed from measured transit counts, cargo volumes, fleet shares, insurance
premia, or canal revenues. We mark at each step what is measured and what is
inferred from it.

We report \(\delta\) as an interval bracketed by episodes rather than as a point, for reasons that
lie in the data. First, aggregate throughput conflates the two
ways in which a disrupted flow adjusts, the detour around the passage, on which the share
\(e\) of the flow survives, and the residual passage through it, which \(\delta\) measures. A fall in transits at a
passage mixes flows that rerouted with flows that were destroyed, but only the residual that continued is \(\delta\). Separating them requires vessel-level
Automatic Identification System (AIS) tracks. The IMF--Oxford PortWatch platform \citep{imf_portwatch}, built on the satellite AIS tracks of cargo ships and tankers at \(28\) chokepoints, lets the displaced tonnage be seen
reappearing on the detour route rather than have its fate inferred. We therefore measure \(\delta\) on vessel-class transit ratios rather than on monthly tonnage. Second, the episodes with a clean \(\delta\) and the episodes with rich data do not overlap. A clean
\(\delta\) needs a terminal gate with no detour, where the whole of what remains is residual transit, while the rich
data come from the reroutable Asia--Europe corridors. Before 2026 the terminal gates had essentially never been
closed, so for them \(\delta\) was bounded but not point-identified.
Third, the residual transit share is commodity-specific. In every episode before 2026 container traffic withdraws while tanker traffic
persists, which is why the intervals are reported by commodity rather than pooled and why the
commodity-differentiated residual transit share is carried as a robustness exercise alongside the uniform baseline. Fourth, the
observed \(\delta\) is conditional on enforcement intensity, which is endogenous
to the episode, and the determined, sustained closure the counterfactual
contemplates sits at the low-\(\delta\) edge of the historical range or beyond
it. The nine historical disruptions of Table~\ref{tab:delta_episodes},
the Red Sea entering twice, once for containers and once for oil,
therefore serve to bracket \(\delta\), commodity by commodity, not to
average it.

\begin{table}[H]
\centering
\caption{Maritime-disruption episodes and the implied residual transit share \(\delta\). The
final column states what each source measures and how \(\delta\) is
inferred from it, and ``LB'' denotes a lower bound. All \(\delta\) values are
intervals inferred by the authors, not figures reported by any source.}
\label{tab:delta_episodes}
\footnotesize
\renewcommand{\arraystretch}{1.25}
\begin{tabularx}{\textwidth}{@{}%
  >{\raggedright\arraybackslash}p{2.75cm}%
  >{\raggedright\arraybackslash}p{2.15cm}%
  >{\raggedright\arraybackslash}p{2.15cm}%
  >{\centering\arraybackslash}p{1.15cm}%
  X@{}}
\toprule
Episode (period) & Gate / corridor & Closure type & \(\delta\) & Basis: measured \(\rightarrow\) inferred \\
\midrule
Suez closure, 1967--75 & Suez Canal & physical, total (8\,yr) & \(\approx 0\) &
canal carried zero; definitional, and the \(e\)-benchmark \citep{feyrer2021distance} \\

Tanker War, 1984--88 & Hormuz / Gulf & war-risk, sustained & \(0.8\text{--}1.0\) &
\(451\) ships hit, strait never closed, under \(2\%\) of Gulf shipping disrupted \(\rightarrow\) LB \citep{usip_tankerwar_2019} \\

Malacca listing, 2005--06 & Malacca & named risk, no closure & \(\approx 1\) &
JWC war-risk listing, traffic uninterrupted; definitional \citep{rsis_malacca_2005,insurancejournal_malacca_2006} \\

Gulf of Oman attacks, 2019 & Hormuz & war-risk, brief & \(0.85\text{--}1.0\) &
premia \(\times 10\), flow maintained \(\rightarrow\) LB \citep{cnbc_hormuz_2019} \\

Ever Given, Mar 2021 & Suez Canal & physical, total (6\,d) & \(\approx 0^{\dagger}\) &
full blockage for six days; definitional \citep{porteconomics_evergiven_2021} \\

Bosphorus insurance rule, 2022 & Turkish Straits & admin.\ friction & \(0.85\text{--}0.95\) &
up to \(27\) tankers queued, then cleared \(\rightarrow\) inferred \citep{insurancejournal_bosphorus_2022} \\

Black Sea blockade \(\to\) BSGI, 2022--23 & Black Sea / Turkish & blockade then corridor & \(0\!\to\!0.6\) &
\(32.9\) Mt moved \citep{un_bsgi_2023} vs.\ our \(\sim\!50\) Mt/yr benchmark \(\rightarrow\) volume ratio \\

Panama drought, 2023--24 & Panama Canal & capacity rationing\(^{\ddagger}\) & n/a &
transits cut \(36\!\to\!22\)/day; a throughput limit the model does not carry, not \(\delta\) \citep{seatrade_panama_2024} \\

Red Sea (boxes), 2023--24 & Bab-el-Mandeb / Suez & war-risk closure & \(0.05\text{--}0.15\) &
container transits \(-84\%\), capacity \(-94\%\) at trough \(\rightarrow\) AIS residual \citep{imf_portwatch} \\

Red Sea (crude \& products), 2023--24 & Bab-el-Mandeb & war-risk closure & \(0.40\text{--}0.55\) &
oil flow \(-54\%\), \(42\text{--}48\%\) of tanker transits persist\(^{\ast}\) \(\rightarrow\) volume/share residual \citep{eia_redsea_2024,imf_portwatch} \\

Determined Hormuz closure & Hormuz & counterfactual & unident.\(^{\S}\) &
no precedent; history gives a lower bound only \citep{eia_hormuz_2024} \\
\bottomrule
\end{tabularx}
\par\smallskip
\footnotesize\textit{Notes:} \(^{\dagger}\)\,\(\delta\approx 0\) but transient
(six days), below any steady-state horizon. \(^{\ddagger}\)\,A water-rationing
event, not a war-risk closure.
\(^{\S}\)\,Every observed Hormuz-area episode left Gulf oil largely flowing, so
the record bounds the residual transit share at Hormuz from below but cannot point-identify it under a determined, sustained closure, the object
the model's Hormuz column contemplates.
\(^{\ast}\)\,Transit counts and capacities are our count of the daily PortWatch series of Bab-el-Mandeb, against the mean of January to October 2023. The container row is taken at the early-2024 trough, and the tanker share of the oil row runs from February and March 2024 to the fourth quarter of 2024. The two rows are therefore not a common-date comparison. In the fourth quarter of 2024 tanker transits stood at \(42\%\) against \(34\%\) for container vessels, and the transiting capacity at \(37\%\) against \(8\%\).
\end{table}

The Houthi campaign against Red Sea shipping from December 2023 gives the cleanest available estimate of \(\delta\), because it is a genuine war-risk closure observed at
vessel-level resolution. Carriers withdrew once war-risk premia rose, exactly the
model's notion of closure, rather than because the passage was physically obstructed. What the sources
measure is unambiguous, and the two measures differ. Container tonnage crossing
the canal fell by $82\%$ between December 2023 and the first half of February
2024, while vessel tonnage rounding the Cape of Good Hope rose by $60\%$ over the
same window \citep{unctad_2024_troubled_waters}. Container-vessel transits of Bab-el-Mandeb fell further, by $84\%$ at the weekly trough of late January 2024, and the container capacity transiting the strait by $94\%$ in February and March, on our count of the daily PortWatch series \citep{imf_portwatch}. It is that transit ratio that measures the residual transit share. The Suez Canal Authority's revenue fell from \textdollar 10.2 billion in 2023 to \textdollar 3.9 billion in 2024, a fall of $61\%$, as the transits it counted halved from $26{,}434$ to $13{,}213$ \citep{sca_2025_statement}. A toll falls one
for one with a transit, and the high-value container traffic that pays the
largest tolls withdrew from the corridor first, so the receipts are a measure of transits weighted toward the container segment. From these measured
quantities we infer \(\delta\) for the container segment as the residual
that still transited, \(\delta_{\text{box}}\in[0.05,0.15]\) at the trough. The
inference is clean here, and only here, because PortWatch shows the trade volume rounding the Cape of Good Hope up by an estimated $74\%$ in the first two months of 2024, while that through Suez fell by half \citep{imf_2024_red_sea}. The decline is detour (\(e\)), so what survives is residual transit (\(\delta\)) rather than destroyed demand. The interval is
triangulated across three independent measurements that converge. The fall of $84$ to $94\%$ in container transits, the \(>\!60\%\) rerouting share, and the $61\%$ annual canal-revenue fall are mutually consistent only with a residual
near one-tenth. We take the steady-state \(\delta\) from the trough plateau. That the residual drifted back to about a third of the 2023 container-vessel count by late 2024 as the war-risk market adapted is itself the signature of \(\delta\)'s dependence on enforcement
intensity, not a contradiction of the trough estimate.

Crude and products behave differently on the same corridor, and the contrast
is what resolving \(\delta\) by commodity delivers. Tanker transits of Bab-el-Mandeb held near $47\%$ of their 2023 level in February and March 2024 and stood at $42\%$ in the fourth quarter, against $34\%$ for container vessels, on our count of the PortWatch series \citep{imf_portwatch}. The strait's oil flow averaged $4.0$ million barrels per day in the first eight months of 2024 against $8.7$ million in 2023, a decline of $54\%$ \citep{eia_redsea_2024}. From the measured tanker share (residual \(0.42\text{--}0.48\)) and the measured volume decline (residual \(0.46\)) we infer \(\delta_{\text{oil}}\in[0.40,0.55]\), so even under active attack the share of oil that keeps moving is four to five times the share of container traffic. Self-insured, war-risk-priced tonnage keeps energy
moving where container lines, answerable to schedule-reliability and
corporate-insurance constraints, withdraw.

Before 2026 the Strait of Hormuz, the most consequential chokepoint, had never been closed, so
history identifies its residual transit share least of all. Its benchmark is enormous, some twenty million
barrels per day, about a fifth of global petroleum-liquids consumption and a
quarter of seaborne oil \citep{eia_hormuz_2024}. But the relevant episodes are attacks, not closures, so they measure \(\delta\) only from below. During the 1984--88 Tanker War some \(451\) ships were hit in the Gulf, yet the strait stayed open, and at its height the war disrupted less than $2\%$ of the shipping through the Gulf. The United States reflagged and escorted Kuwaiti tankers under Operation Earnest Will rather than see the flow stop \citep{usip_tankerwar_2019}. After the 2019 Gulf of Oman attacks, the cost of insuring a transit of the strait rose about tenfold within two months, by one shipping chief executive's account, while business went on as usual \citep{cnbc_hormuz_2019}. What these episodes measure is attack counts and
insurance premia. What we infer is a lower bound of \(0.8\) on the residual transit share at Hormuz
under intense but undetermined pressure, cross-checked by the
absence of any sustained Gulf supply interruption in either episode. The determined, sustained closure of a
terminal gate, the model's most consequential counterfactual, had no
historical analogue before 2026. We therefore treat the baseline \(\delta=0.10\) on the Hormuz
column as a deliberately severe (low residual transit, high loss) calibration and report the
\(\delta\in\{0.01,0.05,0.10,0.20\}\) sensitivity of Online Appendix~\ref{app:robustness}
alongside it.

An administrative-friction episode corroborates that the residual transit share of energy is high and
hard to suppress. In December 2022 Turkey's proof-of-insurance rule at the Bosphorus held up as many as twenty-seven tankers at a time before the backlog began to clear, delaying but never stopping the oil \citep{insurancejournal_bosphorus_2022}. Even a determined
administrative obstruction of a crude corridor slowed the flow without halting
it, pointing to the same conclusion as the Red Sea tanker record, that on the
energy gates the through-passage residual \(\delta\) is large.

We fix the endpoints of the \(\delta\) range, rather than an interior value,
from three further episodes. Physical, total closures drive \(\delta\) to zero, the eight-year Suez
closure of 1967--75, the model's \(e\)-benchmark for the Cape detour
\citep{feyrer2021distance}, and the six-day Ever Given grounding of 2021
\citep{porteconomics_evergiven_2021}, the grounding too brief to register at the
disruption horizon. At the opposite end, a named war-risk without a
closure leaves \(\delta\approx 1\). The Joint War Committee listed the Strait of Malacca as a war-risk area from June 2005 to August 2006, without any closure of the strait \citep{rsis_malacca_2005,insurancejournal_malacca_2006}. The Panama drought of 2023--24, finally, enters the table as a contrast and not as an episode of residual transit. It cut daily transits from
thirty-six to twenty-two by exogenous water rationing
\citep{seatrade_panama_2024}, a throughput limit the model does not carry
rather than residual transit.

The historical record therefore does not support a single universal
\(\delta\). What it supports is a commodity-differentiated one. For the container and general-cargo corridors that
dominate the Asia--Europe gates (Suez, Bab-el-Mandeb, Gibraltar, Malacca, Dover),
a determined war-risk closure leaves \(\delta\in[0.05,0.15]\), centering the baseline \(\delta=0.10\). For the crude, product, and bulk corridors that
dominate the energy gates (Hormuz, the Turkish and Danish Straits, Kerch), the record puts the residual transit share substantially higher, at \(\delta\in[0.40,0.55]\) under
active attack.
The paper adopts the uniform residual transit share \(\delta_r\equiv0.10\) as its baseline,
applying the well-identified residual transit share of the container corridors to every seaborne good.
The commodity-differentiated residual transit share, which sets \(\delta_r=0.45\) on the energy
commodities, the center of the under-attack tanker interval \([0.40,0.55]\), and
\(\delta_r=0.10\) on every other good, the center of the container interval
\([0.05,0.15]\), enters the robustness exercises of
Online Appendix~\ref{app:modelrobust}. Either calibration is admissible without altering any result of Section~\ref{sec:model}. The effective demand floor
behind Lemma~\ref{prop:wellposed} and Online
Appendix~\ref{app:stability} requires only \(\delta_r>0\) on the
benchmark support. What turns that floor into a positive delivered
retention is the interiority the reported equilibria carry. Online Appendix~\ref{app:stability_numerics} reports both retention minima under both calibrations on the data. The uniform baseline is the more
severe of the two calibrations on the energy gates,
since it applies to tankers a box-like enforcement that no episode before 2026 had achieved. The closure of 2026 is the out-of-sample check of that claim (Online Appendix~\ref{app:hormuz_record}).

% ================================================================
\subsection{The 2026 Hormuz closure as an out-of-sample check of the residual transit share}
\label{app:hormuz_record}
% ================================================================

The evidence on the residual transit share in Online Appendix~\ref{app:est_delta} is drawn from
disruptions that predate the model. The closure of the Strait of Hormuz that
began on 28 February 2026 is the first episode at an energy gate to occur after
the calibration was set. It supplies what the calibration needs, namely how long the strait was shut,
how much moved through it while it was shut, and whether energy and other goods
were shut to the same degree. The closure of 2026 informs the parameters of the model and not the losses it reports. Those losses are the unbuffered exposure of Section~\ref{subsec:calibration} of the paper, the cost of a determined and sustained closure that no strategic reserve, no bypass, and no price-induced fall in demand cushions. An actual closure is cushioned by all three, so the output realized during one is not a test of the losses the model reports. We build the record from the daily chokepoint
transit series of IMF PortWatch, which we query directly, from the
strait-specific oil volumes of the IEA Oil Market Reports, and from the
commercial transit monitors that observe the vessels PortWatch cannot see.

The onset of the closure is 28 February 2026, the day
of the strikes, and not the formal Iranian confirmation of 2 March. PortWatch
records $86$ transits on 26 February, $51$ on 28 February, $18$ on 1 March and
$4$ on 2 March. The closure was never complete in the legal sense.
Iran operated a case-by-case carve-out, announced on 26 March, for vessels
linked to China, Russia, India, Iraq and Pakistan, and named vessels transited
throughout \citep{aljazeera_tollbooth_2026}. Against that, the carve-out was quantitatively small in the first phase and large in the second.
Qatar moved no liquefied natural gas out of the Gulf between the end of February and 10 May \citep{bloomberg_qatar_lng_2026}, and the strait carried $1.9$ million barrels a day of crude and condensate in March against $16.1$ before the closure \citep{iea_omr_apr2026}. The memorandum of understanding of 17 June then raised
transits to $20.4$ a day for the rest of that month, before the collapse of
13 July.

Table~\ref{tab:hormuz_record} reports the monthly record on three measures.
The first is the PortWatch transit count, which observes vessels broadcasting
on the automatic identification system. The second corrects that count for the traffic that went dark. Vortexa puts the dark share at $54$ to $65$ percent of outbound laden transits from March to May \citep{vortexa_dark_2026}, and Lloyd's List Intelligence at nearly $70$ percent of tanker transits in the week after the collapse of 13 July \citep{lloyds_hormuz_jul2026}. We apply the correction as the ratio of each commercial monitor's own total to the PortWatch total over the same dates \citep{vortexa_hormuz_2026,windward_dark_2026,bigoceandata_hormuz_2026}. The
third is strait-specific oil volume, taken from the IEA where it is published
and otherwise recovered as Gulf exports less pipeline bypass, a decomposition
that reproduces the April figure to $0.06$ million barrels a day.

\begin{table}[H]
\centering
\caption{The 2026 Strait of Hormuz closure, month by month. Transit counts are
daily means of IMF PortWatch chokepoint transit calls, queried on 15 August
2026, against the February 2026 mean of $69.9$ a day. The corrected column adds the traffic that had gone dark. Oil is crude and condensate through the
strait against the pre-closure $16.1$ million barrels a day, reported by the IEA for March and April \citep{iea_omr_apr2026,iea_omr_may2026} and recovered as Gulf exports less bypass thereafter.
The severance intensity $s$ is one minus the oil share and is the quantity that enters the model.}
\label{tab:hormuz_record}
\small
\setlength{\tabcolsep}{4.5pt}
\begin{tabular}{l rr rr r}
\toprule
 & \multicolumn{2}{c}{Transits per day} & \multicolumn{2}{c}{Strait oil} & \\
\cmidrule(lr){2-3}\cmidrule(lr){4-5}
Month & Observed & Corrected & mb/d & Share & $s$ \\
\midrule
February 2026 (base) & 69.9 & 69.9 & 16.1 & 1.000 & --- \\
March                & 2.8  & 12.4 & 1.9  & 0.118 & 0.882 \\
April                & 5.0  & 22.3 & 2.7  & 0.168 & 0.832 \\
May                  & 3.2  & 14.2 & 1.9  & 0.118 & 0.882 \\
June                 & 10.7 & 32.0 & 6.7  & 0.417 & 0.583 \\
July                 & 8.4  & 21.0 & 6.0  & 0.370 & 0.630 \\
1--15 August         & 4.0  & 10.1 & 4.0  & 0.250 & 0.750 \\
\bottomrule
\end{tabular}
\end{table}

The three measures agree once the dark correction is applied. Integrating the
severance intensity from 28 February to 15 August gives $4.21$
full-closure-equivalent months on strait oil volume and $4.12$ on the corrected
transit count, against $5.04$ on the uncorrected count and $2.69$ on Gulf
exports delivered after bypass. The two corrected measures are the ones that
measure the same object the model closes, namely passage through the gate, and
they place the realized share of calendar 2026 at $0.35$ through 15 August. The
uncorrected count treats dark transits as absent and overstates severance. The
export measure credits the closed gate with the Saudi and Emirati pipelines and
understates it.

The episode gives two estimates of the residual transit share, which differ in what they count. On the PortWatch count, which does not see the dark traffic, the strait carried
$0.065$ of its baseline tanker transits and $0.108$ of its baseline non-energy
transits, pooled over 1 March to 9 August. On strait oil volume it carried about a quarter of its baseline over the same months, the share running from $0.118$ in March to $0.417$ in June. The uniform baseline $\delta_r=0.10$ of Section~\ref{subsec:calibration} thus sits close to the first estimate and below the second, so it is the severe calibration for this
episode as well. Because the country loss is not linear in the residual transit share, the episode is evaluated correctly by solving the model at each month's value rather than by solving once at a pooled value. Online Appendix~\ref{app:hormuz_pipeline} reports the month-by-month solution.

The episode does not support the commodity-differentiated residual transit share at this gate. Online
Appendix~\ref{app:est_delta} sets $\delta_r=0.45$ on the energy commodities on
the evidence that no earlier episode had held tankers to a box-like
enforcement. The 2026 episode held them to less than that, the strait's oil share staying below $0.45$ in every month of Table~\ref{tab:hormuz_record}. On the PortWatch count energy severance
over the episode is $0.935$ on transit counts and $0.927$ on capacity, against
$0.892$ and $0.903$ for containerized, dry bulk, general, and roll-on cargo. On that count
non-energy severance is $0.95$ to $0.97$ times energy severance, so energy was
the more completely severed of the two, not the less.\footnote{A count of broadcasting vessels cannot see a tanker that has gone dark, and Vortexa puts the dark traffic at $54$ to $65$ percent of the outbound laden transits from March to May \citep{vortexa_dark_2026}. The count therefore overstates the severance of energy relative to other cargo, so the ordering of the two classes is less secure than the level.} Within the non-energy
group the ordering runs from roll-on cargo at $0.975$ and containers at $0.947$
down to dry bulk at $0.844$ and general cargo at $0.830$. Solving the
closure at the realized pair, $0.065$ on the energy commodities and $0.108$
elsewhere, moves the Emirates prediction by about two percent relative to the
uniform baseline, so the correction the episode delivers to the model is to
the closure's duration and not to the residual transit share.

The transit series leaves unsettled the closure's share of calendar 2026,
the pre-closure baseline, and the strait's liquefied natural gas flows. It
stops on 15 August 2026 and says
nothing about the remaining four and a half months, so the calendar-year share
is an assumption in its tail. The EIA path of flows increasing from
September gives $0.51$ and a no-recovery path gives $0.60$. The pre-closure baseline is
convention-dependent. PortWatch counts five registered commercial cargo classes at $69.9$ a day, while Lloyd's List Intelligence, whose counts cover cargo-carrying vessels above ten thousand deadweight tonnes \citep{lloyds_hormuz_aug2026}, recorded $3{,}131$ transits in June 2025, about $104$ a day \citep{usni_hormuz_jul2026}. We use PortWatch throughout because its counting universe is documented \citep{imf_portwatch_wp2025}. Its Suez series matches the Suez Canal Authority's own count for 2023 to $1.6$ percent, though it runs $9.5$ percent above the Authority's count for 2024 \citep{sca_2025_statement}. And no source publishes a
monthly liquefied natural gas series through the strait, so the Qatari
prediction rests on the zero-export window and the June recovery rather than on
a continuous series.

% ================================================================
\subsection{Substitution \texorpdfstring{\(\rho\)}{rho} from the production-function literature}
\label{app:est_rho}

Unlike the residual transit share \(\delta\), the substitution parameter \(\rho\) cannot be measured on observed quantities as a ratio. That parameter is the curvature of the production technology across a buyer's sectoral input bundles, the exponent \(\rho\) in
the intermediate input efficiency factor \(\kappa^{(\omega)}_j\) of
Section~\ref{subsec:input_bundle_distortion}, strictly negative when those input bundles are complements. The real-world evidence on it is the body of
structural estimates of how readily a producer substitutes one sectoral input
for another. We take that evidence from the literature that estimates the elasticity of
substitution among intermediate inputs from natural experiments and instrumented price
variation. We translate each estimate into the corresponding \(\rho\) (an
elasticity of substitution \(\sigma\) maps to \(\rho=1-1/\sigma\), so any
\(\sigma<1\) returns \(\rho<0\)). We then make explicit the three features of our setting that
decide which of the many reported elasticities the \(\rho\) of this paper
should match, the short-run horizon, the country-sector granularity of the
inputs, and the country-sector granularity of output. We report \(\rho\) as a range throughout, since it is the structural parameter that the evidence identifies least sharply.

The natural point of departure is \citet{atalay2017}, who estimates the
elasticity of substitution among intermediate inputs from United States industry
input-expenditure shares, instrumenting relative input-price movements with
sector-specific military-procurement shocks. His estimates place that elasticity
below \(0.2\), so the implied \(\rho\) lies below about \(-4\), strongly
complementary and far from the Cobb--Douglas value \(\rho=0\). Atalay separately
estimates the elasticity between value added and the intermediate aggregate, but that is a different nest from the one the intermediate input efficiency factor
penalizes, which is substitution among the sectoral inputs themselves. It bears on the outer-nest curvature \(\varrho\) and not on \(\rho\). The firm-level natural-experiment evidence is sharper about the horizon, and most of it points the same way. \citet{boehm_flaaen_2019} treat the 2011 T\=ohoku earthquake as an exogenous outage of
Japanese suppliers and find that the output of affected affiliates falls almost
one-for-one with their lost imported inputs. The relation is close enough to Leontief
to drive \(\rho\) far below \(-1\), toward the complementary limit. \citet{barrot_sauvagnat_2016}, exploiting natural disasters that strike suppliers, find that the losses of the customers are concentrated where the supplier produces a specific input. \citet{carvalho_etal_2021_covid}, following the same earthquake through the Japanese production network, estimate an elasticity of \(1.2\) between the inputs that a firm buys from its individual suppliers and of \(0.6\) between its primary factors and its intermediate bundle. The network-macro literature that calibrates the elasticity rather than estimating it sets it at \(0.2\) in \citet{baqaee_farhi_supply_2022} and \citet{baqaee_farhi_2024} and at \(0.5\) in \citet{bonadio_etal_2021}, that is \(\rho\in[-4,-1]\).
Table~\ref{tab:rho_estimates} collects these, each with the real-world variation
that identifies it and the \(\rho\) it implies.

\begin{table}[H]
\centering
\caption{Evidence on the elasticity of substitution among intermediate inputs, translated to the CES parameter \(\rho=1-1/\sigma\), or, where the study estimates no elasticity, the sign that its findings imply. The upper rows are identified from real-world variation, the lower block from values that the network-macro literature calibrates. Each entry refers to the nest of intermediate inputs that \(\kappa^{(\omega)}_j\) penalizes, at the level of aggregation of the study.}
\label{tab:rho_estimates}
\footnotesize
\renewcommand{\arraystretch}{1.3}
\begin{tabularx}{\textwidth}{@{}%
  >{\raggedright\arraybackslash}p{3.0cm}%
  >{\raggedright\arraybackslash}X%
  >{\raggedright\arraybackslash}p{2.45cm}%
  >{\centering\arraybackslash}p{1.85cm}@{}}
\toprule
Study & Identifying real-world variation & Nest \& horizon & Implied \(\rho\) \\
\midrule
\citet{atalay2017} & US industry input-expenditure shares; military-procurement instrument & across-intermediate, direct estimate & \(\lesssim -4\) \\
\citet{boehm_flaaen_2019} & 2011 T\=ohoku earthquake; firm-level supplier outage & imported vs.\ domestic, short-run & near-Leontief, \(\ll -1\) \\
\citet{carvalho_etal_2021_covid} & T\=ohoku propagation through the Japanese supply network & across a firm's suppliers, one year & \(\approx0.15\) (\(\sigma=1.2\)) \\
\citet{barrot_sauvagnat_2016} & natural disasters at suppliers & specific inputs, quarters & negative (sign only) \\
\addlinespace
\citet{baqaee_farhi_supply_2022,baqaee_farhi_2024} & network-macro calibration & across intermediates & \(-4\) (\(\sigma=0.2\)) \\
\citet{bonadio_etal_2021} & COVID-19 supply-chain model (calibration) & across intermediates & \(-1\) (\(\sigma=0.5\)) \\
\bottomrule
\end{tabularx}
\end{table}

Read at face value, the short-run estimates on sectoral and on imported inputs sit near \(\rho\le-4\), and the calibrated network-macro values run to \(\rho=-1\). The estimate of \citet{carvalho_etal_2021_covid} lies above the Cobb--Douglas value, but it is taken across the inputs that a firm buys from its individual suppliers, so it mixes the substitution among the suppliers of one input, which our model assigns to the exchange, with the substitution across sectors that \(\rho\) governs. Our baseline \(\rho=-1\) sits at the upper, less-complementary
edge of that macro range, and matching the estimates to the right object (a
short-run, country-sector elasticity rather than a firm-level or micro one) is
what the rest of this subsection supplies. We match the \(\rho\) of this model to an entry in Table~\ref{tab:rho_estimates} on those features
of our setting, and since they do not all push the same way we state each.

The first is that the counterfactual is short run. Substitution
possibilities widen as producers re-engineer their input mix, retool, and qualify
new suppliers, so the elasticity rises with the horizon allowed for adjustment and
the short-run value is the most complementary one. Ports, refineries, supplier
relationships and product specifications are fixed within the disruption window,
so the estimates that bear on our \(\rho\) are those taken on impact and in the first year. \citet{peter_ruane} find, on Indian plant data, elasticities between broad categories of material inputs well above one in the long run and much lower over shorter horizons. A
long-run elasticity would answer a different question from the one the model
poses. The short-run horizon therefore keeps \(\rho\) well below zero.

The second is that inputs enter at the granularity of country-sectors,
collected into sectoral bundles, and \(\rho\) is the elasticity across those sectoral bundles. The parameter \(\rho\) thus governs substitution across sectors alone, the substitutability of, say, refined
petroleum against basic metals in a downstream producer's input bundle. Substitution across sectors is not the substitution among the several country suppliers of a single input, which the
within-sector re-matching \(\tau\) and the exchange of Definition~\ref{def:choke_reroute} already carry. The
literature value we want is therefore the across-broad-sector elasticity, not the
across-origin (Armington) elasticity, which is several times larger and belongs to
\(\tau\) rather than \(\rho\). The input bundle is also coarse, since each of the fifty ICIO sectors pools many narrower products that are themselves mutual
substitutes, so the elasticity between sectoral bundles exceeds the elasticity between
the fine inputs of
the micro studies.  The micro estimates on narrower inputs are on this count a most-complementary bound on our \(\rho\) rather than its value, since the coarseness of the sectoral bundle moves \(\rho\) back toward zero.

The third is that output, too, is at the country-sector level. The producing unit
is not a plant or a firm but an aggregate of many heterogeneous producers, so a single CES technology imposed on that aggregate does not inherit the curvature of
its members. Even if each constituent producer can substitute inputs barely at
all, the country-sector as a whole substitutes more, because the composition of
which producers are active tilts toward those better able to absorb a given
input loss. Aggregation across heterogeneous units thus raises the effective elasticity above the
plant-level one. This aggregation effect is the micro-to-macro result of \citet{oberfield_raval_2021} for
the substitution between capital and labor, and classically of \citet{houthakker_1955}, in which
Leontief plants drawn from a Pareto distribution aggregate to a smooth, substitutable
industry technology. The near-Leontief firm-level relation of
\citet{boehm_flaaen_2019} is on this count also a lower bound, measured on
precisely the units our country-sectors aggregate over, since the aggregation of producers moves \(\rho\) toward zero a second time.

Taken together, the short-run horizon keeps \(\rho\) distinctly below zero, while the two aggregations, of inputs into sectoral
bundles and of output into country-sector producers, raise it above the micro estimates. The evidence then supports an effective short-run country-sector \(\rho\) in roughly
\(\rho\in[-1,-0.25]\), the calibration rectangle of Online Appendix~\ref{app:robustness}. The
baseline \(\rho=-1\) sits at its more-complementary edge, closest to the directly
estimated micro and network-macro elasticities. A reader who weights the aggregation-toward-zero corrections more heavily
would prefer a \(\rho\) nearer \(-0.25\). That spread of defensible values is why \(\rho\) is the parameter we vary most widely and why every main result is reported
across the full rectangle, down to the easy-substitution edge \(\rho=-0.25\) of Online Appendix~\ref{app:robustness}. The production-function literature identifies the sign and the short-run strength of complementarity firmly and locates the
country-sector \(\rho\) within a band. It does not single out a point inside
it.

We set the baseline at the complementary edge \(\rho=-1\) rather than at the deeper micro values
because the choice is mild for the complementarity channel. A less negative exponent penalizes a missing input
less, so the losses the paper reports through this channel are a floor
relative to what the micro estimates would deliver rather than a central
estimate. The unbuffered steady state of Section~\ref{subsec:calibration} also requires the substitution channel to
operate freely once an input is reconstituted. A deeper exponent sharpens the
curvature of the intermediate input efficiency factor at every retention profile, benchmark input bundles
included, so part of the added penalty would fall on small compositional
deviations that a buffered world smooths away. At the baseline both nests
carry an elasticity of substitution \(\sigma=1/(1-\rho)=0.5\), and the leakage \(\beta_j\) of the outer nest breaks the unit gain
that would otherwise admit a continuum. The solved equilibrium is interior across the whole admissible range either way, verified after solving (Online Appendix~\ref{app:stability_numerics}).

The outer-nest curvature \(\varrho\) governs the substitution between the intermediate composite and the primary inputs, and the evidence on it is the elasticity between value added and intermediates. The estimates of \citet{atalay2017} put that elasticity at \(0.8\) to \(0.9\) with his instrument and at \(0.4\) to \(0.8\) on plant-level data, and \citet{carvalho_etal_2021_covid} estimate it at \(0.6\) on Japanese firms in the year after the 2011 earthquake. The network-macro calibrations of \citet{baqaee_farhi_supply_2022} and \citet{baqaee_farhi_2024} set it at \(0.6\) and \(0.5\). These values run from \(0.4\) to \(0.9\), which is to say that \(\varrho\) lies between \(-1.5\) and \(-0.1\) (Table~\ref{tab:calibration}). We set the baseline at \(\varrho=-1\), which gives the outer nest the elasticity of one half that the inner nest carries and that \citet{baqaee_farhi_2024} use. That value lies inside the plant-level estimates and toward the complementary end of the whole span, so a less complementary outer nest would lower the losses, though by little. With \(\varrho\) moved alone across the estimates, the world loss of a closed Hormuz runs from \(2.71\%\) at \(\varrho=-0.125\) to \(2.89\%\) at \(\varrho=-1.5\), against \(2.83\%\) at the baseline, and the loss of Korea runs from \(9.4\%\) to \(10.6\%\), against \(10.2\%\). The harsh corner of Online Appendix~\ref{app:modelrobust} takes \(\varrho\) to \(-4\), which is beyond the estimates and calibrations cited here and is the complementary edge of the range of \(\rho\). With \(\varrho\) moved alone to that value, the world loss of a closed Hormuz is \(3.19\%\).

% ================================================================
\subsection{The unrecovered share \texorpdfstring{\(\tau\)}{tau} and the evidence on supply disruptions}
\label{app:est_tau}

The unrecovered share \(\tau\) is the share of displaced same-sector trade
that is not reconstituted on surviving suppliers within the disruption
horizon. A fraction \(1-\tau\) of a buyer's displaced same-sector demand is re-sourced onto the
suppliers it can still reach, and a fraction \(1-\tau\) of a severed supplier's idle
output is placed on the buyers it can still reach. The residual \(\tau\) is, on each
side, lost (Section~\ref{subsec:rerouting_matrix}). The unrecovered share \(\tau\) therefore measures the substitution within a sector and across its suppliers, distinct
from the other two adjustments. The first is the route detour \(e\),
which salvages a shipment to the same counterparty by a longer voyage and which the
reroutability matrices already capture. The second is the cross-sector complementarity
\(\rho\) of Online Appendix~\ref{app:est_rho}, which governs substitution between sectoral
inputs rather than among the suppliers of one. Unlike \(\rho\), \(\tau\) is a reduced-form recovery rate. No study we know of reports it as such, and the firm-level evidence on supply disruptions bears on it through how far the purchases and the output of a buyer fall when it loses a supplier, how long the fall lasts, and whether the buyer turns to other suppliers. We read that evidence through the short-run horizon, the substitution across suppliers just isolated, and the country-sector granularity. We report \(\tau\) as a range, with baseline \(\tau=0.3\).

In the first months of an outage buyers replace little of what they lose. \citet{boehm_flaaen_2019} find that the US affiliates of Japanese multinationals cut their output roughly one-for-one with their imports from Japan in the months after the 2011 T\=ohoku earthquake, as a Leontief relation between imported and domestic inputs would imply. \citet{barrot_sauvagnat_2016} find that a natural disaster that strikes a supplier lowers the sales growth of its customers by two to three percentage points, and by about five where the supplier produces a differentiated input, while a supplier of a standardized input passes on no significant loss. \citet{khanna_morales_pandalai_2022} find that the Indian firms whose suppliers lay in the districts under the strictest COVID-19 lockdowns of 2020 lost more of their suppliers and cut their input purchases and their sales by more. Where an outage is expected to end, buyers wait for it. \citet{castro_vincenzi_etal_2024} find that the buyers of suppliers struck by floods return to those suppliers once they recover, a few months later, rather than switch to others. Over longer horizons the trade moves. \citet{carvalho_etal_2021_covid} estimate an elasticity of \(1.2\) between the inputs that a Japanese firm buys from its individual suppliers, on sales in the year after the earthquake, which is to say that the inputs of different suppliers are mild substitutes over that horizon. \citet{freund_etal_2022} find that the importers most dependent on Japan before the earthquake cut their imports from Japan by about a third in electronics and by more than a half in autos, relative to other importers, and turned to new suppliers, chiefly in developing and larger economies. Table~\ref{tab:tau_estimates} collects these studies.

\begin{table}[H]
\centering
\caption{Evidence on the replacement of lost inputs. No study reports the unrecovered share \(\tau\) itself. Each row gives what the study finds and over what horizon. }
\label{tab:tau_estimates}
\footnotesize
\renewcommand{\arraystretch}{1.3}
\begin{tabularx}{\textwidth}{@{}%
  >{\raggedright\arraybackslash}p{3.0cm}%
  >{\raggedright\arraybackslash}p{3.1cm}%
  >{\raggedright\arraybackslash}X%
  >{\raggedright\arraybackslash}p{1.5cm}@{}}
\toprule
Study & Disruption & Finding & Horizon \\
\midrule
\citet{boehm_flaaen_2019} & 2011 T\=ohoku earthquake & output of US affiliates of Japanese firms falls about one-for-one with their imports from Japan & months \\
\citet{barrot_sauvagnat_2016} & natural disasters at US suppliers & sales growth of customers falls 2 to 3 percentage points, about 5 for differentiated inputs & quarters \\
\citet{khanna_morales_pandalai_2022} & 2020 COVID-19 lockdowns in India & more separations from suppliers, larger falls in input purchases and in sales & months \\
\citet{castro_vincenzi_etal_2024} & floods at suppliers & buyers return to the flood-hit suppliers rather than switch & months \\
\citet{carvalho_etal_2021_covid} & 2011 T\=ohoku earthquake & elasticity of \(1.2\) between the inputs bought from different suppliers & one year \\
\citet{freund_etal_2022} & 2011 T\=ohoku earthquake & importers dependent on Japan turn to new suppliers & years \\
\bottomrule
\end{tabularx}
\end{table}

The evidence fixes how \(\tau\) moves and not where it lies. The unrecovered share falls with the time allowed for re-sourcing, from close to one in the first months of an outage, when buyers replace little of what they lose, to well below one over the years after the T\=ohoku earthquake, when the trade of Japan's suppliers moved to others. It is also heterogeneous across inputs, low
where the severed good is fungible and re-sourced almost in full and
high where it is specific. Crude oil, the largest single mass in the chokepoint problem, shows why even a traded commodity is not re-sourced in full. Within the model,
deliveries inside one input sector are fungible, and what enters the CES intermediate input efficiency factor is the shortfall in the sector's delivered mass after the unrecovered share of the displaced demand has been written off. The grade specificity of the real world puts content behind
those parameters. Japanese refiners are configured for Arabian Light,
Arabian Medium, Murban, or Kuwait Export Blend, against substitutes
that differ in sulfur, API gravity, and residue balance, and that mismatch is one reason why a re-matched barrel is not a full replacement within the horizon. The model has no separate term for grade mismatch. So any single \(\tau\) is a central value standing in for a
distribution, the same simplification
the uniform baseline \(\delta_r\equiv0.10\) of Online Appendix~\ref{app:est_delta}
makes for a residual transit share that the record shows varying by commodity.

We place our \(\tau\) within this evidence by the horizon, by the adjustment that the studies observe, and by the granularity of the country-sector. The first consideration is the horizon. The recovered share grows as firms locate, qualify and contract alternative suppliers, so \(\tau\) falls with the time allowed. A closure sustained through a year lies between the months in which the studies find little replacement and the years over which the buyers of Japanese inputs moved to other suppliers. A buyer that faces a closure of a year has no reason to wait for its supplier, as the buyers of flood-hit suppliers do.

The second is that \(\tau\) measures the substitution across suppliers alone. The studies observe it together with the other adjustments, since the output of a buyer that loses a supplier reflects at once its re-sourcing and the complementarity of the lost input with the rest of its bundle. We therefore read them for the direction and the horizon of \(\tau\) and not for its level. The model applies the recovered share symmetrically to the two sides of a severed link, whereas the studies observe the buyer side. We impose the same unrecovered share on the severed supplier's search for replacement buyers.

The third is the country-sector granularity, which here pulls in two directions
that roughly cancel. On one hand a country-sector pools many firms and many
same-sector suppliers, so its scope for re-sourcing exceeds a single firm's.
The aggregate recovered share would then exceed the firm-level one, pushing \(\tau\) down. On the other the exchange of Definition~\ref{def:choke_reroute} is confined to the benchmark support, in that it redistributes displaced mass only onto trading
relationships already present at benchmark, never onto a newly formed one. The re-sourcing that the firm-level studies observe, by contrast, includes precisely such new-supplier formation, which pushes \(\tau\) back up. With the two effects offsetting, the firm-level evidence bears on the country-sector \(\tau\) without the systematic aggregation correction that \(\rho\) required.

Taken together the evidence places \(\tau\) below one at the horizon of a year and higher for specific inputs, but it does not single out a value. We adopt the baseline \(\tau=0.3\), an intended seventy-percent reconstitution of displaced same-sector trade (on a fully exposed block the effective demand retains that block's \(\delta_r\)), which assumes that most of the displaced trade finds a new partner within the year and is the more recoverable reading of the evidence. Online Appendix~\ref{app:robustness} varies it over \([0.1,0.5]\). So on the re-matching, as on the residual transit share and the substitution across sectors, the robustness cells that raise the loss, here the less-recoverable \(\tau=0.4\) and \(0.5\), move toward the more pessimistic values the evidence admits.\footnote{The 2026 Hormuz closure permits a first direct estimate of that re-matching. On a product-level panel of Eurostat Comext monthly imports, sea imports of a product from Gulf origins fall by \(0.31\) log points against same-product imports from clear origins after the onset of March 2026. The panel supplies three post-onset months and covers European reporters, while the heavy Gulf dependence sits with the Asian importers, so the recovery slope it returns has a confidence set covering the whole unit interval of \(\tau\). The unrecovered share therefore stays at its calibrated value. The direct estimate is archived in the replication package.}

% ================================================================
\section{The Rationed Exchange and the Fulfillment Fixed Point on the Data}
\label{app:stability_numerics}
% ================================================================

This section reports how the rationed exchange and the fulfillment fixed point behave on the
data at the baseline $\delta=0.10$. We report the geometry of the problem each sector block is
solved on, the sweeps and residuals the Sinkhorn iteration reaches, the support and the column
totals of the effective demand matrix $\mathbf A_\omega$, and the two retention minima the fulfillment fixed point inherits. We then report the iteration that reaches the fixed point and the interiority of the equilibria it returns. Every count is taken on the problem the exchange actually
solves, after the deletion of the rows with a zero residual capacity and of the columns with a
zero residual budget that precedes Lemma~\ref{prop:wellposed} and on the slack-augmented block
wherever the placement deficit $U_r$ is positive. Each closure touches $27$ of the $50$ sector
blocks, so the twelve single closures and the three joint scenarios contribute $405$ solved
blocks in all.

\subsection{The exchange block by block}

Table~\ref{tab:block_geometry} reports the geometry, in which strict feasibility and the sign of the
placement deficit cross on these data, and
both off-diagonal cells are occupied. Of the $405$ solved blocks, $242$
carry a placement deficit and $45$ are boundary projections, $12$ of the
latter on blocks carrying no deficit and $33$ on augmented ones. Strict feasibility is tested on the support graph of each block rather than on the converged
entries, since a boundary projection approaches its zero at a polynomial rate and the cell the
minimizer sets to zero is still positive after any finite number of sweeps. The $45$ boundary
projections zero $2{,}064$ allowed cells in all. Of those, $1{,}865$ are
trading relationships and $199$ are cells of the slack pair, which is a
write-off account rather than a link. No block loses more than $376$
trading relationships this way.

The solved support of every block is connected, in every scenario, so the
maximum component count is one. The reciprocal gauge of
Lemma~\ref{prop:wellposed}(ii) is accordingly one gauge per block on these
data, and the block and the support component coincide as units of
independence. Which support is counted matters here, since the open links on their own split, with $35$ of the $405$ blocks carrying more than one
component before the slack pair is appended, up to $13$ in one Hormuz
block, and $135$ surviving buyer columns isolated on it, each a buyer whose
whole sector-$r$ sourcing the closure severs. The unmet-demand row
reconnects every one of them, which is why the solved support is connected while the open links on their own are not.

% >>> inlined from 04_outputs/tex_tables/tab_block_geometry.tex
\begin{table}[H]
\centering
\caption{Block geometry of the rationed exchange at the baseline
$(\tau,\rho,\varrho,\delta)=(0.3,-1,-1,0.10)$. Each row counts the sector
blocks the closure touches, classified on the problem the exchange solves, namely the unaugmented block where the placement deficit $U_r$ vanishes and the
slack-augmented block where it does not. ``Strict'' and ``Boundary'' divide
those blocks by whether the projection stays interior to the solved support,
$\operatorname{supp}(\mathbf B^{*})=E$, the equivalence of
Lemma~\ref{prop:wellposed}(ii), tested on the support graph of the block. The last column counts blocks
whose solved support has more than one connected component.}
\label{tab:block_geometry}
\small
\begin{tabular}{l r r r r r r}
\toprule
& & \multicolumn{2}{c}{$U_r=0$} & \multicolumn{2}{c}{$U_r>0$} & \\
\cmidrule(lr){3-4}\cmidrule(lr){5-6}
Closure & Blocks & Strict & Boundary & Strict & Boundary & $>1$ comp. \\
\midrule
Hormuz & 27 & 8 & 1 & 16 & 2 & 0 \\
Turkish & 27 & 14 & 0 & 10 & 3 & 0 \\
Danish & 27 & 12 & 1 & 12 & 2 & 0 \\
Suez & 27 & 10 & 0 & 14 & 3 & 0 \\
Bab-el-Mandeb & 27 & 10 & 0 & 14 & 3 & 0 \\
Gibraltar & 27 & 13 & 0 & 11 & 3 & 0 \\
Malacca & 27 & 11 & 2 & 13 & 1 & 0 \\
Panama & 27 & 9 & 1 & 15 & 2 & 0 \\
Taiwan--Luzon & 27 & 12 & 1 & 12 & 2 & 0 \\
Dover & 27 & 9 & 2 & 15 & 1 & 0 \\
Korea & 27 & 10 & 1 & 14 & 2 & 0 \\
Kerch & 27 & 6 & 0 & 18 & 3 & 0 \\
\midrule
$\cset{}^{ME}$ & 27 & 7 & 1 & 17 & 2 & 0 \\
$\cset{}^{EA}$ & 27 & 11 & 2 & 13 & 1 & 0 \\
$\cset{}^{RE}$ & 27 & 9 & 0 & 15 & 3 & 0 \\
\midrule
All fifteen & 405 & 151 & 12 & 209 & 33 & 0 \\
\bottomrule
\end{tabular}
\end{table}
% <<< end 04_outputs/tex_tables/tab_block_geometry.tex

Table~\ref{tab:rate_regimes} reports the sweeps the iteration takes and the residuals it
reaches, separately for the two rate regimes of Lemma~\ref{prop:wellposed}(iii) rather than by the
placement deficit. The strictly feasible blocks take a median of $320$ sweeps and
the boundary projections a median of $191$, and the worst relative column
residual is $2.2\times10^{-6}$ in the first group and
$9.2\times10^{-11}$ in the second. The ordering of the two medians carries
no rate content, since the regimes differ in which blocks they collect
rather than in a common difficulty. What the table does show is that the
loosest residual on these data belongs to a strictly feasible block and not to a boundary one. Seventeen blocks exhaust the sweep budget, all of them strictly feasible. Three of those $17$ are
the only blocks anywhere whose column residual exceeds $10^{-7}$, and
every block that stops on the tolerance instead reaches $10^{-10}$.

% >>> inlined from 04_outputs/tex_tables/tab_rate_regimes.tex
\begin{table}[H]
\centering
\caption{Sinkhorn sweeps to tolerance by rate regime, pooled over the twelve
single closures and the three joint scenarios at the baseline
$\delta=0.10$. One sweep is one column rescaling followed by one row
rescaling, the order the solver runs. ``At budget'' counts blocks that exhaust the sweep allowance,
$5{,}000$ sweeps on an unaugmented block and $20{,}000$ on a block carrying
the slack pair. The residual is the sup-norm error in the column totals relative to the block's largest capacity.}
\label{tab:rate_regimes}
\small
\begin{tabular}{l r r r r c}
\toprule
Rate regime & Blocks & Median & Maximum & At budget & Column residual \\
& & sweeps & sweeps & & (largest) \\
\midrule
Strictly feasible & 360 & 320 & 20000 & 17 & $2.2\times10^{-6}$ \\
Boundary projection & 45 & 191 & 704 & 0 & $9.2\times10^{-11}$ \\
\bottomrule
\end{tabular}
\end{table}
% <<< end 04_outputs/tex_tables/tab_rate_regimes.tex

The support of the effective demand matrix $\mathbf A_\omega$ is contained
in the benchmark support, with equality on the $360$ strictly feasible
blocks and strict containment on the $45$ boundary projections, which zero
$1{,}865$ clear cells no residual transit reaches. Every used sector block of every
buyer's column nonetheless delivers strictly positive mass, at all
$205{,}622$ pairs $(r,j)$ with $\alpha_{rj}>0$ in every scenario, which is
the property Lemma~\ref{lem:choke_kernel} needs and the boundary
projections leave intact. No column total of $\mathbf A_\omega$ exceeds
one beyond the solver tolerance, the largest excess over the fifteen
scenarios being $1.0\times10^{-7}$. The bound binds harder at the block level, where a single sector block of a buyer's column can overshoot its
benchmark share $\alpha_{rj}$ by up to $2.4\times10^{-6}$ in relative
terms, which is the residual the Sinkhorn iteration leaves on its slowest blocks. The shortfalls elsewhere in the same column absorb part of that overshoot.

The computation implements the exact objects of Section~\ref{subsec:rerouting_matrix} to finite
tolerances. The slack accounts are booked above the computed placement deficit by a relative $10^{-3}$, or by the rounding bound of the maximum-flow computation that measures the deficit where that bound is larger. The booking overstates the deficit by three-tenths of one percent across the twelve single closures.

\subsection{Retention at the solved equilibria}

Two retention objects govern the fulfillment fixed point. Table~\ref{tab:retention_minima} reports both at every scenario's
solved equilibrium. Two sets of blocks have to be kept apart in interpreting them. The formal statements take $\mathcal R_j=\{r:\alpha_{rj}>0\}$,
while the implementation holds at full retention any input sector
whose benchmark share falls below $5\times10^{-4}$ of that buyer's total,
so that an input of negligible share that a closure cuts off cannot by itself stop its buyer. The rule covers $47{,}643$ of the $205{,}622$ pairs with $\alpha_{rj}>0$, which make up $0.17\%$ of the input bundle at the median buyer and at most $0.63\%$. It changes little at the baseline calibration, where the residual transit keeps every retention away from zero. With every input sector treated alike, the world loss of each of the fifteen scenarios is higher by between $0.3$ and $0.7$ percent of the loss itself, which is less than a hundredth of a percentage point of world GDP in every scenario. At the Strait of Hormuz the world loss goes from $2.825\%$ to $2.833\%$. No economy's loss moves by as much as $0.05$ percentage points of its GDP, and the three joint indices are unchanged to the third decimal. The table reports the minima over the second, narrower
set, the sectors whose retention moves the intermediate input efficiency factor. The
raw set is quoted alongside where the two differ.

The effective demand retention $\bar\theta^{(\omega)}_{rj}$ is the object
the residual transit floors. Its smallest value is $0.100$, the residual transit share,
which identity~\eqref{eq:standing_retention} forces on a block whose links
the closure exposes in full and which move by sea alone. That floor is reached over the raw set at
three of the fifteen scenarios, the Turkish Straits, the Danish Straits, and the
Russia--Europe scenario, and the raw set comes within $3\times10^{-5}$ of that floor at the two scenarios that close Hormuz. Over the narrower set the smallest value is $0.130$, under a Hormuz closure, the blocks at the floor
carrying negligible benchmark shares. Realized retention
$\theta^{(\omega)}_{rj}(\mathbf h_\omega)$ is the second object. It bottoms out at $0.123$ on the narrower set and $0.096$ on the raw one,
below the effective demand floor by the order-weighted mean fulfillment of
the block's own suppliers and not by anything the residual transit share supplies.
What separates the two on a given block is that block's own
order-weighted mean fulfillment, and what bounds the separation globally
is $\min_i h^{(\omega)}_i=0.472$, attained at the Hormuz closure along
with $\min_i\kappa^{(\omega)}_i=0.444$. Blocks exposed in full are few once the seaborne share is applied.
Six blocks are exposed in full across the fifteen scenarios, two under the Turkish Straits, one under the Danish Straits,
and three under the Russia--Europe scenario, while ten of the twelve gates
carry none. The factorization~\eqref{eq:realized_retention} holds in the
implementation to machine precision over the used
blocks of every scenario. No used block loses its delivered mass at any
scenario, at any of the $205{,}622$ pairs with $\alpha_{rj}>0$.

% >>> inlined from 04_outputs/tex_tables/tab_retention_minima.tex
\begin{table}[H]
\centering
\caption{Effective demand and realized retention at the baseline
$(\tau,\rho,\varrho,\delta)=(0.3,-1,-1,0.10)$, one row per scenario.
$\min\bar\theta^{(\omega)}_{rj}$ and $\min\theta^{(\omega)}_{rj}$ are the
smallest effective demand and realized retention over the used sector blocks
of the solved equilibrium. The last two columns give the smallest fulfillment rate and intermediate input efficiency factor over the $4{,}350$
country-sectors. The two retention
minima are taken over the effective input-bundle set the implementation uses, which holds input sectors below $5\times10^{-4}$ of a buyer's benchmark
total at full retention.}
\label{tab:retention_minima}
\small
\begin{tabular}{l r r r r}
\toprule
Closure & $\min\bar\theta^{(\omega)}_{rj}$ & $\min\theta^{(\omega)}_{rj}$ & $\min_i h^{(\omega)}_i$ & $\min_i\kappa^{(\omega)}_i$ \\
\midrule
Hormuz & 0.130 & 0.123 & 0.472 & 0.444 \\
Turkish & 0.552 & 0.533 & 0.887 & 0.834 \\
Danish & 0.752 & 0.750 & 0.832 & 0.821 \\
Suez & 0.897 & 0.890 & 0.924 & 0.920 \\
Bab-el-Mandeb & 0.897 & 0.890 & 0.924 & 0.920 \\
Gibraltar & 0.819 & 0.819 & 0.938 & 0.924 \\
Malacca & 0.926 & 0.925 & 0.980 & 0.977 \\
Panama & 0.820 & 0.818 & 0.955 & 0.943 \\
Taiwan--Luzon & 0.950 & 0.950 & 0.985 & 0.981 \\
Dover & 0.976 & 0.976 & 0.993 & 0.991 \\
Korea & 0.919 & 0.919 & 0.990 & 0.988 \\
Kerch & 0.906 & 0.906 & 0.993 & 0.990 \\
\midrule
$\cset{}^{ME}$ & 0.130 & 0.123 & 0.478 & 0.450 \\
$\cset{}^{EA}$ & 0.908 & 0.907 & 0.969 & 0.965 \\
$\cset{}^{RE}$ & 0.640 & 0.617 & 0.832 & 0.821 \\
\bottomrule
\end{tabular}
\end{table}
% <<< end 04_outputs/tex_tables/tab_retention_minima.tex

\subsection{Reaching the fixed point, and its interiority}

The implementation reaches the selected equilibrium by a damped iteration with Anderson mixing, where Definition~\ref{def:selection} of the paper iterates the fulfillment rule itself from \(\mathbf h=\mathbf 1\). The two reach the same limit on these data. At the Hormuz closure and at the Middle East scenario, the iteration of the definition ends within \(2\times10^{-7}\) of the solver's fulfillment rate at every country-sector. At the baseline $\delta=0.10$ the damped fulfillment iteration contracts its residual at a median factor between $0.31$ and
$0.48$ per sweep across the twelve closures, and converges in $12$ to $22$
sweeps. These factors are observed rates of the implemented scheme along its own
orbit. No geometric rate is proved for the damped iteration on these data. The obstruction is the supplier-side row sum
$\bar a_\omega$ of $\mathbf A_\omega$ that any $\ell_1$ bound on levels
carries. Lemma~\ref{lem:choke_kernel} holds every column of
$\mathbf A_\omega$ at or below one and leaves the rows free, since a row sum is a seller's market breadth rather than a budget. Here
$\bar a_\omega=19.3$ at every scenario, with $1{,}299$ of the $4{,}350$
rows above one under a Hormuz closure against a median of $0.54$ over the
rows that carry any sales. The maximizing row belongs to a country-sector
no closure of this paper reaches, which is why the figure does not move
across scenarios. What the reported equilibria rest on instead is
Proposition~\ref{prop:equilibrium}, part (ii) for monotone convergence of
the benchmark path and part (iv) for the rate, in the relative norm
whose modulus is an output elasticity of intermediates. That modulus is $0.991$ at
every active country-sector except the zero-technology-share accounting
column of Online Appendix~\ref{oa:data_integrity}, where it is exactly
one.

Under the two-nest fulfillment rule the model returns an
interior equilibrium at every single and joint closure, across the calibration
rectangle and at every residual transit share down to a near-total blockade
$\delta=0.01$. No scenario, corner, or value of the residual transit share drives the iteration to a saturated or trivial state.
The Middle East scenario, the most demanding of the three, remains interior throughout, at $3.00\%$ of world GDP at the baseline uniform residual transit share
$\delta=0.10$. Under a one-nest alternative that passes the delivery
shortfall through intermediates at unit gain, the interior equilibrium can cease to exist as the closure deepens. The leakage $\beta_j$ of the outer nest breaks that unit gain and is what makes a strictly positive fixed point
available at all. It does not by itself guarantee one, since the collapse vector remains a fixed point of the fulfillment rule, and what shows on these data that the selected equilibrium is not a collapse is the ex-post check rather than the interiority condition. The commodity-differentiated residual transit share,
which enters the robustness exercises of Online
Appendix~\ref{app:modelrobust}, also returns an interior
equilibrium for every scenario.

The ex-ante interiority condition~\eqref{eq:choke_cushion} fails at the baseline as well, and not only at the harsh corner of Online Appendix~\ref{app:modelrobust}. It fails at $22$ of
the $4{,}222$ intermediate-buying country-sectors under a Hormuz closure and at $23$
under the Middle East scenario, at three under the Danish Straits and under the Russia--Europe scenario, at two under Suez and under Bab-el-Mandeb,
and at exactly one under each
of the remaining nine scenarios. That last one is structural rather than
economic. The Cypriot air-transport column carries $\beta^{\rm tech}_j=0$
(Online Appendix~\ref{oa:data_integrity}), where the threshold
$(1-\beta_i)^{-1/\varrho}$ equals one against an intermediate input efficiency factor
$\kappa^{(\omega)}_i(\mathbf 1)$ that never exceeds one, so the condition
cannot hold at that coordinate under any closure. On the subset where the
condition does hold, the floor~\eqref{eq:choke_floor} it computes is
$c_\omega=0.016$ under a Hormuz closure, its weakest across
the fifteen scenarios, more than an order of magnitude below the realized
$\min_i h^{(\omega)}_i$ of $0.472$. Every equilibrium this paper reports is
therefore interior by verification rather than by guarantee, at the
baseline as much as at the corner, and $\min_i h^{(\omega)}_i$ is the
quantity that discharges it.

% ================================================================
\section{The Loss by Source and by Channel}
\label{app:addl_decomp}
% ================================================================

This section takes the decomposition of the loss of Section~\ref{subsec:activity_decomposition} of the paper to the data. It divides the loss from a closed Strait of Hormuz by source and by channel for the economies that lose the most and for the world loss of every single closure. It then follows the network-propagation share across all \(87\) economies under that closure and across the twelve closures for the twenty largest economies, and closes with the direct-loss shares of the African economies.

\subsection{The decomposition under a closed Strait of Hormuz and across the twelve closures}
\label{app:decomposition}

The loss from a closed Strait of Hormuz has a different source at the two ends of the corridor. The Gulf exporters lose the orders for their exports, and the Asian energy importers go short of an input. Table~\ref{tab:decomposition_hormuz} decomposes the loss for ten major and Gulf economies and for the world at the baseline calibration. No entry is negative, and the few entries of the direct demand shortfall that are near zero belong to economies that ship essentially nothing through Hormuz.

% >>> inlined from 04_outputs/tex_tables/tab_decomposition_hormuz.tex
\begin{table}[H]
\centering
\caption{Decomposition of country-level GDP loss by source and by channel under closure of the Strait of Hormuz, at the baseline calibration $(\tau,\rho,\varrho)=(0.3,-1,-1)$ with a uniform residual transit share $\delta=0.10$.}
\label{tab:decomposition_hormuz}
\small
\begin{tabular}{l rr rrr rr r}
\toprule
 & \multicolumn{2}{c}{Lost orders} & \multicolumn{3}{c}{Input shortfalls} & \multicolumn{2}{c}{By source} & \\
\cmidrule(lr){2-3}\cmidrule(lr){4-6}\cmidrule(lr){7-8}
Country & Direct & Upstream & Direct & Downstream & Upstream & Orders & Inputs & Network \\
\midrule
ARE & 66 & 16 & 3 & 3 & 11 & 82 & 18 & 30 \\
SAU & 68 & 13 & 4 & 3 & 12 & 81 & 19 & 27 \\
OtherGulf & 51 & 33 & 4 & 3 & 10 & 84 & 16 & 45 \\
IND & 0 & 1 & 13 & 38 & 47 & 2 & 98 & 87 \\
JPN & 0 & 1 & 14 & 39 & 46 & 1 & 99 & 86 \\
KOR & 0 & 1 & 8 & 41 & 50 & 1 & 99 & 91 \\
CHN & 1 & 2 & 5 & 28 & 64 & 2 & 98 & 94 \\
USA & 1 & 4 & 6 & 23 & 67 & 5 & 95 & 94 \\
DEU & 2 & 6 & 3 & 21 & 68 & 8 & 92 & 95 \\
FRA & 1 & 4 & 5 & 28 & 62 & 5 & 95 & 94 \\
\midrule
World & 14 & 7 & 7 & 24 & 48 & 21 & 79 & 79 \\
\bottomrule
\end{tabular}
\par\smallskip
{\footnotesize\textit{Note:} Entries are shares of each economy's GDP loss, in percent, and the last row gives the same shares of the world loss. The first five columns are the entries of the two rows of Table~\ref{tab:loss_decomposition} of the paper, its zero entry apart, and sum to $100$ in every row up to rounding. Direct is the direct demand shortfall under Lost orders and the direct input shortfall under Input shortfalls, and the two Direct columns together are the direct-loss share. The two Upstream columns sum to the upstream network-propagation share, and Downstream is the downstream network-propagation share. Orders and Inputs divide the same loss by source: the lost-orders share and the input-shortfall share. Network is the network-propagation share, the two Upstream columns and Downstream together.\par}
\end{table}
% <<< end 04_outputs/tex_tables/tab_decomposition_hormuz.tex

Most of the network propagation of the three Gulf exporters is the upstream propagation of the lost sales themselves, $13$ to $33$ percentage points of the loss, as the producers that lose their export sales order less from their own suppliers. The Gulf exporters are also buyers through the same strait. Part of the machinery, chemicals, metals, and processed food that their refining, construction, and downstream sectors draw on arrives by the corridor the closure shuts, so part of each Gulf economy's loss is the degradation of its own input bundle. That buying-side loss is small beside the loss of the export market, about $7\%$ of each Gulf economy's total.

For the Asian energy importers the loss begins at the producers that buy from the Gulf. A Japanese or Korean refinery still has its primary inputs and its non-Gulf inputs, but the crude its technology treats as nearly non-substitutable is gone, so the same primary inputs now yield less output. That worse rate of transformation from inputs to output, and not any lost sale, is what a fall in the intermediate input efficiency factor $\kappa^{(\omega)}_i$ measures. Downstream propagation accounts for $28$ to $41\%$ of the loss of the four importers, as the producers that run short deliver less to the producers that buy from them, who go short in their turn. Upstream propagation, the fall in orders that follows once their own producer country-sectors and their trading partners contract, accounts for about half of the loss in Japan, Korea, and India and about two thirds in China. Almost all of that upstream propagation follows from input shortfalls, since the producers that contract are themselves short of inputs. 

The Japanese case isolates the input-shortfall mechanism. Japan meets $97\%$ of its oil demand with imports, and in 2022 more than $90\%$ of its crude-oil imports came from four Persian Gulf suppliers: Saudi Arabia, the UAE, Kuwait, and Qatar \citep{eia_japan_country_brief_2024}. Most of that crude passed through Hormuz. Nothing domestic stands ready to take the place of those barrels within the year. Japan is an island with no pipeline connection to any supplier, and its nuclear fleet, though partly restarted after Fukushima, still supplies only about six percent of its electricity \citep{eia_japan_country_brief_2024}. The adjustments available over a horizon of years, diversifying into LNG, contracting for Russian Far-East crude, letting demand fall, lie outside the short-run window the model covers. In the extended ICIO matrix the dependence of Japan appears as a heavy weight on Gulf-supplied intermediates in the columns of its refining (C19), chemicals (C20), basic metals (C24A and C24B), and downstream manufacturing sectors. When Hormuz closes, the Gulf rows of $\mathbf A_\omega$ collapse, the realized retention ratios $\theta^{(\omega)}_{rj}(\mathbf h)$ in those Japanese columns fall sharply, and the intermediate input efficiency factor $\kappa^{(\omega)}_i$ at Japanese refining and downstream suppliers contracts accordingly. The producers that buy from the Gulf bear the direct input shortfall, about $14\%$ of Japan's loss, and the producers to which they and other suppliers short of inputs then deliver less bear downstream propagation, a further $39\%$. Korea and India follow the same logic, at $10.2$ and $6.4\%$ of GDP, and lose more than Japan on import baskets that draw about as heavily on the Gulf. Japan, Korea, and India divide the loss across sources and channels in nearly the same proportions.

Table~\ref{tab:decomposition_world} carries the two decompositions to the world loss of each of the twelve single closures. At Hormuz the network-propagation share coincides with the input-shortfall share, at $79\%$. The two shares divide the loss differently, one by its channel and the other by its source, so they are $6$ to $12$ percentage points apart at every other gate.

% >>> inlined from 04_outputs/tex_tables/tab_decomposition_world.tex
\begin{table}[H]
\centering
\caption{Network propagation and input shortfalls in the world loss of each single-chokepoint closure, at the baseline calibration $(\tau,\rho,\varrho)=(0.3,-1,-1)$ with a uniform residual transit share $\delta=0.10$.}
\label{tab:decomposition_world}
\small
\begin{tabular}{l rrr r}
\toprule
 & \multicolumn{3}{c}{Network propagation} & Input \\
\cmidrule(lr){2-4}
Chokepoint & Upstream & Downstream & Total & shortfalls \\
\midrule
Strait of Hormuz & 55 & 24 & 79 & 79 \\
Panama Canal & 63 & 15 & 78 & 70 \\
Strait of Gibraltar & 62 & 15 & 77 & 69 \\
Suez Canal & 61 & 16 & 77 & 70 \\
Bab-el-Mandeb & 61 & 16 & 77 & 70 \\
Turkish Straits & 60 & 17 & 77 & 71 \\
Danish Straits & 61 & 16 & 77 & 70 \\
Strait of Malacca & 62 & 16 & 78 & 72 \\
Taiwan--Luzon Corridor & 66 & 15 & 82 & 72 \\
Korea Strait & 63 & 16 & 79 & 73 \\
Strait of Dover & 60 & 16 & 76 & 68 \\
Kerch Strait & 65 & 13 & 78 & 66 \\
\bottomrule
\end{tabular}
\par\smallskip
{\footnotesize\textit{Note:} Upstream, Downstream, and Total are the upstream and the downstream network-propagation share of the world loss, $\eta^{\mathrm{up}}_W$ and $\eta^{\mathrm{dn}}_W$, and their sum $\eta_W$, in percent, and Total equals the sum of the other two up to rounding. Input shortfalls is the input-shortfall share of the world loss, in percent. The closures are ordered by world loss.\par}
\end{table}
% <<< end 04_outputs/tex_tables/tab_decomposition_world.tex

% ----------------------------------------------------------------
\subsection{The network-propagation share across economies and closures}
\label{app:propccdf}

\begin{figure}[H]
\centering
\includegraphics[width=\linewidth]{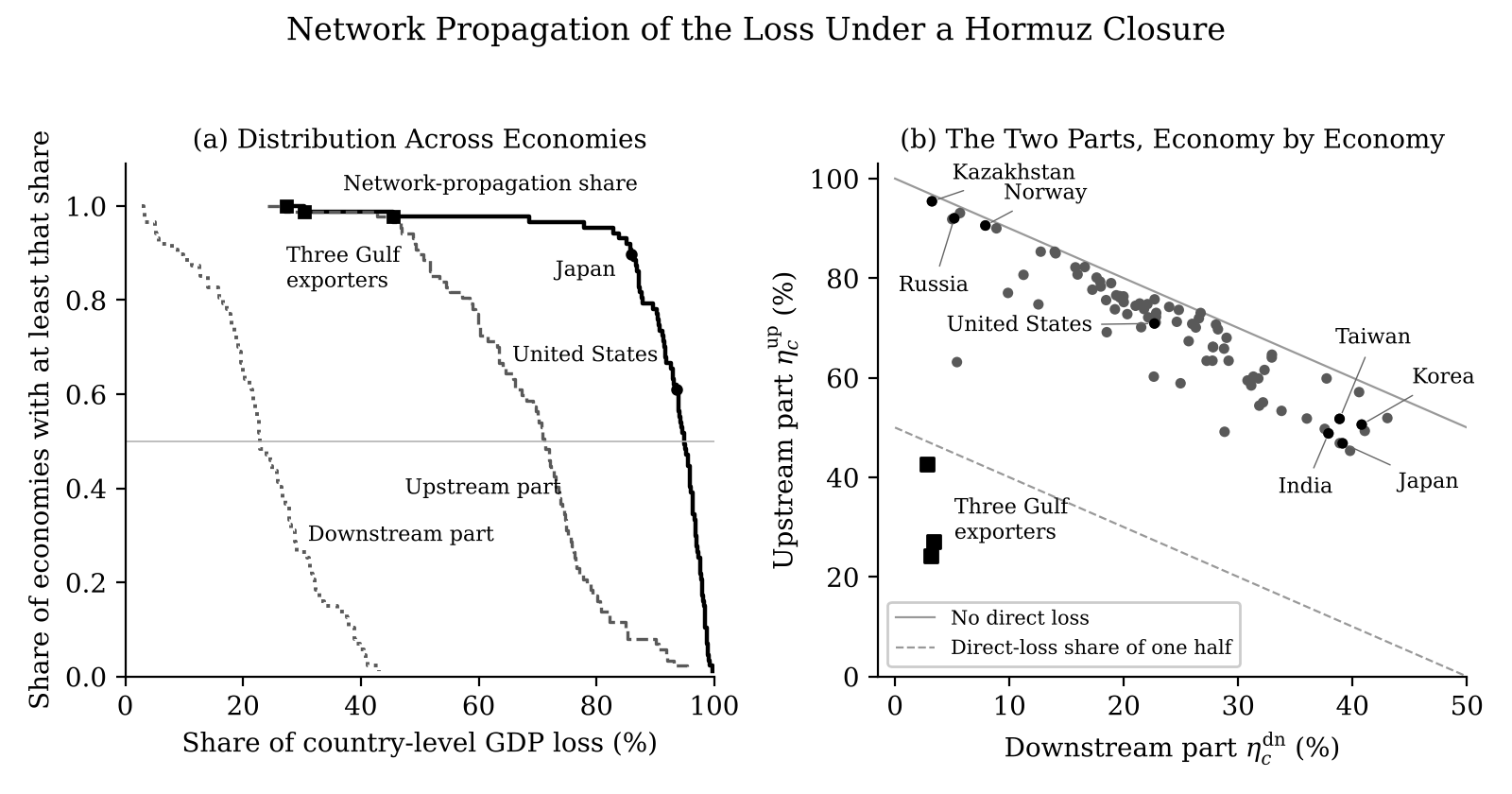}
\caption{Network propagation of the country-level GDP loss under a closed Strait of Hormuz, across all $87$ economies. Panel (a) plots the CCDF of the network-propagation share \(\eta_c\) (solid), of its upstream part \(\eta^{\mathrm{up}}_c\) (dashed), and of its downstream part \(\eta^{\mathrm{dn}}_c\) (dotted), and the thin horizontal line marks the median. Panel (b) plots the upstream part against the downstream part, one point per economy. An economy on the solid diagonal has no direct loss, and on the dashed diagonal the two parts sum to half of the loss. Squares are the three Gulf exporters, and labeled points are economies named in the text.}
\label{fig:propagation_ccdf}
\end{figure}

Figure~\ref{fig:propagation_ccdf} draws the network propagation of the loss from a closed Strait of Hormuz across all $87$ economies. Its first panel plots the complementary cumulative distribution function (CCDF) of the network-propagation share \(\eta_c\), the share of economies whose \(\eta_c\) is at least a given fraction, together with the CCDFs of its upstream part \(\eta^{\mathrm{up}}_c\) and its downstream part \(\eta^{\mathrm{dn}}_c\). Its second panel sets the two parts of every economy against each other. The three Gulf exporters take $55$ to $73\%$ of their loss directly, whereas the direct-loss share of every other economy is below a third, and the median economy takes $95\%$ of its loss through the network. Since the network-propagation share also counts the propagation among the producer country-sectors of the country itself (Section~\ref{subsec:country_world_losses} of the paper), a high share says that the producer country-sectors which the closure itself reaches bear a small part of what the economy loses. Upstream propagation is the larger part in every economy, at a median of $71\%$ of the loss, and it is largest in Kazakhstan, Brunei, and Russia, at $92$ to $95\%$. Downstream propagation has a median of $23\%$, and it comes to about two fifths of the loss in Korea, Japan, India, and Taiwan, which buy through the strait intermediate inputs worth $2$ to $5\%$ of their GDP. By source, the three Gulf exporters are the only economies whose loss originates mostly in lost orders, whereas the median economy owes $94\%$ of its loss to input shortfalls.

In the first panel the curves of the upstream and the downstream part fall at much the same rate, whereas the curve of the network-propagation share is nearly flat up to a share of $80\%$, which $82$ of the $87$ economies reach, and then falls to zero over the last twenty points. The middle half of the economies spans $16$ points of the upstream part and $12$ of the downstream part, and only $7$ of the network-propagation share, which is the sum of the two parts in every economy. The second panel shows why the sum varies so much less than its parts, which is that the two parts are not large in the same economies. Since the direct-loss share is small in every economy but the three Gulf exporters, the two parts of an economy add up to nearly the same total everywhere, so the economies lie in a narrow band under the line on which the two parts make up the whole loss. An economy with a larger downstream part therefore has a smaller upstream part almost point for point, and the correlation between the two parts is $-0.90$ across the economies other than the three Gulf exporters. Which is to say that economies differ far more in the side from which the network brings them their loss than in how much of their loss it brings.

Table~\ref{tab:propagation_top10} lists the ten named economies with the largest upstream network-propagation share \(\eta^{\mathrm{up}}_c\) under the Hormuz closure and the ten with the largest downstream network-propagation share \(\eta^{\mathrm{dn}}_c\), each with both of its shares and its proportional loss \(\lambda_c\). The two lists differ in how much the economies on them trade through the strait. Nine of the ten economies on the upstream list sell and buy through it intermediate inputs that together come to less than $1\%$ of their GDP, and the tenth, Angola, $1.1\%$. Seven of the ten economies on the downstream list buy through the strait intermediate inputs worth more than $1\%$ of their GDP, as much as $5\%$ in Korea and $6\%$ in Jordan, and the other three, Bangladesh, Cambodia, and S\~ao Tom\'e and Pr\'incipe, $0.5\%$ or less. Downstream propagation is below $17\%$ of the loss in every economy on the upstream list, and upstream propagation is $45$ to $60\%$ of the loss in every economy on the downstream list.

% >>> inlined from 04_outputs/tex_tables/tab_propagation_top10_hormuz.tex
\begin{table}[H]
\centering
\caption{The ten named economies with the largest upstream network-propagation share and the ten with the largest downstream network-propagation share under closure of the Strait of Hormuz, at the baseline calibration $(\tau,\rho,\varrho)=(0.3,-1,-1)$ with a uniform residual transit share $\delta=0.10$.}
\label{tab:propagation_top10}
\small
\begin{tabular}{l rrr @{\qquad} l rrr}
\toprule
\multicolumn{4}{c}{Largest upstream share} & \multicolumn{4}{c}{Largest downstream share} \\
\cmidrule(r){1-4}\cmidrule(l){5-8}
Economy & Upstream & Downstream & GDP loss & Economy & Upstream & Downstream & GDP loss \\
\midrule
KAZ & 95.4 & 3.2 & 3.1 & GRC & 51.9 & 43.0 & 3.2 \\
BRN & 93.1 & 5.7 & 16.1 & PHL & 49.3 & 41.1 & 4.3 \\
RUS & 92.0 & 5.2 & 1.8 & KOR & 50.6 & 40.8 & 10.2 \\
AGO & 91.8 & 5.0 & 5.9 & KHM & 57.1 & 40.6 & 1.7 \\
NOR & 90.6 & 7.9 & 1.5 & BGD & 45.3 & 39.8 & 1.5 \\
NGA & 90.0 & 8.9 & 2.2 & JPN & 46.8 & 39.1 & 5.5 \\
AUS & 85.3 & 12.7 & 3.2 & JOR & 46.9 & 38.9 & 7.4 \\
CHL & 85.3 & 14.0 & 1.5 & TWN & 51.7 & 38.9 & 7.8 \\
CHE & 82.2 & 16.6 & 0.8 & IND & 48.8 & 37.9 & 6.4 \\
LUX & 82.2 & 15.8 & 0.9 & STP & 59.9 & 37.7 & 0.5 \\
\bottomrule
\end{tabular}
\par\smallskip
{\footnotesize\textit{Note:} Upstream and Downstream are the upstream network-propagation share $\eta^{\mathrm{up}}_c$ and the downstream network-propagation share $\eta^{\mathrm{dn}}_c$, in percent of the economy's GDP loss. GDP loss is the proportional loss $\lambda_c$, in percent of the economy's benchmark GDP. The left panel ranks the economies by the upstream share and the right panel by the downstream share. The seven rest-of-world blocks are left out of the ranking.\par}
\end{table}
% <<< end 04_outputs/tex_tables/tab_propagation_top10_hormuz.tex

\begin{figure}[H]
\centering
\includegraphics[width=0.85\linewidth]{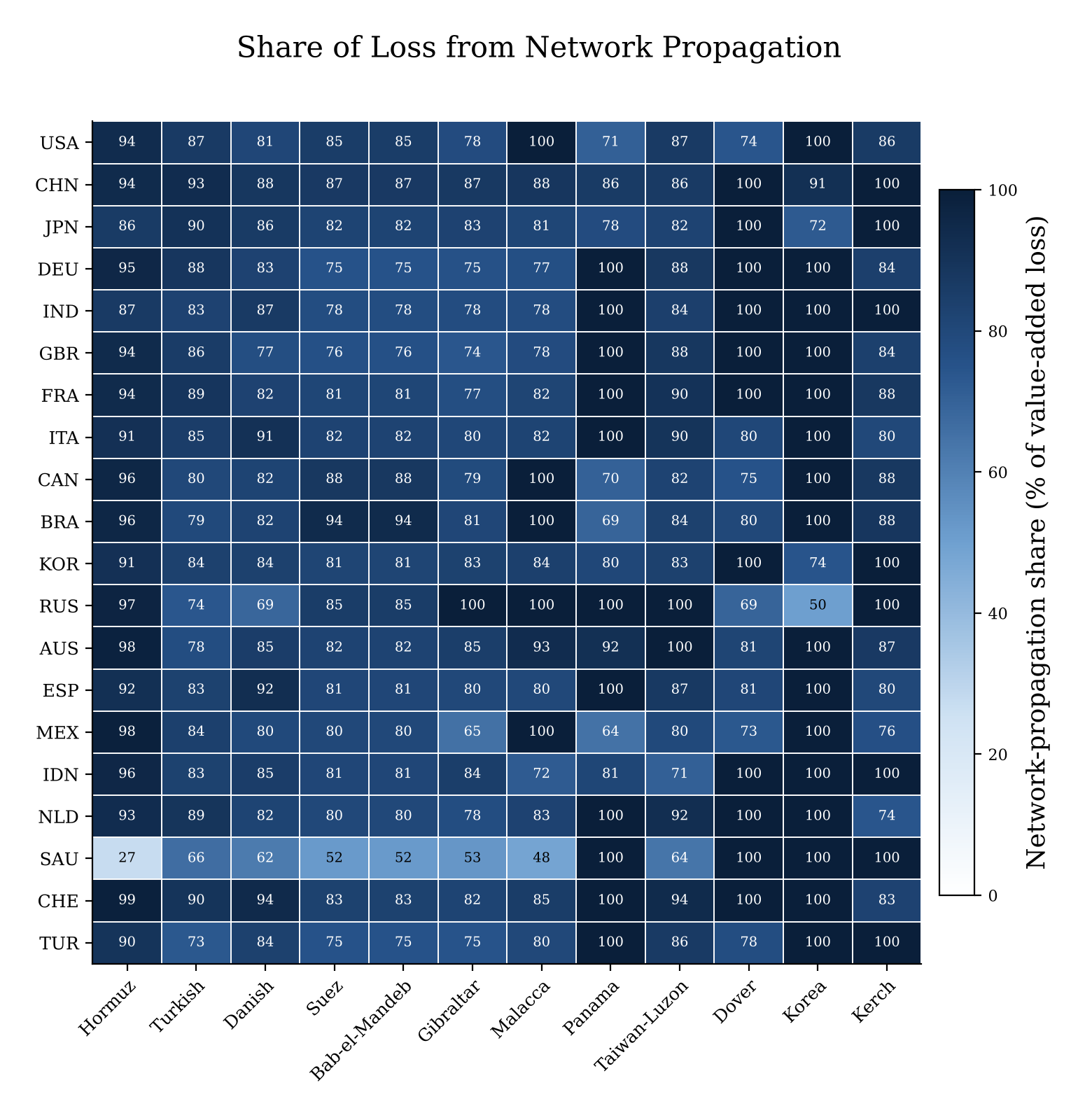}
\caption{Network-propagation share \(\eta_c\) of country-level GDP loss for the
top-twenty economies by GDP, across the twelve single-chokepoint
closures.}
\label{fig:propagation_share}
\end{figure}

Figure~\ref{fig:propagation_share} carries the network-propagation share \(\eta_c\) to all twelve single-chokepoint closures, for the twenty largest economies by GDP. Dark cells mark an economy that takes its loss through the network, and light cells an economy that takes it directly. Along the Hormuz column every one of the twenty but Saudi Arabia takes at least $86\%$ of its loss through the network. The United States, the least exposed major economy of Section~\ref{subsec:master_matrix_results} of the paper, takes $94\%$ of its loss there through the network, and Germany, the United Kingdom, France, and Switzerland between $94$ and $99\%$ of theirs. On the other columns the United States takes between $71$ and $100\%$ of its loss through the network. Its share is lowest on Panama ($71\%$), where its Atlantic-to-Pacific flows of refined products, gases, and bulk chemicals are only partly saved by the Cape Horn detour. Since the shale revolution, little US intermediate input transits any listed strait. Russia is likewise dark everywhere except the Turkish Straits ($74\%$), the Danish Straits ($69\%$), Dover ($69\%$), and the Korea Strait ($50\%$), gates its seaborne exports cross, since most of its oil and gas leaves by pipeline or through Arctic and Far-East ports. Brazil, Mexico, Canada, and Indonesia belong to the same regime for the same reason, bulk or Atlantic-basin trade that bypasses most of the chokepoint set. Each takes more of its loss directly on the closure that touches a route it actually uses (Mexico, Canada, and Brazil on Panama, Indonesia on Malacca).

% ----------------------------------------------------------------
\subsection{African direct-loss shares}
\label{app:africa}

\begin{figure}[H]
\centering
\includegraphics[width=0.85\linewidth]{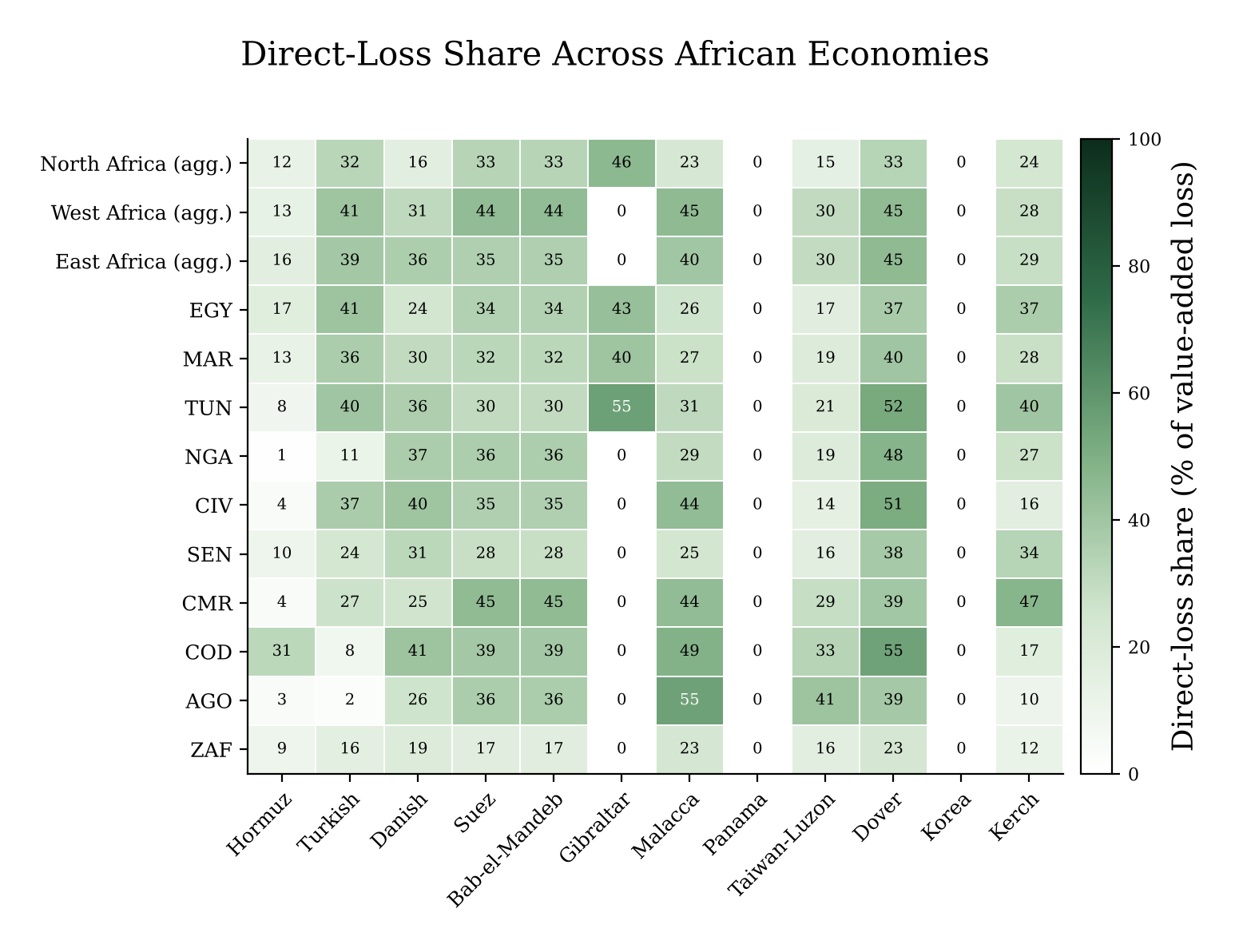}
\caption{Direct-loss share \(1-\eta_c\) of country-level GDP loss for thirteen
African economies, across the twelve single-chokepoint closures. The direct-loss share is
one minus the network-propagation share, the part of the loss that arises at the economy's own producer country-sectors from the shortfalls of demand and of inputs that the closure itself causes.}
\label{fig:direct_impact_africa}
\end{figure}

Figure~\ref{fig:direct_impact_africa} reports the direct-loss share \(1-\eta_c\), the complement of the network-propagation share, for thirteen African economies across the twelve closures. The three ROW Africa aggregates take $32$ to $44\%$ of their loss directly on the Turkish, Suez, and Bab-el-Mandeb columns and $23$ to $45\%$ on Malacca, where the median economy of the top twenty of Figure~\ref{fig:propagation_share} takes $16$ to $19\%$. Their direct-loss share is $12$ to $16\%$ on the Hormuz column, where their loss arrives mostly through buyers that contract, and at or near zero on the Panama and Korea columns. The shares of the West African aggregate and of Cameroon on the Suez and Bab-el-Mandeb columns are overstated, since the routing sends their trade with the Mediterranean through the Red Sea (Section~\ref{sec:step1_def} of the paper).

Among the individually identified economies, Tunisia and Morocco take $32$ to $60\%$ of their loss directly on the Mediterranean gates, and Egypt $37$ to
$49\%$ there. Cameroon takes $55\%$ of its loss directly on Suez and Bab-el-Mandeb, and the Democratic Republic of the Congo $32\%$ on Hormuz.  Angola carries high direct-loss shares on both the Malacca ($67\%$) and Taiwan--Luzon
($61\%$) columns, the highest pair among the Sub-Saharan oil exporters, because Angolan crude shipments to Chinese refineries
cross the Strait of Malacca and reach Taiwanese petrochemical buyers by
through-routes. South Africa carries the most propagation-driven African
profile, at a direct-loss share of $34\%$ on Malacca and on Dover, of
about a fifth to a quarter on the Turkish, Danish, Suez, Bab-el-Mandeb, and Taiwan--Luzon columns, of $9\%$ on Hormuz,
and near zero elsewhere. This profile resembles a diversified Western
European economy more than its commodity-narrow neighbors.

The direct-loss share of the African economies is high because their trade is narrow on both sides. On the export side, an African economy typically sells one or two primary commodities: crude from the Gulf of Guinea, copper and cobalt from the Congo basin, ores and agricultural staples elsewhere. It sells them to a diversified world market through a single corridor. When that corridor closes, the buyers at the far end re-source within sector with comparative ease, since other suppliers of crude or copper abound off the corridor. The African seller, by contrast, finds few off-corridor same-sector buyers onto which the rationed exchange can push its displaced output, so the unrecovered share $\tau$ writes the remainder off. On the import side the same concentration appears in mirror image. The fuel, chemicals, machinery, and processed food that African downstream sectors use as inputs arrive from a handful of suppliers along the same few corridors. The within-sector re-matching that re-places seventy percent of displaced trade for diversified economies places far less here, the open links being too few to carry it. The intermediate input efficiency factor $\kappa^{(\omega)}_i$ falls steeply where the surviving suppliers cannot match the lost input mix.

The concentration that the model finds in $\mathbf Z^*$ has an independent counterpart in the trade statistics. UNCTAD classifies more than $\nicefrac{4}{5}$ of African economies as
commodity-dependent, with over $60\%$ of merchandise-export revenue from primary
commodities \citep{unctad_commodity_dependence_2023}. Most of the continent's
long-haul trade moves on a small number of liner services that funnel through the
Suez--Bab-el-Mandeb corridor toward Europe and through Indian-Ocean hubs toward Asia. Nothing in the model is specific to Africa. The exposure comes from concentrated supply that meets diversified demand along a few routes, evaluated at the African rows and columns of $\mathbf Z^*$.

% ================================================================
\section{The Structure of Exposure}
\label{app:addl_structure}
% ================================================================

This section collects the additional results behind Section~\ref{sec:structure} of the paper: the estimation of the tail coefficients, the largest falls in output under each closure, the concentration of exposure across corridors, and the named exporters and importers at three corridors. One construction
convention governs every cross-chokepoint statistic, that of Section~\ref{subsec:mean_tail} of the paper. Worst-case
statistics accordingly run over the twelve gates, to which a duplicated
column is invisible, and mean-based statistics run over the set \(\mathcal K^{\mathrm d}\) of the eleven distinct corridors, the twelve gates with Bab-el-Mandeb removed as a
duplicate of Suez.

\subsection{Estimation detail for the tail coefficients}
\label{app:zipfdetail}

The tail statements of Section~\ref{subsec:heavy_tail} rest on a threshold
fit rather than on a rank-size regression. For each sample we choose the
threshold $x_{\min}$ that minimizes the Kolmogorov--Smirnov distance
between the empirical distribution above it and a fitted power law. The power-law exponent is then the maximum-likelihood value
$\hat\alpha=1+n_{\mathrm{tail}}/\sum\log(x_i/x_{\min})$ over that tail. We
assess the fit by the semiparametric bootstrap of
\citet{clauset2009powerlaw}, resampling the body empirically and the tail
from the fitted law over $1{,}000$ synthetic samples. We then compare the
power law with a lognormal and with an exponential over the same range by
normalized likelihood ratio. The ratio reported throughout is the
normalized statistic $R/(\sqrt{n}\sigma)$, which is the one that is
asymptotically standard normal. The raw log-likelihood sum is not a
$z$-score, and treating it as one overstates the separation.

The country cross-section places $x_{\min}$ at a loss of $0.8\%$ of value
added, leaving $76$ of the $80$ named economies, with $\hat\alpha=1.93$ and a bootstrap
$p$-value of $0.01$. The power law is therefore rejected on this window.
The normalized likelihood ratio against a lognormal is $-1.30$ ($p=0.19$), which leans to the
lognormal without separating the two laws. Against an
exponential the ratio is $1.02$ ($p=0.31$), so the likelihood ratio does not set the
exponential aside on these $76$ observations either. The comparison laws are also
unbounded while the loss is bounded by one, so the fits and their bootstraps are descriptions of the window rather than tail models, and a formal test would truncate each null at one.

The country-sector cross-section places $x_{\min}$ at a proportional activity loss of $0.091$, leaving $452$ country-sectors, with $\hat\alpha=2.91$ and a
bootstrap $p$-value of $0.14$. A power law therefore survives on this window, and the exponential is set aside at the five percent level, with a normalized likelihood ratio of
$2.30$ ($p=0.02$). The ratio against a lognormal is $-1.77$ ($p=0.08$).
The lognormal is therefore ahead of the power law at the country-sector level at the ten percent level, and the country level rejects the power law outright.
The loss $\lambda$ cannot exceed one, the worst country-sector (Emirati crude extraction under a Hormuz closure) already stands at $0.91$, and a bounded
variable cannot carry a power law into its endpoint.

A rank-size slope over a chosen
window is a description of that
window and not a tail index, since the threshold fits
place the exponent of the density at $\hat\alpha\approx1.9$ at the country level and $2.9$ at the country-sector level, where the rank-size slope on ranks $5$ to $40$ is $1.39$. The evidence
supports only the weaker statement that worst-case exposure is more dispersed than an exponential benchmark, with a coefficient of variation of $1.39$, and concentrated in a few economies, with no
claim about an infinite second moment.

\subsection{The largest falls in output under each closure}
\label{app:sector_declines}

Table~\ref{tab:sector_declines_top} lists, for each of the twelve single closures and the three joint scenarios, the number of country-sectors whose output falls by more than a tenth and by more than half, and the five largest falls, each as a share of the country-sector's benchmark output. The falls are the proportional activity declines \(\lambda_i(\cset{})\) of Section~\ref{subsec:heavy_tail} of the paper, measured between the benchmark and the post-closure fixed point of each scenario, and every country-sector with positive benchmark activity is counted. Hormuz and the Middle East scenario stand apart. Under every other closure at most \(98\) country-sectors lose a tenth of their output and none loses half. The replication package lists the fall of every country-sector under every scenario (\texttt{04\_outputs/experiments/sector\_output\_declines.csv}).

% >>> inlined from 04_outputs/tex_tables/tab_sector_declines_top.tex
\begin{table}[H]
\centering
\caption{The five largest falls in output under each closure, at the baseline calibration $(\tau,\rho,\varrho,\delta)=(0.3,-1,-1,0.10)$. The second and third columns count the country-sectors whose output falls by more than $10\%$ and by more than $50\%$ of its benchmark, and the remaining columns name the five largest falls, in percent of benchmark output. Economy and sector codes are those of the extended input--output table of Online Appendix~\ref{sec:data}.}
\label{tab:sector_declines_top}
\small
\begin{tabular}{l r r p{9.6cm}}
\toprule
Closure & $>10\%$ & $>50\%$ & The five largest falls, in percent of benchmark output \\
\midrule
Hormuz & 321 & 10 & ARE B06 (91), SAU B06 (69), OtherGulf B06 (63), ARE B09 (62), SAU B09 (62) \\
Turkish & 48 & 0 & TUR B07 (35), TUR C24B (27), OtherCentralAsia B09 (27), BGR B07 (25), BGR C24B (24) \\
Danish & 43 & 0 & SWE B06 (32), FIN C24B (27), FIN B07 (26), SWE C19 (24), FIN C24A (21) \\
Suez & 0 & 0 & TUN C19 (10), JOR B08 (10), GRC C19 (9), GRC B06 (9), OtherNorthAfrica B09 (9) \\
Bab-el-Mandeb & 0 & 0 & TUN C19 (10), JOR B08 (10), GRC C19 (9), GRC B06 (9), OtherNorthAfrica B09 (9) \\
Gibraltar & 3 & 0 & TUN C302T309 (12), TUN C22 (10), TUN B06 (10), AUT B06 (10), TUN C24B (9) \\
Malacca & 0 & 0 & ARE B06 (4), SGP C19 (3), COD C24B (3), COD C19 (3), KOR C19 (3) \\
Panama & 2 & 0 & CHL B07 (11), COL B06 (10), CRI C24B (9), PER B07 (9), CRI C24A (9) \\
Taiwan--Luzon & 0 & 0 & TWN B07 (10), TWN C24B (7), TWN C20 (6), TWN C24A (4), TWN B08 (4) \\
Dover & 0 & 0 & SWE B06 (2), EST C24A (1), BLR B05 (1), SWE B07 (1), LTU C24A (1) \\
Korea & 0 & 0 & RUS B05 (4), RUS B07 (4), RUS C24B (2), KOR C24A (2), RUS C16 (2) \\
Kerch & 0 & 0 & UKR C24A (3), UKR A01 (3), UKR B08 (3), UKR B05 (2), UKR C20 (2) \\
\midrule
$\cset{}^{ME}$ & 370 & 11 & ARE B06 (91), SAU B06 (69), OtherGulf B06 (64), ARE B09 (62), SAU B09 (62) \\
$\cset{}^{EA}$ & 1 & 0 & TWN B07 (12), TWN C24B (9), TWN C20 (8), TWN C19 (7), TWN B06 (7) \\
$\cset{}^{RE}$ & 98 & 0 & TUR B07 (35), SWE B06 (32), TUR C24B (28), FIN C24B (28), OtherCentralAsia B09 (27) \\
\bottomrule
\end{tabular}
\end{table}
% <<< end 04_outputs/tex_tables/tab_sector_declines_top.tex

\subsection{The concentration of exposure across corridors}
\label{app:tailcomp}

We measure the shape of an economy's exposure profile across the distinct corridors by an index
normalized so that the number of corridors sets neither its minimum nor its maximum. Let \(w_{ck}:=\lambda_c(\cset{}_k)/\sum_{k'\in\mathcal K^{\mathrm d}}\lambda_c(\cset{}_{k'})\)
be the share of country \(c\)'s total exposure carried by corridor
\(k\in\mathcal K^{\mathrm d}\), and write \(n:=\lvert\mathcal K^{\mathrm d}\rvert\) for the number of
distinct corridors. Country \(c\)'s exposure concentration is the normalized Herfindahl
index of those shares,
\begin{equation}
H_c
\;:=\;
\frac{\sum_{k\in\mathcal K^{\mathrm d}} w_{ck}^{2}\;-\;\nicefrac1n}
{1-\nicefrac1n}
\;\in\;[0,1].
\label{eq:herfindahl}
\end{equation}
The index runs from $0$, exposure spread evenly across the distinct
corridors, to $1$, a single gate carrying the whole exposure, whatever
the number of corridors. Across the $80$ named economies it spans $0.10$
to $0.72$ with a median near $0.25$, and twenty-eight named economies carry
$H_c\ge0.30$. Table~\ref{tab:herfindahl} lists them, each with its binding gate, and the seven pooled rest-of-world aggregates. Twenty-four of the twenty-eight
are bound by Hormuz, led by the United Arab Emirates at $0.72$ and by Pakistan, Saudi Arabia, and New Zealand at $0.69$ to $0.71$.
The other four are Sweden and Finland behind the Danish Straits, at $0.38$ and $0.36$,
Bulgaria behind the Turkish Straits, and Colombia behind the Panama Canal. The multi-country ROW aggregates other than the Gulf block sit near the middle of the named economies,
with OtherNorthAfrica, OtherSouthAsia, OtherCentralAsia, OtherWestAfrica,
and OtherRestOfWorld falling between $0.16$ and $0.28$ and OtherEastAfrica at $0.39$.
OtherGulf, whose members share a single sea
exit, stands fifth among all $87$ economies, at $0.66$. The Turkish Straits are the binding gate of OtherCentralAsia
and Hormuz that of every other
aggregate.

\begin{table}[H]
\centering
\caption{Corridor concentration of single-gate exposure. The upper block
lists the twenty-eight named economies with \(H_c\ge0.30\) and the lower
block the seven pooled rest-of-world aggregates, each with the
normalized index \eqref{eq:herfindahl} and its binding
gate, the corridor carrying the largest single-closure loss. 
The gates are the Strait of Hormuz (H), the Danish Straits (D), the Panama Canal (P), and the
Turkish Straits (T). Baseline calibration \((\tau,\rho,\varrho)=(0.3,-1,-1)\) and
\(\delta=0.10\), OECD ICIO 2025, year 2019.}
\label{tab:herfindahl}
\small
\begin{tabular}{lcc lcc lcc}
\toprule
\multicolumn{9}{c}{Named economies with $H_c\ge0.30$}\\
\midrule
 & $H_c$ & Gate & & $H_c$ & Gate & & $H_c$ & Gate\\
\midrule
ARE & 0.72 & H & ZAF & 0.56 & H & HKG & 0.38 & H\\
PAK & 0.71 & H & JOR & 0.51 & H & SWE & 0.38 & D\\
SAU & 0.69 & H & MMR & 0.49 & H & CHN & 0.37 & H\\
NZL & 0.69 & H & AUS & 0.46 & H & FIN & 0.36 & D\\
IND & 0.66 & H & TWN & 0.43 & H & NGA & 0.34 & H\\
JPN & 0.65 & H & LAO & 0.43 & H & BGR & 0.34 & T\\
BRN & 0.64 & H & SGP & 0.41 & H & BGD & 0.31 & H\\
THA & 0.60 & H & MYS & 0.41 & H & COL & 0.31 & P\\
KOR & 0.58 & H & IDN & 0.39 & H & & & \\
PHL & 0.57 & H & AGO & 0.39 & H & & & \\
\midrule
\multicolumn{9}{c}{Pooled aggregates}\\
\midrule
OtherGulf & 0.66 & H & OtherWestAfrica & 0.27 & H & OtherSouthAsia & 0.25 & H\\
OtherEastAfrica & 0.39 & H & OtherCentralAsia & 0.25 & T & OtherNorthAfrica & 0.16 & H\\
OtherRestOfWorld & 0.28 & H & & & & & & \\
\bottomrule
\end{tabular}
\end{table}

\subsection{Named exporters and importers at three corridors}
\label{app:asym_three}

Table~\ref{tab:asym_three} names the leading net exporters and net importers of three of the corridors of Table~\ref{tab:asymmetry_by_chokepoint} of the paper, ranked by disrupted trade mass. The three corridors run from a wide division of the loss to a narrow one.
Under closure of the Strait of Hormuz the crude and refined products that leave the Gulf are
bought largely by the refining and petrochemical sectors of East and South Asia. The Gulf side loses roughly four times the buyers' proportional loss. Even so, the buyers lose about $6\%$ of GDP, because the Asian refiners that the Gulf can no longer supply go short of inputs and pass the shortfall on to their own customers. At the
Panama Canal, where South American ores, crude oil, and farm products and Chinese manufactures
are bound mostly for North American buyers, the exporting side loses more than twice the
buyers' loss. At the Turkish Straits, whose cargo is Caspian and Russian crude with chemicals,
steel, and refined products behind it, the division narrows to near parity. The ratios of Table~\ref{tab:asym_three} are unweighted means over the economies it names, about four at Hormuz, two at Panama, and one at the Turkish Straits. Table~\ref{tab:asymmetry_by_chokepoint} of the paper weights each economy by its disrupted trade mass and gives four and a half, two and a half, and one and a half at the same three corridors. The comparison therefore does not turn on which economies are named.

\begin{table}[H]
\centering
\caption{Exporter-side against importer-side GDP losses at three corridors, at the baseline calibration. Each side lists the
corridor's leading net economies by disrupted intermediate-trade mass,
with the unweighted mean of their proportional GDP losses.}
\label{tab:asym_three}
\small
\begin{tabular}{l l r l r c}
\toprule
Corridor & Net exporters & Loss (\%) & Net importers & Loss (\%) & Ratio \\
\midrule
Hormuz & OtherGulf, SAU, ARE & 26 & CHN, IND, JPN, KOR & 6.3 & $\approx4\times$ \\
Panama Canal & BRA, CHN, CHL & 1.4 & USA, MEX, CAN & 0.6 & $\approx2\times$ \\
Turkish Straits & KAZ, RUS, CHN & 2.0 & TUR, ITA, DEU & 1.9 & $\approx1\times$ \\
\bottomrule
\end{tabular}
\end{table}

% ================================================================
\section{Joint Chokepoint Closures}
\label{app:addl_joint}
% ================================================================

This section collects the material behind Section~\ref{sec:joint} of the paper: the political background of the three scenarios, the trade that the joint closures displace, and the additivity index of all sixty-six pairwise closures.

% ----------------------------------------------------------------
\subsection{Political background of the three scenarios}
\label{app:scenariopol}

The three joint scenarios of Section~\ref{subsec:joint_scenarios} are drawn
from recognizable international-political fault lines, and the mechanisms
that would close their gates together differ from one scenario to the next.

A regional Middle East escalation can connect the Hormuz and Red Sea
theaters through alliance politics, proxy conflict, naval deployments, or
retaliation against commercial shipping. The 2026 war has begun to realize
the connection. The closure followed American and Israeli strikes on
Tehran that began in late February 2026, and Iran declared the strait
closed on 2~March. An April ceasefire reopened the strait briefly, a
renewed declaration followed the American blockade of Iranian ports in
mid-April, and Iran declared a complete closure on 11~June. Negotiated
corridors and attacks on vessels using alternative routes have alternated
since \citep{aljazeera_2026_hormuz,lloyds_hormuz_2026}. \citet{aawsat_2026_iran_houthi} reports an instruction to the Houthi movement to stand ready to close the Red Sea route, and
\citet{aljazeera_2026_blockade} the blockade declaration of July
2026. All Suez traffic enters or leaves the Red
Sea through Bab-el-Mandeb. The 2023--24 campaign took most of the container traffic
and three fifths of the receipts from the canal without any action at the
canal itself (Online Appendix~\ref{app:est_delta}).
The blockade of July 2026 has reproduced the coupling. Mainstream tanker
transits of Bab-el-Mandeb fell by about two fifths within a week of the
declaration, under renewed attacks on shipping, while canal transit
counts have so far held \citep{lloyds_redsea_2026}. The economic event need not be a
legally declared blockade. Once war-risk premia rise enough, or once major
carriers refuse to transit the affected corridors, the corridor ceases to
be part of the commercial network for a large share of firms.

A Taiwan contingency need not remain geographically narrow. A naval
response, export controls, insurance withdrawal, or precautionary rerouting
would draw the wider Indo-Pacific into the scenario, and the three East
Asian chokepoints enter the closure set together because they belong to the
same strategic--economic system. The 2022 exercises declared six closure zones encircling
Taiwan, several overlapping commercial shipping lanes, and shipping and
flights rerouted around them for the drills' duration. The Eastern Theater Command named the blockade of key ports and areas among the subjects of Joint Sword-2024B, of 14~October 2024 \citep{reuters_2024_jointsword}. Justice Mission-2025, of 29--30 December
2025, rehearsed the blockade with some two hundred aircraft sorties over its two days and some twenty-eight naval and coast-guard ships around the island on each day. Its declared areas approached the island's twenty-four-nautical-mile contiguous zone and in places appear to have crossed it \citep{gti_2026_justice_mission}.

The three Russia-adjacent gates close through different mechanisms. The
Turkish Straits are governed by the Montreux Convention, which gives Turkey
a distinctive legal role during war. The Danish Straits sit inside a Baltic
security environment in which the risks come less from formal closure than
from sabotage and increased naval inspection. After declining to
renew the Black Sea grain corridor on 17~July 2023, the Russian defense
ministry announced that from 20~July vessels sailing to Ukrainian ports
would be treated as potential carriers of military cargo
\citep{reuters_2023_blacksea}. Turkey applied the Montreux Convention on 28~February 2022 and warned all countries, Black Sea states or not, against sending warships through the Bosphorus and the Dardanelles \citep{reuters_2022_montreux}. Russia seized three Ukrainian naval vessels at the Kerch Strait, the sole passage into the Sea of Azov, in November 2018 \citep{bbc_2018_kerch}. In the Baltic, sabotage destroyed the Nord Stream
pipelines in September 2022, eleven cables suffered damage in the fifteen
months to January 2025, and NATO launched the Baltic Sentry patrol
mission in response \citep{defensenews_2025_baltic}. A
Russia--Europe escalation can impair maritime access at all three gates at
once.

\subsection{The trade that the joint closures displace}
\label{app:joint_displaced}

Section~\ref{subsec:joint_empirics} of the paper sets the three indices beside two measures that involve no solved equilibrium. The first is the trade that a closure set \(\cset{}\) displaces, \(Z^{\rm disp}(\cset{}):=\sum_{i,j}(1-\delta)\,\mu_{ij}\,d_{ij}(\cset{})\,Z_{ij}\), i.e., the displaced sales \(Z^{\rm disp}_i\) of \eqref{eq:choke_rowtarget} of the paper summed over the sellers, with the seaborne shares \(\mu_{ij}\) of Section~\ref{sec:seashare} of the paper. It depends on the routes, the detours, the seaborne shares, and the trade table, and with a uniform residual transit share its ratio across closure sets depends on nothing else. The trade that a joint closure displaces falls short of the sum of what its single closures displace by the trade that two of its gates would both displace. We call that shortfall, as a share of the sum, the `duplication share' of the closure set,
\begin{equation}
\chi(\cset{})
\;:=\;
1-\frac{Z^{\rm disp}(\cset{})}
{\displaystyle\sum_{k\in\cset{}}Z^{\rm disp}(\cset{}_k)}.
\label{eq:duplication_share}
\end{equation}
The duplication share is zero for gates in pure parallel, which stop different shipments, and one half for two gates in pure series, whose joint closure is the closure of one. Table~\ref{tab:joint_displaced} reports the trade that each joint closure displaces against the sum of what its three single closures displace.

% numbers: 04_outputs/experiments/joint_displaced_trade.json (02_scripts/experiments/joint_displaced_trade.py)
\begin{table}[H]
\centering
\caption{The trade that each joint closure displaces against the sum of what its single closures displace, in \% of world GDP, with a uniform residual transit share $\delta=0.10$. The ratio is computed on unrounded values, and $S(\cset{})$ is the index of the solved model at the baseline calibration.}
\label{tab:joint_displaced}
\small
\begin{tabular}{l r r r r}
\toprule
Scenario & Single closures, sum & Joint closure & Ratio & $S(\cset{})$ \\
\midrule
$\cset{}^{ME}$ & 1.43 & 1.08 & 0.76 & 0.89 \\
$\cset{}^{EA}$ & 0.35 & 0.34 & 0.99 & 0.98 \\
$\cset{}^{RE}$ & 0.52 & 0.51 & 0.97 & 0.97 \\
\bottomrule
\end{tabular}
\end{table}

The ratio is \(0.99\) for the East Asian scenario and \(0.97\) for the Russia--Europe scenario, against indices of \(0.98\) and \(0.97\), so their duplication shares are \(0.01\) and \(0.03\). On unrounded values the ratio exceeds the index by \(0.004\) in the East Asian scenario and by less than \(0.001\) in the Russia--Europe scenario. The ratio of the Middle East scenario is \(0.76\), a duplication share of \(0.24\), against an index of \(0.89\). A unit of trade displaced at Hormuz costs the world about three times as much as a unit displaced at a Red Sea gate, \(2.83\%\) of world GDP against a displaced trade of \(0.88\%\) at Hormuz and \(0.26\%\) against \(0.27\%\) at either Red Sea gate. The trade that the two Red Sea gates duplicate therefore weighs less in the losses of the scenario than in the trade that it displaces.

The second measure concerns the inputs that two closures take from the same buyer. For an ordered pair of gates \((k,k')\) of a scenario we take the share \(\ell_{rj}(\cset{}_{k'})/\alpha_{rj}\) of buyer \(j\)'s purchases of input sector \(r\) that the closure of gate \(k'\) displaces. We average that share over the input sectors and the buyers, with the trade that the closure of gate \(k\) displaces there as weights. The average is at most \(1.3\%\) across the ordered pairs of the East Asian scenario and at most \(2.7\%\) across those of the Russia--Europe scenario. Between the Turkish and the Danish Straits, the two large gates of the Russia--Europe scenario, it is \(1.5\%\) in one order and \(1.6\%\) in the other. By Online Appendix Proposition~\ref{prop:joint_pair} the interaction of two closures at a buyer is proportional to the unrecovered share of these displaced purchases, which is why it leaves no visible mark on the indices of the two scenarios. Across the ordered pairs of the Middle East scenario the average reaches \(8.3\%\), for the trade that either Red Sea gate displaces set against the closure of Hormuz.

\subsection{The additivity index of the sixty-six pairwise closures}
\label{app:pairwise_grid}

\begin{figure}[H]
\centering
\includegraphics[width=0.85\linewidth]{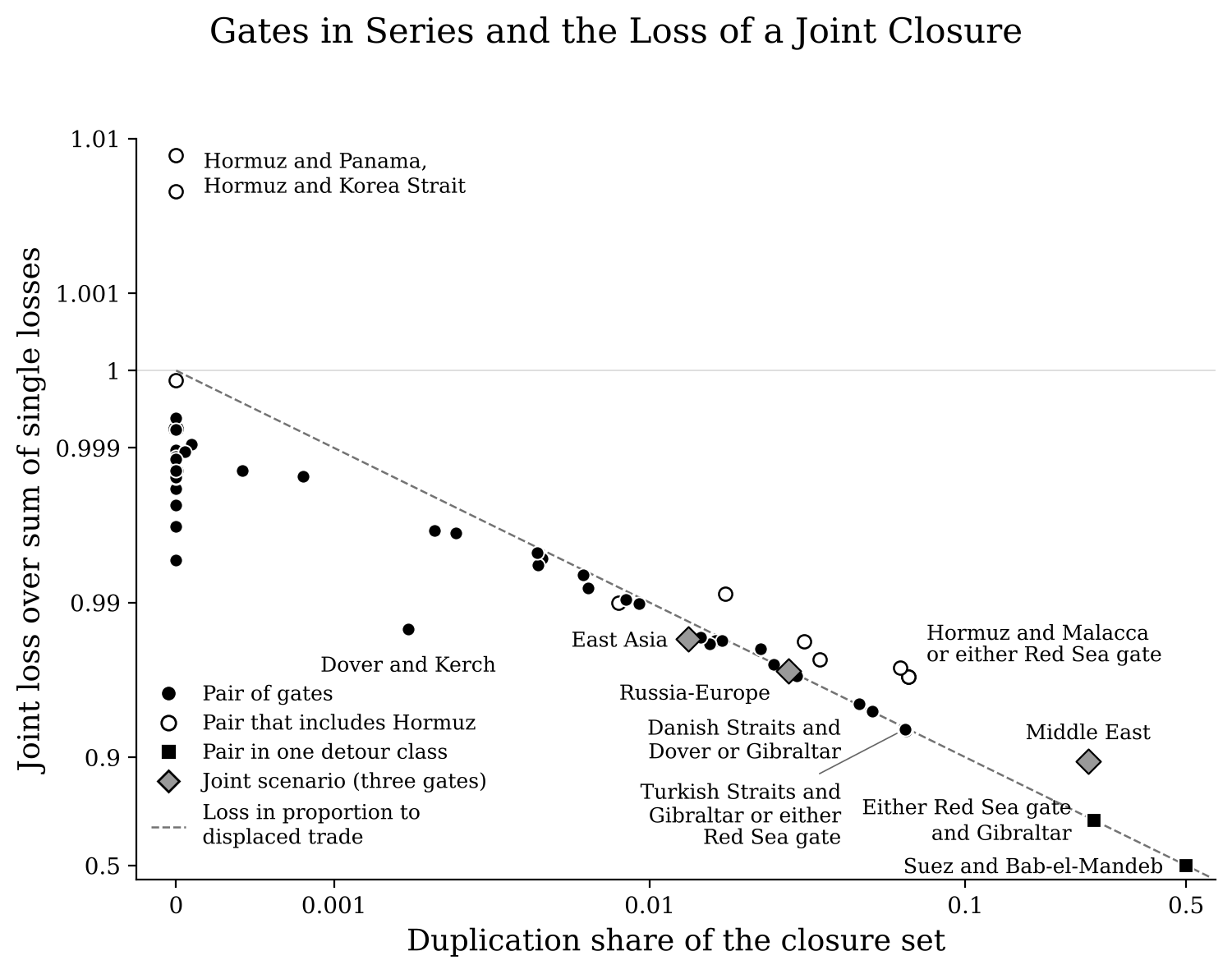}
\caption{The index \(S\) of the \(66\) pairwise closures and of the three joint scenarios against their duplication share \(\chi\) \eqref{eq:duplication_share}, at the baseline calibration. Both axes are logarithmic, the vertical one in the distance \(1-S\) of the index from one, except within \(0.001\) of \(\chi=0\) and of \(S=1\), where they are linear. On the dashed diagonal \(S=1-\chi\). Squares are the three pairs within the detour class of the Cape of Good Hope, two of which coincide, open circles are the eleven pairs that include Hormuz, and diamonds are the three scenarios.}
\label{fig:seriesness}
\end{figure}

The three joint scenarios of the paper are three points in a much larger space of closure sets.\footnote{The general problem this
raises is a combinatorial optimization problem on the network, in which one
chooses which subset of chokepoints to close and predicts the loss from the
combinatorial structure of that subset. The problem is closest in spirit to the
network interdiction, minimum cut, and shortest path problems, which ask
how removing edges blocks or lengthens a flow
\citep{ford_fulkerson_1956,wood_1993_interdiction,israeli_wood_2002}.} To cover more of that space, we solve the model at the baseline calibration for all \(66\) pairwise closures of the twelve gates. Unless Hormuz is one of its two gates, the index of a pair is one minus its duplication share \eqref{eq:duplication_share}, to within \(0.013\). The \(55\) pairs without Hormuz thus fall on the line \(S=1-\chi\) of Figure~\ref{fig:seriesness}, from the pairs whose gates stop different shipments, at an index of one, to Suez and Bab-el-Mandeb, which stop the same shipments, at one half. The exceptions are the six pairs with Hormuz whose duplication share exceeds \(0.01\), which lie above the line by as much as \(0.036\), and the Middle East scenario, which lies above it by \(0.14\). Across all \(66\) pairs the correlation of the index with one minus the duplication share is \(0.99\).

% numbers: 04_outputs/experiments/seriesness.json (02_scripts/experiments/seriesness.py)

The line is where the loss of a closure set is in proportion to the trade that it displaces. If every dollar of displaced trade cost the world the same, the index of a pair, its joint loss over the sum of its two single losses, would equal the trade that its joint closure displaces over the sum of what its two single closures displace, which is one minus its duplication share. At every gate but Hormuz a dollar of displaced trade costs the world between \(0.86\) and \(1.07\) dollars of GDP, so the pairs without Hormuz fall on the line. Which is to say that the routes, the detours, and the trade table give the index of such a pair to within \(0.013\) before the model is solved. At Hormuz a dollar of displaced trade costs \(3.2\) dollars. The loss from the trade that the two gates of a pair with Hormuz duplicate is then a smaller part of the sum of their single losses, which the costly Hormuz trade inflates, than the duplicated trade is of the sum of what they displace. The joint loss of such a pair therefore falls short of the sum of its single losses by less than its duplication share, and the pair lies above the line.

Along the line the pairs sort by the geography of Section~\ref{subsec:superadditivity} of the paper. The pairs within the detour class of the Cape of Good Hope are the most sub-additive. Suez with Bab-el-Mandeb sits at the pure-duplication value \(S=0.5\), and either Red Sea gate with Gibraltar at \(0.74\). Next come pairs of different classes that share shipments. The Turkish Straits sit at \(0.93\) with either Red Sea gate and with Gibraltar, and the Danish Straits at \(0.95\) with Gibraltar and with Dover, since every shipment through Dover also clears the Danish Straits. Pairs whose gates share few shipments sit near additivity, and \(50\) of the \(66\) pairs lie within \(0.02\) of one. The East Asian and the Russia--Europe scenarios lie on the line among them.

At the other end of the line, \(24\) pairs have no benchmark link that both of their gates cross, so their duplication share is zero. Twenty-two of them cost between \(0.01\) and \(0.5\%\) less than the sum of their two single closures. The other two, Hormuz with Panama and Hormuz with the Korea Strait, cost \(0.8\) and \(0.5\%\) more, and they are the only pairs of the \(66\) that cost more than that sum. The gates of either pair carry crude oil and gas to the same buyers in Korea, Japan, and Taiwan over different routes, which is the configuration that compounds (Online Appendix Proposition~\ref{prop:joint_pair}), and the closure of Hormuz has already left these buyers far short of that input.

% ----------------------------------------------------------------
\section{Robustness and Parametric Variation}
\label{app:robustness}
% ----------------------------------------------------------------

We vary, in turn, the parameters, the detour distances, the input--output table, and the maintained structure of the model. For each variation we ask what becomes of the paper's four headline findings: the top rank of Hormuz with the cross-chokepoint ordering, the heavy tail of cross-country exposure, the exporter--importer asymmetry, and the comparison of each joint closure with the sum of its single closures. Three further exercises measure what three conventions of the baseline do to the level of the losses, namely that it credits no bypass pipeline, charges the trade of land neighbors outside Europe to the sea route, and leaves out final goods.

The variations move the level of the losses and leave their structure in place. Hormuz remains the top-loss gate and exposure remains concentrated in every variation that can assess them. The Middle East scenario remains sub-additive, with an index between \(0.78\) and \(0.95\), and the most sub-additive of the three scenarios under every variation of the parameters, the input--output vintage, and the structure of the model. The exporter--importer asymmetry holds at all twelve gates wherever it is assessed, save at the Turkish Straits, the Danish Straits, and the Korea Strait in some vintages, the gates the model already places closest to symmetry, and at the Turkish Straits in one cell of the grouped unrecovered shares. None of this makes the country-sector loss vector parameter-invariant, since both its level and its direction move with the unrecovered share \(\tau\), the curvatures \((\rho,\varrho)\), and the residual transit share \(\delta\).

We report first the world loss on a $(\tau,\rho)$ grid (Online
Appendix~\ref{app:robustness_heatmap}), the four findings at the corners of the rectangle
(Online Appendix~\ref{app:paramvar}), and the sensitivity to the residual transit share (Online
Appendix~\ref{app:robustness_delta}). We then vary the detour distances (Online
Appendix~\ref{app:robustness_detour}) and the reference year and edition of the input--output
table (Online Appendix~\ref{app:yearrobust}). A structural exercise then varies what the exchange and the fulfillment rule themselves maintain (Online Appendix~\ref{app:modelrobust}). We last credit the Gulf crude pipelines at a closed Hormuz (Online Appendix~\ref{app:hormuz_pipeline}), take the trade of land neighbors off the sea routes (Online Appendix~\ref{app:land_neighbors}), and count the final goods that a closure stops (Online Appendix~\ref{app:final_goods}). Every computation this section reports returns an interior equilibrium, and Online Appendix~\ref{app:stability_numerics} documents the numerics behind them.

% ----------------------------------------------------------------
\subsection{World loss as a function of $(\tau,\rho)$}
\label{app:robustness_heatmap}
% ----------------------------------------------------------------

Table~\ref{tab:robust_grid} reports the world GDP loss from closing the
Strait of Hormuz over $\tau\in\{0.1,0.2,0.3,0.4,0.5\}$ at the three complementarity
values $\rho\in\{-0.25,-0.5,-1\}$ that bracket and include the baseline, all at the
uniform baseline residual transit share $\delta=0.10$. The loss rises monotonically in both the unrecovered share and the complementarity, from $2.39\%$ at the corner of low unrecovered share and easy substitution $(\tau,\rho)=(0.1,-0.25)$ to $2.83\%$ at the baseline
$(0.3,-1)$ and on to $3.02\%$ at $(0.5,-1)$. Every cell is an interior equilibrium, since the intermediate input efficiency factor $\min_i\kappa^{(\omega)}_i$ stays at or above $0.44$
across the grid, so no cell saturates, and it remains near $0.45$ even at $\tau=0.7$. The aggregate scale of the loss, by contrast, moves with the parameters, by about a quarter from the mildest cell of the grid to the harshest.

\begin{table}[H]
\centering
\caption{World GDP loss from closing the Strait of Hormuz across
$\tau\in[0.1,0.5]$ at $\rho\in\{-0.25,-0.5,-1\}$, uniform residual transit share $\delta=0.10$. Entries are $L_W(\{\text{Hormuz}\})$ in \% of world GDP,
and the parenthetical is the minimum intermediate input efficiency factor
$\min_i\kappa^{(\omega)}_i$ at the solution. The baseline calibration $(\tau,\rho,\varrho)=(0.3,-1,-1)$ is in bold.}
\label{tab:robust_grid}
\small
% >>> inlined from 04_outputs/tex_tables/tab_robust_grid.tex
\begin{tabular}{lccccc}
\toprule
& $\tau=0.1$ & $\tau=0.2$ & $\tau=0.3$ & $\tau=0.4$ & $\tau=0.5$\\
\midrule
$\rho=-0.25$ & 2.39 (0.46) & 2.48 (0.46) & 2.57 (0.46) & 2.66 (0.47) & 2.75 (0.47)\\
$\rho=-0.50$ & 2.47 (0.46) & 2.56 (0.46) & 2.65 (0.46) & 2.74 (0.46) & 2.83 (0.46)\\
$\rho=-1.00$ & 2.65 (0.44) & 2.73 (0.44) & \textbf{2.83} (0.44) & 2.92 (0.44) & 3.02 (0.45)\\
\bottomrule
\end{tabular}
% <<< end 04_outputs/tex_tables/tab_robust_grid.tex
\end{table}

\subsection{The four findings at the corners of the rectangle}
\label{app:paramvar}

Table~\ref{tab:corner_findings} re-solves the twelve single closures and the three joint
scenarios at the four corners of the rectangle \(\tau\in[0.2,0.5]\), \(\rho\in[-1,-0.25]\), with
the outer-nest curvature at \(\varrho=-1\) and the uniform residual transit share at \(\delta=0.10\), and sets
each of the paper's four findings beside its baseline value.

The Strait of Hormuz is the top-loss gate at every corner, and it leads the runner-up by a
factor of $5.2$ to $12.9$, against $8.9$ at the baseline. The Panama Canal holds second place
at every corner, as it does at the baseline, and the order of the twelve gates is the same at
all four corners as at the baseline.

The heavy tail and the exporter--importer asymmetry are properties of the cross-section of
country losses rather than of the world aggregate, and both hold at every corner. The rank-size
slope on the fixed window of ranks $5$ to $40$ runs from $1.17$ to $1.72$ against $1.39$ at the
baseline, a comparable summary of concentration and not a tail index (Online
Appendix~\ref{app:zipfdetail}), and the five most exposed economies carry between $30$ and $36$
percent of total exposure. The exporting side loses more than the importing side at all twelve
gates at every corner. The Hormuz ratio runs from $4.5$ to $5.6$ against $4.6$ at the baseline,
and it is larger at the easy-substitution edge $\rho=-0.25$, where the supply shortfalls of
the importers are smaller. The ranking of economies by their Hormuz exposure holds a Spearman
correlation of at least $0.996$ with its baseline counterpart at every corner, and the ranking by worst-case exposure a correlation
between $0.95$ and $0.99$.

The additivity pattern of the joint closures holds at every corner as well. The Middle East scenario stays
sub-additive, with \(S(\cset{}^{ME})\) between \(0.84\) and \(0.92\), and the East Asian and
Russia--Europe scenarios stay close to additive, with \(S(\cset{}^{EA})\) at \(0.98\) and
\(S(\cset{}^{RE})\) at \(0.97\) at every corner. The Middle East index is lower at the two corners of
high unrecovered share, \(0.84\) and \(0.85\), than at the two of low unrecovered share, \(0.91\) and \(0.92\).

% >>> inlined from 04_outputs/experiments/tab_corner_findings.tex
\begin{table}[H]
\centering
\caption{The four findings at the corners of the calibration rectangle, at the outer-nest curvature
$\varrho=-1$ and the uniform residual transit share $\delta=0.10$. The lead is the ratio of the Hormuz world
loss to the world loss of the runner-up gate. The rank-size slope $\hat\zeta$ is taken over ranks
$5$ to $40$ of the worst-case exposure of the named economies. The asymmetry column gives the ratio
of the exporting side's loss to the importing side's loss at Hormuz and, in parentheses, the number
of gates at which the exporting side loses more. The last row is the baseline calibration.}
\label{tab:corner_findings}
\small
\setlength{\tabcolsep}{5.5pt}
\begin{tabular}{l r l r r r r r r}
\toprule
$(\tau,\rho)$ & Hormuz $L_W$ & Runner-up & Lead & $\hat\zeta$ & Hormuz asym. & $S(\cset{}^{ME})$ & $S(\cset{}^{EA})$ & $S(\cset{}^{RE})$ \\
 & (\% world GDP) & & & & ratio (gates) & & & \\
\midrule
$(0.2,-0.25)$ & 2.48 & Panama & 11.7 & 1.26 & 5.2$\times$ (12) & 0.91 & 0.98 & 0.97 \\
$(0.2,-1)$ & 2.73 & Panama & 12.9 & 1.17 & 4.5$\times$ (12) & 0.92 & 0.98 & 0.97 \\
$(0.5,-0.25)$ & 2.75 & Panama & 5.2 & 1.72 & 5.6$\times$ (12) & 0.84 & 0.98 & 0.97 \\
$(0.5,-1)$ & 3.02 & Panama & 5.7 & 1.67 & 4.9$\times$ (12) & 0.85 & 0.98 & 0.97 \\
\midrule
$\mathbf{(0.3,-1)}$ & 2.83 & Panama & 8.9 & 1.39 & 4.6$\times$ (12) & 0.89 & 0.98 & 0.97 \\
\bottomrule
\end{tabular}
\end{table}
% <<< end 04_outputs/experiments/tab_corner_findings.tex

\subsection{Sensitivity to the residual transit share}
\label{app:robustness_delta}

Table~\ref{tab:paramvar_delta} reports the world GDP loss for Hormuz and the three joint
scenarios, and the three joint indices, as the uniform residual transit share is varied, holding
$(\tau,\rho,\varrho)=(0.3,-1,-1)$. World loss falls monotonically as the residual transit share rises,
because a larger residual pass-through leaves more of the benchmark flow intact. The Hormuz
single closure runs from $3.29\%$ at a near-total blockade $\delta=0.01$ through the baseline
$2.83\%$ at $\delta=0.10$ to $2.38\%$ at $\delta=0.20$. The dependence on the residual transit share is
mild for the small East Asian shock and largest for the Middle East scenario, which loses $3.47\%$
at the near-total blockade and $3.00\%$ at the baseline. The United Arab Emirates is the worst-hit economy under the Hormuz closure at every residual transit share, with a loss of $34.8\%$ of its GDP at $\delta=0.01$, $32.3\%$ at the baseline, and $29.3\%$ at $\delta=0.20$.

The additivity pattern of the joint closures of Section~\ref{sec:joint} holds at every residual transit share. The index \(S(\cset{}^{ME})\) of the Middle East scenario stays between \(0.89\) and \(0.90\), and the indices
\(S(\cset{}^{EA})\) and \(S(\cset{}^{RE})\) at \(0.98\) and \(0.97\) throughout. The exporting side loses
more than the importing side at all twelve gates at every residual transit share in the table.

Every scenario returns an interior equilibrium at every residual transit share, down to $\delta=0.01$. The
effective demand floor behind Lemma~\ref{prop:wellposed} keeps the orders well defined at
every strictly positive residual transit share, since the slack pair keeps every closure well posed on the open links and the residual transit keeps every disrupted cell alive. The interiority of the
equilibrium is a separate matter, delivered by part (iii) of Proposition~\ref{prop:equilibrium} where its condition holds and by the ex-post interiority checks of Online Appendix~\ref{app:stability_numerics} elsewhere, since the effective demand floor does not by itself keep
the selected fulfillment vector off zero.

% >>> inlined from 04_outputs/experiments/tab_attenuation_findings.tex
\begin{table}[H]
\centering
\caption{World GDP loss (percent) and the joint indices as the uniform residual transit share \(\delta\)
varies at the baseline unrecovered share and substitution
\((\tau,\rho,\varrho)=(0.3,-1,-1)\), for the Strait of Hormuz and the three
joint scenarios. The lower block gives the additivity index of each joint scenario.  The baseline column
\(\delta=0.10\) is in bold.}
\label{tab:paramvar_delta}
\small
\begin{tabular}{lcccc}
\toprule
Scenario & \(\delta=0.01\) (near-total) & \(\delta=0.05\) & \(\mathbf{\delta=0.10}\) (baseline) & \(\delta=0.20\) \\
\midrule
Hormuz (single) & 3.29 & 3.07 & \textbf{2.83} & 2.38 \\
Middle East scenario & 3.47 & 3.25 & \textbf{3.00} & 2.54 \\
East Asian scenario & 0.40 & 0.38 & \textbf{0.36} & 0.32 \\
Russia--Europe scenario & 0.54 & 0.52 & \textbf{0.49} & 0.44 \\
\midrule
$S(\cset{}^{ME})$ & 0.90 & 0.90 & \textbf{0.89} & 0.89 \\
$S(\cset{}^{EA})$ & 0.98 & 0.98 & \textbf{0.98} & 0.98 \\
$S(\cset{}^{RE})$ & 0.97 & 0.97 & \textbf{0.97} & 0.97 \\
\bottomrule
\end{tabular}
\end{table}
% <<< end 04_outputs/experiments/tab_attenuation_findings.tex

% ----------------------------------------------------------------
\subsection{Sensitivity of world loss to reroutability}
\label{app:robustness_detour}
% ----------------------------------------------------------------

The reroutability rule of Online Appendix~\ref{app:reroutability} derives every reroutability $e^{(k)}_{cc'}$ from one geographic quantity, the extra
distance $\Delta_k$ that the detour around chokepoint $k$ adds. That
distance is a calibration input like the residual transit share $\delta$. We vary it in the same way. Scaling every bypassable gate's detour distance by a
factor $\vartheta$ carries the reroutability to
\[
e'\;=\;\frac{e}{e+\vartheta\,(1-e)},
\]
so the halved geometry $\vartheta=0.5$ gives $e'=2e/(1+e)$ and the lengthened
geometry $\vartheta=1.5$ gives $e'=2e/(3-e)$. A terminal gate has
$\Delta_k=\infty$ and $e^{(k)}_{cc'}=0$, which this map fixes at zero for
every $\vartheta$, so the four terminal gates are invariant by construction.
Table~\ref{tab:esens} re-solves four single closures and the three joint
scenarios on both variant geometries at the baseline
$(\tau,\rho,\varrho,\delta)=(0.3,-1,-1,0.10)$.

\begin{table}[H]
\centering
\caption{World GDP loss (\%) under a $\pm50\%$ scaling of the detour
distance $\Delta_k$ of every bypassable chokepoint, at the baseline
$(\tau,\rho,\varrho,\delta)=(0.3,-1,-1,0.10)$. Terminal gates carry
$e^{(k)}_{cc'}=0$ and are invariant under the scaling.}
\label{tab:esens}
\small
\begin{tabular}{l ccc}
\toprule
Closure & $\Delta_k\times0.5$ & Baseline & $\Delta_k\times1.5$ \\ \midrule
Hormuz (terminal) & 2.83 & 2.83 & 2.83 \\
Suez & 0.16 & 0.26 & 0.34 \\
Malacca & 0.09 & 0.17 & 0.24 \\
Taiwan--Luzon & 0.08 & 0.14 & 0.20 \\ \midrule
Middle East joint & 2.93 & 3.00 & 3.05 \\
East Asian joint & 0.21 & 0.36 & 0.48 \\
Russia--Europe joint & 0.49 & 0.49 & 0.49 \\
\bottomrule
\end{tabular}
\end{table}

The bypassable gates move with
the detour geometry, though less than proportionally. The loss of the Suez closure runs from $0.16\%$ on the halved geometry to $0.34\%$ on the lengthened one against a baseline of $0.26\%$, and the loss of the East Asian scenario runs from $0.21\%$ to $0.48\%$ against a baseline of $0.36\%$.
The Middle East scenario moves by less than a thirtieth of its loss in
either direction, because the Strait of Hormuz carries most of that
scenario's loss and is terminal. The Russia--Europe scenario does not move at
all, since the Turkish Straits, the Danish Straits, and the Kerch Strait
are all terminal.

The detour geometry accordingly matters least where the aggregate loss is
largest. That geometry matters most where a detour exists and carries real mass.
Closing the Lombok and Sunda passages together with the route east of
Taiwan, which sets $e^{(k)}_{cc'}=0$ on the Malacca and Taiwan--Luzon
supports, raises the East Asian scenario from $0.36\%$ to $3.51\%$, with
Vietnam losing $19.2\%$ of GDP and Taiwan $16.9\%$. What holds the East
Asian scenario small is therefore the existence of the Lombok--Sunda and east-of-Taiwan detours rather than their calibrated length.

The East Asian scenario also depends on the resolution of the routing. The regional rule leaves
the intra--East-Asian pairs other than the two overrides outside the chokepoint incidence
(Online Appendix~\ref{app:reroutability}). We can instead assign the whole East Asian block at
point resolution, entering the mixed-origin CHN--JPN and CHN--KOR pairs at half to full weight
on the Korea Strait alone or on both East Asian gates. That assignment moves the world GDP loss
under the East Asian scenario closure set $\cset{}^{EA}$ of Section~\ref{sec:joint} to between
about $0.44\%$ and $0.58\%$, against the reported $0.36\%$. The joint index stays at $0.98$ when the pairs
enter the Korea Strait alone, and falls to $0.94$ and to $0.85$ when they enter both East Asian gates at half and at full weight.

% ----------------------------------------------------------------
\subsection{Robustness across input--output vintages}
\label{app:yearrobust}
% ----------------------------------------------------------------

The paper's four findings are claims about the world trade network
rather than about the one input--output table we calibrate to. The baseline measures the closures against the 2019 benchmark, the last
pre-pandemic normal year (Section~\ref{sec:mapping}). To test whether the
findings depend on that vintage, we rebuild the extended intermediate-flow
matrix $\mathbf Z^*$ for every reference year the OECD ICIO 2025 edition
carries, 2011 through 2022. Each vintage holds the maritime route geometry,
the chokepoint incidence, the seaborne shares of the 2019 benchmark, and the calibration
$(\tau,\rho,\varrho,\delta)=(0.3,-1,-1,0.10)$ fixed. Only the benchmark
trade flows change across the twelve vintages. We re-solve the twelve single
closures and the three joint scenarios on each year's network.
Table~\ref{tab:year_robustness} reports the outcome.

Three of the four findings hold in every year, and the fourth, the exporter--importer asymmetry, holds in every year at Hormuz and at most of the other gates. The Strait of Hormuz is the top-loss
gate in all twelve years, with second place held by Gibraltar in 2011 through 2014 and by the Panama Canal from 2015 on.
The worst-hit economy is a Gulf oil exporter in every year, Saudi Arabia in 2011 through 2013 and the United Arab Emirates in the other nine, with a
loss between $27.7\%$ and $42.0\%$ of its GDP. The cross-country
distribution stays concentrated, with the top five economies carrying
between $30$ and $35$ percent of total exposure in every vintage. The
country-level rank-size slope over the rank $5$--$40$ window runs between
$1.07$ and $1.69$, comparable across vintages because the window is fixed, and not a tail index. The
exporter--importer asymmetry at Hormuz stays large, between $3.5$ and
$5.6$ times, and the exporting side loses more than the importing side at ten to
twelve of the twelve gates each year. The additivity pattern of the joint closures holds in every year as well.
The Middle East scenario is sub-additive in every year, with $S(\cset{}^{ME})$ between $0.88$ and $0.91$, while
$S(\cset{}^{EA})$ holds at $0.98$
and $S(\cset{}^{RE})$ between $0.97$ and $0.98$. Every scenario settles into an interior
equilibrium. The minimum intermediate input efficiency factor runs between $0.17$ and $0.50$ across the
twelve years and is $0.44$ in the 2019 baseline. The
minimum fulfillment rate is the larger of the two objects at every
scenario, and it is $0.472$ in the 2019 baseline.

What moves across years is the level, not the structure. The Hormuz world
loss runs from $1.97\%$ in the 2020 pandemic trough to $5.22\%$ in the
high-oil-price early 2010s, tracking the energy-trade mass the corridor
carries. The trade that the strait carries is the size of the shock, which is why the level of the loss moves with it from one year to the next. That trade does not say how much the world loses for each dollar of it, or on which economies the loss falls. The sign of the exporter--importer asymmetry flips mostly at the gates the model
already places closest to symmetry. Those near-symmetric gates are the Turkish Straits in 2016, 2017, 2020, and 2022
and the Danish Straits in the 2022 energy crisis. The flips elsewhere are at the Korea Strait in 2011 through 2013.
The sign never flips at Hormuz, Suez and Bab-el-Mandeb, Gibraltar, Panama, Malacca, the Taiwan--Luzon Corridor, Dover, or Kerch.

% >>> inlined from 04_outputs/tex_tables/tab_year_robustness.tex
\begin{table}[H]
\centering
\caption{Year-robustness of the four findings across ICIO reference years $2011$--$2022$ (OECD ICIO 2025 edition), at the baseline calibration $(\tau,\rho,\varrho)=(0.3,-1,-1)$ with a uniform residual transit share $\delta=0.10$. The columns give the Hormuz world loss, the loss of the United Arab Emirates under a closed Hormuz, the country-level rank-size slope $\hat\zeta$ over ranks $5$ to $40$, the exporter--importer asymmetry ratio at Hormuz, and the three joint indices. The year $2019$ is the paper's baseline. The last two rows recompute the 2019 and 2015 benchmarks on the independently constructed ICIO 2023 edition, which does not name the United Arab Emirates.}
\label{tab:year_robustness}
\small
\begin{tabular}{r rr r r rrr}
\toprule
Year & Hormuz $L_W$ & UAE-Hormuz & $\hat\zeta$ & Hormuz asym. & $S(\cset{}^{ME})$ & $S(\cset{}^{EA})$ & $S(\cset{}^{RE})$ \\
 & (\% world GDP) & (\% GDP) & & ratio & & & \\
\midrule
2011 & 4.87 & 38.7 & 1.16 & 3.8$\times$ & 0.91 & 0.98 & 0.98 \\
2012 & 5.22 & 37.0 & 1.07 & 3.6$\times$ & 0.91 & 0.98 & 0.98 \\
2013 & 4.80 & 35.8 & 1.12 & 3.5$\times$ & 0.91 & 0.98 & 0.97 \\
2014 & 4.28 & 35.0 & 1.12 & 3.5$\times$ & 0.91 & 0.98 & 0.97 \\
2015 & 3.10 & 29.6 & 1.29 & 3.8$\times$ & 0.90 & 0.98 & 0.97 \\
2016 & 2.60 & 27.7 & 1.31 & 4.0$\times$ & 0.89 & 0.98 & 0.97 \\
2017 & 2.89 & 32.4 & 1.34 & 4.3$\times$ & 0.90 & 0.98 & 0.97 \\
2018 & 3.26 & 31.9 & 1.34 & 4.4$\times$ & 0.90 & 0.98 & 0.97 \\
\textbf{2019} & 2.83 & 32.3 & 1.39 & 4.6$\times$ & 0.89 & 0.98 & 0.97 \\
2020 & 1.97 & 33.9 & 1.40 & 5.6$\times$ & 0.88 & 0.98 & 0.97 \\
2021 & 2.55 & 35.1 & 1.48 & 5.2$\times$ & 0.88 & 0.98 & 0.97 \\
2022 & 3.48 & 35.7 & 1.69 & 4.4$\times$ & 0.89 & 0.98 & 0.98 \\
\midrule
2019, 2023 ed. & 2.46 & -- & 1.44 & 5.0$\times$ & 0.88 & 0.98 & 0.97 \\
2015, 2023 ed. & 2.28 & -- & 1.37 & 4.9$\times$ & 0.86 & 0.98 & 0.98 \\
\bottomrule
\end{tabular}
\end{table}
% <<< end 04_outputs/tex_tables/tab_year_robustness.tex

The findings survive a change of ICIO edition as well as of reference year, as the last two
rows of Table~\ref{tab:year_robustness} show.
We rebuild the 2019 and 2015 benchmarks from the OECD ICIO 2023 edition,
which revises the underlying data, resolves 45 sectors in place of 50, and
names four fewer economies, and re-solve the closures on each. The 2023
edition is not a re-dating of the 2025 one but a separate estimate of
the world economy, built from an earlier vintage of national-accounts
and supply--use data, so agreement across the two editions tests the accounting layer beneath the model, not just the
choice of reference year. The Strait of Hormuz stays the top-loss gate, the exposure distribution stays concentrated, with a country-level rank-size slope of $1.37$ to $1.44$ on the fixed comparison window, and the Hormuz asymmetry stays large at $4.9$ to
$5.0$ times. The exporting side loses more than the importing side at all twelve gates on the 2019 benchmark and at eleven on the 2015 one, where the sign flips at the Turkish Straits. The Middle East scenario stays the most sub-additive of the three scenarios, with $S(\cset{}^{ME})$ at $0.88$ on the 2019 benchmark and $0.86$ on the 2015 one, against $0.97$ to $0.98$ for the other two indices. The
Hormuz loss comes out lower than in the 2025 edition, by about
$13\%$ for the 2019 benchmark and $27\%$ for 2015.
The United Arab
Emirates is not named in the 2023 edition and enters only through the
rest-of-world disaggregation, so its individual Hormuz loss is not
comparable across editions. Among the economies that the edition names, Saudi Arabia is the worst hit on both benchmarks, losing $21.7\%$ of
its GDP on the 2019 benchmark and $22.9\%$ on the 2015 one.

% ----------------------------------------------------------------
\subsection{Robustness to the maintained structure of the exchange and the fulfillment rule}
\label{app:modelrobust}
% ----------------------------------------------------------------

The checks so far vary parameter values and data vintages within a fixed model. We now vary the maintained structure itself. The first two variations take up the fixed residual transit of the exchange (Definition~\ref{def:choke_reroute}) and its symmetric unrecovered share. The next three take up the absence of any constraint on the re-matched flows beyond their row and column totals,
the proportional fulfillment rule (Definition~\ref{def:fulfillment}), and
the common unrecovered share across shipped commodities. A closing exercise moves
all four calibrated parameters at once to their loss-maximizing corner.
Table~\ref{tab:robustness_new_summary} collects the outcome. The
heavy tail and the top rank of Hormuz survive every
variation that can assess them. The exporter--importer asymmetry reverses at one gate, the Turkish Straits, in one cell of the grouped unrecovered shares. The Middle East scenario is the most sub-additive of the three scenarios in every cell of every variation, with an index between $0.78$ and $0.95$, and the indices of the East Asian and Russia--Europe scenarios lie between $0.95$ and $0.99$ throughout.

Each experiment reproduces its reference computation to a relative $10^{-12}$ when its new
ingredient is set to the neutral value. The reference is the adopted baseline for the recovery tilts, the alternative fulfillment rule, the grouped unrecovered shares, and the harsh corner. The row and column totals of the exchange are met to a
relative $10^{-9}$ or better on every block except the small group that
exhausts the sweep budget, where the loosest residual is
$2.2\times10^{-6}$ (Online Appendix~\ref{app:stability_numerics}). Every
reported fixed point is interior and fully converged, and the per-scenario convergence diagnostics ship with the
replication package.

% >>> inlined from 04_outputs/robustness_new/tab_robustness_new_summary.tex
\begin{table}[H]
\centering
\caption{Summary of the structural robustness computations. Each row reports the range of the Hormuz world loss across the variation's cells, in percent of world GDP, and whether each of three headline findings survives in every cell. Every variation is compared with the adopted baseline, whose Hormuz loss is $2.83\%$. The three findings are that the Strait of Hormuz remains the top-loss gate, that country-level exposure stays concentrated (rank-size slope $\hat\zeta<2$ over ranks 5--40), and that the exporter--importer asymmetry is positive at all 12 gates. The last three columns give the range of each joint index across the variation's cells, against $0.89$, $0.98$, and $0.97$ at the baseline. Dashes mark findings a variation's scenario set cannot assess, since the per-link caps solve one exchange variant and the harsh corner solves Hormuz and the three joint scenarios only.}
\label{tab:robustness_new_summary}
\footnotesize
\setlength{\tabcolsep}{2.5pt}
\begin{tabular}{l c c c c c c c c}
\toprule
Variation & Cells & Hormuz $L_W$ & Hormuz & Exposure & Asym.\ $>0$ & $S(\cset{}^{ME})$ & $S(\cset{}^{EA})$ & $S(\cset{}^{RE})$ \\
 & & range (\%) & top gate & concentrated & all gates & & & \\
\midrule
Rescaled residual transit & 1 & 1.01 & yes & yes & yes & 0.78 & 0.98 & 0.97 \\
Asymmetric recovery & 5 & 2.83--2.94 & yes & yes & yes & 0.86--0.89 & 0.98 & 0.97--0.98 \\
Per-link caps & 1 & 3.88 & yes & -- & yes & 0.80 & 0.98 & 0.97 \\
Alternative fulfillment rule & 1 & 1.28 & yes & yes & yes & 0.89 & 0.96 & 0.95 \\
Grouped unrecovered shares & 2 & 1.55--2.79 & yes & yes & \textbf{no} & 0.86--0.90 & 0.98 & 0.97 \\
Harsh corner & 6 & 2.83--7.91 & -- & -- & -- & 0.88--0.95 & 0.98--0.99 & 0.97--0.98 \\
\bottomrule
\end{tabular}
\end{table}
% <<< end 04_outputs/robustness_new/tab_robustness_new_summary.tex

We first let the exchange rescale the residual transit. At the baseline the residual transit of a disrupted link is held fixed (Definition~\ref{def:choke_reroute}), so displaced trade must be re-placed over an open route. In the exchange of \citet{bhatt2026_sanctions} the residual on a severed link is scaled along with every other surviving flow, so displaced trade can return across the closed passage. Under that exchange the world loss of a closed Strait of Hormuz is $1.01\%$ of world GDP against $2.83\%$ at the baseline, and the United Arab Emirates loses $15.5\%$ of its GDP against $32.3\%$. The strait is a terminal gate, so no detour carries the trade that its closure severs, and an exchange that may scale the residual transit up sends displaced trade back through the closed strait. At the other eleven gates the two exchanges return world losses that agree to within $0.0003$ percentage points of world GDP.

The findings stand under the exchange that rescales the residual transit. The Strait of Hormuz remains the top-loss gate, with a lead of $3.2$ over the Panama Canal against $8.9$ at the baseline, and the order of the twelve gates is that of the baseline. The rank-size slope over ranks $5$ to $40$ is $1.46$ against $1.39$. The exporting side loses more than the importing side at all twelve gates, and the Hormuz ratio widens from $4.6\times$ to $8.1\times$, since the rescaling lowers the loss of the importers by more than that of the exporters. Korea, for instance, loses $2.4\%$ of its GDP against $10.2\%$. The ranking of the economies by their Hormuz exposure holds a Spearman correlation of $0.98$ with its baseline counterpart. The ranking by worst-case exposure holds a correlation of $0.81$, because Hormuz is the binding gate of $35$ of the $87$ economies under that exchange against $63$ at the baseline. The Middle East scenario is more sub-additive, with $S(\cset{}^{ME})$ at $0.78$, and the other two indices stay at $0.98$ and $0.97$. We hold the residual transit fixed for the reason that Section~\ref{subsec:rerouting_matrix} of the paper gives. A closure is imposed by a third party or by nature, so what still crosses the passage is a datum of the closure and not something that the trading pair can adjust.

We then let the two sides of the exchange, the sellers' capacities and the buyers' budgets,
recover at different rates. The unrecovered share $\tau$ enters the budgets and capacities of
Definition~\ref{def:choke_reroute} at one common value, so equal recovery opportunity is a
maintained symmetry of the exchange, recorded in Section~\ref{subsec:asymmetry} of the paper,
and the measured exporter--importer asymmetry is conditional on it. We replace $\tau$ by a
buyer-side and a seller-side unrecovered share $(\tau^{\rm buy},\tau^{\rm sell})$. The budgets
become $t^{(\omega)}_{rj}(1-\beta_j)s_j=(1-\beta_j)s_j(\alpha_{rj}-\tau^{\rm buy}\,\ell_{rj})$
and the capacities $g^{(\omega)}_i=s_i-f_i-\tau^{\rm sell}\,Z^{\rm disp}_i$. The residual transit
stays fixed, and the unexposed flows are rationed to these capacities and budgets net of it, as at
the baseline. The block totals then differ by $(\tau^{\rm buy}-\tau^{\rm sell})D_r$, with $D_r$
the sector-$r$ block's total displaced flow.

The slack pair of the rationed exchange absorbs that difference. At the baseline the two
accounts of the pair carry one common target, the placement deficit $U_r$ of the block. With side-specific shares the open links leave more unplaced on the side that recovers more. Write
$F_r$ for the largest total that the open links of the block can carry from its sellers to its
buyers within those capacities and budgets. The unsold-output column then takes what the
sellers' capacities leave beyond $F_r$ and the unmet-demand row what the buyers' budgets leave
beyond $F_r$, both net of the residual transit, so the two targets differ by
$(\tau^{\rm buy}-\tau^{\rm sell})D_r$ and the smaller of them is the placement deficit of the
block. A block with no placement deficit carries a single account, a column of unsold output
when $\tau^{\rm buy}>\tau^{\rm sell}$ and a row of unmet demand when
$\tau^{\rm sell}>\tau^{\rm buy}$. Both accounts are deleted after the exchange. Unsold output is
thereby written off, and unmet demand arrives as a delivered-mass shortfall that the realized
retention ratios $\theta^{(\omega)}_{rj}$ pass to the CES penalty.

We re-solve all fifteen scenarios on the five mean-preserving tilts
\[
(\tau^{\rm buy},\tau^{\rm sell})\in\bigl\{(0.3,0.3),\,(0.2,0.4),\,(0.4,0.2),\,(0.25,0.35),\,(0.35,0.25)\bigr\}.
\]
The exporting side loses more than the
importing side at all twelve gates in every cell, including the seller-favorable tilt
$(0.4,0.2)$ built to erode the gap. The Hormuz ratio stays between $4.6\times$ and $4.8\times$,
against $4.6\times$ at the baseline. The gate at which the two sides come closest remains the
Danish Straits, where the ratio falls from $1.3\times$ at the baseline to $1.2\times$ under the
tilt $(0.2,0.4)$. The country ranking barely moves, with a Spearman rank correlation of the
exposure vector against the baseline of $0.998$ or higher in every cell. The asymmetry is
therefore not an artifact of imposing equal recovery on the two sides of the exchange. The
Strait of Hormuz remains the top-loss gate in every cell, and the joint indices move little, the Middle East index
lying between $0.86$ and $0.89$ and the other two between $0.97$ and $0.98$.

Both tilt directions raise the world loss at every one of the fifteen scenarios. Under a Hormuz
closure the loss rises from $2.83\%$ of world GDP at the baseline to between $2.87\%$ and
$2.94\%$. The side that recovers less sets how much of the displaced trade is re-placed, so the rest of what the other side could have placed is written off. A mean-preserving spread of the
unrecovered share is therefore loss-increasing.

We next impose hard per-link caps on the re-matched flows, an exercise that brackets the adopted baseline from above by making the capacity discipline of the exchange exact. The exchange constrains row and column totals but not individual cells, so a severed exporter can lean on a thin open link harder than
a per-link physical capacity would permit. The exporter-side cushioning is therefore an upper bound. Imposing the hard caps
$Z^{(\omega)}_{ij}\le\check Z_{ij}$ cell by cell, the benchmark flow on clear links and the residual transit on disrupted links, closes the gap from
the other side. Because the disrupted transactions matrix $\check{\mathbf Z}$ sits exactly at the caps while the capacities and the budgets weakly exceed the capped totals, the projection of the exchange places every cell at its cap, with the two slack accounts absorbing the remainders. The exchange then reconstitutes nothing, $\mathbf A_\omega=\check{\mathbf A}$, and, in the exchange that rescales the residual transit, the capped equilibrium
coincides with the uncapped equilibrium at $\tau=1$, an equivalence we verify to machine precision.

Under a Hormuz closure the world loss rises from $2.83\%$ to $3.88\%$ of world GDP, the UAE loss from $32\%$ to
$45\%$ of GDP, and the Saudi loss from $23\%$ to $32\%$. The distance
between the uncapped and capped bounds is the value of re-matching along
surviving relationships, and the adopted baseline of $2.83\%$ sits between the $1.01\%$ returned by the exchange that rescales the residual transit and the capped $3.88\%$. The
asymmetry survives at the no-cushioning bound as well, positive at all
twelve gates with a Hormuz ratio of $5.2\times$. The Middle East scenario is more sub-additive at this bound,
with $S(\cset{}^{ME})$ at $0.80$, and the other two indices stay at $0.98$ and $0.97$.

We then replace the fulfillment rule, which at the baseline sets deliveries proportional to the output
ratio, $h_i=G^{(\omega)}_i$ (Definition~\ref{def:fulfillment}).
The alternative rule rations demand against effective capacity. With
effective demand
$N_i(\mathbf x)=f_i+[\mathbf A_\omega(\mathbf I-\boldsymbol\beta)\mathbf x]_i$
and effective capacity $G^{(\omega)}_i(\mathbf h)\,s_i$, the fulfillment rate is $h_i=\min\{1,\,G^{(\omega)}_i(\mathbf h)\,s_i/N_i(\mathbf x)\}$ and
activity is $x_i=h_i\,N_i(\mathbf x)$, with $\mathbf A_\omega$ and both
CES nests unchanged. The pair $(\mathbf x,\mathbf h)$ is no longer
triangular, and the jointly damped iteration converges monotonically to a
$10^{-10}$ residual in every scenario. Losses are measured against the
alternative rule's own fixed point with no chokepoint closed, so the null scenario returns exactly zero loss.

In Table~\ref{tab:robustness_new_B}, which reports the outcome, world losses fall below the baseline at every scenario, to between $34$ and $61$ percent of it, $1.28\%$
against $2.83\%$ for Hormuz. The baseline rule is therefore the upper
envelope at the world level. Exposure stays concentrated, the rank-size slope standing at $1.41$ against $1.39$ at the baseline on the same window, and the Hormuz asymmetry widens
to $12\times$. The joint indices move little. The
index $S(\cset{}^{ME})$ stays at $0.89$, and $S(\cset{}^{EA})$ and $S(\cset{}^{RE})$ fall from $0.98$ and $0.97$ to $0.96$ and $0.95$. The Middle East scenario is therefore the most sub-additive of the three under both rules.

% >>> inlined from 04_outputs/robustness_new/B_alt_closure/tab_B.tex
\begin{table}[H]
\centering
\caption{Alternative fulfillment rule at the baseline calibration $(\tau,\rho,\varrho,\delta)=(0.3,-1,-1,0.10)$. Deliveries follow demand-capacity rationing $h_i=\min\{1,G^{(\omega)}_i s_i/N_i(x)\}$ and $x_i=\min\{N_i(x),G^{(\omega)}_i s_i\}$, with effective capacity $G^{(\omega)}_i(\mathbf h)\,s_i$, in place of the baseline proportional rule $h_i=G^{(\omega)}_i$. $\mathbf A_\omega$ and both CES nests are unchanged. A few small country-sector cells lose more than under the baseline rule, documented in the replication package. $\rho_{\rm rank}$ is the Spearman rank correlation of the 87-economy exposure vector against the baseline rule.}
\label{tab:robustness_new_B}
\small
\begin{tabular}{l rr r r}
\toprule
Scenario & $L_W$ alternative & $L_W$ baseline & $S(\cset{})$ alternative & $\rho_{\rm rank}$ \\
 & (\% world GDP) & (\% world GDP) & (baseline) & \\
\midrule
Hormuz & 1.28 & 2.83 & -- & 0.959 \\
Turkish & 0.10 & 0.26 & -- & 0.964 \\
Danish & 0.09 & 0.24 & -- & 0.938 \\
Gibraltar & 0.11 & 0.28 & -- & 0.956 \\
Suez & 0.10 & 0.26 & -- & 0.952 \\
Bab-el-Mandeb & 0.10 & 0.26 & -- & 0.952 \\
Panama & 0.12 & 0.32 & -- & 0.982 \\
Taiwan--Luzon & 0.05 & 0.14 & -- & 0.964 \\
Malacca & 0.07 & 0.17 & -- & 0.970 \\
Korea & 0.02 & 0.05 & -- & 0.963 \\
Dover & 0.00 & 0.01 & -- & 0.981 \\
Kerch & 0.00 & 0.01 & -- & 0.987 \\
\midrule
$\cset{}^{ME}$ & 1.33 & 3.00 & 0.89 (0.89) & 0.957 \\
$\cset{}^{EA}$ & 0.13 & 0.36 & 0.96 (0.98) & 0.977 \\
$\cset{}^{RE}$ & 0.19 & 0.49 & 0.95 (0.97) & 0.965 \\
\bottomrule
\end{tabular}
\end{table}
% <<< end 04_outputs/robustness_new/B_alt_closure/tab_B.tex

We next let the unrecovered share differ across shipped commodities. The recovery
evidence behind $\tau$ (Online Appendix~\ref{app:est_tau}) places fungible
commodities at fast, near-complete re-sourcing and specific,
relationship-intensive inputs at the slow end of the recovery band. We give
the energy and raw-material sectors (B05--B09 and C19) $\tau^{\rm sell}=0.20$, the
machinery and specialized-equipment sectors (C26--C29, C301, C302T309)
$\tau^{\rm sell}=0.45$, and every other seaborne good the baseline $0.30$. The grouping takes crude oil and gas at the fungible end, as traded commodities, and sets aside the grade specificity of Online Appendix~\ref{app:est_tau}, so these cells are the favorable case for the energy gates. The exchange is block-diagonal in the seller's sector, so a per-sector
$\tau^{\rm sell}$ preserves the block consistency of the capacities and the budgets. The
symmetric slack pair is engaged only on blocks whose structural max-flow
deficit demands it, so deficit-free blocks engage no slack. We cross the grouped unrecovered shares with the uniform baseline residual transit share
$\delta=0.10$ and with the commodity-differentiated residual transit share ($\delta=0.45$ on B05, B06, and C19).

Three of the four findings stand in both cells. Hormuz keeps the top rank with the Panama Canal second, as at the baseline, the rank-size concentration summary
stands at $1.38$ and $1.74$ on the same rank window, and the
joint indices stay close to their baseline values, with $S(\cset{}^{ME})$ at $0.90$ and $0.86$ and the other two at $0.98$ and $0.97$ in both cells. The exporter--importer asymmetry is the exception. With
easy energy re-sourcing alone, the asymmetry stays positive at all twelve gates, with the Hormuz ratio at $4.6\times$ and the Hormuz loss at $2.79\%$ of
world GDP.
Raising the residual transit share of energy cuts the Hormuz loss to $1.55\%$ and raises the Hormuz ratio to $6.6\times$, but the asymmetry reverses at the Turkish Straits, where the ratio falls to $0.93$,
and at Suez and Bab-el-Mandeb the ratio falls to $1.1$. No economy
edges out the UAE as the worst-hit economy in either cell. The higher residual transit share of energy takes $45$ percent off the Hormuz loss and no more than $15$ percent off the loss at any other gate. The dominance of Hormuz therefore rests on energy. It holds insofar as energy trade is hard to
re-source within sector and does not pass through a determined closure, and the closure of 2026 held the strait's oil to about a quarter of its benchmark (Online
Appendix~\ref{app:hormuz_record}). Under the commodity-differentiated residual transit share Hormuz keeps the top rank, but its lead over the runner-up, the Panama Canal, falls from a factor of nine to a factor of about five.

We close at the harsh corner of the calibration. The rectangle checks vary
$(\tau,\rho)$ jointly but hold the outer-nest curvature at $\varrho=-1$ and the
residual transit share at separate values. We now move all four parameters at once to the
loss-maximizing corner the documented ranges admit,
$(\tau,\rho,\varrho,\delta)=(0.5,-4,-4,0.01)$, together with its
one-coordinate relaxations. The corner deepens a failure that is already
present at the center, where the interiority condition~\eqref{eq:choke_cushion} fails at a handful of country-sectors (Online
Appendix~\ref{app:stability_numerics}). At $\varrho=-4$ the
threshold $(1-\beta_i)^{-1/\varrho}$ of the condition rises toward one and the ex-ante
check fails at up to $124$ of the $4{,}222$ intermediate-buying
country-sectors. The failures
are ex ante only. Every cell beyond the center settles into an interior equilibrium with
$\min_i h_i\ge0.116$, converged to a $10^{-10}$ residual with empirical
contraction factors between $0.41$ and $0.54$, and no cell collapses to the zero fixed point.

The Hormuz world loss rises from $2.83\%$ at the center to $7.91\%$ at the
corner, with Saudi Arabia at $69.1\%$ of GDP and the UAE at $48.5\%$, and the country ranking holds with a
Spearman rank correlation against the center of $0.97$ or higher. The additivity pattern of the joint
closures holds in every cell. The index $S(\cset{}^{ME})$ lies between $0.88$ and $0.95$ and is the lowest of the three in every cell, $S(\cset{}^{EA})$ stays at $0.98$ to $0.99$,
and $S(\cset{}^{RE})$ at $0.97$ to $0.98$.
The corner therefore scales the losses and maintains a high,
but imperfect, country-rank correlation, and it leaves the additivity pattern of the joint closures in place.
Interiority itself survives the harshest admissible
calibration, verified ex post at every cell. The a priori guarantee of
Proposition~\ref{prop:equilibrium}(iii) is not what the reported
equilibria rest on anywhere in this paper, so the interiority reported here is a solved property of the
equilibrium throughout.

% ----------------------------------------------------------------
\subsection{Crediting the Gulf crude pipelines}
\label{app:hormuz_pipeline}
% ----------------------------------------------------------------

The chokepoint incidence of Section~\ref{sec:mapping} of the paper routes
every seaborne flow out of the Persian Gulf through the strait, so the crude that crosses Arabia by pipeline and is loaded at a port outside the strait
is not credited at a closed Hormuz. That is the severe convention everywhere in the paper. We measure how much it matters at Hormuz, because the strait is the
one gate at which a large overland bypass exists and the 2026 closure
mobilized it.

The record of the 2026 closure gives the size of the bypass, since at the end of March the Saudi East--West line
to Yanbu carried $4.4$ million barrels a day and the Emirati line to
Fujairah $1.6$, against a pre-closure total under $4$, while Kuwait, Qatar, Bahrain, and Iran had no comparable bypass (Online
Appendix~\ref{app:hormuz_record}). Table~\ref{tab:hormuz_pipeline}
re-solves the closure with that crude re-origined off the strait onto the
routes the routing rule assigns it, a share of $0.238$ of the two
economies' crude sales at the gate, calibrated to the $2$ million barrels a
day the closure actually mobilized against the strait's $16.1$ million
pre-closure. Nothing downstream of the severance changes, so the two
columns of the table differ in the shock and in nothing else. The
re-origined share enters the rationed exchange as open trade, so the exchange is free to scale it up. After the re-matching the effective
demand restores more than four fifths of the two economies' benchmark crude
sales at the gate, where the pipelines and the residual transit together carried under a third. The credited
column therefore assumes a bypass with no throughput ceiling, which makes it
the floor of the loss with the pipelines in use, whereas the baseline column is the ceiling of that loss.

On those terms the credit halves the world loss, from $2.83\%$ of
world GDP to $1.42\%$, and takes the Emirates from $32.3$ to $16.2\%$ of
its own GDP and Saudi Arabia from $23.1$ to $11.3\%$. The lowest intermediate input efficiency factor rises from $0.44$ to $0.64$.
The bypass rescales the shock without taking the closure out of the
model's nonlinear regime.

Panel~B of Table~\ref{tab:hormuz_pipeline} takes the model from the year-long closure of the paper to the closure as it was measured. Weighted day by day over the severance path of Online Appendix~\ref{app:hormuz_record}, the closure costs $1.44\%$ of world GDP in calendar 2026 with no pipeline credited and $0.73\%$ with the pipelines credited. The United Arab Emirates loses $18.7$ and $8.8\%$ of its GDP on the two counts, and Korea $4.7$ and $1.9\%$. These losses are unbuffered like the others, so they are not a forecast of what the closure of 2026 cost.

A pipeline bypass differs from the maritime detour classes of
Section~\ref{sec:model} of the paper in what it costs. A
longer voyage costs distance, and a pipeline costs throughput, which the credited computation does not charge. Closing
Bab-el-Mandeb alongside the strait shows how little the route of the bypass
matters once its throughput is free, since
crude leaving Yanbu on the Red Sea must cross that gate to reach Asia. Only
$28\%$ of the Saudi crude that crosses the strait in the benchmark goes to the western basin the line can
still serve through Suez. Yet blocking the eastbound leg raises the
joint-closure world loss only from $1.61\%$ to $1.66\%$, and the Saudi loss
does not rise, standing at $11.0\%$ with the leg blocked against $11.5\%$ with the leg open. The two losses are $3.00\%$ and $23.2\%$ with no pipeline
at all. In the exchange the Saudi line's sales to the western basin scale
up, and the Asian buyers it can no longer reach re-source from the
suppliers the two closures leave open, so the bypass survives the
joint closure built to break it. A bypass entered instead as a uniform
residual transit share would credit Qatar and Kuwait with relief they did not receive,
which is why the credit is taken line by line from the throughput of the individual pipelines rather
than entered as a parameter.

\begin{table}[H]
\centering
\caption{The Hormuz closure with and without the Gulf crude pipelines
credited, at the baseline calibration $(\tau,\rho,\varrho)=(0.3,-1,-1)$ with
$\delta=0.10$. Panel~A reports a complete closure sustained for a year, the
convention of Section~\ref{subsec:calibration} of the paper. Panel~B day-weights the model over the measured 2026 severance path of
Online Appendix~\ref{app:hormuz_record} on the EIA forward assumption, entered as a severance intensity that holds its August level to the end of that month and falls by $0.15$ a month from September. Country
entries are $\lambda_c$, in percent of the economy's own GDP. }
\label{tab:hormuz_pipeline}
\small
\begin{tabular}{l cc}
\toprule
 & No pipeline & Pipeline credited \\
  & (baseline) & (share $0.238$) \\
\midrule
\multicolumn{3}{l}{\textit{Panel A. Complete closure sustained for a year}}\\
\addlinespace
World loss $L_W$ & $2.83\%$ & $1.42\%$ \\
Lowest intermediate input efficiency factor & $0.444$ & $0.637$ \\
Active country-sectors below $\kappa=0.90$ & $3.8\%$ & $1.0\%$ \\
United Arab Emirates, $\lambda_c$ & $32.3\%$ & $16.2\%$ \\
Saudi Arabia, $\lambda_c$ & $23.1\%$ & $11.3\%$ \\
\addlinespace
\multicolumn{3}{l}{\textit{Panel B. Calendar 2026 on the measured severance path}}\\
\addlinespace
World loss $L_W$ & $1.44\%$ & $0.73\%$ \\
United Arab Emirates & $18.7\%$ & $8.8\%$ \\
Saudi Arabia & $12.9\%$ & $5.9\%$ \\
OtherGulf block & $12.6\%$ & $10.5\%$ \\
Japan & $2.51\%$ & $0.87\%$ \\
Korea & $4.70\%$ & $1.89\%$ \\
India & $3.05\%$ & $1.44\%$ \\
China & $1.66\%$ & $0.83\%$ \\
\bottomrule
\end{tabular}
\end{table}

% ----------------------------------------------------------------
\subsection{Trade between land neighbors outside Europe}
\label{app:land_neighbors}
% ----------------------------------------------------------------

A pair of economies in which neither is a reporter takes the seaborne share \(\mu_{ij}\) that the exporter's goods show on arrival in the Union, and that share tells sea from air but not from land (Online Appendix~\ref{app:seashare}). The trade of two such economies that share a land border, and that do not both lie inside the continental network of Europe and Central Asia, is therefore charged to the sea route between them almost in full. The $87$ economies form $79$ pairs of this kind. A pair shares a land border when the GeoDist database records its two economies as contiguous \citep{mayer_zignago_2011}, and a block borders an economy when one of its members does. The construction already takes two of the pairs, China with Kazakhstan and with the Central Asian block, to move overland. The routing sends the trade of $25$ of the other $77$ pairs through at least one gate. The trade of these pairs is $3.2$ percent of the seaborne trade on the routes of the Strait of Hormuz, $9.1$ percent at the Turkish Straits, $4.4$ percent at the Suez Canal and Bab-el-Mandeb, $9.2$ percent at the Strait of Malacca, and $7.8$ percent at the Taiwan--Luzon Corridor. The routes of the other six gates carry none of it.

No customs record gives the mode of transport on most of these pairs. The Trade-and-Transport Dataset of UNCTAD and the World Bank \citep{unctad_tt_2025} holds a record of the mode only where the importer reports it. The importers that reported it for 2019 were $65$ of the $170$ in the dataset, and China, India, Vietnam, Pakistan, Saudi Arabia, and Jordan were not among them. For the other importers the dataset predicts the mode from a model of route choice. We therefore correct by hand the pairs on which geography leaves no doubt, and we bound the error on the rest.

Two computations re-solve the twelve single closures and the three joint scenarios at the baseline calibration, with the seaborne shares of the baseline changed on these pairs alone. The first computation sets the seaborne share to zero on four pairs, and we call its shares the hand-set shares. On three of the four pairs one side is landlocked and reaches the other overland, the ground on which the construction treats China and Kazakhstan. These three pairs are the Central Asian block with the OtherGulf block, in which it borders Iran, the Central Asian block with OtherSouthAsia, in which it borders Afghanistan, and Laos with China. The fourth pair is Jordan with Saudi Arabia. Crude oil and gas are $72$ percent of the intermediate goods that Saudi Arabia sells to Jordan in the benchmark, and the routing sends them through the Strait of Hormuz, Bab-el-Mandeb, and the Suez Canal. The crude is in fact shipped from the Saudi port of Yanbu on the Red Sea to Aqaba \citep{petra_jordan_oil_2026}, a voyage that crosses none of the twelve gates, and the two economies share a land border. The second computation sets the seaborne share to zero on all $77$ pairs. It gives a lower bound on the world losses and not a correction, since a block borders an economy when one of its members does, and the lower bound removes the trade that the economy has by sea with the other members. Pakistan borders Iran, for instance, and the lower bound takes off the Strait of Hormuz the oil and gas that Pakistan buys from the other members of the OtherGulf block.

Table~\ref{tab:land_neighbors} reports the outcome. The hand-set shares lower the world loss of a Hormuz closure from $2.825$ to $2.814$ percent of world GDP, and the lower bound lowers it to $2.788$ percent. In proportion the falls are largest at the Turkish Straits, where the world loss falls by $4$ percent of its baseline value under the hand-set shares and by $10$ percent at the lower bound. The world losses at the Strait of Malacca, the Taiwan--Luzon Corridor, and the Suez Canal fall by $8$, $7$, and $5$ percent at the lower bound. The world losses at the other six gates do not change. The Strait of Hormuz remains the top-loss gate, and the ordering of the gates changes in one place, at the lower bound, where the Turkish Straits fall from sixth to seventh behind the Danish Straits. The three additivity indices move by less than $0.004$. In both computations the exporting side still loses more than the importing side at each of the twelve gates, and the Hormuz ratio stays at $4.6$. The rank-size slope \(\hat\zeta\) is $1.39$ under the hand-set shares and $1.43$ at the lower bound, against $1.39$ at the baseline.

Among the economies, the hand-set shares matter for Jordan. Its loss under the Hormuz closure falls from $7.4$ to $1.7$ percent of its GDP, and its loss under the Middle East scenario from $7.8$ to $2.2$ percent. Jordan stands eighth of the $87$ economies by total exposure at the baseline and forty-fifth under the hand-set shares, and Turkey takes its place among the nine most exposed economies of Section~\ref{subsec:master_matrix_results} of the paper. The loss of the Central Asian block at the Turkish Straits falls from $3.3$ to $2.3$ percent, and no other cell of the master exposure matrix moves by as much as a percentage point. At the lower bound Pakistan's loss under the Hormuz closure falls from $5.3$ to $3.5$ percent and Turkey's loss at the Turkish Straits from $5.1$ to $4.4$ percent. The cells above $5$ percent of GDP number $15$ at the baseline, $14$ under the hand-set shares, and $12$ at the lower bound, where all twelve lie in the Hormuz column. The Spearman rank correlation \(\rho_{\rm rank}\) of the $87$-economy exposure vector against the baseline is at least $0.99$ in each of the fifteen scenarios under the hand-set shares and at least $0.97$ at the lower bound. The four headline findings of the paper therefore survive both computations, the ordering of the gates with the one exchange of places. The place of Jordan among the most exposed economies does not survive the hand-set shares.

\begin{table}[H]
\centering
\caption{Losses with the trade of land neighbors taken off the sea routes, at the baseline calibration $(\tau,\rho,\varrho)=(0.3,-1,-1)$ with $\delta=0.10$. The first column is the baseline of the paper, the second sets the seaborne share to zero on the four hand-set pairs, and the third sets it to zero on all $77$ pairs of land neighbors. Panel~A lists the single closures whose world loss changes and the three joint scenarios. Panel~C lists the cells of the master exposure matrix that fall by half a percentage point or more in the third column, without the cells of Bab-el-Mandeb, which repeat those of the Suez Canal.}
\label{tab:land_neighbors}
\small
\begin{tabular}{l ccc}
\toprule
 & Baseline & Hand-set shares & Lower bound \\
\midrule
% >>> inlined from 04_outputs/experiments/overland_trade/tab_land_neighbors.tex
\multicolumn{4}{l}{\textit{Panel A. World loss $L_W$, percent of world GDP}}\\
\addlinespace
Strait of Hormuz & 2.825 & 2.814 & 2.788 \\
Suez Canal and Bab-el-Mandeb & 0.265 & 0.260 & 0.251 \\
Turkish Straits & 0.259 & 0.248 & 0.234 \\
Strait of Malacca & 0.169 & 0.169 & 0.155 \\
Taiwan--Luzon Corridor & 0.144 & 0.143 & 0.133 \\
$\cset{}^{ME}$ & 2.997 & 2.984 & 2.953 \\
$\cset{}^{EA}$ & 0.360 & 0.360 & 0.336 \\
$\cset{}^{RE}$ & 0.490 & 0.479 & 0.466 \\
\addlinespace
\multicolumn{4}{l}{\textit{Panel B. Additivity index}}\\
\addlinespace
$S(\cset{}^{ME})$ & 0.893 & 0.895 & 0.897 \\
$S(\cset{}^{EA})$ & 0.983 & 0.983 & 0.983 \\
$S(\cset{}^{RE})$ & 0.972 & 0.972 & 0.971 \\
\addlinespace
\multicolumn{4}{l}{\textit{Panel C. Loss $\lambda_c$ of an economy at a single closure, percent of its GDP}}\\
\addlinespace
Jordan, Strait of Hormuz & 7.40 & 1.68 & 1.37 \\
Pakistan, Strait of Hormuz & 5.28 & 5.28 & 3.47 \\
OtherCentralAsia, Turkish Straits & 3.30 & 2.28 & 2.25 \\
OtherSouthAsia, Strait of Hormuz & 4.24 & 4.20 & 3.42 \\
OtherGulf, Strait of Hormuz & 22.58 & 22.45 & 21.80 \\
Jordan, Suez Canal & 1.46 & 0.84 & 0.70 \\
Turkey, Turkish Straits & 5.08 & 5.07 & 4.41 \\
Egypt, Strait of Hormuz & 1.77 & 1.74 & 1.13 \\
OtherSouthAsia, Turkish Straits & 1.30 & 0.76 & 0.75 \\
Vietnam, Taiwan--Luzon Corridor & 1.14 & 1.14 & 0.63 \\
% <<< end 04_outputs/experiments/overland_trade/tab_land_neighbors.tex
\bottomrule
\end{tabular}
\end{table}

% ----------------------------------------------------------------
\subsection{Counting the final goods that a closure stops}
\label{app:final_goods}
% ----------------------------------------------------------------

The paper holds final demand at its benchmark, so the goods that a closed
passage carries to households, governments, and investors are not counted
among its losses. The input--output table records those goods by the
country-sector that sells them and the economy that receives them, which
makes a rough count of what they would add possible. We take the shipments
to final users from the household, non-profit, government, and
fixed-investment columns of the table. A final shipment is displaced to
the same extent as the seller's intermediate sales on the same sea route,
at the seaborne share of the seller's goods on that route and at the
baseline residual transit share. The displaced final shipments are sales lost to the
producers that shipped them, so we lower the final demand of those
producers and solve each of the twelve single closures again, measuring
the losses against the benchmark of the paper. Nothing is added at the
receiving end, where the buyer is a final user that the model does not
carry, and no producer gains the sales that another loses. We run the
count twice. In the first run the sellers write off the unrecovered share
\(\tau\) of their displaced final shipments, the share they write off of
displaced intermediate trade, and in the second run they write off all of
it. The count is rough in one further respect. The table does not resolve
the final goods that the seven rest-of-world blocks sell and receive. We divide them among the blocks in the proportions of their intermediate
trade.

Table~\ref{tab:final_goods} sets the two runs beside the figures of the
paper. Final goods raise every world loss, which is the sense in which the
losses of the paper are a lower bound. The addition is small at the Strait
of Hormuz, whose loss rises from \(2.83\%\) of world GDP to \(2.88\%\) in
the first run and to \(3.02\%\) in the second, because the strait carries
mostly inputs. The other gates carry more finished goods, so their losses rise by between a tenth and a quarter in the first run and by between
\(30\) and \(85\) percent in the second. Hormuz remains by far the costliest gate in both
runs.

The exporter--importer asymmetry of Section~\ref{subsec:asymmetry} of the
paper survives the count. The exporting side loses more than the importing side
at all twelve gates in both runs. Both sides lose more than they do in the
paper, however, because the economies that buy inputs through a gate often
sell finished goods through it. The importing side of the paper's comparison
sells two-thirds of the final goods that the Strait of Hormuz carries,
China alone a sixth of them, and more than four-fifths of those that the
three East Asian gates carry, China, Japan, and Korea the greater part.
The ratio of the exporting side's loss to the importing side's loss accordingly
widens at five gates, Hormuz, Panama, the Danish Straits, Dover, and Kerch, and
narrows at the other seven. The ratio narrows most at Malacca and at the Korea Strait,
where it falls from \(3.2\) and \(2.4\) to \(2.9\) and \(2.3\) in
the first run and to \(2.4\) and \(2.0\) in the second.

\begin{table}[H]
\centering
\caption{World GDP loss and exporter--importer loss ratio at each single closure when the final goods that the closure stops are counted as sales lost to their sellers, at the baseline calibration $(\tau,\rho,\varrho)=(0.3,-1,-1)$ with $\delta=0.10$. The first column of each panel is the figure of the paper, which holds final demand at its benchmark. In the second column the sellers write off the unrecovered share $\tau$ of their displaced final shipments, and in the third column all of them. The two sides of each gate are assigned as in the paper, by net disrupted intermediate trade.}
\label{tab:final_goods}
\small
\begin{tabular}{l ccc ccc}
\toprule
 & \multicolumn{3}{c}{World loss $L_W$ (\% of world GDP)} & \multicolumn{3}{c}{Exporting-side over importing-side loss} \\
\cmidrule(lr){2-4}\cmidrule(lr){5-7}
Chokepoint & Paper & Share $\tau$ & All & Paper & Share $\tau$ & All \\
\midrule
% >>> inlined from 04_outputs/experiments/tab_final_goods.tex
Strait of Hormuz & 2.825 & 2.883 & 3.017 & 4.6 & 4.7 & 4.8 \\
Panama Canal & 0.318 & 0.399 & 0.588 & 2.4 & 2.5 & 2.7 \\
Strait of Gibraltar & 0.276 & 0.340 & 0.489 & 1.6 & 1.6 & 1.6 \\
Suez Canal & 0.265 & 0.318 & 0.441 & 1.5 & 1.5 & 1.5 \\
Bab-el-Mandeb & 0.265 & 0.318 & 0.441 & 1.5 & 1.5 & 1.5 \\
Turkish Straits & 0.259 & 0.297 & 0.386 & 1.5 & 1.4 & 1.3 \\
Danish Straits & 0.241 & 0.288 & 0.398 & 1.3 & 1.3 & 1.4 \\
Strait of Malacca & 0.169 & 0.191 & 0.244 & 3.2 & 2.9 & 2.4 \\
Taiwan--Luzon Corridor & 0.144 & 0.161 & 0.202 & 2.4 & 2.4 & 2.4 \\
Korea Strait & 0.054 & 0.059 & 0.070 & 2.4 & 2.3 & 2.0 \\
Strait of Dover & 0.011 & 0.014 & 0.020 & 5.2 & 6.0 & 7.2 \\
Kerch Strait & 0.005 & 0.006 & 0.009 & 35 & 57 & 100 \\
% <<< end 04_outputs/experiments/tab_final_goods.tex
\bottomrule
\end{tabular}
\end{table}

\section{Code, Data, and Replication}
\label{app:replication}
All code and results behind the paper are public. The source repository
(\url{https://bitbucket.org/VipinVeetil/chokepoints}) holds the model, the full
data-processing pipeline, and the scripts that regenerate every number, table,
and figure in the paper. The code is released under the MIT license. The repository also carries the numerical verifications that accompany the proofs. Those
verifications check lemmas and propositions of Online Appendix~A on exact
small economies. The calibrated
model inputs (the input--output operators for the $87$ economies and $50$ sectors
of our 2019 benchmark) are archived on Zenodo
(DOI:~\href{https://doi.org/10.5281/zenodo.21263266}{10.5281/zenodo.21263266})
under a CC~BY~4.0 license. The code and these inputs store the detour geometry as lost shares, so the array they call \texttt{E} holds \(1-e^{(k)}_{cc'}\), the complement of the reroutability of the paper. We do not redistribute the two primary
sources, the GTAP~10 Data Base and the OECD Inter-Country Input--Output tables, which their providers distribute under their own terms.
But the repository documents how to obtain them and rebuild the inputs from
scratch. To lower the cost of replication further still, we distribute the model
itself as the \href{https://pypi.org/project/maritime-choke/}{\texttt{maritime-choke}}
package on the Python Package Index. The single command
\href{https://pypi.org/project/maritime-choke/}{\texttt{pip install maritime-choke}}
installs the solver and all its dependencies. On first use it fetches the
calibrated inputs automatically. A user therefore needs no prior configuration
of a scientific-computing environment to reproduce our results or run scenarios
of their own. The package also carries an interactive
interface, launched with \texttt{maritime-choke run}, that opens in a web
browser. There a user can close any subset of the twelve chokepoints, vary the
structural parameters, and solve the model forward, reading the resulting
GDP losses off a world map disaggregated by country and by sector. Every
output is written to disk. The model can also be run without any installation through a Streamlit app
(\url{https://maritime-chokepoints.streamlit.app}), which solves the same scenarios in the browser. These tools let researchers and policymakers study
disruptions well beyond the ones we consider here.

% ================================================================
\section{Notation}
\label{app:notation}
% ================================================================

Every symbol used in the paper is cataloged here, organized by role
rather than by order of introduction. A
single line of text suffices for each symbol because each is given a
unique meaning in the paper, and the few cases in which an established
letter is used in more than one sense are flagged explicitly.

\subsection*{Sets, indices and dimensions}

\begin{longtable}[l]{@{}p{2.6cm}p{12.2cm}@{}}
\toprule
Symbol & Meaning \\ \midrule
\endhead
$\mathcal C$ & Set of countries in the OECD ICIO baseline, with $C:=|\mathcal C|=81$. \\
$\mathcal C^*$ & Extended country set with seven ROW-block disaggregations, with $C^*:=|\mathcal C^*|=87$. \\
$\mathcal B$ & Set of the seven constructed rest-of-world blocks, so $\mathcal C^*=(\mathcal C\setminus\{\text{ROW}\})\cup\mathcal B$; a block is indexed $b$ (Online Appendix~\ref{sec:data}). \\
$\mathcal S$ & Set of ICIO sectors, with $S:=|\mathcal S|=50$. \\
$\mathcal M$ & Seaborne goods sectors $\subset\mathcal S$, with $|\mathcal M|=27$. \\
$\mathcal{CS}$ & Country-sector index set $\mathcal C\times\mathcal S$, with $|\mathcal{CS}|=CS$. \\
$i,j$ & Country-sector indices, with $i=(c(i),r(i))$ as in Section~\ref{sec:model}. \\
$c(i)$ & Country of country-sector $i$. \\
$r(i)$ & Sector of country-sector $i$. \\
$r$ & Sector index, $r\in\mathcal S$, with $\mathcal N_r:=\{i:\,r(i)=r\}$ the set of country-sectors in sector $r$. \\
$\mathcal K$ & Set of maritime chokepoints, with $K:=|\mathcal K|=12$. \\
$k$ & Chokepoint index. \\
$\mathcal K^{\mathrm d}$ & Set of the eleven distinct corridors, the twelve gates with Bab-el-Mandeb removed as a duplicate of Suez (Section~\ref{subsec:mean_tail}). \\
$\cset{}$ & Closure set, the set of chokepoints shut together,
$\cset{}\subseteq\mathcal K$, with $\cset{}_k:=\{k\}$ the closure of gate
$k$ alone. \\
$\omega=(\cset{},\mathbf E)$ & The scenario: a closure set paired with the route geography it acts on, the label post-closure objects carry. Determined by $\cset{}$, since $\mathbf E$ is exogenous. \\
$\cset{}^{ME},\cset{}^{EA},\cset{}^{RE}$ & The three politically coherent joint scenarios of Section~\ref{sec:joint}. \\
$\mathcal D,\;\mathcal D_n$ & Detour-equivalence partition of $\mathcal K$ and its classes \eqref{eq:detour_classes}. \\
$\mathcal R_j$ & Sectoral support of column $j$, $\{r:\alpha_{rj}>0\}$. \\
$\mathcal P(c,c')$ & Set of maritime routes from $c$ to $c'$, of which $p^*(c,c')$ is the least-cost element. \\
$\mathcal G_k$ & Geographic boundary of chokepoint $k$, a latitude--longitude polygon; distinct from the fulfillment rule $\mathcal G_\omega$. \\
$\mathcal{C}_k^{\text{sell}},\mathcal{C}_k^{\text{buy}}$ & Seller and buyer country sets of chokepoint $k$. \\
$\mathcal{S}_k^{\text{sell}},\mathcal{S}_k^{\text{buy}}$ & Seller and buyer sector sets of chokepoint $k$. \\
\bottomrule
\end{longtable}

\subsection*{Trade flows and benchmark network}

\begin{longtable}[l]{@{}p{2.6cm}p{12.2cm}@{}}
\toprule
Symbol & Meaning \\ \midrule
\endhead
$\mathbf Z,\;Z_{ij}$ & Benchmark intermediate-flow matrix and its entries (OECD ICIO 2025 edition, reference year 2019). \\
$\mathbf Z^*$ & Extended intermediate-flow matrix at $C^*\times S$ resolution. \\
$(1-\beta_j)s_j$ & Buyer $j$'s benchmark intermediate spending, $\sum_i Z_{ij}$; the paper carries the expression rather than a separate symbol. \\
$\mathbf I$ & Identity matrix, at whatever dimension the surrounding expression fixes. \\
$\mathbf A,\;a_{ij}$ & Input-share matrix, $a_{ij}=Z_{ij}/(1-\beta_j)s_j$, column-stochastic. \\
$\mathbf A^*$ & Extended input-share matrix. \\
$\mathbf Z_\omega,\;Z^{(\omega)}_{ij}$ & Transactions matrix after the exchange: the residual transit $\delta_r d_{ij}Z_{ij}$ of $\check{\mathbf Z}$ is held fixed, and the unexposed flow $(1-d_{ij})Z_{ij}$ is projected onto the capacities and budgets \eqref{eq:choke_rowtarget}--\eqref{eq:choke_coltarget} net of the row and column sums of the residual transit, in biproportional form with finite positive scalings where that projection is positive on every open link and as a limit of such forms where it does not; deficit blocks carry a symmetric slack pair (unsold-output column, unmet-demand row). \\
$\mathsf B(\cdot)$ & The rationed exchange of Definition~\ref{def:choke_reroute}, written $\mathsf B(\check{\mathbf Z};\mathbf g,\mathbf t)$ for capacities $\mathbf g$ and budgets $\mathbf t$; distinct from the block set $\mathcal B$. \\
$\mathbf A_\omega,\;a^{(\omega)}_{ij}$ & Effective demand matrix under scenario $\omega=(\cset{},\mathbf E)$, recovered as $Z^{(\omega)}_{ij}/(1-\beta_j)s_j$ \eqref{eq:choke_reroute}. \\
$\boldsymbol\beta,\;\beta_j$ & Leakage parameters, with $(1-\beta_j)s_j=\sum_i Z_{ij}$ buyer $j$'s intermediate spending; collected in the diagonal matrix $\boldsymbol\beta$. \\
$\mathbf s,\;s_i$ & Benchmark gross activity. The severance intensity $s$ of Table~\ref{tab:hormuz_record} is a separate use of the letter. \\
$\mathbf x$ & Trial activity vector of the fixed-point operator $\mathcal T_\omega$, $\mathbf x\in[\mathbf 0,\mathbf s]$ \eqref{eq:choke_operator}. The benchmark is $\mathbf s$ and the equilibrium $\mathbf s_\omega$. \\
$\mathbf s_\omega,\;s^{(\omega)}_{i}$ & Post-closure gross activity, written $s^{(\cset{})}_{i}$ when scenario-indexed. \\
$\mathbf f,\;f_i$ & Non-intermediate (final $+$ other) demand. \\
$\boldsymbol\gamma,\;\gamma_i$ & Benchmark GDP content per unit of activity, so $\boldsymbol\gamma^{\!\top}\mathbf s$ is world GDP. \\

\bottomrule
\end{longtable}

\subsection*{Chokepoint disruption}

\begin{longtable}[l]{@{}p{2.6cm}p{12.2cm}@{}}
\toprule
Symbol & Meaning \\ \midrule
\endhead
$\mathbbm 1^{(k)}_{ij}$ & Transit indicator, $\in\{0,1\}$ \eqref{eq:three_filters}. \\
$\boldsymbol\Gamma$ & $K\times M$ incidence matrix \eqref{eq:incidence_def}. \\
$e^{(k)}_{cc'},\;\mathbf E_k,\;\mathbf E$ & Reroutability (the share of the flow from country $c$ to country $c'$ that survives the detour around the closed chokepoint $k$, the lost share being $1-e^{(k)}_{cc'}$), reroutability matrix of chokepoint $k$, and reroutability profile, the collection of the $K$ matrices. \\
$d_{ij}(\cset{})$ & Disrupted fraction of link $i\to j$ under $\cset{}$ \eqref{eq:partial_choke_disruption}. \\
$\mu_{ij}$ & Seaborne share of link $i\to j$, the share of its benchmark flow that moves by ship (Online Appendix~\ref{app:seashare}); the closure displaces $(1-\delta)\mu_{ij}d_{ij}Z_{ij}$.  \\
$\ell_{rj}(\cset{})$ & Displaced sector-$r$ input share at buyer $j$ \eqref{eq:partial_disrupted_mass}. \\
$p^{(k)}_{ij}$ & Share of the flow on link $i\to j$ that the closure of gate $k$ alone would cost, $\bigl(1-e^{(k)}_{c(i)c(j)}\bigr)\mathbbm 1^{(k)}_{ij}$. Gate $k$ exposes the link when $a_{ij}\,p^{(k)}_{ij}>0$ (Online Appendix~\ref{app:joint_pair}). \\
$\ell_r(\cset{}'),\;\ell(\cset{}')$ & Displaced sector-$r$ input share of the common buyer $j^*$ of an isolated pair of gates under the closure set $\cset{}'$, and its sum over the input sectors, with $\alpha_r:=\alpha_{rj^*}$ and $\beta:=\beta_{j^*}$ (Online Appendix~\ref{app:joint_pair}). \\
$Z^{\rm disp}_i$ & Seller $i$'s displaced sales in flow units, $\sum_j(1-\delta)d_{ij}Z_{ij}$ \eqref{eq:choke_rowtarget}. \\
$t^{(\omega)}_{rj}$ & Buyer $j$'s surviving sector-$r$ budget as a share of its spending, $\alpha_{rj}-\tau\ell_{rj}$ \eqref{eq:choke_coltarget}, carrying the flow $t^{(\omega)}_{rj}(1-\beta_j)s_j$. \\
$g^{(\omega)}_i$ & Seller $i$'s surviving capacity, $s_i-f_i-\tau Z^{\rm disp}_i$ \eqref{eq:choke_rowtarget}. \\
$Y^{(\omega)}_{ij},\;\mathbf Y_\omega$ & Residual transit $\delta_{r(i)}d_{ij}Z_{ij}$ of a disrupted link and its matrix \eqref{eq:choke_split}: the residual that still crosses the closed passage, held fixed by the exchange. In the computations it also holds the non-seaborne flow $(1-\mu_{ij})d_{ij}Z_{ij}$ and the seaborne residual is $\delta_{r(i)}\mu_{ij}d_{ij}Z_{ij}$. \\
$\mathring Z_{ij},\;\mathring g_i,\;\mathring t_{rj}$ & Unexposed flow $(1-d_{ij})Z_{ij}$ and the residual capacities and budgets, the capacities and budgets net of the residual transit (Online Appendix~\ref{app:stability}). \\
$U_r,\;U_{rj}$ & Placement deficit of a sector-$r$ block \eqref{eq:wellposed_deficit} (Lemma~\ref{prop:wellposed}) and its unmet-demand share absorbed at buyer $(r,j)$ (Lemma~\ref{lem:choke_kernel}); zero on deficit-free blocks. \\
$u_i,\;v_j$ & Seller and buyer scalings of the exchange \eqref{eq:choke_reroute}, acting on the unexposed flows of each disrupted block (augmented by the slack pair on deficit blocks): multipliers on seller $i$'s capacity and buyer $j$'s budget, finite where the projection stays interior to the support and fixed only up to one reciprocal gauge per connected component of it (Definition~\ref{def:choke_reroute} of the paper). \\
$\mathbbm 1^{(k)}_{cc'}$ & Routing indicator: $1$ if $p^*(c,c')$ crosses $\mathcal G_k$. \\
$\Delta_k$ & Detour distance (nm) for bypassable chokepoint $k$. \\
$D_{cc'}$ & Benchmark voyage distance from $c$ to $c'$ (nm); distinct from the effective intermediate demand $D^{(\omega)}_i$. \\
$\vartheta$ & Factor by which the detour distance of every bypassable gate is scaled in the reroutability sensitivity (Online Appendix~\ref{app:robustness_detour}). \\
\bottomrule
\end{longtable}

\subsection*{Model parameters and equilibrium}

\begin{longtable}[l]{@{}p{2.6cm}p{12.2cm}@{}}
\toprule
Symbol & Meaning \\ \midrule
\endhead
$\sigma$ & Elasticity of substitution of either CES nest, $\sigma=1/(1-\rho)$; distinct from the standard deviation in the normalized likelihood ratio. \\
$\mathbf r^{B}$ & Benchmark accounting residual $\mathbf s-\mathbf f-\mathbf M_0\mathbf s$ of the assembled table. \\
$\tau$ & Unrecovered share (scalar or per input sector), with $(1-\tau)$ of disrupted mass reconstituted on clear suppliers. \\
$\delta_r,\;\underline\delta,\;\bar\delta$ & Residual transit share of seaborne sector $r$ and its uniform floor and ceiling, $0<\underline\delta\le\delta_r\le\bar\delta<1$ (Assumption~\ref{ass:attenuation}). Well-posedness of the exchange needs only $\delta_r>0$. The uniform baseline is $\delta_r\equiv0.10$ on all seaborne goods, and the commodity-differentiated residual transit share ($0.45$ on \texttt{B05}/\texttt{B06}/\texttt{C19}) enters the robustness exercises of Online Appendix~\ref{app:modelrobust}. \\
$\rho$ & CES substitution parameter, $\rho\in[\underline\rho,\overline\rho]\subset(-\infty,0)$. \\
$\alpha_{rj}$ & Benchmark sector-$r$ input share of column $j$, $\alpha_{rj}=\sum_{i\in\mathcal N_r}a_{ij}$, with $\sum_r\alpha_{rj}=1$. \\
$\kappa^{(\omega)}_j(\mathbf h;\rho)$ & Intermediate input efficiency factor \eqref{eq:choke_ces_kappa}. \\
$\mathbf G_\omega(\mathbf h)$ & Diagonal matrix of output ratios $G^{(\omega)}_j$ \eqref{eq:choke_Kmatrix}. \\
$\theta^{(\omega)}_{rj}(\mathbf h)$ & Realized retention ratio at column $j$, sector $r$ \eqref{eq:choke_tilde_f}. \\
$\bar\theta^{(\omega)}_{rj}$ & Effective demand retention, the same ratio at full fulfillment, $\bar\theta^{(\omega)}_{rj}:=\theta^{(\omega)}_{rj}(\mathbf 1)$, floored by the residual transit \eqref{eq:standing_retention}. The bar marks evaluation at $\mathbf h=\mathbf 1$, not a benchmark total. \\
$e^{(\omega)}_{rj},\;\upsilon^{(\omega)}_{rj}$ & Exposed share of the block, $\bigl(\sum_{i\in\mathcal N_r}d_{ij}a_{ij}\bigr)/\alpha_{rj}$, and the share of it the unmet-demand account absorbs \eqref{eq:standing_retention}. Distinct from the reroutability $e^{(k)}_{cc'}$ and from the output elasticity of intermediates $\widetilde e_j$. \\
$\mathcal S^{(\omega)}_{rj},\;\pi^{(\omega)}_{ij}$ & The block's effective demand support $\{i\in\mathcal N_r:a^{(\omega)}_{ij}>0\}$ and the order weights it induces, over which realized retention is the order-weighted mean fulfillment \eqref{eq:realized_retention}. \\
$\widehat{\mathbf K},\;E,\;\mathbf B^{*}$ & The kernel the exchange actually solves on, after the deletion of the rows with a zero residual capacity and of the columns with a zero residual budget and after the slack augmentation where the block carries a placement deficit; its support $E:=\operatorname{supp}(\widehat{\mathbf K})$, distinct from the reroutability profile $\mathbf E$; and the re-matched flows the projection returns. Strict feasibility is $\operatorname{supp}(\mathbf B^{*})=E$ (Lemma~\ref{prop:wellposed}(ii)). \\
$h_i,\;\mathbf h,\;\mathbf h_\omega$ & Fulfillment rate, fulfillment vector, and equilibrium fulfillment vector. \\
$\bar e_*$ & Envelope modulus of the reduced iteration, $\max_i\widetilde e_i(t_0\kappa^{(\omega)}_i(\mathbf h_\omega))<1$ (Proposition~\ref{prop:equilibrium}(iv)). \\
$s_i-f_i$ & Supplier $i$'s benchmark intermediate sales, $s_i-f_i=\sum_j Z_{ij}$. \\
$\mathbf M_0,\;\mathbf M_\omega$ & Propagation matrices of the benchmark and of the scenario, $\mathbf M_0=\mathbf A(\mathbf I-\boldsymbol\beta)$ and $\mathbf M_\omega=\mathbf A_\omega(\mathbf I-\boldsymbol\beta)$, whose entry in row $i$ and column $j$ is the intermediate demand that a unit of buyer $j$'s activity places on seller $i$ (Section~\ref{subsec:short_run_activity}). \\
$D^{(\omega)}_i(\mathbf x)$ & Supplier $i$'s total effective intermediate demand under $\omega$, $[\mathbf M_\omega\mathbf x]_i$. \\
$G_j,\;\varrho$ & Outer CES aggregator $G_j(\kappa)=[\beta_j+(1-\beta_j)\kappa^{\varrho}]^{1/\varrho}$ and its curvature $\varrho\in[\underline\varrho,\overline\varrho]\subset(-\infty,0)$ \eqref{eq:choke_outer}. \\
$G^{(\omega)}_j(\mathbf h)$ & Output ratio: the outer aggregator $G_j$ evaluated at $\kappa^{(\omega)}_j(\mathbf h)$ \eqref{eq:choke_lambda}. \\
$\widetilde e_j(\kappa)$ & Output elasticity of intermediates, $d\log G_j/d\log\kappa\in[1-\beta_j,1)$ \eqref{eq:choke_shadow}. \\
$\mathcal G_\omega$ & Fulfillment rule with coordinates $G^{(\omega)}_i=G_i(\kappa^{(\omega)}_i)$ \eqref{eq:choke_fulfillment}; a consistent fulfillment vector is a fixed point of $\mathcal G_\omega$. \\
$c_i,\;c_\omega$ & Interiority floor of the benchmark-selected equilibrium under the interiority condition \eqref{eq:choke_cushion}--\eqref{eq:choke_floor}. \\
$\mathcal T_\omega$ & Post-closure fixed-point operator \eqref{eq:choke_operator}. \\
$\bar a_\omega$ & Maximum supplier-side row sum of $\mathbf A_\omega$, $\max_j\sum_i a^{(\omega)}_{ji}$, a seller's market breadth rather than a budget: Lemma~\ref{lem:choke_kernel} bounds the columns of $\mathbf A_\omega$ and leaves the rows free (Online Appendix~\ref{app:stability_numerics}). \\
$\zeta^{(\omega)}_i,\,\xi^{(\omega)}_i$ & Direct input shortfall and downstream propagation, the two parts of the supply shortfall of producer $i$ \eqref{eq:choke_zeta}--\eqref{eq:choke_xi}. \\
\bottomrule
\end{longtable}

\subsection*{Country and world losses}

\begin{longtable}[l]{@{}p{2.6cm}p{12.2cm}@{}}
\toprule
Symbol & Meaning \\ \midrule
\endhead
$\boldsymbol\Delta\mathbf s,\;\Delta s^{(\cset{})}_{i}$ & Loss vector $\mathbf s-\mathbf s_\cset{}$ and its entries. \\
$\Delta\mathrm{GDP}_c(\cset{})$ & Country $c$'s GDP loss in dollars \eqref{eq:country_loss_choke}. \\
$\lambda_c(\cset{})$ & Country $c$'s proportional GDP loss \eqref{eq:country_loss_share_choke}. \\
$\lambda_i(\cset{}),\;\lambda_i^{\max}$ & Country-sector $i$'s proportional activity loss $\Delta s^{(\cset{})}_{i}/s_i$ and its worst-case exposure $\max_{k\in\mathcal K}\lambda_i(\cset{}_k)$ (Section~\ref{subsec:heavy_tail}). \\
$L_W(\cset{})$ & World GDP loss as a share of benchmark world GDP \eqref{eq:world_loss_choke}. \\

$\boldsymbol\Psi$ & $C^*\times K$ master exposure matrix of proportional GDP losses, rows $=$ economies, columns $=$ chokepoints. \\

$\mathbf s^{\ordonly}_\omega$ & Activity with lost orders alone $(\mathbf I-\mathbf M_\omega)^{-1}\mathbf f$ \eqref{eq:orders_alone}, the activity of the model with the output ratio matrix held at the identity ($\kappa\equiv 1$).  \\
$L_W^{\,\ordonly}(\cset{})$ & World GDP loss through lost orders, which is the world loss of the economy with lost orders alone (Online Appendix~\ref{app:loss_identity}). The lost-orders share of the world loss is $L_W^{\,\ordonly}/L_W$ and the input-shortfall share is $1-L_W^{\,\ordonly}/L_W$ (Section~\ref{subsec:country_world_losses} of the paper). The same two shares of the GDP loss of a country are its lost-orders share and its input-shortfall share. \\
$\eta_c,\;\eta^{\mathrm{up}}_c,\;\eta^{\mathrm{dn}}_c$ & Network-propagation share of country $c$'s GDP loss and its upstream and downstream parts \eqref{eq:network_shares}; $1-\eta_c$ is the direct-loss share. The same shares of the world loss carry the subscript $W$, and the direct loss of a closure is $(1-\eta_W)L_W$. \\
$\nu_W(\cset{})$ & Network multiplier of closure set $\cset{}$, the world loss per unit of its direct loss, $1/(1-\eta_W)$ \eqref{eq:network_multiplier}. \\
$\lambda_X,\;\lambda_M$ & Mean proportional GDP loss of the exporting and of the importing side of a gate, weighted by disrupted trade mass (Section~\ref{subsec:asymmetry}); $\lambda^{\mathrm{dir}}_X,\;\lambda^{\mathrm{dir}}_M$ are the same means of the direct losses $(1-\eta_c)\lambda_c$, and $\nu_X=\lambda_X/\lambda^{\mathrm{dir}}_X$, $\nu_M=\lambda_M/\lambda^{\mathrm{dir}}_M$ the network multipliers of the two sides \eqref{eq:asymmetry_split}. \\
$\bar U_j(\cset{})$ & Column $j$'s absorbed unmet demand as a share of its intermediate spending, $\bigl[\sum_r U_{rj}\bigr]/\bigl[(1-\beta_j)s_j\bigr]$. \\
$q_j(\omega)$ & Fall in the column of buyer $j$ in the propagation matrix, $(1-\beta_j)-[\mathbf 1^{\!\top}\mathbf M_\omega]_j$ \eqref{eq:dissipative_shortfall}, i.e., the intermediate orders per unit of its activity that the buyer no longer places with any seller. \\
$S(\cset{})$ & Additivity index $L_W(\cset{})/\sum_{k\in \cset{}}L_W(\cset{}_k)$ \eqref{eq:superadd_index}, read only where its denominator is strictly positive. \\
$Z^{\rm disp}(\cset{}),\;\chi(\cset{})$ & Trade that closure set $\cset{}$ displaces, $\sum_{i,j}(1-\delta)\mu_{ij}d_{ij}(\cset{})Z_{ij}$, and its duplication share $1-Z^{\rm disp}(\cset{})/\sum_{k\in\cset{}}Z^{\rm disp}(\cset{}_k)$ \eqref{eq:duplication_share}, the share of what the single closures of $\cset{}$ displace that two of its gates would both take. \\
\bottomrule
\end{longtable}

\subsection*{Cross-country statistics}

\begin{longtable}[l]{@{}p{2.6cm}p{12.2cm}@{}}
\toprule
Symbol & Meaning \\ \midrule
\endhead
$\lambda_c^{\max}$ & Worst-case exposure of country $c$, $\max_{k\in\mathcal K}\lambda_c(\cset{}_k)$. \\
$\bar\lambda_c$ & Mean exposure of country $c$: the average of $\lambda_c(\cset{}_k)$ over the eleven distinct corridors (Section~\ref{subsec:mean_tail}). \\
$\hat\alpha$ & Maximum-likelihood power-law exponent above the Kolmogorov--Smirnov-minimizing threshold $x_{\min}$, with the fit assessed by the bootstrap of \citet{clauset2009powerlaw} (Online Appendix~\ref{app:zipfdetail}). \\
$\hat\zeta$ & Rank-size slope of worst-case exposure over ranks $5$ to $40$ of the named economies, a description of that window and not a tail index (Online Appendix~\ref{app:zipfdetail}). Distinct from the direct input shortfall $\zeta^{(\omega)}_i$. \\
$\rho_{\rm rank}$ & Spearman rank correlation between two exposure vectors, distinct from the substitution parameter $\rho$ (Online Appendix~\ref{app:paramvar}). \\
$H_c$ & Normalized Herfindahl concentration of country $c$'s exposure
shares across the distinct corridors $\mathcal K^{\mathrm d}$, so
$H_c\in[0,1]$ \eqref{eq:herfindahl}. \\
\bottomrule
\end{longtable}

\noindent Decorations carry fixed meanings for a generic symbol $x$. A caron
$\check x$ marks a disrupted quantity taken before the exchange, a ring
$\mathring x$ the free part of such a quantity, a breve $\breve x$
an equation-consistent state before the reporting map, and a hat $\hat x$ an empirical estimate or an explicitly reported
numerical state. In the online appendices a hat also marks the
counterpart of $x$ in a reduced or redirected problem. A tilde $\widetilde x$ distinguishes a variant of a letter that
already carries another meaning, as the output elasticity of intermediates
$\widetilde e_j$ does against the reroutability $e^{(k)}_{cc'}$. A bar
$\bar x$ marks a bound, a benchmark total, or a mean where stated, and an
underline $\underline x$ a lower bound. A superscript or subscript
$\omega$ marks an object of the post-closure economy. An upright
superscript names a part or a side of a quantity, as in $Z^{\rm disp}_i$ and $\tau^{\rm sell}$.

\end{document}